\pdfoutput=1
\documentclass[twocolumn,prd,groupaddress,preprintnumbers,nofootinbib]{revtex4}
\usepackage{graphicx}
\usepackage{epsfig}
\usepackage{bm}
\usepackage{latexsym,amssymb,amsmath,amsfonts,amssymb,txfonts,pxfonts,wasysym,float}
\usepackage{color}

\usepackage{bm}
\usepackage{epsfig}

\usepackage{mathrsfs}
\usepackage{slashbox}
\usepackage{slashed}
\usepackage[makeroom]{cancel}
\newcommand{\postscript}[2]{\setlength{\epsfxsize}{#2\hsize}
   \centerline{\epsfbox{#1}}}
\usepackage[titletoc,toc]{appendix}
\usepackage{multirow}
\usepackage{slashbox}
\usepackage[usenames,dvipsnames]{xcolor}
\definecolor{orange}{cmyk}{0,0.5,1,0}
\definecolor{rossoCP3}{cmyk}{0,.88,.77,.40}
\definecolor{graa}{rgb}{0.8,0.8,0.8}
\definecolor{blaa}{rgb}{0.2,0.2,0.6}
\usepackage{dsfont}
\newcommand{\be}{\begin{equation}}
\newcommand{\ee}{\end{equation}}
\newcommand\ylm{Y_{l m}}

\newcommand{\mbh}{M_{\text{BH}}}

\newcommand{\bea}{\begin{eqnarray}}
\newcommand{\eea}{\end{eqnarray}}

\def\tfrac#1#2{{\textstyle\frac{#1}{#2}}}

\begin{document}

\title{\color{rossoCP3} Lectures on Astronomy, Astrophysics, and Cosmology}

\author{Luis A. Anchordoqui}
\affiliation{Department of Physics and Astronomy,  Lehman College, City University of New York, NY 10468, USA \\
Department of Physics, Graduate Center, City University  of New York, 365 Fifth Avenue, NY 10016, USA\\
Department of Astrophysics, American Museum of Natural History, Central Park West  79 St., NY 10024, USA
}

\date{Spring 2016}
\begin{abstract}
  \noindent This is a written version of a series of lectures aimed at
  undergraduate students in astrophysics/particle theory/particle
  experiment. We summarize the important progress
  made in recent years towards understanding high energy astrophysical
  processes and we survey the state of the art regarding the
  concordance model of cosmology.
\end{abstract}


\maketitle

\section{Across the Universe}

A look at the night sky provides a strong impression of a changeless
universe. We know that clouds drift across the Moon, the sky rotates
around the polar star, and on longer times, the Moon itself grows and
shrinks and the Moon and planets move against the background of stars.
Of course we know that these are merely local phenomena caused by 
motions within our solar system. Far beyond the planets, the stars appear
motionless. Herein we are going to see that this impression of
changelessness is illusory.  

\begin{figure}
 \postscript{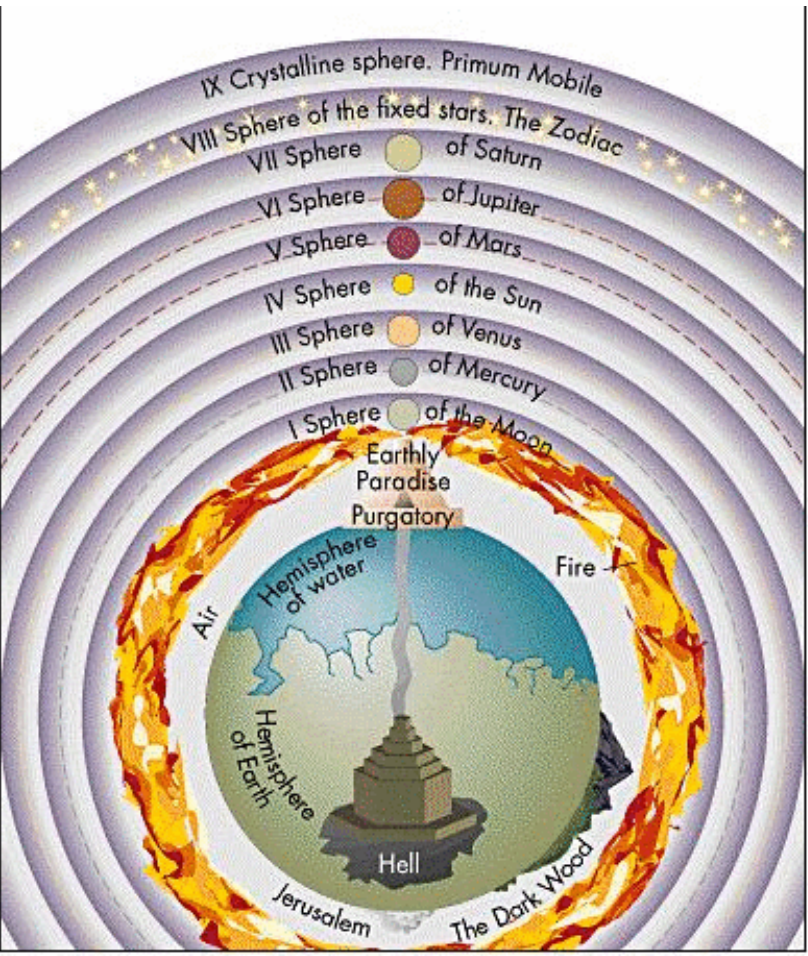}{0.8}
 \caption{Celestial spheres of ancient cosmology.}
\label{celestial-spheres}
\end{figure}

\subsection{Nani gigantum humeris insidentes}

According to the ancient cosmological belief, the stars, except for a
few that appeared to move (the planets), where fixed on a sphere
beyond the last planet; see Fig.~\ref{celestial-spheres}.  The universe was self
contained and we, here on Earth, were at its center. Our view of the
universe dramatically changed after Galileo's first telescopic
observations: we no longer place ourselves at the center and we view
the universe as vastly larger~\cite{Galileo:1610,Galileo,Copernicus}. 

In the early 1600s,  Kepler proposed three laws that described the motion of planets in a sun-centered solar system~\cite{Kepler:1619}.  
The laws are:
\begin{enumerate}
\item Planets orbit the Sun in ellipses, with the Sun in one of the two focuses.
\item The line connecting the Sun and a planet sweeps out equal area in equal time.
\item The “harmonic law” states the squared orbital period ${\cal T}$ of
planets measured in years equals to the third power of their major
axis measured in astronomical units, $({\cal T}/yr)^2 = (a/{\rm
  AU})^3$. 
\end{enumerate}
Newton used later the harmonic law to derive the $1/r^2$ dependence of the gravitational force~\cite{Newton:1687}. We will follow the opposite way and discuss how Kepler's laws follow from Newton's law for gravitation.
We begin by recalling how a two-body problem can be reduced to a 
one-body problem in the case of a central force. Denoting the position
and the masses of the two objects by $m_i$ and $r_i$, with $i = 1,2$
the equations of motion are found to be
\begin{equation}
m_1 \ddot {\vec r}_1 = - f(|\vec r_1 - \vec r_2|) (\vec r_1 - r_2) \,,
\label{samy-1}
\end{equation}
and
\begin{equation}
m_2 \ddot {\vec r}_2= + f(|\vec r_1 - \vec r_2|) (\vec r_1 - r_2) \, .
\label{samy-2}
\end{equation}
In other words, the center-of-mass (c.m.) of the system 
\begin{equation}
\vec R = \frac{m_1 \vec{r}_1 + m_2 \vec {r}_2}{m_1 +m_2} \, .
\label{Rc.m.}
\end{equation}
moves freely. Now, multiplying (\ref{samy-1}) by $m_2$ and
(\ref{samy-2}) by $m_1$ and substracting the two equation we obtain
\begin{equation}
\mu \ddot {\vec r} = f(r) \vec r \,,
\label{samy-3}
\end{equation}
where
\begin{equation}
\mu = \frac{m_1 m_2}{m_1 + m_2} \, .
\end{equation}
We can then solve a one-body problem for the reduced mass $\mu$ moving
with the distance $r = |\vec r_1 - \vec r_2|$ in the gravitational field of the mass $M = m_1 +
m_2$.

We can now derive the second law (a.k.a. the area law). Consider the
movement of a body under the influence  of a central force (\ref{samy-3}). Since
$\vec r \times \vec r =0$, the vectorial multiplication of (\ref{samy-3}) by $\vec
r$ leads to
\begin{equation}
\mu \vec r \times \ddot {\vec r} = 0\,,
\end{equation}
that looks already similar to a conservation law. Since
\begin{equation}
\frac{d}{dt} (\vec r \times \dot {\vec r}) = \dot {\vec r} \times \dot
{\vec r} + \vec r \times \ddot {\vec r} \,,
\end{equation} 
the first term in the right-hand-side is zero and we obtain the
conservation of angular momentum $\vec L = \mu \vec r \times \dot
{\vec r}$ for the motion in a cental potential
\begin{equation}
\mu \, \vec 
r \times \ddot {\vec r} = \frac{d}{dt} (\mu \vec r \times
\dot {\vec r}) = \frac{d}{dt} \vec L =0 \, .
\end{equation}
There are two immediate consequences: First, the motion is always in
the plane perpendicular to $\vec L$. Second, the area swept out by the
vector $\vec r$ is
\begin{equation}
d \vec A = \frac{1}{2} \vec r \times \vec v \, dt = \frac{1}{2 \mu} d
\vec L \,,
\label{samy1}
\end{equation}
and thus also constant.

We now turn to demonstrate the first law. We introduce the unit vector
$\hat r = \vec r/r$ and rewrite the definition of the angular momentum
$\vec L$ as
\begin{eqnarray}
\vec L & = & \mu \vec r \times \dot{\vec r} = \mu r \hat r \times
\frac{d}{dt} (r \hat r) \nonumber\\
& = & \mu r \hat r \times \left(\dot r \hat r + r
  \frac{d \hat r}{dt} \right) = \mu r^2 \hat r \times \frac{d \hat
r}{dt} \, .
\end{eqnarray}
The first term in the parenthesis vanishes, because of $\hat r \times
\hat r = 0$. Next we take the cross product of the gravitational acceleration,
\begin{equation}
\vec a = - \frac{GM}{r^2} \hat r \,,
\end{equation}
with the angular momentum
\begin{eqnarray}
\vec a \times \vec L & = & - \frac{GM}{r^2} \hat r \times \left(\mu r^2
  \hat r \times \frac{d\hat r}{dt} \right) \nonumber \\
& = & - GM \mu \hat r \times \left(\hat r \times \frac{d \hat r}{dt}
\right) \, ,
\end{eqnarray}
where $G = 6.674 \times 10^{-11}~{\rm N}\, {\rm m}^2 \, {\rm
  kg}^{-2}$~\cite{Beringer:1900zz}. 
The identity from vector analysis, $\vec A \times (\vec B \times \vec
C) = (\vec A \cdot \vec C) \vec B - (\vec A \cdot \vec B) \vec C$,
leads to
\begin{equation}
\vec a \times \vec L = - GM \mu \left[\left( \hat r \cdot \frac{d\hat
      r}{dt} \right) \hat r - (\hat r \cdot \hat r) \frac{d \hat
  r}{dt} \right] \, .
\end{equation}
Since $\hat r$ is a unit vector, we have $\hat r \cdot \hat r =1$ and
$d(\hat r \cdot \hat r)/dt = 0$, hence
\begin{equation}
\vec a \times \vec L = GM \mu \frac{d\hat r}{dt} \, .
\end{equation}
Since $\vec L$ and $GM\mu$ are constant, we can write this as
\begin{equation}
\frac{d}{dt} (\vec v \times \vec L ) = \frac{d}{dt} (GM\mu \hat r) \,
.
\label{samy2}
\end{equation} 
Integration of (\ref{samy2}) leads to
\begin{equation}
\vec v \times \vec L = GM\mu \hat r + \vec C \,,
\end{equation}
where the integration constant $\vec C$ is a constant vector.
Taking now the dot product with $\vec r$, we have
\begin{equation}
\vec r \cdot (\vec v \times \vec L) = GM \mu r \hat r \cdot \hat r +
\vec r \cdot \vec C \, .
\end{equation}
Applying next the identity $\vec A \cdot (\vec B \times \vec C) =
(\vec A \times  \vec B) \cdot \vec C$, it follows
\begin{eqnarray}
(\vec r \times \vec v) \cdot \vec L & = & GM \mu r + r C \cos \vartheta
\nonumber \\
& = & G M \mu r \left(1 + \frac{C \cos \vartheta}{GM \mu} \right) \, ,
\end{eqnarray}
where $\vartheta$ is the angle between $\vec r$ and $\vec C$.
Expressing $\vec r \times \vec v$ as $\vec L/\mu$, defining $e =
C/(GMμ)$ and solving for $r$, we obtain finally the equation for a
conic section, which is Kepler's first law:
\begin{equation}
r = \frac{L^2/\mu^2}{GM (1 + e \cos \vartheta)} \, .
\end{equation}
Using (\ref{elipse-eq}) we obtain angular momentum
\begin{equation}
L = \mu \sqrt{GMa(1-e^2)} \, .
\label{samy3}
\end{equation}

To obtain the harmonic law we integrate the second law in the form of
(\ref{samy1}) over one orbital period ${\cal T}$,
\begin{equation}
A = \pi ab = \frac{L}{2 \mu} {\cal T} \, .
\end{equation}
Squaring and solving for ${\cal T}$ , it follows
\begin{equation}
{\cal T}^2 = 4 \pi^2 \frac{(a b \mu)^2}{L^2} \, .
\end{equation}
Using (\ref{elipse-rel}) and (\ref{samy3}) for the angular momentum $L$, we obtain Kepler's harmonic law,
\begin{equation}
{\cal T}^2 = \frac{4 \pi^2}{G (m_1 + m_2) } a^3 \, .
\label{Kepler-h}
\end{equation}

\vspace{0.2in}

{\bf EXERCISE 1.1}~The planet Neptune, the most distant gas giant from
the Sun, orbits with a semimajor axis $a = 30.066~{\rm AU}$ and an
eccentricity $e = 0.01$. Pluto, the next large world out from the Sun
(though much smaller than Neptune) orbits with $a = 39.48~{\rm AU}$
and $e = 0.250$.  {\it (i)}~To correct number of significant figures
given the precision of the data in this exercise, how many years does
it take Neptune to orbit the Sun? {\it (ii)}~How many years does it
take Pluto to orbit the Sun? {\it (iii)}~Take the ratio of the two
orbital periods you calculated in parts {\it (i)} and {\it (ii)}. You
will see that it is very close to the ratio of two small integers;
which integers are these? Thus the two planets regularly come close to
one another, in the same part of their orbits, which allows them to
have a maximum gravitational influence on each other's orbits. This is
an example of an orbital resonance (other examples in the solar system
can be found among the moons of Jupiter, and between the moons and
various features of the rings of Saturn). {\it (iv)}~What is the
aphelion distance of Neptune's orbit? Express your answer in AU. {\it
  (v)}~What are the perihelion and aphelion distances of Pluto’s
orbit? Is Pluto always farther from the Sun than Neptune?\\

{\bf EXERCISE 1.2}~A satellite in geosynchronous orbit (GEO)
orbits the Earth once every day. A satellite in geostationary orbit
(GSO) is a satellite in a circular GEO  in the Earth's equatorial
plane. Therefore, from the point of view of an observer on Earth's
surface, a satellite in GSO seems always to {\it hover} in the same
point in the sky. For example, the satellites used for satellite TV
are in GSO so that satellite dishes can be stationary and need not
track their motion through the sky. Take a look; you will notice all
satellite dishes on people's houses point towards the Equator, that is
South. How far above Earth's equator (i.e., above the Earth's surface)
is a satellite in GSO? Express your answer in kilometers, and in Earth
radii.\\

{\bf EXERCISE 1.3}~The space station Mir traveled 3.6 billion kilometers
during its life. Its circular orbit was 200~km above the surface of
the Earth. {\it (i)}~How many years was it in orbit? {\it (ii)}~How many times
did Mir circle the Earth per day (i.e., 24 hours)? {\it (iii)}~Can you put a
satellite into such an orbit that it circles the Earth 20 times per
day?\\

The astronomical distances are so large
that we specify them in terms of the time it takes the light to travel
a given distance. For example, one light second $ = 3 \times 10^8 {\rm m} = 300,000~{\rm
  km}$, one light minute = $1.8 \times 10^7~{\rm km}$, and one light year
\begin{equation}
1~{\rm ly}  =  9.46 \times 10^{15}~{\rm m} 
  \approx  10^{13}~{\rm km}. 
\label{unoEQ}
\end{equation}
For specifying distances to the Sun and the Moon, we usually use
meters or kilometers, but we could specify them in terms of light. The
Earth-Moon distance is 384,000~km, which is 1.28 ls. The Earth-Sun
distance is $150,000,000~{\rm km}$; this is
equal to 8.3~lm. Far out in the solar system, Pluto is
about $6 \times 10^9~{\rm km}$ from the Sun, or $6 \times 10^{-4}$ ly.
The nearest star to us, Proxima Centauri, is about 4.2~ly away.
Therefore, the nearest star is 10,000 times farther from us that the
outer reach of the solar system.

\subsection{Stars and galaxies}

On clear moonless nights, thousands of stars with varying degrees of
brightness can be seen, as well as the long cloudy strip known as the
Milky Way. Galileo first observed with his telescope that the Milky
Way is comprised of countless numbers of individual stars. A half
century later Wright suggested that the Milky Way was a flat disc of
stars extending to great distances in a plane, which we call the
Galaxy~\cite{Wright:1750}.

Our Galaxy has a diameter of 100,000~ly and a thickness of roughly
2,000~ly. It has a bulging central {\it nucleus} and spiral arms. Our Sun,
which seems to be just another star, is located half  way from the
Galactic center to the edge, some $26,000$~ly from the center. The Sun
orbits the Galactic center approximately once every 250 million years
or so, so its speed is
\begin{equation}
v  =  
\frac{2\pi\ \ \  26,000 \times 10^{13}~\rm km}{2.5 \times 10^8~{\rm yr}\ 
3.156 \times 10^7~{\rm s/yr} } 
   = 200~{\rm km/s} \,.
\end{equation}
The total mass of all the stars in the Galaxy can be estimated using
the orbital data of the Sun about the center of the Galaxy. To do so,
assume that most of the mass is concentrated near the center of the
Galaxy and that the Sun and the solar system (of total mass $m$) move
in a circular orbit around the center of the Galaxy (of total mass
$M$), 
\begin{equation}
\frac{GMm}{r^2} = m \frac{v^2}{r}\,\, ,
\label{dmeq}
\end{equation}
where $a = v^2/r$ is the centripetal acceleration. All in all,
\begin{equation}
M = \frac{r\,v^2}{G} \approx 2 \times 10^{41}~{\rm kg}\, \,.
\end{equation}
Assuming all the stars in the Galaxy are similar to our Sun ($M_\odot
\approx 
2 \times 10^{30}~{\rm kg}$), we conclude that there are roughly
$10^{11}$ stars in the Galaxy.

In addition to stars both within and outside the Milky Way, we can see
with a telescope many faint cloudy patches in the sky which were once
all referred to as {\it nebulae} (Latin for clouds).  A few of these,
such as those in the constellations of Andromeda and Orion, can
actually be discerned with the naked eye on a clear night. In the XVII
and XVIII centuries, astronomers found that these objects were getting
in the way of the search for comets. In 1781, in order to provide a
convenient list of objects not to look at while hunting for comets,
Messier published a celebrated catalogue~\cite{Messier:1781}. Nowadays astronomers
still refer to the 103 objects in this catalog by their Messier
numbers, e.g., the Andromeda Nebula is M31.

Even in Messier's time it was clear that these extended objects are
not all the same. Some are star clusters, groups of stars which are so
numerous that they appeared to be a cloud. Others are glowing clouds
of gas or dust and it is for these that we now mainly reserve the word
nebula. Most fascinating are those that belong to a third category:
they often have fairly regular elliptical shapes and seem to be a
great distance beyond the Galaxy. Kant seems to have been the first to
suggest that these latter might be circular discs, but appear
elliptical because we see them at an angle, and are faint because they
are so distant~\cite{Kant:1755}. At first it was not universally
accepted that these objects were extragalactic (i.e. outside our
Galaxy). The very large telescopes constructed in the XX century
revealed that individual stars could be resolved within these
extragalactic objects and that many contain spiral arms. Hubble did
much of this observational work in the 1920's using the 2.5~m
telescope on Mt.  Wilson near Los Angeles, California. Hubble
demostrated that these objects were indeed extragalactic because of
their great distances~\cite{Hubble:b}. The distance to our nearest
spiral galaxy, Andromeda, is over 2 million ly, a distance 20 times
greater than the diameter of our Galaxy. It seemed logical that these
nebulae must be galaxies similar to ours. Today it is thought that
there are roughly $4 \times 10^{10}$ galaxies in the observable
universe -- that is, as many galaxies as there are stars in the
Galaxy.

\section{Distance Measurements}

We have been talking about the vast distance of the objects in the
universe. We now turn to discuss different
methods to estimate these distances. 

\subsection{Stellar Parallax}

\begin{figure}
 \postscript{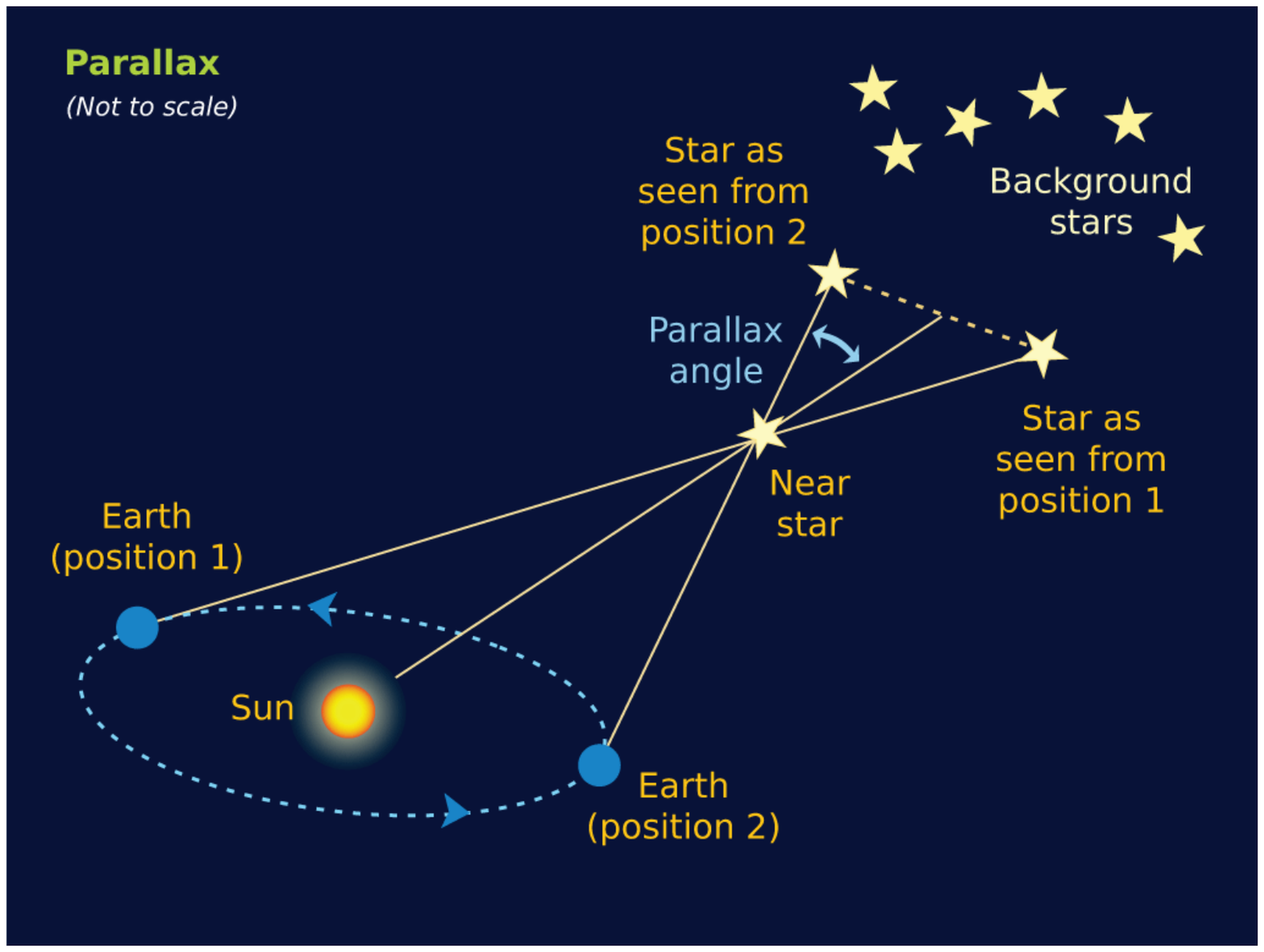}{0.98}
 \caption{The parallax method of measuring a star's distance.}
\label{parallax}
\end{figure}

One basic method to measure distances to nearby stars employs simple
geometry and stellar parallax. Parallax is the apparent displacement of an
object because of a change in the observer's point of view. One way to
see how this effect works is to hold your hand out in front of you and
look at it with your left eye closed, then your right eye closed. Your
hand will appear to move against the background. By stellar parallax we mean the
apparent motion of a star against the background of more distant
stars, due to Earth's motion around the Sun; see Fig.~\ref{parallax}. The sighting angle of a
star relative to the plane of Earth's orbit (usually indicated by
$\theta$) can be determined at two different times of the year separated
by six months. Since we
know the distance $d$ from the Earth to the Sun, we can determine the
distance $D$ to the star.  For example, if the angle $\theta$ of a
given star is measured to be $89.99994^\circ,$ the parallax angle is
$p \equiv \phi = 0.00006^\circ.$ From trigonometry, $\tan \phi = d/D$,
and since the distance to the Sun is $d= 1.5 \times 10^8~{\rm km}$ the
distance to the star is
\begin{equation}
D =  \frac{d}{\tan \phi} \approx \frac{d}{\phi} = 
\frac{1.5 \times 10^{8}~{\rm km}}{1 \times 10^{-6}} = 
1.5 \times 10^{14}~{\rm km} \,\,,
\end{equation}
or about 15~ly.

Distances to stars are often specified in terms of parallax angles
given in seconds of arc: 1 second (1'') is 1/60 of a minute (1') of
arc, which is 1/60 of a degree, so 1'' = 1/3600 of a degree. The
distance is then specified in parsecs (meaning {\em par}allax angle in
{\em sec}onds of arc), where the parsec is defined as $1/\phi$ with
$\phi$ in seconds. For example, if $\phi = 6 \times 10^{-5\, \circ}$, we
would say the the star is at a distance $D = 4.5~{\rm pc}$.

The angular resolution of the Hubble Space Telescope (HST) is about
1/20 arcs. With HST one can measure parallaxes of about 2 milli arc seconds (e.g., 1223 Sgr).  This corresponds to a distance of about 500~pc. Besides, there are stars with radio emission for which
observations from the Very Long Baseline Array (VLBA) allow accurate
parallax measurements beyond 500~pc. For example, parallax
measurements of Sco X-1 are $0.36\pm 0.04$~milli arc seconds which puts it
at a distance of 2.8~kpc. Parallax can be used to determine the
distance to stars as far away as about 3~kpc from Earth.  Beyond that
distance, parallax angles are two small to measure and
more subtle techniques must be employed.\\

{\bf EXERCISE 2.1}~One of the first people to make a very accurate
measurement of the circumference of the Earth was Eratosthenes, a
Greek philosopher who lived in Alexandria around 250~B.C. He was told
that on a certain day during the summer (June 21) in a town called
Syene, which was 4900 stadia (1 stadia = 0.16 kilometers) to the south
of Alexandria, the sunlight shown directly down the well shafts so
that you could see all the way to the bottom. Eratosthenes knew that
the sun was never quite high enough in the sky to see the bottom of
wells in Alexandria and he was able to calculate that in fact it was
about 7 degrees too low. Knowing that the sun was 7 degrees lower at
its highpoint in Alexandria than in Syene and assuming that the sun's
rays were parallel when they hit the Earth, Eratosthenes was able to
calculate the circumference of the Earth using a simple proportion:
C/4900 stadia = 360 degrees/ 7 degrees. This gives an answer of
252,000 stadia or 40,320 km, which is very close to today's
measurements of 40,030 km. Assume the Earth is flat and determine the
parallax angle that can explain this phenomenon. Are the results
consistent with the hypothesis that the Earth is flat?

\subsection{Stellar luminosity}

\begin{figure}
 \postscript{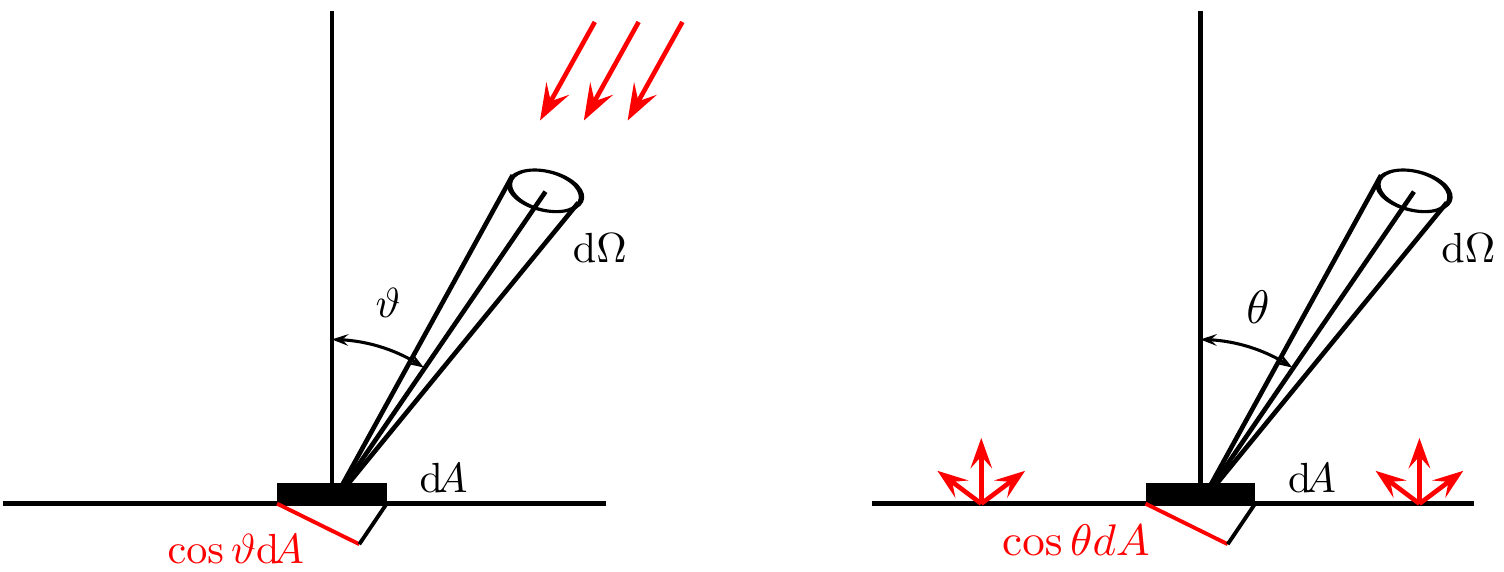}{0.98}
 \caption{{\bf Left.} A detector with surface element $dA$ on Earth
   measuring radiation coming from a direction with zenith angle
   $\vartheta$ (left).  {\bf Right.} An imaginary detector of area 
   $dA$ on the surface of a star measuring radiation emitted in the
   direction $\theta$~\cite{Kachelriess}.}
\label{brightness}
\end{figure}

In 1900, Planck found empirically the distribution
\begin{equation}
B_\nu \, d\nu = \frac{2h\nu^3}{c^2} \,  \left[ {\rm exp} \left
    (\frac{h\nu}{kT} \right) - 1 \right]^{-1} \, d\nu
\label{Planck}
\end{equation}
describing the amount of energy emitted into the frequency interval
$[\nu, \nu + d\nu]$ and the solid angle $d\Omega$ per unit time and
area by a body in thermal equilibrium~\cite{Planck:1901tja}. The {\it
  intrinsic} (or surface) brightness $B_\nu$ depends only on the
temperature $T$ of the blackbody (apart from the natural constants
$k$, $c$ and $h$).  The dimension of $B_\nu$ in the cgs system of
units is
\begin{equation}
[B_\nu] = \frac{\rm erg}{{\rm Hz} \, {\rm cm^2} \, {\rm s} \, {\rm sr}} \, .
\end{equation}
In general the amount of energy per frequency interval $[\nu,  \nu +
d\nu]$ and solid angle $d\Omega$ crossing the perpendicular area
$A_\perp$ per time is called the specific (or differential) intensity~\cite{Rybicki:1979} \begin{equation}
I_\nu = \frac{dE}{d\nu d\Omega dA_\perp dt} \, ;
\end{equation}
see Fig.~\ref{brightness}. For the special case of the blackbody
radiation, the specific intensity at the emission surface is given by
the Planck distribution, $I_\nu = B_\nu$. Stars are fairly good
approximations of blackbodies.

Integrating (\ref{Planck}) over all frequencies and possible solid
angles gives the emitted flux $F$ per surface area $A$. The angular
integral consists of the solid angle $d\Omega = \sin \theta d \theta d
\phi$ and the factor $\cos \theta$ taking into account that only the
perpendicular area $A_\perp = A \cos \theta$ is
visible~\cite{Anchordoqui:2015uww}. The flux emitted by a star is
found to be
\begin{equation}
F = \pi \int_0^\infty d \nu B_\nu = \frac{2 \pi}{c^2 h^3} (kT)^4
\int_0^\infty \frac{x^3 \, dx}{e^x-1} = \sigma T^4,
\label{Fnueve}
\end{equation} 
where $x = h \nu/(kT)$,
\begin{equation}
\sigma = \frac{2 \pi^5 k^4}{15 c^2 h^3} = 5.670 \times 10^{-5}~\frac{\rm
  erg}{\rm cm^2 \, K^4 \, s}  
\end{equation}
is the Stefan-Boltzmann constant~\cite{Stefan:1879,Boltzmann:1884}, and where we used $\int_0^\infty
x^3 \ [e^x - 1]^{-1} dx = \pi^4/15$.

A useful parameter for a star or galaxy is its luminosity. 
The total luminosity $L$ of a star is given by the product of its
surface area  and the radiation emitted per area 
\begin{equation}
L = 4 \pi R^2 \sigma T^4 \, .
\label{Luminosity}
\end{equation} 
Careful analyses of nearby stars have shown that the absolute
luminosity for most of the stars depends on the mass: {\em the more
  massive the star, the greater the luminosity}.

Consider a thick spherical source of radius $R$, with constant
intensity along the surface, say a star. An observer at a distance $r$
sees the spherical source as a disk of angular radius $\vartheta_c
=R/r$. Note that since the source is optically thick the observer only
sees the surface of the sphere. Because the intensity is constant
over the surface there is a symmetry along the $\varphi$ direction such
that the solid solid angle is given by $d\Omega = 2 \pi \sin \vartheta d
\vartheta$. By looking at Fig.~\ref{brightness} it is straightforward
to see that the flux observed at $r$ is given by
\begin{eqnarray}
F (r) & = & \int I \cos \vartheta d\Omega = 2 \pi I \int_0^{\vartheta_c} \sin \vartheta
\cos \vartheta d \vartheta \nonumber \\
& = & \left. \pi I \cos^2 \vartheta\right|^0_{\vartheta_c} = \pi I \sin^2 \vartheta_c =
\pi I (R/r)^2 \, .
\label{Fdoce}
\end{eqnarray}
At the surface of the star $R =r$ and we recover (\ref{Fnueve}). Very
far away, $r \gg R$, and (\ref{Fdoce}) yields $F = \pi \vartheta_c^2 I
= I \Omega_{\rm source}$; see Appendix~\ref{appA}. The validity of the
inverse-square law $F \propto 1/r^2$ at a distance $r > R$ outside of
the star relies on the assumptions that no radiation is absorbed and
that relativistic effects can be neglected. The later condition
requires in particular that the relative velocity of observer and
source is small compared to the speed of light. All in all, the total
(integrated) flux at the surface of the Earth from a given
astronomical object with total luminosity $L$ is found to be
\begin{equation}
F_{\rm observed \,  @ \, Earth} = {\cal F} = \frac{L}{4 \pi d_L^2},
\label{L}
\end{equation}
where $d_L$ is the distance to the object. 

Another important parameter of a star is its surface temperature,
which can be determined from the spectrum of electromagnetic
frequencies it emits. The wavelength at the peak of the spectrum,
$\lambda_{\rm max},$ is related to the temperature by Wien's displacement law~\cite{Wien:1894}
\begin{equation}
\lambda_{\rm max} T = 2.9 \times 10^{-3}~{\rm m \, K} \,\,.
\end{equation}

We can now use Wien's law and the Steffan-Boltzmann equation (power
output or luminosity $\propto AT^4$) to determine the temperature and
the relative size of a star. Suppose that the distance from Earth to
two nearby stars can be reasonably estimated, and that their apparent
luminosities suggest the two stars have about the same absolute
luminosity, $L$.  The spectrum of one of the stars peaks at about
700~nm (so it is reddish).  The spectrum of the other peaks at about
350~nm (bluish). Using Wien's law, the temperature of the reddish star
is $T_{\rm r} \simeq 4140~{\rm K}$.  The temperature of the bluish
star will be double because its peak wavelength is half, $T_{\rm b}
\simeq 8280~{\rm K}$.  The power radiated per unit of area from a star
is proportional to the fourth power of the Kelvin temperature
(\ref{Luminosity}).  Now the temperature of the bluish star is double
that of the redish star, so the bluish must radiate 16 times as much
energy per unit area. But we are given that they have the same
luminosity, so the surface area of the blue star must be 1/16 that of
the red one. Since the surface area is $4 \pi R^2$, we conclude that
the radius of the redish star is 4 times larger than the radius of the
bluish star (and its volume 64 times larger)~\cite{FA}.

\begin{figure*}
 \postscript{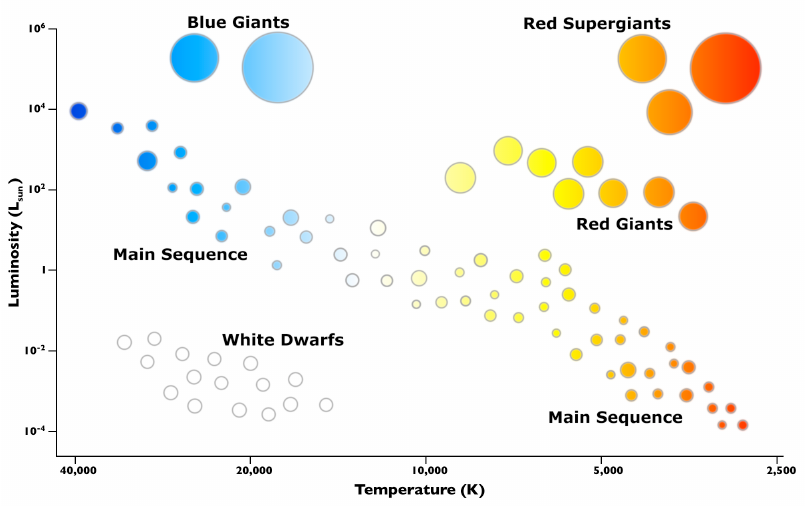}{0.98}
 \caption{ HR diagram. The vertical axis depicts the
   inherent brightness of a star, and the horizontal axis the surface
   temperature increasing from right to left~\cite{Shirley}.}
\label{9}
\end{figure*}

An important astronomical discovery, made around 1900, was that for
most of the stars, the color is related to the absolute luminosity and
therefore to the mass. A useful way to present this relationship is
by the so-called Hertzsprung-Russell (HR) diagram~\cite{HR}. On the HR
diagram, the horizontal axis shows the temperature $T$, whereas the
vertical axis the luminosity $L$, each star is represented by a point
on the diagram shown in Fig.~\ref{9}. Most of the stars fall along the
diagonal band termed the main sequence. Starting at the lowest right, we
find the coolest stars, redish in color; they are the least luminous
and therefore low in mass. Further up towards the left we find hotter
and more luminous stars that are whitish like our Sun. Still farther
up we find more massive and more luminous stars, bluish in color.
There are also stars that fall outside the main sequence. Above and to
the right we find extremely large stars, with high luminosity but with
low (redish) color temperature: these are called red giants. At the
lower left, there are a few stars of low luminosity but with high
temperature: these are white dwarfs.

Suppose that a detailed study of a certain star suggests that it most
likely fits on the main sequence of the HR diagram. The observed flux
is ${\cal F} = 1 \times 10^{-12}$~W m$^{-2}$, and the peak
wavelength of its spectrum is $\lambda_{\rm max} \approx 600$~nm.  We
can first find the temperature using Wien's law and then estimate the
absolute luminosity using the HR diagram; namely, $T \approx 4800~{\rm
  K}$.  A star on the main sequence of the HR diagram at this
temperature has absolute luminosity of about $L \approx
10^{26}$~W. Then, using (\ref{L}) we can estimate its distance from
us, $d_L = 3 \times 10^{18}~{\rm m}$
or equivalently 300~ly.\\

{\bf EXERCISE 2.2}~About 1350 J of energy strikes the atmosphere of the Earth from
the Sun per second per square meter of area at right angle to the
Sun's rays. What is {\it (i)} the observed flux from the Sun ${\cal F}_\odot$ and
{\it (ii)}~its absolute luminosity $L_\odot$. {\it (iii)}~What is the average Solar flux density measured at Mars?
{\it (iv)} If the approximate efficiency of the solar panels (with
area of 
$1.3~{\rm m^2}$) on the
Martian rover Spirit  is 20\%, then how
many Watts could the fully illuminated panels generate? \\

{\bf EXERCISE 2.3}~Suppose the MESSENGER spacecraft, while orbiting
Mercury, decided to communicate with the Cassini
probe, now exploring Saturn and its moons. When Mercury is closest to
Saturn in their orbits, it takes 76.3 minutes for the radio signals
from Mercury to reach Saturn. A little more than half a mercurian year
later, when the 2 planets are furthest apart in their orbits, it takes
82.7 minutes. {\it (i)}~What is the distance between Mercury and the
Sun? Give answers in both light-minutes and astronomical
units. Assume that the planets have circular orbits.
{\it (ii)}~What is the distance between Saturn and the Sun?\\

{\bf EXERCISE 2.4}~The photometric method to search for extrasolar
planets is based on the detection of stellar brightness variations,
which result from the transit of a planet across a star's disk. If a
planet passes in front of a star, the star will be partially eclipsed
and its light will be dimmed. Determine the reduction in the {\it
  apparent} surface brightness $I$ when Jupiter passes in front of the
Sun. \\

{\bf EXERCISE 2.5}~The angular resolution of a telescope (or other
optical system) is a measure of the smallest details which can be
seen. Because of the distorting effects of earth's atmosphere, the
best angular resolution which can be achieved by optical telescopes
from earth's surface is normally about 1~arcs. This is why much
clearer images can be obtained from space. The angular resolution of
the HST is about 0.05 arcs, and the smallest angle that can be
measured accurately with HST is actually a fraction of one resolution
element.  {\it (i)}~ Cepheid variable stars are very
important distance indicators because they have large and well-known
luminosities. What is the distance of a Cepheid variable star whose
parallax angle is measured to be $0.005 \pm 0.001~{\rm arcs}$? {\it (ii)}~The faintest stars that can be detected with the
HST have apparent brightnesses which are $4 \times 10^{21}$ times
fainter than the Sun. How far away could a star like the
Sun be, and still be detected with the HST? Express your answer in
light years. {\it (iii)}~How far away could a Cepheid variable with
20,000 times the luminosity of the Sun be, and still be detected with
the HST?  Express your answer in light years.  \\

{\bf EXERCISE 2.6}~The discovery of the dwarf planet {\it Eris} in
2005 threw the astronomical community into a tizzy and made
international headlines; it is slightly larger than Pluto and brought
up interesting questions about what the definition of a planet
is. Eventually, this resulted in the controversial demotion of Pluto
from the 9th planet of the Solar System to just one of a number of
dwarf planets. Throughout, assume that Eris is spherical and is
observed at opposition (i.e., the Earth lies on the straight line
connecting the Sun and Eris). {\it (i)}~In five hours, Eris is
observed to move 7.5 arcseconds relative to the background stars as
seen from Earth. Because Eris is much further from the Sun than is the
Earth, it is moving quite a bit slower around the Sun than the Earth,
so this apparent motion on the sky is essentially entirely parallax
due to the Earth's motion. Calculate the speed with which the Earth
goes around the Sun, in kilometers/second. Use this information and
the small-angle formula to calculate the distance from the Earth to
Eris. Express your result in AU. Compare with the semi-major axis of
Pluto's orbit (which you will need to look up). {\it (ii)}~Eris shines
in two ways: from its reflected light from the Sun (which will be
mostly visible light), and from its blackbody radiation from absorbed
sunlight (which will mostly come out as infrared light). The albedo of
Eris (i.e., the fraction of the sunlight incident on Eris that is
reflected) is very high, about 85\%. This suggests that Eris is
covered by a layer of shiny ice; spectroscopy tells us that the ice is
composed of frozen methane, CH$_4$.  Derive an expression for the
brightness of Eris which depends on its distance from the Sun $d$, and
its radius $r$. First, calculate the amount of sunlight reflected by
Eris per unit time (i.e., its luminosity in reflected light); express
your answer in terms of $d$, $r$, the albedo $\mathfrak{a}$, and the
luminosity of the Sun $L_\odot$. {\it (iii)}~We detect only a tiny
fraction of this light reflected by Eris. Calculate the brightness,
via the inverse square law, of Eris as perceived here on Earth. {\it
  (iv)}~The measured brightness of Eris is $2.4 \times 10^{-16}$
Joules meters$^{-2}$ second$^{-1}$. Use this information to determine
the radius $r$ of Eris.  This is what led to the controversy of what a
planet is: if Pluto is considered a planet, then certainly Eris should
be as well. We further elaborate about this controversy in the
solution of the exercise. {\it (v)}~Calculate the angular size of Eris
(i.e., the angle the diameter of Eris makes on the sky). Compare this
to the resolution of the HST; will
you be able to resolve Eris (i.e., will it look like a point of light
or a finite-size object in a telescope)? {\it (vi)}~You go ahead and
observe Eris with Hubble, and find that it has a moon orbiting it. 
Observations with Hubble show that this moon (called Dysnomia) makes
an almost circular orbit around Eris with a period of 15.8 Earth
days. The semi-major axis of the orbit subtends an angle of $0.53''$ as
seen from Earth. Calculate the semi-major axis in kilometers, and
calculate the mass of Eris in kilograms. Compare with the mass of
Pluto ($1.3 \times 10^{22}~{\rm kg}$). Is Eris more massive?\\

{\bf EXERCISE 2.7}~A perfect blackbody at temperature $T$ has the shape of an
oblate ellipsoid, its surface being given by the equation
\begin{equation}
 {x^2\over a^2} + {y^2\over a^2} + {z^2 \over b^2} = 1 \,,
\end{equation}
with $a > b$. {\it (i)}~Is the luminosity of the blackbody isotropic?
Why?  {\it (ii)}~Consider an observer at a distance $d_L$ from the
blackbody, with $d_L \gg a$. What is the direction of the observer for
which the maximum amount of flux will be observed (keeping the
distance $d_L$ fixed)?  Calculate what this maximum flux is. {\it
  (iii)}~Repeat the same exercise for the direction for which the
minimum flux will be observed, for fixed $d_L$. {\it (iv)}~If the two
observers who see the maximum and minimum flux from distance $d_L$ can
resolve the blackbody, what is the {\it apparent}  brightness, $I$, that each one
will measure? {\it (v)}~Write down an expression for the total
luminosity emitted by the black body as a function of $a$, $b$ and
$T$. {\it (vi)}~Now, consider a  galaxy with a perfectly oblate
shape, which contains only a large number $N$ of stars, and no gas or
dust.  To make it simple, assume that all stars have radius $R$ and
surface temperature $T$. Answer again the questions {\it (i-v)} for
the galaxy, assuming $NR^2 \ll ab$. Are there any
differences from the case of a blackbody? Explain why. {\it (vii)}~Imagine that
there were a very compact galaxy that did not obey the condition $NR^2
\ll ab$. Would the answer to the previous question be modified? Do you
think such a galaxy could be stable?\\

{\bf EXERCISE 2.8}~The HR diagram is usually plotted in logarithmic
coordinates ($\log L$ vs. $\log T$, with the temperature increasing to
the left). Derive the slope of a line of constant radius in the
logarithmic HR diagram.

\section{Doppler Effect}

There is observational evidence that stars move at speeds ranging up
to a few hundred kilometers per second, so in a year a fast moving
star might travel $\sim 10^{10}$~km. This is $10^3$ times less than
the distance to the closest star, so their apparent position in the
sky changes very slowly. For example, the relatively fast moving star
known as Barnard's star is at a distance of about $56 \times
10^{12}$~km; it moves across the line of sight at about 89~km/s, and
in consequence its apparent position shifts (so-called ``proper
motion'') in one year by an angle of 0.0029 degrees. The HST has
  measured proper motions as low as about 1 milli arc second per year. In
  the radio (VLBA), relative motions can be measured to an accuracy of
  about 0.2 milli arc second per year. The apparent position in the sky
of the more distant stars changes so slowly that their proper motion
cannot be detected with even the most patient observation. However,
the rate of approach or recession of a luminous body in the line of
sight  can be measured much more accurately
than its motion at right angles to the line of
sight. The technique makes use of a familiar property of any sort of
wave motion, known as Doppler effect~\cite{Doppler}.

When we observe a sound or light wave from a source at rest, the time
between the arrival wave crests at our instruments is the same as the
time between crests as they leave the source. However, if the source
is moving away from us, the time between arrivals of successive wave
crests is increased over the time between their departures from the
source, because each crest has a little farther to go on its journey
to us than the crest before. The time between crests is just the
wavelength divided by the speed of the wave, so a wave sent out by a
source moving away from us will appear to have a longer wavelength 
than if the source were at rest. Likewise, if the source is moving
toward us, the time between arrivals of the wave crests is decreased
because each successive crest has a shorter distance to go, and the
waves appear to have a shorter wavelength. A nice analogy was put
forward by Weinberg~\cite{Weinberg:1977ji}. He compared
the situation with a travelling man that has to send a letter home
regularly once a week during his travels: while he is travelling away
from home, each successive letter will have a little
farther to go than the one before, so his letters will arrive a little
more than a week apart; on the homeward leg of his journey, each
succesive letter will have a shorter distance to travel, so they will
arrive more frequently than once a week.

The Doppler effect began to be of enormous importance to astronomy
in 1968, when it was applied to the study of individual spectral
lines. In 1815,  Fraunhofer first realized that when
light from the Sun is allowed to pass through a slit and then through
a glass prism, the resulting spectrum of colors is crossed with
hundreds of dark lines, each one an image of the slit~\cite{Fraunhofer}. The dark lines
were always found at the same colors, each corresponding to a
definite wavelength of light. The same dark spectral lines were also
found in the same position in the spectrum of the Moon and brighter
stars. It was soon realized that these dark lines are produced by the
selective absorption of light of certain definite wavelengths, as
light passes from the hot surface of a star through its cooler outer
atmosphere. Each line is due to absorption of light by a specific
chemical element, so it became possible to determine that the elements
on the Sun, such as sodium, iron, magnesium, calcium, and chromium,
are the same as those found on Earth.

In 1868, Sir  Huggins was able to show that the dark lines in
the spectra of some of the brighter stars are shifted slightly to the
red or the blue from their normal position in the spectrum of the Sun~\cite{Huggins}.
He correctly interpreted this as a Doppler shift, due to the motion of
the star away from or toward the Earth. For example, the wavelength of
every dark line in the spectrum of the star Capella is longer than the
wavelength of the corresponding dark line in the spectrum of the Sun
by $0.01\%,$ this shift to the red indicates that Capella is receding
from us at $0.01\% \,c$ (i.e., the radial velocity of Capella is about
30~km/s).

\begin{figure}
 \postscript{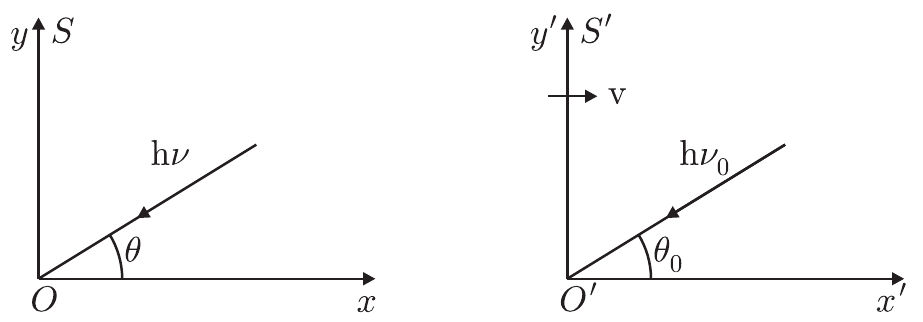}{0.98}
 \caption{A source of light waves moving to the right, relative to
   observers in the $S$ frame, with velocity $v$. The frequency is
   higher for observers on the right, and lower for observers on the
   left~\cite{Sarkar}.}
\label{Doppler1}
\end{figure}

Consider two inertial frames, $S$ and $S'$, moving with relative
velocity $v$ as shown in Fig.~\ref{Doppler1}. Assume a light source
(e.g. a star) at rest in $S'$ emits light of frequency $\nu_0$ at an
angle $\theta_0$ with respect to the observer $O'$. Let 
\begin{equation}
p^\mu = \left( \frac{h\nu}{c}, -\frac{h \nu}{c} \cos \theta , -
  \frac{h \nu}{c} \sin \theta, 0 \right)
\end{equation}
be the momentum 4-vector for the photon as seen in $S$ and 
\begin{equation}
p_0^\mu = \left( \frac{h\nu_0}{c}, -\frac{h \nu_0}{c} \cos \theta_0 , -
  \frac{h \nu_0}{c} \sin \theta_0, 0 \right)
\end{equation}
in $S'$. To get the 4-momentum relation from $S'
\to S$, apply the inverse Lorentz transformation~\cite{Lorentz:1904}
\begin{eqnarray}
\frac{h \nu}{c} & = & \gamma \left[ \frac{h \nu_0}{c} + \beta \left(-
    \frac{h\nu_0}{c} \cos \theta_0 \right) \right] \nonumber \\ 
-\frac{h \nu}{c} \cos \theta & = & \gamma \left( -\frac{h \nu_0}{c} \cos \theta_0 + \beta 
    \frac{h\nu_0}{c}  \right) \nonumber \\ 
\frac{h\nu}{c} \sin \theta & = & \frac{h \nu_0}{c} \sin \theta_0 \, .
\end{eqnarray}
The first expression gives
\begin{equation}
\nu = \nu_0 \gamma ( 1 - \beta \cos \theta_0) \,,
\label{dop1}
\end{equation}
which is the relativistic Doppler formula. 

For observational astronomy (\ref{dop1}) is not useful because both $\nu_0$ and
$\theta_0$ refer to the star's frame, not that of the observer. Apply
instead the direct Lorentz transformation $S \to S'$ to the photon
energy to obtain 
\begin{equation}
\nu_0 = \gamma \nu ( 1 + \beta \cos \theta) \, .
\end{equation}
This equation gives $\nu_0$ in terms of quantities measured by the
observer. It is sometimes written in terms of wavelengths:  $\lambda = \lambda_0 \gamma (1 +
\beta \cos \theta)$. (For details see
e.g.~\cite{Anchordoqui:2015xca}.)\\

{\bf EXERCISE 3.1} Consider the inertial frames $S$ and $S'$ shown in
Fig.~\ref{Doppler1}. Use the inverse Lorentz transformation to show that
the relation between angles is given by
\begin{equation}
\cos \theta = \frac{\beta - \cos \theta_0}{\beta \cos \theta_0 -1} \, .
\label{dopplerang}
\end{equation}

\vspace{0.2in}

There are three special cases: {\it (i)} $\theta_0 =0$, which gives 
\begin{equation}
\nu = \nu_0 \sqrt{(1-\beta)/(1+\beta)} \, .
\end{equation}
In the non-relativistic limit we have
$\nu = \nu_0 (1- \beta)$. This corresponds to a source moving away
from the observer. Note that $\theta = 0$. {\it (ii)}~$\theta_0 =
\pi$, which gives 
\begin{equation}
\nu = \nu_0 \sqrt{(1+\beta)/(1-\beta)} \, . 
\end{equation}
Here the
source is moving towards the observer. Note that $\theta = \pi$. {\it
  (iii)}~$\theta_0 = \pi/2$, which gives 
\begin{equation}
\nu = \nu_0 \gamma \, . 
\end{equation}
This last is the transverse Doppler effect -- a second order relativistic
effect. It can be thought of as arising from the dilation of time in
the moving frame.\\

{\bf EXERCISE 3.2}~Suppose light is emitted isotropically in a star's
rest frame $S'$, i.e. $dN/d\Omega_0 = \varkappa$, where $dN$ is the
number of photons in the solid angle $d\Omega_0$ and $\varkappa$ is a
constant. What is the angular distribution in the inertial frame
$S$?\\

{\bf EXERCISE 3.3}~Show that for $v\ll c$, the Doppler shift in wavelength 
is 
\begin{equation}
\frac{\lambda' - \lambda}{\lambda} \approx \frac{v}{c} \, . 
\label{doppler-shift}
\end{equation}
To avoid confusion, it should be
  kept in mind that $\lambda$ denotes the wavelength of the light if
  observed near the place and time of emission, and thus presumably
  take the values measured when the same atomic transition occurs in
  terrestrial laboratories, while $\lambda'$ is the wavelength of the
  light observed after its long journey to us. If $\lambda'-\lambda > 0$ then
  $\lambda' > \lambda$ and we speak of a redshift; if $\lambda' -
  \lambda < 0$ then
  $\lambda' < \lambda$, and we speak of a blueshift.\\

{\bf EXERCISE 3.4}~Through some coincidence, the Balmer lines from single ionized
helium in a distant star happen to overlap with the Balmer lines from
hydrogen in the Sun. How fast is that star receding from us? [Hint:
the wavelengths from single-electron energy level transitions are inversely
proportional to the square of the atomic number of the nucleus.]\\

\begin{figure*}[tbp] 
\begin{minipage}[t]{0.5\textwidth}
\postscript{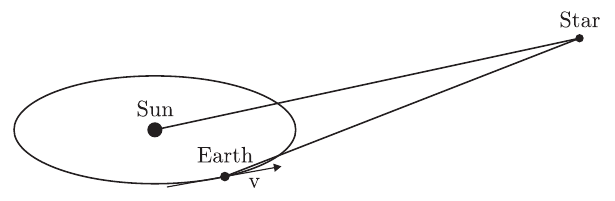}{0.99} 
\end{minipage}
\hfill
\begin{minipage}[t]{0.48\textwidth}
\postscript{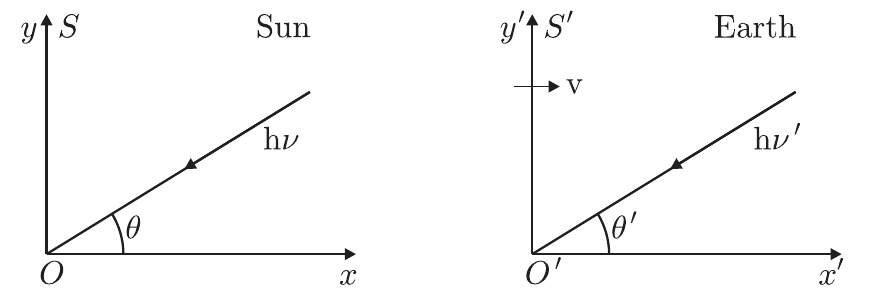}{0.99} 
\end{minipage}
\caption{Schematic representation of stellar aberration~\cite{Sarkar}.}
\label{Doppler2}
\end{figure*}

{\bf EXERCISE 3.5}~Stellar aberration is the apparent motion of a star
due to rotation of the Earth about the Sun. Consider an incoming
photon from a star with 4-momentum $p^\mu$. Let $S$ be the Sun's frame
and $S'$ the Earth frame moving with velocity $v$ as shown in
Fig.~\ref{Doppler2}. Define the angle of aberration $\alpha$  by $\theta' =
\theta - \alpha$ and show that $\alpha \approx \beta \sin \theta$.\\

{\bf EXERCISE 3.6}~HD 209458 is a star in the constellation Pegasus
very similar to our Sun ($M = 1.1 M_\odot$ and $R = 1.1 R_\odot$),
located at a distance of about $150~{\rm ly}$.  In 1999, two teams
working independently discovered an extrasolar planet orbiting the
star using the so-called {\it radial velocity planet search
method}~\cite{Mazeh:2000pa,Queloz:2000xp}.  Note that a star with a
planet must move in its own small orbit in response to the planet's
gravity. This leads to variations in the speed with which the star
moves toward or away from Earth, i.e. the variations are in the radial
velocity of the star with respect to Earth. The radial velocity can be
deduced from the displacement in the parent star's spectral lines due
to the Doppler shift. If a planet orbits the star, one should have a
periodic change in that rate, except for the extreme case in which the
plane of the orbit is perpendicular to our line of sight. Herein we
assume that the motions of the Earth relative to the Sun have already
been taken into account, as well as any long-term steady change of
distance between the star and the sun, which appears as a median line
for the periodic variation in radial velocity due to the star's
{\it wobble} caused by the orbiting planet. The observed Doppler shift
velocity of HD 209458 is found to be $K = V \sin i = 82.7 \pm 1.3~{\rm
  m/s}$, where $i = 87.1^\circ \pm 0.2^\circ$ is the inclination of
the planet's orbit to the line perpendicular to the
line-of-sight.~\cite{Wittenmyer:2005nz}. Soon after the discovery,
separate teams were able to detect a transit of the planet across the
surface of the star making it the first known transiting extrasolar
planet~\cite{Charbonneau:1999nk,Henry:2000gr}. The planet received the
designation HD 209458b. Because the planet transits the star, the star
is dimmed by about 2\% every $3.52447 \pm 0.00029$~days. Tests
allowing for a non-circular Keplerian orbit for HD 209458 resulted in
an eccentricity indistinguishable from zero: $e = 0.016 \pm 0.018$.
Consider the simplest case of a nearly circular orbit and find: {\it
  (i)}~the distance from the planet to the star; {\it (ii)} the mass $m$
of the planet; {\it (iii)} the radius $r$ of the planet.

\section{Stellar Evolution}

The stars appear unchanging. Night after night the heavens reveal no
significant variations. Indeed, on human time scales, the vast
majority of stars change very little. Consequently, we cannot follow
any but the tiniest part of the life cycle of any given star since
they live for ages vastly greater than ours. Nonetheless, herein 
 we will follow the process of stellar evolution from
the birth to the death of a star, as we have theoretically
reconstructed it.

\subsection{Stellar nucleosynthesis}

There is a general consensus that stars are born when gaseous
clouds (mostly hydrogen) contract due to the pull of gravity. A huge
gas cloud might fragment into numerous contracting masses, each mass
centered in an area where the density is only slightly greater than
at nearby points. Once such {\it globules} formed, gravity would cause
each to contract in towards its center-of-mass. As the particles of
such protostar accelerate inward, their kinetic energy increases. When
the kinetic energy is sufficiently high, the Coulomb repulsion between
the positive charges is not strong enough to keep hydrogen nuclei
appart, and nuclear fussion can take place. In a star like our Sun,
the ``burning'' of hydrogen occurs when four protons fuse to form a
helium nucleus, with the release of $\gamma$ rays, positrons and
neutrinos.\footnote{The word ``burn'' is put in quotation marks
  because these high-temperature fusion reactions occur via a {\em
    nuclear} process, and must not be confused with ordinary burning
  in air, which is a {\em chemical} reaction, occurring at the {\em
    atomic} level (and at a much lower temperature).}

The energy output of our Sun is believed to be due principally to the
following sequence of fusion reactions:
\begin{equation}
_1^1 \!{\rm H} + _1^1 \!\!{\rm H}  \to _1^2 \!\!{\rm H} + 2\, e^+ + 
2\, \nu_e \hspace{0.9cm} (0.42~{\rm MeV})\,,
\end{equation}
\begin{equation}
_1^1 \!{\rm H} + _1^2 \!\!{\rm H}  \to _2^3 \!\!{\rm He} + \gamma \hspace{2.2cm} (5.49~{\rm MeV})\,,
\end{equation}
and 
\begin{equation}
_2^3 {\rm He} + _2^3 \!\!{\rm He}  \to _2^4 \!\!{\rm He} + _1^1 \!\!{\rm H} + _1^1 \!\!{\rm H} \hspace{1.3cm} (12.86~{\rm MeV})\,,
\end{equation}
where the energy released for each reaction (given in parentheses)
equals the difference in mass (times $c^2$) between the initial and
final states. Such a released energy is carried off by the outgoing
particles.  The net effect of this sequence, which is called the
$pp$-cycle, is for four protons to combine to form one $_2^4$He
nucleus, plus two positrons, two neutrinos, and two gamma rays:
\begin{equation}
4 \, _1^1\!{\rm H} \to _2^4\!\!{\rm He} + 2 e^+ + 2 \nu_e + 2 \gamma \,\,.
\end{equation}
Note that it takes two of each of the first two reactions to produce
the two $_2^3$He for the third reaction. So the total energy
released for the net reaction is 24.7~MeV. However, each of the two
$e^+$ quickly annihilates with an electron to produce $2 m_e c^2 =$
1.02~MeV; so the total energy released is 26.7~MeV. The first
reaction, the formation of deuterium from two protons, has very low
probability, and the infrequency of that reaction serves to limit the
rate at which the Sun produces energy. These reactions requiere a
temperature of about $10^7$~K, corresponding to an average kinetic
energy ($kT$) of 1~keV.\\ 

{\bf EXERCISE 4.1}~Approximately $10^{38}$ neutrinos are produced by
the $pp$ chain in the Sun every second. Calculate the number of
neutrinos from the Sun that are passing through your brain every
second.\\

In more massive stars, it is more likely that the energy output comes
principally from the carbon (or CNO) cycle, which comprises the
following sequence of reactions:
\begin{equation}
^{12}_{\phantom{1}6}\!{\rm C} + ^1_1\!\!{\rm H} \to ^{13}_{\phantom{1}7}\!\!{\rm N} + \gamma \,,
\end{equation}
\begin{equation}
\phantom{^{12}_{\phantom{1}6}\!{\rm C} +++} ^{13}_{\phantom{1}7}\!{\rm N} \to ^{13}_{\phantom{1}6}\!\!{\rm C} + e^+ 
+ \nu \,,
\end{equation}
\begin{equation}
^{13}_{\phantom{1}6}\!{\rm C} + ^1_1\!\!{\rm H} \to ^{14}_{\phantom{1}7} \!\!{\rm N} + \gamma \,,
\end{equation}
\begin{equation}
^{14}_{\phantom{1}7}\!{\rm N} + ^1_1\!\!{\rm H} \to ^{15}_{\phantom{1}8} \!\!{\rm O} + \gamma \,,
\end{equation}
\begin{equation}
\phantom{^{12}_{\phantom{1}6}\!{\rm C} +++}  ^{15}_{\phantom{1}8} \!{\rm O} \to ^{15}_{\phantom{1}7} \!\!{\rm N} + e^+ 
+ \nu \,,
\end{equation}
\begin{equation}
^{15}_{\phantom{1}7} \!{\rm N} + ^1_1 \!\!{\rm H} \to ^{12}_{\phantom{1}6} \!\!{\rm C} + ^{4}_2 \!\!{\rm He} \, .
\end{equation}
It is easily seen that no carbon is consumed in this cycle (see first
and last equations) and that the net effect is the same as the $pp$
cycle. The theory of the $pp$ cycle and the carbon cycle as the source
of energy for the Sun and the stars was first worked out by Bethe
in 1939~\cite{Bethe:1939bt}.

The fusion reactions take place primarily in the core of the star,
where $T$ is sufficiently high. (The surface temperature is of course
much lower, on the order of a few thousand K.) The tremendous release
of energy in these fusion reactions produces an outward pressure
sufficient to halt the inward gravitational contraction; and our
protostar, now really a young star, stabilizes in the main sequence.

To a good approximation the stellar structure on the main sequence can
be described by a spherically symmetric system in hydrostatic
equilibrium. This requires that rotation, convection, magnetic fields,
and other effects that break rotational symmetry have only a minor
influence on the star. This assumption is in most cases very well
justified.

We denote by $M(r)$ the mass enclosed inside a sphere with radius $r$
and density $\rho(r)$
\begin{equation}
M(r) = 4\pi \int_0^r dr' \ {r'}^2 \ \rho(r')
\end{equation}
or
\begin{equation}
\frac{dM(r)}{dr} = 4 \pi r^2 \rho(r) \, .
\label{star-continuity}
\end{equation}
An important application of (\ref{star-continuity}) is to express
physical quantities not as function of the radius $r$ but of the
enclosed mass $M(r)$. This facilitates the computation of the stellar
properties as function of time, because the mass of a star remains
nearly constant during its evolution, while the stellar radius can
change considerably. 

 A radial-symmetric mass distribution $M(r)$
produces according Gauss law the same gravitational acceleration, as
if it would be concentrated at the center $r = 0$. Therefore the
gravitational acceleration produced by $M(r)$ is
\begin{equation}
g(r) = - \frac{G M(r)}{r^2} \, .
\end{equation}
If the star is in equilibrium, this acceleration is balanced by a
pressure gradient from the center of the star to its surface. Since
pressure is defined as force per area, $P = F/A$, a pressure change
along the distance $dr$ corresponds to an increment
\begin{eqnarray}
dF & = & dA P - (P + dP) dA \nonumber \\
 & = & - \underbrace{dA dP}_{\rm force} = - \underbrace{\rho(r) d A
 dr}_{\rm mass} \underbrace{a(r)}_{\rm acceleration}
\end{eqnarray}
 of the force $F$ produced by the pressure gradient $dP$. For
increasing $r$, the gradient $dP < 0$ and the resulting force $dF$ is
positive and therefore directed outward. Hydrostatic equilibrium, $g(r)
= -a(r)$, requires then
\begin{equation}
\frac{dP}{dr} = \rho(r) g(r) = - \frac{G M(r) \ \rho(r)}{r^2} \, .
\label{staruno}
\end{equation} 
If the pressure gradient and gravity do not balance each other, the
layer at position $r$ is accelerated,
\begin{equation}
a(r) = \frac{GM(r)}{r^2} + \frac{1}{\rho(r)} \frac{dP}{dr} \, .
\end{equation}
In general, we need an equation of state, $P = P(\rho,T,Y_i)$, that
connects the pressure $P$ with the density $\rho$, the (not yet) known
temperature $T$ and the chemical composition $Y_i$ of the star. For an
estimate of the central pressure $P_c = P (0)$  of a star in hydrostatic
equilibrium, we integrate (\ref{staruno}) and obtain with $P (R) \approx 0$,
\begin{equation}
P_c = \int_0^R \frac{dP}{dr} dr = G \int_0^M  dM \frac{M}{4 \pi r^4} \,,
\end{equation}
where we used the continuity equation (\ref{star-continuity}) to
substitute $dr = dM/(4\pi r^2\rho)$ by $dM$. If we replace furthermore $r$ by the
stellar radius $R \geq r$, we obtain a lower limit for the central
pressure,
\begin{eqnarray}
P_c & = & G \int_0^M dM \frac{M}{4 \pi r^4} \nonumber \\
 & > & G \int_0^M dM \frac{M}{4 \pi R^4} = \frac{M^2}{8 \pi R^4}  \, .
\end{eqnarray}
Inserting values for the Sun, it follows
\begin{equation}
P_c > \frac{M^2}{8 \pi R^4} = 4 \times 10^8~{\rm bar}
\left(\frac{M}{M_\odot}\right)^2 \, \left(\frac{R_\odot}{R} \right)^4
\, .
\label{Pclimit}
\end{equation}
The value obtained integrating the hydrostatic equation using the
``solar standard model'' is $P_c = 2.48 \times 10^{11}~{\rm bar}$,
i.e. a factor 500 larger.\\

{\bf EXERCISE 4.2}~Calculate the central pressure $P_c$ of a star in hydrostatic equilibrium as a function of its mass $M$ and radius $R$ for
{\it (i)} a constant mass density, $\rho(r) = \rho_0$ and {\it (ii)}~ a linearily decreasing mass density, $\rho(r) = \rho_c [1 - (r/R)]$. \\

Exactly where the star falls along the main sequence depends on its
mass. The more massive the star, the further up (and to the left) it
falls in the HR diagram. To reach the main sequence requires perhaps
30 million years and the star  is expected to
remain there 10 billion years ($10^{10}$~yr). Although most of stars
are billions of years old, there is evidence that stars are actually
being born at this moment in the Eagle Nebula. 

As hydrogen fuses to form helium, the helium that is formed is denser
and tends to accumulate in the central core where it was formed. As
the core of helium grows, hydrogen continues to fuse in a shell around
it. When much of the hydrogen within the core has been consumed, the
production of energy decreases at the center and is no longer
sufficient to prevent the huge gravitational forces from once again
causing the core to contract and heat up. The hydrogen in the shell
around the core then fuses even more fiercely because of the rise in
temperature, causing the outer envelope of the star to expand and to
cool. The surface temperature thus reduced, produces a spectrum of
light that peaks at longer wavelength (reddish). By this time the star
has left the main sequence. It has become redder, and as it has grown
in size, it has become more luminous. Therefore, it will have moved to
the right and upward on the HR diagram. As it moves upward, it enters
the red giant stage. This model then explains the origin of red giants
as a natural step in stellar evolution. Our Sun, for example, has been
on the main sequence for about four and a half billion years. It will
probably remain there another 4 or 5 billion years. When our Sun
leaves the main sequence, it is expected to grow in size (as it
becomes a red giant) until it occupies all the volume out to roughly
the present orbit of the planet Mercury.

If the star is like our Sun, or larger, further fusion can occur. As
the star's outer envelope expands, its core is shrinking and heating
up. When the temperature reaches about $10^8$~K, even helium nuclei,
in spite of their greater charge and hence greater electrical
repulsion, can then reach each other and undergo fusion:
\begin{equation}
^4_2 {\rm He} + ^4_2 \!{\rm He} \to ^8_4 \!{\rm Be} +
\gamma \hspace{1.2cm}  (-91.8~{\rm keV}) \, .
\end{equation}
Once beryllium-8 is produced a little faster than it decays (half-life
is $6.7 \times 10^{-17}~{\rm s}$), the number of beryllium-8 nuclei in
the stellar core increases to a large number. Then in its core there
will be many beryllium-8 nuclei that can fuse with another helium
nucleus to form carbon-12, which is stable:
\begin{equation}
^4_2 {\rm He} + ^8_4 \!{\rm Be} \to ^{12}_{\phantom{1}6} \!{\rm
  C} + \gamma \hspace{1.2cm}  (7.367~{\rm MeV}) \,  .
\end{equation}
The net energy release of the triple-$\alpha$ process is
7.273~MeV. Further fusion reactions are possible, with $^4_2$He
fusing with $^{12}_{\phantom{1}6}$C to form $^{16}_{\phantom{1}
  8}$O.  Stars spend approximately a few thousand to 1 billion years
as a red giant. Eventually, the helium in the core runs out and fusion
stops.  Stars with $0.4 M_\odot < M < 4 M_\odot$ are fated to end up
as spheres of carbon and oxygen. Only stars with $M > 4 M_\odot$
become hot enough for fusion of carbon and oxygen to occur and higher
$Z$ elements like $^{20}_{10}$Ne or $^{24}_{12}$Mg can be made.

As massive ($M > 8 M_\odot$) red supergiants age, they
produce ``onion layers'' of heavier and heavier elements in their
interiors. A star of this 
mass can contract under gravity and heat up even further,
($T= 5 \times 10^9$~K), producing nuclei as heavy as $^{56}_{26}$Fe and
$^{56}_{28}$Ni. However, the average binding energy 
per nucleon begins to decrease beyond the iron group of isotopes.
Thus, 
the formation of heavy nuclei from
lighter ones by fusion ends at the iron group.  Further fusion would
require energy, rather than release it.
As a consequence, a core of
iron builds up in the centers of massive supergiants. 

A star's lifetime as a giant or supergiant is shorter than its main
sequence lifetime (about 1/10 as long). As the star's core becomes hotter, and the fusion reactions powering
it become less efficient, each new fusion fuel is used up in a shorter
time. For example, the stages in the life of a $25 M_\odot$ star are as
follows::
hydrogen fusion lasts 7 million years, hellium fusion lasts 500,000
years, carbon fusion lasts 600 years, neon fusion lasts 1 year, oxygen
fusion lasts 6 months, and sillicon fusion lasts 1 day. The star core
in now pure iron. The process of creating heavier nuclei
from lighter ones, or by absorption of neutrons at higher $Z$ (more on
this below) is
called nucleosynthesis.

\subsection{White dwarfs and Chandrasekhar limit}

At a distance of 2.6~pc Sirius is the fifth closest stellar system to
the Sun. It is the brightest star in the Earth's night sky.  Analyzing
the motions of Sirius from 1833 to 1844, Bessel concluded that it had
an unseen companion, with an orbital period $T \sim 50~{\rm
  yr}$~\cite{Bessel}. In 1862, Clark discovered this companion, Sirius
B, at the time of maximal separation of the two components of the
binary system (i.e. at
apastron)~\cite{Clark:1863}. Complementary follow up observations
  showed that the mass of Sirius B equals approximately that of the
  Sun, $M \approx M_\odot$. Sirius B's peculiar properties were not
  established until the next apastron by
  Adams~\cite{Adams:1915}. He noted that its high temperature ($T\simeq 25,
  000~{\rm K}$)  together with its small
  luminosity ($L= 3.84 \times 10^{26}~{\rm W}$) require an extremely small
  radius and thus a large density. From Stefan-Boltzmann
  law we have
\begin{equation}
\frac{R}{R_\odot} = \left(\frac{L}{L_\odot} \right)^{1/2} \left(
  \frac{T}{T_\odot}\right)^2 \approx 10^{-2} \, .
\end{equation}
Hence, the mean density of Sirius B is a factor $10^6$ higher than
that of the Sun; more precisely,  $\rho = 2 \times 10^6~{\rm g/cm}^3$.

A lower limit for the central pressure of Sirius B follows from (\ref{Pclimit})
\begin{equation}
P_c > \frac{M^2}{8 \pi R^4} = 4 \times 10^{16}~{\rm bar} \, .
\label{siriuspressureL}
\end{equation}
Assuming the pressure is dominated by an ideal gas the central
temperature is found to be
\begin{equation}
T_c = \frac{P_c}{nk} \sim 10^2 T_{c,\odot} \approx 10^9~{\rm K} \, .
\end{equation}
For such a high $T_c$, the temperature gradient $dT/dr$
in Sirius B would be a factor $10^4$ larger than in the Sun. This
would in turn require a larger luminosity  and a larger energy
production rate than that of main sequence stars. 

Stars like Sirius B are called white dwarfs. They have very long
cooling times, because of their small surface luminosity. This type of
stars is rather numerous. The mass density of main-sequence stars in
the solar neighborhood is $0.04M_\odot/{\rm pc}^3$ compared to
$0.015M_\odot /{\rm pc}^3$ in white dwarfs. The typical mass of white
dwarfs lies in the range $0.4 \alt M/M_\odot \alt 1$, peaking at $0.6
M_\odot$. No further fusion energy can be
obtained inside a white dwarf. The star loses internal energy by radiation, decreasing
in temperature and becoming dimmer until its light goes out.  

For a classical gas, $P = nkT$, and thus in the limit of zero
temperature,  the pressure inside a star also goes to  zero. How can a
star be stabilized after the fusion processes and thus energy
production stopped?  The solution to this puzzle is that the main
source of pressure in such compact stars has a different origin.

According to Pauli's exclusion principle no two fermions can occupy the
same quantum state~\cite{Pauli}.  In statistical mechanics,
Heisenberg's uncertainty principle $\Delta x \Delta p \geq
\hslash$~\cite{Heisenberg:1927zz} together with Pauli's principle imply
that each phase-space volume, $\hslash^{-1} \, dx \, dp$, can only be
occupied by  one fermionic state.

A (relativistic or non-relativistic) particle in a box of volume $L^3$
collides per time interval $\Delta t = L/v_x$ once with the $yz$-side of the
box, if the $x$ component of its velocity is $v_x$. Thereby it exerts the
force $F_x = \Delta p_x/\Delta t = p_xv_x/L$. The pressure produced by $N$ particles is
then $P = F/A = Np_xv_x/(LA) = np_xv_x$.  For an isotropic distribution,
with $\langle v^2\rangle = \langle v_x^2\rangle + \langle v_y^2\rangle
+ \langle v_z^2 \rangle = 3 \langle v_x^2 \rangle$, we have 
\begin{equation}
P = \tfrac{1}{3} n v p \, .
\label{lagarde}
\end{equation}
Now, if we take $\Delta x = n^{-1/3}$ and $\Delta p \approx \hslash􏰤/\Delta
x \approx \hslash n^{1/3}$, combined with the non-relativistic
expression  $v=p/m$,  the pressure of a degenerate
fermion gas is found to be
\begin{equation}
P \approx nvp \approx \frac{\hslash^2 n^{5/3}}{m} \, .
\label{fortyniners}
\end{equation}
(\ref{fortyniners}) implies $P \propto \rho^{5/3}$, where $\rho$ is the density.
For relativistic particles, we can obtain an estimate for the pressure inserting $v = c$,
\begin{equation}
P \approx n c p \approx c \hslash n^{4/3} \,,
\end{equation}
which implies $P \propto \rho^{4/3}$.  It may be worth noting at this
juncture that
{\it (i)}~both the non-relativistic and the relativistic pressure laws
are polytropic equations of state, $P = K \rho^\gamma$; {\it (ii)}~a
non-relativistic degenerate Fermi gas has the same adiabatic index 
($\gamma = 5/3$) as an ideal gas, whereas a relativistic degenerate Fermi
gas has the same adiabatic index ($\gamma = 4/3$) as radiation; {\it
  (iii)}~in the non-relativistic limit the pressure is inversely
proportional to the fermion mass, $P \propto 1/m$, and so for
non-relativistic systems  the degeneracy will first become important 
to electrons.\\

{\bf EXERCISE 4.3}~Estimate the average energy of electrons in Sirius
B from the equation of state for non-relativistic degenerate fermion 
gas, 
\begin{equation}
P =  \frac{(3\pi^2)^{2/3}}{5} \frac{\hslash^2}{m}
n^{5/3} \,, 
\end{equation}
and calculate the Lorentz factor of the electrons. Give a short qualitative statement about the validity of the non-relativistic equation of state for white dwarfs with a density of Sirius B and beyond. \\

Next, we compute the pressure of a degenerate non-relativistic
electron gas inside Sirius B and check if it is consistent with the lower
limit for the central pressure derived in (\ref{siriuspressureL}).  The
only bit of information needed is the value of $n_e$, which can be
written in terms of the density of the star, the atomic mass of the
ions making up the star, and the number of protons in the ions
(assuming the star is neutral):
\begin{equation}
n_e = \frac{\rho}{ \mu_e \ m_p} 
\end{equation}
where $\mu_e \equiv A/Z$ is the average number of nucleon per free
electron. For metal-poor stars $\mu_e = 2$, and so from (\ref{fortyniners}) we obtain
\begin{eqnarray}
P & \approx & \frac{h^2 n_e^{5/3}}{m_e} \nonumber \\
& \approx & \frac{(1.05 \times 10^{27}~{\rm
  erg \, s})^2}{9.11 \times 10^{-28}~{\rm g}} \left( \frac{10^6~{\rm
      g/cm}^3}{2 \times 1.67 \times 10^{-24}~{\rm g}} \right)^{5/3}
\nonumber \\
& \approx & 10^{23}~{\rm dyn/cm}^2 \, .
\end{eqnarray}
Since  $10^{6}~{\rm dyn/cm}^2 = 1~{\rm bar}$, we have
$P = 10^{17}~{\rm bar}$, which is consistent with the lower limit
derived in (\ref{siriuspressureL}).

We can now relate the mass of the star to its radius by combining the
lower limit on the central pressure
$P_c \sim GM^2/R^4$
 and the polytropic equation of state 
$P = K \rho^{5/3} \sim K
(M/R^3)^{5/3} =K M^{5/3}/R^5$. It  follows that 
\begin{equation}
\frac{GM^2}{R^4} = \frac{KM^{5/3}}{R^5} \,,
\end{equation}
or equivalently
\begin{equation}
R = \frac{M^{(10-12)/6}}{K} = \frac{1}{K M^{1/3}} \, .
\end{equation}
If the small differences in chemical composition can be neglected, then 
there is unique relation between the mass and the radius of  white dwarfs.
Since the star's radius  decreases with increasing mass, 
there must be a maximal mass allowed.

To derive this maximal mass we first assume the pressure can be described by a non-relativistic degenerate Fermi
gas. The total kinetic energy of the star is $U_{\rm
  kin} = N p^2/(2m_e)$, where $n \sim N/R^3$ and $p \sim \hslash n^{1/3}$. Thus
\begin{equation}
U_{\rm kin} \sim N \frac{\hslash^2 n^{2/3}}{2m_e} \sim \frac{\hslash^2
N^{(3+2)/3}}{2m_eR^2} = \frac{\hslash^2 N^{5/3}}{2 m_e R^2} \, .
\end{equation}
For the potentail gravitational energy, we use the approximation
$U_{\rm pot} = \alpha GM^2/R$, with $\alpha =1$. Hence
\begin{equation}
U(R) = U_{\rm kin} + U_{\rm pot} \sim \frac{\hslash^2 N^{5/3}}{2 m_eR^2}
  - \frac{GM^2}{R} \, .
\end{equation}
For small $R$, the positive term dominates and so there exists a stable minimum
$R_{\rm min}$  for each $M$.

However, if the Fermi gas inside the star becomes relativistic, then
$U_{\rm kin} = N cp$, or
\begin{equation}
U_{\rm kin} \sim N c \hslash n^{1/3} \sim \frac{c \hslash N^{4/3}}{R} 
\end{equation}
and
\begin{equation}
U(R) = U_{\rm kin} + U_{\rm pot} \sim \frac{c \hslash N^{4/3}}{R} -
\frac{GM^2}{R} \, .
\label{morelagarde}
\end{equation}
Now both terms scale like $1/R$. For a fixed chemical composition, the
ratio $N/M$ remains constant. Therefore, if  $M$ is increased the negative term increases
faster than the first one. This implies there exists a
critical $M$ so that $U$ becomes negative, and can be made arbitrary small
by decreasing the radius of the star: {\it the star collapses}.  This
critical mass is called Chandrasekhar mass $M_{\rm Ch}$.  It can be obtained by
solving (\ref{morelagarde}) for $U=0$. Using $M=N_N m_N$ we have $c \hslash N_{\rm
  max}^{4/3} = G N_{\rm max}^2 m_N^2$, or,  with $m_N \simeq m_p$,
\begin{equation}
N_{\rm max} \sim \left(\frac{c \hslash}{G m_p^2} \right)^{3/2} \sim
  \left(\frac{M_{\rm Pl}}{m_p} \right)^3 \sim 2 \times 10^{57} \, .
\end{equation}
This leads to
\begin{equation}
M_{\rm Ch} = N_{\rm max} m_p \sim 1.5 M_{\odot} \, .
\label{Chandra}
\end{equation}
The Chandrasekhar mass derived ``professionally'' is found to be
$M_{\rm Ch} \simeq 1.46 M_\odot$~\cite{Longair:1994wu}.\\

{\bf EXERCISE 4.4}~Derive approximate Chandrasekhar mass limits in
units of solar mass by setting the central pressures of exercise 4.1  equal to the
relativistic degenerate electron pressure,
\begin{equation}
P= \frac{(3 \pi^2)^{1/3} \hslash c}{4} n^{4/3} \, .
\end{equation}
Compare the estimates with the exact limit. \\

The critical size can be determined by imposing two conditions: that the gas
becomes relativistic, $U_{\rm kin} \alt N m_e c^2$, and $N =
N_{\rm max}$,
\begin{equation}
N_{\rm max} m_e c^2 \agt \frac{c \hslash N_{\rm max}^{4/3}}{R}  \, .
\end{equation}
This leads to
\begin{equation}
m_e c^2 \agt \frac{c \hslash}{R} \left(\frac{c \hslash}{G m_N^2} 
\right)^{1/2} \,,
\end{equation}
or  equivalently
\begin{equation}
R \agt \frac{\hslash}{m_e c} \left( \frac{c \hslash}{G m_N^2}
\right)^{1/2} \sim 5 \times 10^8~{\rm cm} \, .
\label{RChandra}
\end{equation}
which is in agreement with the radii found for white dwarf stars.

\subsection{Supernovae}

Supernovae are massive explosions that take place at the end of a
star's life cycle. They can be triggered by one of two basic
mechanisms: (I) the sudden re-ignition of nuclear fusion in a
degenerate star, or (II) the sudden gravitational collapse of
the massive star's core. 

In a type I supernova, a degenerate white dwarf accumulates sufficient
material from a binary companion, either through accretion or via a
merger. This material raise its core temperature to then trigger
runaway nuclear fusion, completely disrupting the star.  Since the
white dwarf stars explode crossing the Chandrasekhar limit, $M >
M_\odot$, the release total energy should not vary so much.  Thus one
may wonder if they are possible standard candles.\\

{\bf EXERCISE 4.5}~Type Ia supernovae have been observed in some
distant galaxies. They have well-known luminosities and at their peak
$L_{\rm Ia} \approx 10^{10} L_\odot$. Hence, we can use
them as standard candles to measure the distances to very remote
galaxies. How far away could a type Ia supernova be, and still be detected with  HST?\\

\begin{figure*}
 \postscript{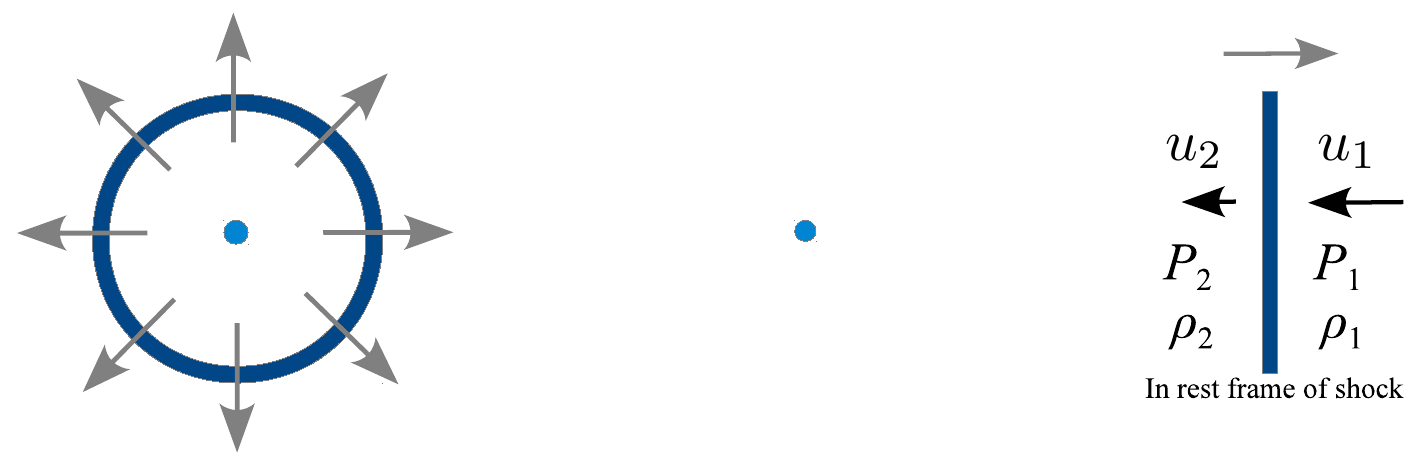}{0.85}
 \caption{{\bf Left.} The sudden release of a large amount of energy
   into a background fluid of density $\rho_1$ creates a strong
   spherical shock wave, emanating from the point where the energy was
   released. {\bf Right.} Jump conditions across normal shock
   waves. If the shock moves to the right with velocity $u_{\rm sh}$,
   then in the rest-frame of the shock the background gas streams with
   velocity $u_1 = - u_{\rm sh}$ to the left, and comes out of the
   shock with a higher density $\rho_2$, higher pressure $P_2$, and
   with a lower velocity $u_2$. Conservation of momentum requires $P_1
   + \rho_1 u_1^2 = P_2 + \rho_2 u_2^2$, see Appendix~\ref{appB}. For
   the case at hand, $P_1 \ll P_2$ and so $P_2 \sim \rho_1 u_1^2$.}
\label{fig_shock}
\end{figure*}

In type II supernovae the core of a $M \agt 8
M_\odot$  star undergoes sudden gravitational collapse. These stars have an
onion-like structure with a degenerate iron core. When the core is
completely fused to iron, no further processes releasing energy are
possible. Instead, high energy collisions 
break apart iron into helium and
eventually into protons and neutrons,
\begin{equation}
^{56}_{26} {\rm Fe} \to 13 \, ^{4}_{2} {\rm He} + 4\,n 
\end{equation}
and 
\begin{equation}
^{4}_{2} {\rm He} \to 2\,p + 2\,n \,\,.
\end{equation}
This removes the thermal
  energy necessary to provide pressure support and the star collapses.
 When the star begings to contract the density increases and the free
  electrons are forced together with protons to form neutrons via
  inverse beta decay, 
\begin{equation}
e^- + p \to n + \nu_e \,\, ;
\end{equation}
even though neutrinos do not interact easily with matter, at these
extremely high densities, they exert a tremendous outward
pressure. The outer layers fall inward when the iron core collapses,
forming an enormously dense neutron star~\cite{Oppenheimer:1939ne}. If
$M \alt M_{\rm Ch}$, then the core stops
collapsing because the neutrons start getting packed too tightly.
Note that $M_{\rm Ch}$ as derived in (\ref{Chandra}) is valid for both neutrons and
electrons, since the stellar mass is in both cases given by the sum of
the nucleon masses, only the main source of pressure (electrons or
neutrons) differs. The critical size follows from (\ref{RChandra}) by substituting
$m_e$ with $m_N$,
\begin{equation}
R \agt \frac{\hslash}{m_N c} \left(\frac{c\hslash}{G m_N^2} 
\right)^{1/2}  \sim 3
  \times 10^{5}~{\rm cm} \, .
\end{equation}
Since already Sirius B was difficult to detect, the question arises if
and how these extremely small stars can be observed. When core density
reaches nuclear density, the equation of state stiffens suddenly and
the infalling material is ``reflected.'' Both the neutrino outburst and the
outer layers that crash into the core and rebound cause the entire
star outside the core to be blown apart.  The released energy goes
mainly into neutrinos (99\%), kinetic energy (1\%); only 0.01\% into
photons.

Much of the modeling of supernova explosions and their remnants
derives from the nuclear bomb research program. Whenever a supernova
goes off a large amount of energy $E$ is injected into the ``ambient
medium'' of uniform density $\rho_1$. In the initial phase of the
expansion the impact of the external medium will be small, because the
mass of the ambient medium that is overrun and taken along is still
small compared with the ejecta mass. The supernova remnant is said to
expand adiabatically. After some time a strong spherical shock front
(a ``blast wave'') expands into the ambient medium, and the mass swept
up by the outwardly moving shock significantly exceeds the mass of the
initial ejecta, see Fig.~\ref{fig_shock}. The ram pressure, $P_2 \sim \rho_1 u_{\rm sh}^2$, of
the matter that enters the shock wave is much larger than the ambient
pressure $P_1$ of the upstream medium, and any radiated energy is much
smaller than the explosion energy $E$. This regime, during which the
energy remains constant is known as the Sedov--Taylor
phase~\cite{Sedov,Taylor:1950a,Taylor:1950b}.  The mass of the swept up
material is of order $M(t) \sim \rho_1 r^3(t)$, where
$r$ is the radius of the shock. The fluid velocity behind the
shock will be of order the mean radial velocity of the shock, $u_{\rm
  sh} (t) \sim r(t)/t$ and so the kinetic energy is
\begin{equation}
E_{\rm kin} = \frac{1}{2} M u_{\rm sh}^2 \sim \rho_1 r^3
\frac{r^2}{t^2} = \rho_1
\frac{r^5}{t^2} \, .
\end{equation}
What about the thermal energy in the bubble created by the explosion?
This should be of order
\begin{equation}
E_{\rm therm} = \frac{3}{2} P_2 V \sim P_2 r^3 \sim \rho_1
u_{\rm sh}^2 r^3 \sim \rho_1 \frac{r^5}{t^2} \, .
\end{equation}
This suggests that the thermal energy is of the same order as the kinetic energy, and scales in the same fashion with time. Therefore
\begin{equation}
E = E_{\rm kin} + E_{\rm therm} \sim \rho_1 \frac{r^5}{t^2} \, ,
\end{equation}
yielding
\begin{equation}
r (t) \sim \left(\frac{E t^2}{\rho_1} \right)^{1/5} \, .
\label{lagarde-sedov-taylor}
\end{equation}
The expanding shock wave slows as it expands
\begin{equation}
u_{\rm sh} = \frac{2}{5} \left(\frac{E}{\rho_1 t^3}\right)^{1/5} = \frac{2}{5} \left(\frac{E}{\rho_1}\right)^{1/2}
r^{-3/2} \, .
\end{equation}
This means that the blask wave decelerates and dissapears after some
time. The expanding supernova remnant then passes from its
Taylor-Sedov phase to its ``snowplow'' phase. During the snowplow 
phase, the matter of the ambient interstellar medium is swept up by
the expanding dense shell, just as snow is swept up by a coasting
snowplow.\\

\begin{figure*}
 \postscript{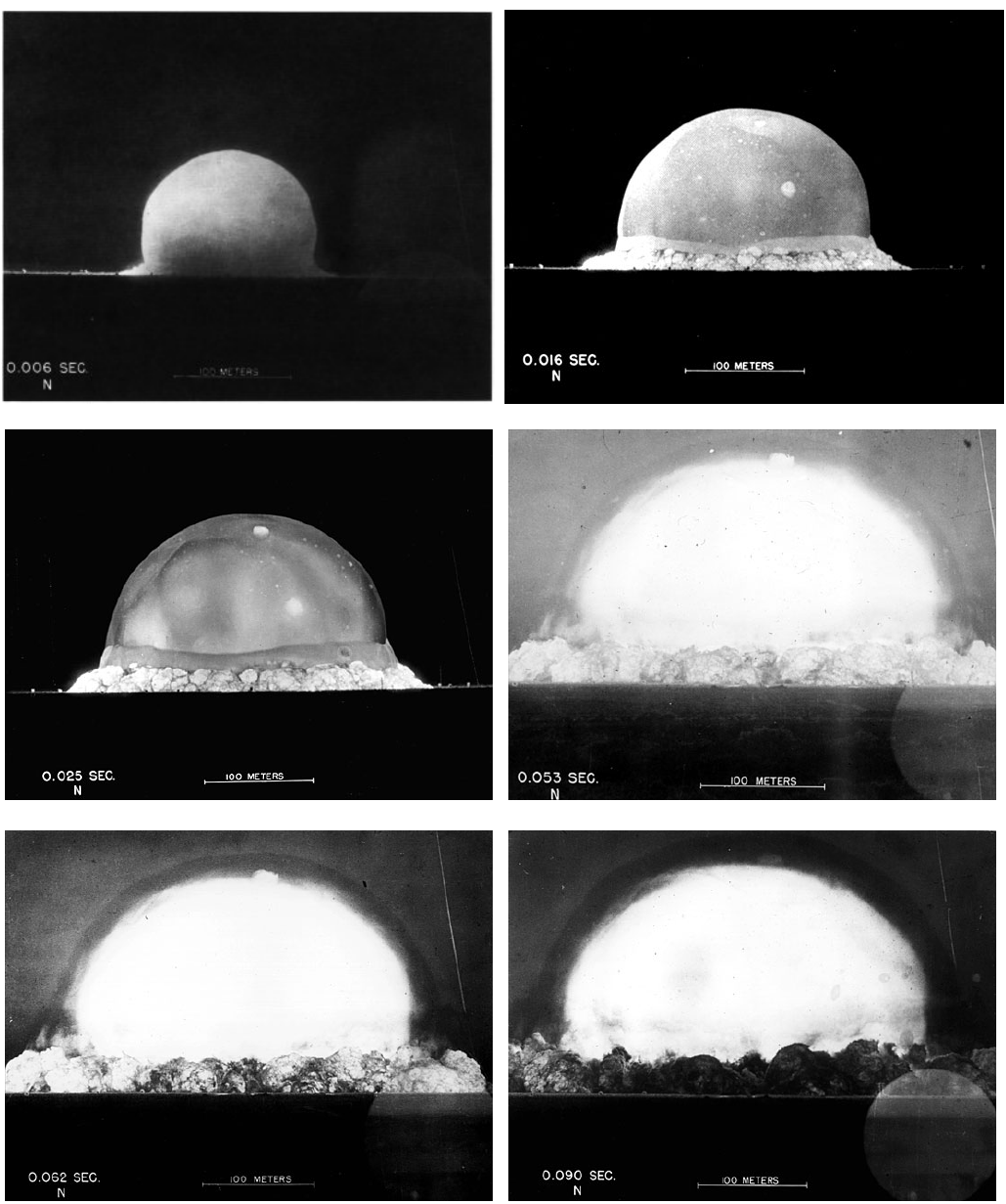}{0.99}
 \caption{Trinity test of July 16, 1945.}
\label{trinity}
\end{figure*}

{\bf EXERCISE 4.6}~Estimate the energy of the first detonation of a
nuclear weapon (code name Trinity) from the time dependence of the
radius of its shock wave. Photographs of the early stage of the
explosion are shown in Fig.~\ref{trinity}. The device was placed on the top
of a tower, $h = 30~{\rm m}$ and the explosion took place at about
$1100~{\rm m}$
above sea level.  {\it (i)}~Explain the origin of the thin layer above the
bright ``fireball'' that can be seen in the last three pictures ($t \geq
0.053~{\rm s}$). Is the shock front behind or ahead of this layer? Read the
radius of the shock front from the figures and plot it as a function
of time after the explosion. The time and length scale are indicated
in the lables of the figures. {\it (ii)}~Fit (by eye or numerical
regression) a line to the radius vs. time dependence of the shock
front in a log-log representation, $\ln(r) = a+b \ln(t)$. Verifiy that $b$
is compatible with a Sedov-Taylor expansion. Then fix $b$ to the
theoretical expectation, re-evaluate $a$ and estimate the energy of the
bomb in tons of TNT equivalent. [Hint: ignore the initial (short)
phase of free expansion.]\\

If the final mass of a neutron star is less than $M_{\rm Ch}$ its
subsequent evolution is thought to be similar to that of a white
dwarf. In 1967, an unusual object emitting a radio signal with period
$T = 1.377~{\rm s}$ was detected at the Mullard Radio Astronomy
Observatory. By its very nature the object was called ``pulsar.'' Only
one year later, Gold argued that pulsars are rotating neutron
stars~\cite{Gold}. He predicted an increase on the pulsar period
because of electromagnetic energy losses. The slow-down of the Crab
pulsar was indeed discovered in 1969~\cite{Boynton}.

If the mass of the neutron star is greater than $M_{\rm Ch}$, then the
star collapses under gravity, overcoming even the neutron exclusion
principle~\cite{Oppenheimer:1939ue}.  The star eventually collapses to
the point of zero volume and infinite density, creating what is known
as a ``singularity''~\cite{Penrose:1964wq,Hawking:1965mf,Hawking:1966sx,Hawking:1966jv,Hawking:1967ju,Hawking:1969sw}. As the density increases,
the paths of light rays emitted from the star are bent and eventually
wrapped irrevocably around the star. Any emitted photon is trapped
into an orbit by the intense gravitational field; it will never
leave it. Because no light escapes after the star reaches this
infinite density, it is called a black hole.

\section{Warping Spacetime}

A hunter is tracking a bear. Starting at his camp, he walks one mile
due south. Then the bear changes direction and the hunter follows it
due east. After one mile, the hunter loses the bear's track. He turns
north and walks for another mile, at which point he arrives back at
his camp. What was the color of the bear?

An odd question. Not only is the color of the bear unrelated to the
rest of the question, but how can the hunter walk south, east and
north, and then arrive back at his camp? This certainly does not work
everywhere on Earth, but it does if you start at the North pole.
Therefore the color of the bear has to be white. A surprising
observation is that the triangle described by the hunter's path has two right angles in
the two bottom corners, and so the sum of all three angles is
greater than $180^\circ$. This implies the metric space is curved.

What is meant by a curved space? Before answering this question, we recall that
our normal method of viewing the world is via Euclidean plane
geometry, where the line element of the $n$-dimensional space is given by
\begin{equation}
ds^2 = \sum_{i = 1}^n  dx_i^2 \, .
\end{equation}
Non-Euclidean geometries which involve curved spaces have
been independently imagined by Gauss~\cite{Gauss},
B\'olyai~\cite{Bolyai}, and Lobachevsky~\cite{Lobachevsky}. To
understand the idea of  a metric space herein we will greatly simplify the
discussion by considering only  2-dimensional surfaces. For
2-dimensional metric spaces, the so-called first and second fundamental
forms of differential geometry uniquely determine how to measure
lengths, areas and angles on a surface, and how to describe the shape
of a parameterized surface.

\subsection{2-dimensional metric spaces}

The parameterization of a surface maps points $(u, v)$ in the domain
to points $\vec \sigma (u,v)$ in space:
\begin{equation}
\vec \sigma (u,v) = \left( \begin{array}{c} 
x (u,v) \\  y(u,v) \\ z(u,v) \end{array} \right) \, .
\end{equation}
Differential geometry is the local analysis of how small changes in
position $(u, v)$ in the domain affect the position on the surface
$\vec \sigma (u, v)$, the first derivatives $\vec \sigma_u(u,v)$ and
$\vec \sigma_v(u,v)$, and the surface normal $\hat n(u,v)$.

The first derivatives, $\vec \sigma_u(u, v)$ and $\vec \sigma_v(u,
v)$, are vectors that span the tangent plane to the surface at point
$\vec \sigma (u, v)$. The surface normal at point
$\vec \sigma$
is defined as the unit vector normal to the tangent plane at point
$\vec \sigma$
and is computed using the cross product of the partial derivatives of
the surface parameterization,
\begin{equation}
\hat n (\vec \sigma) = \frac{\vec \sigma_u \times \vec \sigma_v}{|| \vec
  \sigma_u \times \vec \sigma_v ||} \, .
\end{equation}
The tangent vectors and the surface normal define an
orthogonal coordinate system at point $\vec \sigma (u, v)$ on the surface, which is the framework for describing the local shape of the surface.

Geometrically, $d\vec \sigma$ is a differential vector quantity that
is tangent to the surface in the direction defined by $du$ and $dv$.
The first fundamental form, I,  which measures the distance of neighboring points on the surface with parameters
$(u, v)$ and $(u + du, v + dv)$, is given by the inner
product of $d\vec \sigma$ with itself 
\begin{eqnarray}
{\rm I} & \equiv & ds^2  =   d \vec \sigma \cdot d \vec \sigma = (\vec \sigma_u du +
\vec \sigma_v dv) \cdot  (\vec \sigma_u du + \vec \sigma_v dv)
\nonumber \\
& = & (\vec \sigma_u \cdot \vec \sigma_u) du^2 + 2 (\vec \sigma_u
\cdot \vec \sigma_v) du dv + (\vec \sigma_v \cdot \vec \sigma_v) dv^2
\nonumber \\
& = & E du^2 + 2 F du dv + G dv^2 \,,
\end{eqnarray}
where 
$E$, $F$ and $G$ are the first fundamental coefficients. The
coefficients have some remarkable properties. For example, they can
be used to calculate the surface area. Namely, the area bounded by
four vertices $\vec \sigma (u,v), \vec \sigma (u + \delta u,v), \vec
\sigma (u, v + \delta v), \vec \sigma (u + \delta u , v + \delta v)$
can be expressed in terms of the first fundamental form with the
assistance of Lagrange identity
\begin{eqnarray}
\sum_{i=1}^{n-1} \sum_{j =
  i+1}^n (a_i b_j - a_j b_i)^2  & = &\left(\sum_{k=1}^n a_k^2\right)
\left(\sum_{k=1}^n b_k^2\right) \nonumber \\
& - & \left(\sum_{k=1}^n a_k b_k \right)^2 \,,  
\end{eqnarray}
which applies to any two sets $\{a_1, a_2, \cdots , a_n\}$ and $\{b_1, b_2, \cdots, b_n\}$ of real numbers.
The classical area element is found to be 
\begin{equation}
\delta A = |\vec \sigma_u \ \delta u \times \vec \sigma_v \ \delta v| =
\sqrt{EG - F^2} \ \delta u  \ \delta v \, ,
\end{equation}
or in differential form
\begin{equation}
dA =  \sqrt{EG - F^2} \ du \ dv \, .
\label{areafactor}
\end{equation}
Note that the expression under the square root in (\ref{areafactor}) is precisely $|\vec \sigma_u \times \vec \sigma_v|$ and so it is strictly positive at the regular points.

The key to the second fundamental form, II, is the unit normal vector.
The second fundamental form coefficients at a given point in the
parametric $uv$-plane are given by the projections of the second
partial derivatives of $\vec \sigma$ at that point onto the normal vector and
can be computed with the aid of the dot product as follows: $e = \vec
\sigma_{uu} \cdot \hat n$, $f = \vec \sigma_{uv} \cdot \hat n$, and $g
= \vec \sigma_{vv} \cdot \hat n$. The second fundamental form, 
\begin{equation}
{\rm II}  =  e \, du^2 + 2f \, du \, dv + g \, dv^2 \, ,
\end{equation} 
can be used to characterize the local shape of the folded surface. 

The concept of curvature, while intuitive for a plane curve (the
reciprocal of the radius of curvature), requires a more comprehensive
definition for a surface. Through a point on a surface any number of
curves may be drawn with each having a different curvature at the
point. We have seen that at any point on a surface we can find $\hat
n$ which is at right angles to the surface; planes containing the
normal vector are called normal planes. The intersection of a normal
plane and the surface will form a curve called a normal section and
the curvature of this curve is the normal curvature $\kappa$. For most points
on most surfaces, different sections will have different curvatures;
the  minimum and maximum values of these are called the principal
curvatures, denoted by $\kappa_1$ and $\kappa_2$. The Gaussian
curvature is defined by the product of the two principal curvatures $K
=\kappa_1 \kappa_2$. It may be calculated using the first and second
fundamental coefficients. At each grid point where these values are
known two matrices are defined. The matrix of the first fundamental
form,
\begin{equation}
{\rm I} = \left(\begin{array}{cc} E & F \\ F & G \end{array} \right),
\end{equation}
 and the matrix of the second fundamental form,
\begin{equation}
{\rm II} = \left(\begin{array}{cc} e & f \\ f & g \end{array} \right)
\, .
\end{equation}
The Gaussian curvature is given by
\begin{equation}
 K = \frac{{\rm det} \, {\rm II}}{{\rm det} \,  {\rm I}} \, .
\end{equation}
 
As an illustration, consier a half-cylinder of radius $R$ oriented along the $x$
axis. At a particular point on the surface, the scalar curvature can
have different values depending on direction. In the direction of the
half-cylinder's axis (parallel to the $x$ axis), the surface has zero scalar
curvature, $\kappa = 0$. This is the smallest curvature value at any point
on the surface, and therefore $\kappa_1$ is in this direction. For a
curve on the half-cylinder's surface parallel to the $(y,z)$ plane, the
cylinder has uniform scalar curvature. In fact this curvature is the
greatest possible on the surface, so that $\kappa_2 = 1/R$ is in this
direction. For a curve on the surface not in one of these directions,
the scalar curvature is greater than $\kappa_1$ and less than
$\kappa_2$. The Gaussian curvature is $K = 0$.

2-dimensional metric spaces can be classified according to the 
Gaussian curvature into elliptic ($K>0$), flat ($K=0$), and hyperbolic
($K<0$). Triangles which lie on the surface of an elliptic geometry
will have a sum of angles which is greater than $180^\circ$. Triangles
which lie on the surface of an hyperbolic geometry will have a sum of
angles which is less than $180^\circ$.\\

{\bf EXERCISE 5.1}~The unit sphere can be parametrized as
\begin{equation}
\vec \sigma (u,v) = \left( \begin{array}{c} \cos u \, \sin v\\ \sin u \, \sin v\\ \cos v \end{array} \right)
\label{unitS}
\end{equation}
where $(u,v) \in [0,\, 2\pi) \times [0,\, \pi]$. {\it (i)}~Find the
distance of neighboring points on the surface with parameters $(u, v)$
and $(u + du, v + dv)$, a.k.a. the line element $ds^2$. {\it
  (ii)}~Find the surface area. {\it  (iii)}~Find the Gaussian curvature. \\

{\bf EXERCISE 5.2} The tractrix is a curve with the following nice
interpretation: Suppose a dog-owner takes his pet along as he goes
for a walk ``down'' the $y$-axis. He starts from the origin, with his dog
initially standing on the $x$-axis at a distance $r$ away from the
owner. Then the tractrix is the path followed by the dog if he
``follows his owner unwillingly'', i.e., if he constantly pulls against
the leash, keeping it tight. This means mathematically that the leash
is always tangent to the path of the dog, so that the length of the
tangent segment from the tractrix to the $y$-axis has constant length
$r$. The tractrix has a well-known surface of
revolution called the pseudosphere which, for $r=1$, can be parametrized as
\begin{equation}
\vec \sigma (u,v) = \left( \begin{array}{c} {\rm sech}\;\! u \, \cos
    v\\ {\rm sech}\;\! u \, \sin v\\ u - {\rm tanh} \;\! u \end{array}
\right) \,,
\label{pseudosphere}
\end{equation}
with $u \in (-\infty, \infty)$ and $v \in [0, 2\pi)$.  {\it (i)}~Find the
line element. {\it (ii)}~Find the surface area. {\it
  (iii)}~Find the Gaussian curvature.  \\

A curve $\gamma$ with parametr $t$ on a surface $\vec \sigma (u,v)$ is
called a {\it geodesic} if at every point $\gamma (t)$ the
acceleration vector $\ddot {\vec \gamma} (t)$ is either
zero or parallel to its unit normal $\hat n$. \\

{\bf EXERCISE 5.3} Show that a geodesic $\gamma (t)$ on a surface
$\vec \sigma$ has constant speed.\\

{\bf EXERCISE 5.4} A curve $\gamma$ on a surface $\vec \sigma$ is a geodesic if
and only if for any part $\gamma (t) = \vec \sigma (u(t), v(t))$
contained in a surface patch $\vec \sigma$, the following two
equations are satisfied:
\begin{equation}
\frac{d}{dt} (E \dot u + F \dot v) = \frac{1}{2} (E_u \dot u^2 + 2 F_u
\dot u \dot v + G_u \dot v^2) \,,
\label{geodesic1}
\end{equation}
\begin{equation}
\frac{d}{dt} (F \dot u + G \dot v) = \frac{1}{2} (E_v \dot u^2 + 2 F_v
\dot u \dot v + G_v \dot v^2) \,,
\label{geodesic2}
\end{equation}
where $E du^2 + 2 F du dv + G dv^2$ is the first fundamental form of
$\vec \sigma$. (\ref{geodesic1}) and (\ref{geodesic2}) are called the
geodesic equations. They are nonlinear and solvable analytically on
rare occasions only.\\

{\bf EXERCISE 5.5}~Show that if $\gamma$ is a geodesic on the unit
sphere $S^2$, then $\gamma$ is part of a great circle.  Consider the patch
under the parametrization 
\begin{equation}
\vec \sigma (\theta, \phi) = \left(\begin{array}{c} \cos \theta
\cos \phi \\ \cos \theta \sin
\phi \\ \sin \theta \end{array} \right).
\label{esfera}
\end{equation}
[{\it Hint:} A great circle (a.k.a. orthodrome) of a sphere is the intersection of the sphere and a plane which passes through the center point of the sphere.]\\

The scalar curvature (or  Ricci scalar) is the simplest curvature
invariant of an $n$-dimensional hypersurface. To each point on the
hypersurface, it assigns a single real number determined by the
intrinsic geometry of the hypersurface near that point.  It provides
one way of measuring the degree to which the geometry determined by a
given metric might differ from that of ordinary Euclidean
$n$-space. In two dimensions, the scalar
curvature is twice the Gaussian curvature, $R = 2K$, and completely
characterizes the curvature of a surface. In more than two dimensions,
however, the curvature of hypersurfaces involves more than one
functionally independent quantity.

\subsection{Schwarzschild metric}

Consider a freely falling spacecraft in the gravitational field of a
radially symmetric mass distribution with total mass $M$. Because the
spacecraft is freely falling, no effects of gravity are felt inside.
Then, the spacetime coordinates from $r \to \infty$ should be valid
inside the spacecraft. Let us call these coordinates $\vec
\Sigma_\infty(t_\infty,x_\infty,y_\infty,z_\infty)$, with $x_\infty$
parallel and $y_\infty,
z_\infty$ transversal to movement. The spacecraft has velocity
$v$ at the distance $r$ from the mass $M$, measured in the coordinate
system $\vec \Sigma=(t,r,\theta,\phi)$ in which the mass $M$ is at rest
at $r=0$. As long as the {\it gravitational field is weak},  to first order approximation that the laws of special
relativity hold~\cite{Einstein:1905ve}, and we can use a Lorentz
transformation~\cite{Lorentz:1904}  to relate $\vec \Sigma$ at rest and $\vec
\Sigma_\infty$ moving with $v = \beta c$. We will define shortly what
``weak'' means in this context. For the moment, we presume that
effects of gravity are small if the velocity of the spacecraft, which
was at rest a $r \to \infty$, is still small $v \ll c$. Should this be 
the case, we have
\begin{eqnarray}
dt_\infty & = & dt \ \sqrt{1 - \beta^2} \nonumber \\
dx_\infty & = & \frac{dr}{\sqrt{1-\beta^2}} \nonumber \\
 dy_\infty & = & r \ d\theta \nonumber \\
dz_\infty & = & r \ \sin \theta \ d \phi \, .
\end{eqnarray}
The infinitesimal distance between two spacetime events is given by
the Minkowskian line element~\cite{Minkowski}
\begin{equation}
ds^2 = g_{\mu \nu} dx^\mu dx^\nu = c^2 dt_{\infty}^2 - dx_\infty^2 -
dy_\infty^2 - dz_\infty^2 \, ,
\end{equation}
which, for the case at hand, becomes
\begin{equation}
ds^2 = (1 -\beta^2) c^2 dt^2 - \frac{dr^2}{1-\beta^2} + r^2 (d\theta^2 + \sin^2
d\phi^2) \, .
\label{SchMink}
\end{equation}
Herein we follow the notation of~\cite{Anchordoqui:2015xca}:
Greek indices $(\mu, \nu, \cdots)$ run from 0 to 3 and Latin indices
$(i,j,\cdots)$ from 1 to 3.

We now turn to determine $\beta$ from measurable quantities
of the system: $M$ and $r$. Consider the energy of the spacecraft with
rest mass $m$,
\begin{equation}
(\gamma - 1) mc^2 - \frac{G \gamma m M}{r} = 0 \,,
\end{equation}
where the first term is the kinetic energy and the second the
Newtonian expression for the potential energy. Note that here we have made  the
crucial assumption that gravity couples not only to the mass of the
spacecraft but also to its total energy. Dividing by $\gamma mc^2$ gives
\begin{equation}
\left(1 - \frac{1}{\gamma} \right) - \frac{GM}{rc^2} = 0 \, .
\label{Sch1}
\end{equation}
Introducing $\alpha = GM/c^2$ we can re-write (\ref{Sch1}) as
\begin{equation}
\sqrt{1 - \beta^2} = 1 - \frac{\alpha}{r} \,,
\label{Sch2}
\end{equation}
where  $\gamma = (1 -\beta^2)^{-1/2}$.  (\ref{Sch2}) leads to 
\begin{equation}
1 - \beta^2 = 1 - \frac{2 \alpha}{r} + \frac{\alpha^2}{r^2} \approx 1
- \frac{2 \alpha}{r} \, ;
\end{equation}
in the last step, we neglected the term $(\alpha/r)^2$, since we
attempt only at an approximation for large distances, where gravity is
still weak.  Inserting this expression into (\ref{SchMink}), we obtain the
metric describing the gravitational field produced by a radially 
symmetric mass distribution,
\begin{equation}
ds^2 = \left(1 - \frac{2 \alpha}{r}\right)  c^2 dt^2 - \left(1 - \frac{2 \alpha}{r}\right)^{-1} dr^2
 - r^2 d\Omega^2 \, ,
\label{Sch-metric}
\end{equation} 
where $d\Omega^2 = d \theta^2 + \sin^2 \theta d\phi^2$.  Wickedly,
this agrees with the exact result found by
Schwarzschild~\cite{Schwarzschild:1916uq} by solving Einstein's vacuum
field equations of general relativity~\cite{Einstein:1916vd}.

As in special relativity, the line element $ds^2$ determines the time
and spatial distance between two spacetime events. The time measured
by an observer in the instantaneous rest frame, known as the proper
time $d\tau$, is given by $d\tau =
ds/c$~\cite{Anchordoqui:2015xca}. In particular, the time difference
between two events at the same point is obtained by setting $dx^i =
0$. If we choose two static observers at the position $r$ and $r'$,
then we find with $dr = d\phi = d\theta = 0$,
\begin{equation}
\frac{d \tau(r)}{d\tau(r')} = \frac{\sqrt{g_{00}(r)} \, dt}{\sqrt{g_{00}
  (r')} \, dt} = \sqrt{\frac{g_{00}(r)}{g_{00}
  (r')}} \, .
\end{equation}
The time intervals $d\tau(r')$ and $d\tau (r)$ are different and thus
the time measured by clocks at different distances $r$ from the mass
$M$ will differ too. In particular, the time $\tau_\infty$ measured by
an observer at infinity will pass faster than the time experienced in
a gravitational field,
\begin{equation}
\tau_\infty = \frac{\tau(r)}{\sqrt{1 - 2 \alpha/r}} < \tau (r) \, .
\end{equation}
Since frequencies are inversely proportional to time, the frequency or
energy of a photon traveling from $r$ to $r'$ will be affected by the
gravitational field as 
\begin{equation}
\frac{\nu(r')}{\nu(r)} = \sqrt{\frac{1 - 2\alpha /r}{1 - 2 \alpha r'}}
\, .
\end{equation}
Therefore, an observer at $r' \to \infty$ will receive photons, which were
emitted with frquency $\nu$ by a
source at position $r$, redhsifted to 
frequency $\nu_\infty$, 
\begin{equation}
\nu_\infty = \sqrt{1 - \frac{2GM}{r c^2}} \ \nu(r) \, .
\end{equation}
Note that the photon frequency  is
redshifted by the gravitational field.  The size of this effect is of order
$\Phi/c^2$, where $\Phi = -GM/r$ is the Newtonian gravitational potential. We are
now in position to specify more precisely what weak gravitational
fields means. As long as $|\Phi|/c^2 \ll 1$, the deviation of 
\begin{equation}
g_{00} = 1 - \frac{2GM}{rc^2} \approx 1 - 2 \frac{\Phi(r)}{c^2}
\end{equation} 
from the Minkowski  value $g_{00} = 1$ is small, and
Newtonian gravity is a sufficient approximation.

What is the meaning of  $r = 2\alpha?$ At
\begin{equation}
R_{\rm Sch} = \frac{2GM}{c^2} = 3~{\rm km} \frac{M}{M_\odot} \,,
\label{Sch3km}
\end{equation}
the Schwarzschild coordinate system (\ref{Sch-metric}) becomes
ill-defined. However, this does not mean necessarily that at $r =
R_{\rm Sch}$ physical quantities like tidal forces become infinite. As
a matter of fact, all scalar invariants are finite, e.g.  $R=0$ and
$\mathfrak{K} = 12 R_{\rm Sch}/r^6$.  Here $R$ is the Ricci scalar and
$\mathfrak{K}$ the Kretschmann scalar~\cite{Kretschmann}, a quadratic
scalar invariant used to find the true singularities of a
spacetime. The Schwarzschild's scalar invariants can only be found by
long and troublesome calculation that is beyond the scope of this
course; for a comprehensive discussion see
e.g.~\cite{Weinberg:1972kfs,Misner:1974qy}. Before proceeding we
emphasize again that, whether or not the singularity is moved to the
origin, only depends on the coordinate frame used, and has no physical
significance whatsoever; see Appendix~\ref{appC} for an example.

If the gravitating mass is concentrated inside a radius smaller than
$R_{\rm Sch}$ then we cannot obtain any information about what is
going on inside $R_{\rm Sch}$, and we say $r = R_{\rm Sch}$ defines an
event horizon. An object smaller than its Schwarzschild radius, is
called a black hole.  In Newtonian gravity, only the enclosed mass
$M(r)$ of a spherically symmetric system contributes to the
gravitational potential outside $r$. Therefore, we conclude the Sun is
not a black hole, becasue for all values of $r$ the enclosed mass is
$M(r) < rc^2/(2G)$. The Schwarzschild black hole is fully characterized
by its mass $M$. To understand this better, we consider next what
happens to a photon crossing the event horizon as seen from an
observer at $r \to \infty$.

Light rays are characterized by $ds^2 = 0$. Consider a light ray
traveling in the 
radial direction, that is to say $d\phi  = d\theta = 0$. The Schwarzschild metric (\ref{Sch-metric}) becomes
\begin{equation}
\frac{dr}{dt} = \left(1 - \frac{2\alpha}{r} \right) c \, .
\end{equation}
As seen from far away a light ray approaching a massive star will
travel slower and slower as it comes closer to the Schwarzschild
radius. In fact, for an observer at infinity the signal will reach $r
= R_{\rm Sch}$ only asymptotically, for $t \to \infty$. Similarly, the
communication with a freely falling spacecraft becomes impossible as
it reaches $r = R_{\rm Sch}$. A more detailed analysis shows that
indeed, as seen from infinity, no signal can cross the surface at $r =
R_{\rm Sch}$.  The factors $(1 - 2\alpha/r)$ in (\ref{Sch-metric}) control the bending of
light, a phenomenon known as gravitational
lensing.  The first observation of light deflection was performed by
noting the change in position of stars as they passed near the Sun on
the celestial sphere. The observations were performed in May 1919
during a total solar eclipse, so that the stars near the Sun (at that
time in the constellation Taurus) could be observed~\cite{Dyson}.\\

{\bf EXERCISE 5.6}~In addition to the time dilation due to an object
moving at a finite speed that we have learned about in special
relativity, we have seen that there is an effect in general relativity,
termed ``gravitational redshift,'' caused by gravity itself. To
understand this latter effect, consider a photon escaping from the
Earth's surface to infinity. It loses energy as it climbs out of the
Earth's gravitational well. As its energy $E$ is related to its
frequency $\nu$ by Planck's formula $E = h \nu$, its frequency must
therefore also be reduced, so observers at a great distance $r \to
\infty$ must see clocks on the surface ticking at a lower frequency as
well. Therefore an astronaut orbiting the Earth ages differently from
an astronomer sitting still far from the Earth for two reasons; the
effect of gravity, and the time dilation due to motion. In this
problem, you will calculate both these effects, and determine their
relative importance.  {\it (i)}~The escape speed from an
object of mass $M$ if you are a distance $r$ from it is given by
\begin{equation}
v_{\rm escape} = \sqrt{\frac{2GM}{r}} \, .
\end{equation}
That is, if you are moving this fast, you will not fall back to the
object, but will escape ￼its gravitational field
entirely. Schwarzschild's solution to Einstein's field equations of
general relativity shows that a stationary, non-moving clock at a
radius $r \geq R_\oplus$ from the Earth will tick at a rate that is
\begin{equation}
\sqrt{1 - \frac{1}{c^2} \frac{2 GM_\oplus}{r} } = \sqrt{1 -
  \frac{v^2_{\rm escape}}{c^2}} 
\label{Sch-escape}
\end{equation}
times as fast as one located far away from the Earth (i.e. at $r \to
\infty$). Note how much this expression looks like the equivalent
expression from special relativity for time dilation. Here $R_\oplus$
is the radius of the Earth, and $M_\oplus$ is its mass. Using
(\ref{Sch-escape}) calculate the rate at which a stationary clock at a
radius $r$ (for $r > R_\oplus$) will tick relative to one at the
surface of the Earth. Is your rate greater or less than 1? If greater
than 1, this means the high altitude clock at $r > R_\oplus$ ticks
faster than one on the surface; if less than one, this means the high
altitude clock ticks slower than one on the surface. {\it (ii)}~Now
consider an astronaut orbiting at $r > R_\oplus$. What is her orbital
velocity as a function of $r$? Because she is moving with respect to a
stationary observer at radius $r$, special relativity says that her
clock is ticking slower. Calculate the ratio of the rate her clock
ticks to that of a stationary observer at radius $r$. (Note that for
circular motion, the acceleration in the spaceship travelling in a
circle is not zero, so the spaceship is not in a single frame of
inertia.) {\it (iii)}~Determine an expression for the ratio of the
rate at which the orbiting astronaut's clock ticks to a stationary
clock on the surface of the Earth, as a function of the radius $r$ at
which she orbits. You may ignore the small velocity of the clock on
the surface of the Earth due to the Earth's rotation. {\it (iv)}~Using
$\sqrt{1 - x} \approx 1 -  x/2 + \cdots$, $(1-x)^{-1} \approx
1 + x + \cdots$, and $(1-x) (1 -y) = 1 - x - y + xy \approx 1 -( x+y)$,
all valid for $x\ll 1$ and $y\ll1$, derive an expression of the form $1
- \delta$  for the relative rate of a clicking clock on the surface of
the Earth and the orbiting astronaut. Demonstrate that $\delta \ll
1$. {\it (v)}~Calculate the radius $r$ at which the clock of the
orbiting astronaut ticks at the same rate as a stationary one on the
surface of the Earth; express your result in Earth radii and
kilometers. Will an astronaut orbiting at a smaller radius age more or
less than one who stayed home? Thus, do astronauts on the Space
Shuttle (orbiting 300~km above the Earth's surface) age more or less
than one staying home?\\

{\bf EXERCISE 5.7}~``A full set of rules [of Brockian Ultra Cricket,
as played in the higher dimensions] is so massively complicated that
the only time they were all bound together in a single volume they
underwent gravitational collapse and became a black hole''~\cite{Adams:1982hh}.
 A quote like this is crying out for a calculation. In this
problem, we will answer Adams challenge, and determine just how
complicated these rules actually are.  An object will collapse into a
black hole when its radius is equal to the radius of a black hole of
the same mass; under these conditions, the escape speed at its surface
is the speed of light (which is in fact the defining characteristic of
a black hole). We can rephrase the above to say that an object will
collapse into a black hole when its density is equal to the density of
a black hole of the same mass. {\it (i)}~Derive an expression for the
density of a black hole of mass $M$. Treat the volume of the black
hole as the volume of a sphere of radius given by the Schwarzschild
radius. As the mass of a black hole gets larger, does the density grow
or shrink? {\it (ii)}~Determine the density of the paper making up the
Cricket rule book, in units of kilograms per cubic meter. Standard
paper has a surface density of 75~g per square meter, and a
thickness of 0.1~mm. {\it (iii)}~Calculate the mass (in solar
masses), and radius (in AU) of the black hole with density equal to
that of paper. {\it (iv})~How many pages long is the Brockian Ultra
Cricket rule book? Assume the pages are standard size ($8.5'' \times
11''$). For calculational simplicity, treat the book as spherical (a
common approximation in this kind of problem). What if the rule book
were even longer than you have just calculated? Would it still
collapse into a black hole? \\

{\bf EXERCISE 5.8}~Black holes provide the ultimate laboratory for
studying strong-field gravitational physics. The tides near black
holes can be so extreme that a process informally called
``spaghettification'' occurs in which a body falling towards a black
hole is strongly stretched due to the difference in gravitational
force at different locations along the body (this is called a tidal
effect). In the following, imagine that you are falling into a $3
M_\odot$ black hole. {\it (i)}~What is the Schwarzschild radius of
this black hole (in km)? {\it (ii)}~You are 1.5~m tall and 70~kg in
mass and are falling feet first. At what distance from the black hole
would the gravitational force on your feet exceed the gravitational
force on your head by 10~kN? Express this distance in km and in
Schwarzschild radii of the black hole. {\it (iii)}~To appreciate if
this amount of force is enough to ``spaghettify'' and kill you,
imagine that you are suspended from a ceiling of your room (on Earth)
with a steel plate tied to your feet. Calculate the mass of the plate
(in kg) that will give you a nice tug of 10~kN (you can ignore the
weight of your body here). Do you think this pull will kill you? {\it
  (iv)}~Now consider a trip toward the supermassive black hole at the
center of our Galaxy, which has an estimated mass of 4 million
$M_\odot$. How does this change the distance at which you will be
``spaghettified'' by the differential gravity force of 10~kN? Express
your answer in km and in Schwarzschild radii of the black
hole. {\it(v)}~Find the smallest mass of the black hole for which you
would not die by ``spaghettification'' before falling within its event
horizon.\\

{\bf EXERCISE 5.9}~In the Schwarzschild metric $r$ is a comoving
coordinate, not a real physical distance. Rather, integrals over $ds$
constitute physical distances. In the following we will take slices of
spacetime at a constant time ($dt = 0$). {\it (i)}~Compute the
physical circumference, $C$, at a given coordinate distance $R$ from
the center of a black hole of mass $M$ at $\theta = \pi/2$. {\it
  (ii)}~Compute the physical distance $R_{\rm phys}$ from the center of the black hole out to
the coordinate distance $R$ (assume $R > R_{\rm Sch}$ and take the absolute value of $g_{rr}$).
[{\it Hint:} The following facts may be helpfull: 
\begin{equation}
\int_0^1 \sqrt{\frac{\xi}{1-\xi}} d \xi = \frac{\pi}{2} \,,
\label{SCHint1}
\end{equation}
and
\begin{equation}
\int_1^\alpha \sqrt{\frac{\xi}{1 -\xi}} d \xi = \ln(\sqrt{\alpha -1} +
  \sqrt{\alpha} )+ \sqrt{\alpha -1} \sqrt{\alpha} \,,
\label{SCHint2}
\end{equation}
where $\alpha >1$ is constant.] {\it (iii)}~Now use your answers to
part {\it (i)} and part {\it (ii)} to compute $\Pi$ where $C = 2 \Pi
R_{\rm phys}$. {\it (iv)}~Plot $\Pi$ as a function of $\xi \equiv
R/R_{\rm Sch}$ for $\xi \in  [1, 10^3]$ (use log axes for the x
axis). What happens with $\Pi$ as $\xi \to \infty$?

\subsection{Eddington luminosity and black hole growth}

Binary X-ray sources are places to find strong black hole
candidates~\cite{Cameron,Wilson}.  A companion star is a perfect
source of infalling material for a black hole. As the matter falls or
is pulled towards the black hole, it gains kinetic energy, heats up
and is squeezed by tidal forces. The heating ionizes the atoms, and
when the atoms reach a few million degrees Kelvin, they emit
X-rays. The X-rays are sent off into space before the matter crosses
the event horizon, and so we can detect this X-ray emission.  Another sign of the
presence of a black hole is random variation of emitted X-rays.  The
infalling matter that emits X-rays does not fall into the black hole
at a steady rate, but rather more sporadically, which causes an
observable variation in X-ray intensity. Additionally, if the X-ray
source is in a binary system, the X-rays will be periodically cut off
as the source is eclipsed by the companion star.  

Cygnus X-1 is one of the strongest X-ray sources we can detect from
Earth~\cite{Giacconi} and the first widely thought to be a black hole,
after the detection of its rapid X-ray variability~\cite{Oda} and the
identification of its optical countempart with the blue supergiant
star HDE 226868~\cite{Webster,Bolton}.  The $X$-ray emission is powered
mainly by accretion from the strong stellar wind from HDE 226868~\cite{Petterson}.
While the disk of accreting matter is incredibly bright on its own,
Cygnus X-1 has another source of light: a pair of jets perpendicular
to the disk erupt from the black hole carrying part of the infalling
material away into  the interstellar space \cite{Stirling:2001xb}.

Consider a steady spherically symmetrical accretion. We assume the
accreting material to be mainly hydrogen and to be fully
ionized. Under these circumstances, the radiation exerts a force
mainly on the free electrons through Thomson scattering, since the
scattering cross section for protons is a factor $(m_e/m_p)^2$
smaller, where $m_e/m_p = 5 \times 10^{-4}$ is the ratio of the
electron and proton masses~\cite{Thomson}. If $F$ is the radiant energy flux (${\rm
  erg} \, {\rm s}^{-1} {\rm cm}^{-2}$) and $\sigma_T = 6.7 \times
10^{-25}~{\rm cm}^2$ is the Thomson cross section, then the outward
radial force on each electron equals the rate at which it absorbs
momentum,
\begin{equation}
F_{\rm
  out} = \frac{\sigma_T F}{c} \, . 
\end{equation}
The attractive electrostatic Coulomb force between the electrons and
protons means that as they move out the electrons drag the protons
with them. In effect, the radiation pushes out electron-proton pairs
against the total gravitational force 
\begin{equation}
F_{\rm in} = \frac{GM}{r^2} (m_p + m_e)
\end{equation}
 acting on each pair at a radial distance $r$ from the center. If the luminosity
of the accreting source is $L$ (${\rm erg} \, {\rm s}^{-1}$), we have   
\begin{equation}
F = \frac{L}{4\pi r^2}
\end{equation}
 by spherical
symmetry, so the net inward force on an electron-proton pair is
\begin{equation}
F_{\rm net} = \left( GM m_p - \frac{L \sigma_T}{4 \pi c} \right)
  \frac{1}{r^2} \, .
\end{equation}
There is a limiting luminosity for which this expression vanishes,
called the Eddington limit~\cite{Eddington}
\begin{equation}
L_{\rm Edd} = \frac{4 \pi G M m_p}{\sigma_T} \simeq 1.3 \times 10^{38}
\left(\frac{M}{M_\odot} \right)~{\rm erg} \ {\rm s}^{-1} \, .
\end{equation}
At greater luminosities the outward pressure of radiation would exceed the inward gravitational attraction and accretion would be halted.

 \begin{figure*}
 \postscript{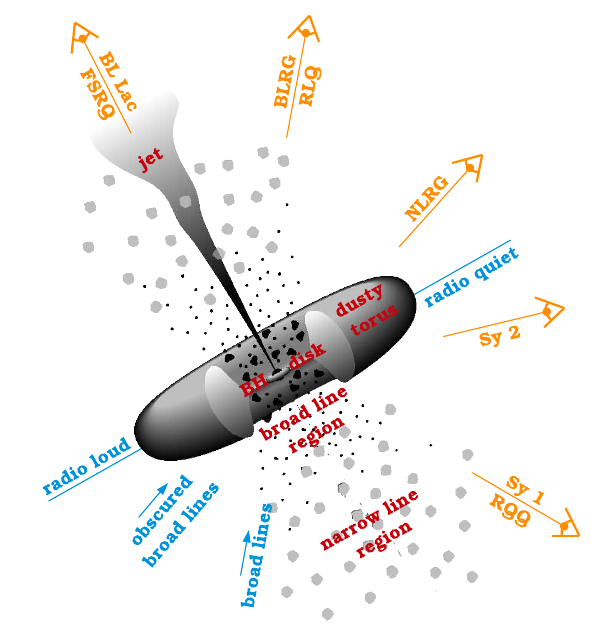}{0.8}
 \caption{Unification scheme of AGN. The acronyms for the different
   sub-classes of AGN are as follows: Fanaroff-Riley radio galaxies
   (FR I/II), narrow line radio galaxy (NLRG), broad line radio galaxy
   (BLRG), radio-loud quasar (RLQ), radio quiet quasar (RQQ), flat
   spectrum radio quasar (FSRQ), and Sefeyrt galaxies (Sy 1/2)~\cite{Biteau:2013nua}.}
\label{fig:AGN}
\end{figure*}

Active galactic nuclei (AGNs) are galaxies that harbor compact masses
at the center exhibiting intense non-thermal emission that is often
variable, which indicates small sizes (light months to light years).
The luminosity of an accreting black hole is proportional to the rate
at which it is gaining mass.  Under favorable conditions, the
accretion leads to the formation of a highly relativistic collimated
jet. The formation of the jet is not well constrained, but it is
thought to change from magnetic-field-dominated near the central
engine to particle (electron and positron, or ions and electrons)
dominated beyond pc distances. The AGN taxonomy, controlled by the
dichotomy between radio-quiet and radio-loud classes, is represented
in Fig.~\ref{fig:AGN}. The appearance of an AGN depends crucially on
the orientation of the observer with respect to the symmetry axis of
the accretion disk~\cite{Urry:1995mg}. In this scheme, the difference
between radio-loud and radio-quiet AGN depends on the presence or
absence of radio-emitting jets powered by the central nucleus, which
in turn may be speculated to depend on: {\it (i)}~black hole rotation;
{\it(ii)}~low
power or high power, as determined by the mass-accretion rate $\dot M
c^2/L_{\rm Edd}$~\cite{Dermer:2016jmw}.\\

\begin{figure}
 \postscript{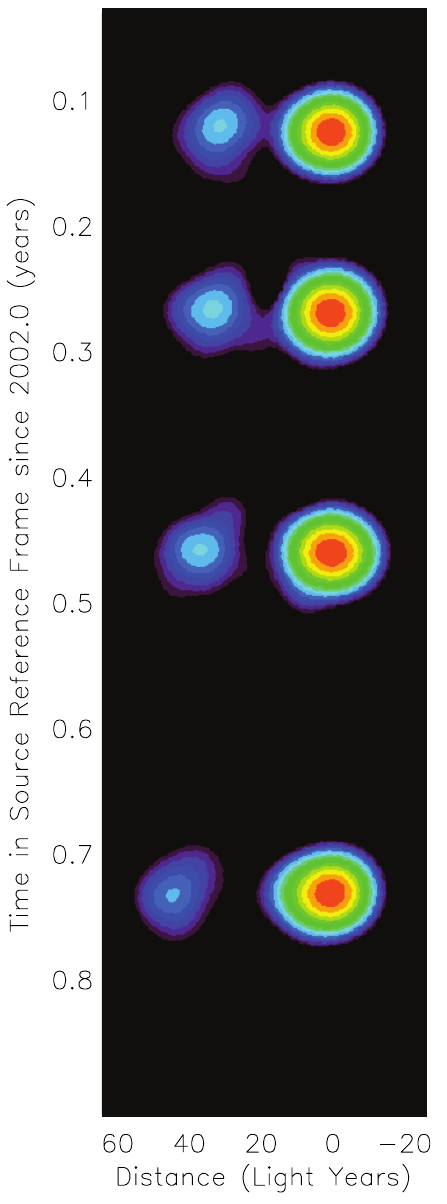}{0.8}
\caption{Mosaic of images of 0827+243 at 22 GHz~\cite{Piner:2005px}.}
\label{quasar}
\end{figure}

{\bf EXERCISE 5.10}~The pictures in Fig.~\ref{quasar} show  a time sequence of radio
observations of the quasar 0827+243. The core of the quasar is the
bright object at a distance of 0~ly and a fainter blob of plasma is
moving away from it. {\it (i)}~What is the apparent velocity of the
motion of the plasma blob? {\it (ii)}~Derive the apparent transverse
velocity of an object ejected from a source at velocity $v$ at an angle
$\theta$ with respect to the line of sight between the source and the
observer. {\it (iii)}~Which angle maximizes the apparent transverse velocity? What is accordingly the minimal
Lorentz-factor of the plasma blob observed in 0827+243?

\section{Expansion of the Universe}

The observations that we will discuss in this section reveal that the
universe is in a state of violent explosion, in which the galaxies
are rushing appart at speeds approaching the speed of light. Moreover,
we can extrapolate this explosion backwards in time and conclude that
all the galaxies must have been much closer at the same time in the
past -- so close, in fact, that neither galaxies nor stars nor even
atoms or atomic nuclei could have had a separate existence.

\subsection{Hubble's law}

The XVI century finally saw what came to be a watershed in the
development of Cosmology. In 1543 Copernicus published his
treatise ``De Revolutionibus Orbium Celestium'' (The Revolution of
Celestial Spheres) where a new view of the world is presented: the
heliocentric model~\cite{Copernicus}. 

It is hard to underestimate the importance of this work: it challenged
the age long views of the way the universe worked and the
preponderance of the Earth and, by extension, of human beings.  The
realization that we, our planet, and indeed our solar system (and even
our Galaxy) are quite common in the heavens and reproduced by myriads
of planetary systems, provided a sobering (though unsettling) view of
the universe. All the reassurances of the cosmology of the Middle Ages
were gone, and a new view of the world, less secure and comfortable,
came into being. Despite these ``problems'' and the many critics the
model attracted, the system was soon accepted by the best minds of the
time such as Galileo.

The simplest and most ancient of all astronomical observations is that
the sky grows dark when the Sun goes down. This fact was first noted
by  Kepler, who, in the XVII century, used it as evidence for
a finite universe. In the XIX century, when the idea of an unending,
unchanging space filled with stars like the Sun was widspread in
consequence of the Copernican revolution, the question of the dark
night sky became a problem. To clearly ascertain this problem, we recall
that if absorption is neglected, the aparent luminosity of a star of
absolute luminosity $L$ at a distance $r$ will be $b=L/4\pi r^2.$ If
the number density of such stars is a constant $n$, then the number of
stars at distances $r$ between $r$ and $r+ dr$ is $dN = 4 \pi n r^2
dr$, so the total radiant energy density due to all stars is
\begin{eqnarray}
\rho_{\rm s} & = & \int b \,\, dN = \int_{\! 0}^\infty 
\left(\frac{L}{4 \pi r^2} \right) \, 4\pi\, n\,r^2 dr \nonumber \\ 
 & = &  L n \int_{\! 0}^\infty dr \, .
\label{olbersp}
\end{eqnarray}
The integral diverges, leading to an infinite energy density of starlight!

In order to avoid this paradox, both de Ch\'eseaux
(1744)~\cite{Cheseaux} and  Olbers (1826)~\cite{Olbers}
postulated the existence of an interstellar medium that absorbs the
light from very distant stars responsible for the divergence of the
integral in (\ref{olbersp}).  However, this resolution of the
paradox is unsatisfactory, because in an eternal universe the
temperature of the interstellar medium would have to rise until the
medium was in thermal equilibrium with the starlight, in which case it
would be emitting as much energy as it absorbs, and hence could not
reduce the average radiant energy density. The stars themselves are of
course opaque, and totally block out the light from sufficiently
distant sources, but if this is the resolution of the so-called
``Olbers paradox'' then every line of segment must terminate at the
surface of a star, so the whole sky should have a temperature equal to
that at the surface of a typical star.\\

{\bf EXERCISE 6.1}~{\it (i)}~In a forest there are $n$ trees per
hectare, evenly spaced.  The thickness of each trunk is $D$. What is
the mean distance that you have an unobstructed view into the woods,
i.e. the mean free path? {\it (ii)}~How is this related to the Olbers
paradox?

In 1929, Hubble discovered that the spectral lines of galaxies were
shifted towards the red by an amount proportional to their
distances~\cite{Hubble}.  If the redshift is due to the Doppler
effect, this means that the galaxies move away from each other with
velocities proportional to their separations.  The importance of this
observation is that it is just what we should predict according to the
simplest possible picture of the flow of matter in an expanding
universe.

\begin{figure}
 \postscript{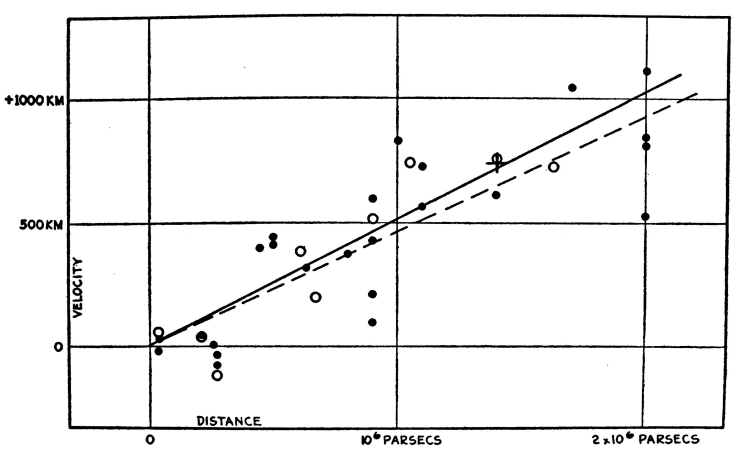}{0.9}
\caption{Hubble's original plot of the relation between redshift
  (vertical axis) and distance (horizontal axis). Note that in the
  vertical axis he actually plots $cz$ rather than $z$, and that the units are accidentally written as km rather than km/s~\cite{Hubble}.}
\label{Hubble-old}
\end{figure}

\begin{figure}
 \postscript{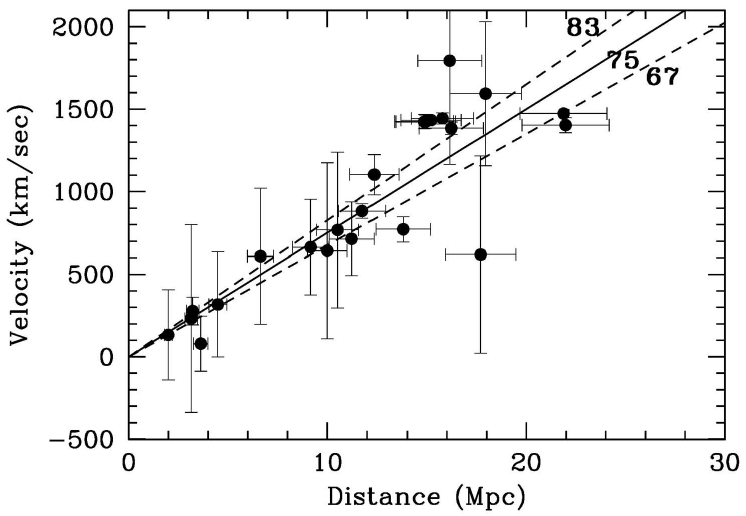}{0.9}
\caption{A more modern version of Hubble's plot, showing $cz$ versus
  distance. In this case, the galaxy distances have been determined
  using Cepheid variable stars as standard candles~\cite{Freedman:2000cf}.}
\label{Hubble-new}
\end{figure}

The {\it redshift} parameter is defined as the traditional shift in
wavelength of a photon emitted by a distant galaxy at time $t_{\rm
  em}$ and observed on Earth today
\begin{equation}
z = \frac{\lambda_{\rm obs}}{\lambda_{\rm em}} - 1 = \frac{\nu_{\rm
    em}}{\nu_{\rm obs}} - 1 , 
\label{zredshit}
\end{equation}
Although measuring a galaxy's redshift is relatively easy, and can be done with high
precision, measuring its distance is difficult. Hubble knew $z$ for
nearly 50 galaxies, but had estimated distances for only 20 of
them. Nevertheless, from a plot of redshift versus distance
(reproduced in Fig.~\ref{Hubble-old}) he found the famous linear
relation now known as the Hubble's law:
\begin{equation}
z = \frac{H_0}{c} r \,,
\end{equation}
where $H_0$ is a constant (now called the Hubble constant).
Since in the study of Hubble all the redshift were small, $z < 0.04$,
he was able to use the classical non-relativistic realtion for small
velocities $(v \ll c)$. From (\ref{doppler-shift}) the Doppler redshift 
is $z \approx v/c$ and Hubble's law takes the form
\begin{equation}
v = H_0 \,  r \, . 
\end{equation}
Since the Hubble constant $H_0$ can be found by dividing velocity by
distance, it is customarily written in the rather baroque units of
${\rm km} \,  {\rm s}^{-1} \, {\rm  Mpc}^{-1}$. From
Fig.~\ref{Hubble-old} it follows that $H_0 = 500~{\rm km} \,  {\rm
  s}^{-1} \, {\rm  Mpc}^{-1}$. However, it turned out that Hubble was severely underestimating the distances to galaxies.
In Fig.~\ref{Hubble-new} we show a more recent determination of the Hubble constant
from nearby galaxies, using HST data~\cite{Freedman:2000cf}. By combining results of different research
groups, the present day Hubble expansion rate is $H_0 = 70^{+5}_{-3}~{\rm km}\, 
{\rm s}^{-1}\, {\rm Mpc}^{-1}$.\\

\begin{figure}
 \postscript{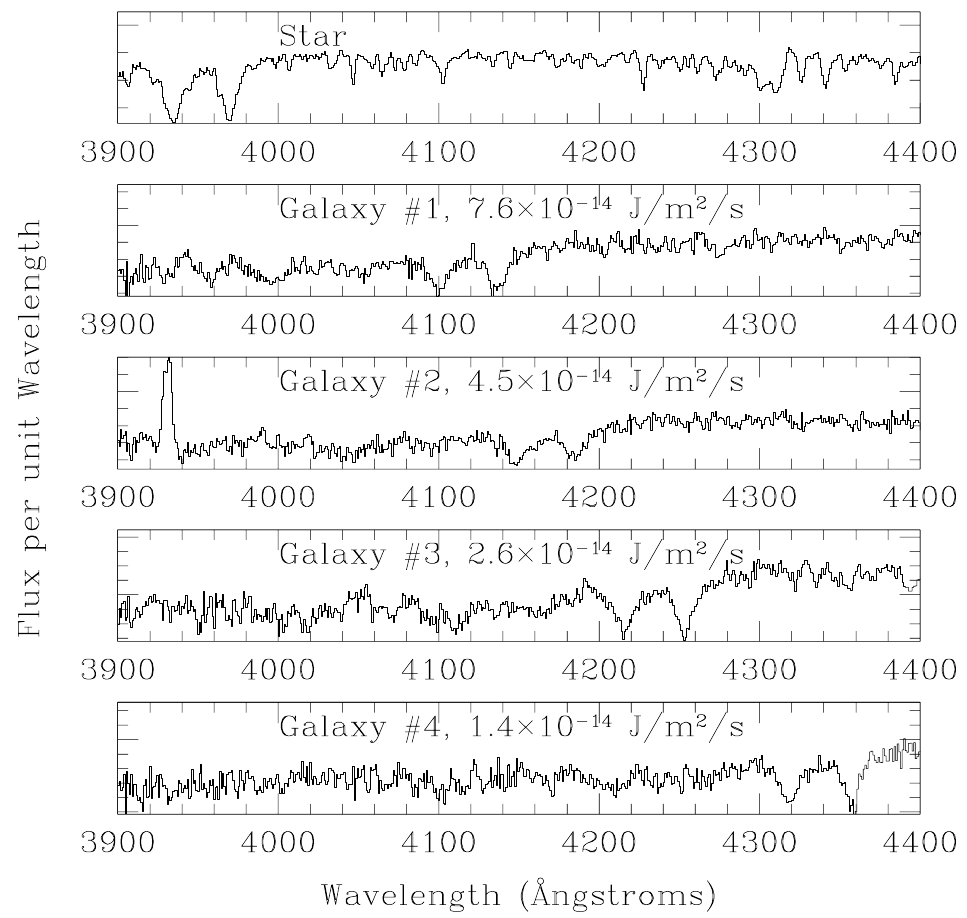}{0.9}
\caption{Spectra measured by the SDSS~\cite{Chyba}.}
\label{Hubble-problem}
\end{figure}

{\bf EXERCISE 6.2}~The Sloan Digital Sky Survey (SDSS) is a survey
that mapped positions and distances of a million galaxies using a
dedicated 2.5~m telescope in New Mexico~\cite{SDSS}. In this exercise, you will
use data from this survey to calculate $H_0$. In
Fig.~\ref{Hubble-problem} we show the spectrum of a star in our galaxy
and spectra of four distant galaxies, as measured by the SDSS. For
each of the galaxies, we indicate the measured brightness in units of
Joules per square meter per second. Assume that each of them has the
same luminosity as that of the Milky Way ($L_{\rm MW} =
10^{11}~L_\odot$, or $L_{\rm MW} = 4 \times 10^{37}~{\rm J/s}$). {\it
  (i)}~Determine the distance to each of the four galaxies, using the
inverse-square law relation between brightness and
luminosity. Express your answers both in meters and in megaparsecs,
and give two significant figures. {\it (ii)}~The spectrum of each of
these objects shows a pair of strong absorption lines of calcium,
which have rest wavelength $\lambda_0 = 3935~{\rm \AA}$ and $3970~{\rm \AA}$, respectively. The wavelengths of these lines in the galaxies
have been shifted to longer wavelengths (i.e., redshifted), by the
expansion of the universe. As a guide, the spectrum of a star like the
Sun is shown in the upper panel; the calcium lines are at zero
redshift. Measure the redshift of each galaxy. That is, calculate the
fractional change in wavelength of the calcium lines. [{\it Hint:}~The
tricky part here is to make sure you are identifying the right lines
as calcium. In each case, they are a close pair; for Galaxy \#2, they
are the prominent absorption dips between 4100~\AA and 4200~\AA.
Measure the redshift for both of the calcium lines in each galaxy (in each case the two lines should give the same redshift, of
course!). Give your final redshift to two significant figures. Do the
intermediate steps of the calculation without rounding; rounding too
early can result in errors. {\it (iii)}~Given the redshifts, calculate
the velocity of recession for each galaxy, and in each case use the
distances to estimate the Hubble constant, in units of kilometers per
second per Megaparsec. You will not get identical results from each of
the galaxies, due to measurement uncertainties (but they should all be
in the same ballpark), so average the results of the four galaxies to
get your final answer.\\

Now a point worth noting at this juncture is that galaxies do not
follow Hubble's law exactly.  In addition to the expansion of the
universe, galaxy motions are affected by the gravity of specific,
nearby structures, such as the pull of the Milky Way and Andromeda
galaxies on each other.  Each galaxy therefore has a peculiar
velocity, where peculiar is used in the sense of ``individual,'' or
``specific to itself.''  Thus, the recession velocity of a galaxy is
really
\begin{equation}
v = H_0 d + v_{\rm pec}, 
\label{vpeculiar}
\end{equation}
where $v_{\rm pec}$ is
the peculiar velocity of the galaxy along the line of sight.  If
peculiar velocities could have any value, then this would make
Hubble's law useless.  However, peculiar velocities are typically only
about 300~km/s, and they very rarely exceed 1000~km/s.  Hubble's
law therefore becomes accurate for galaxies that are far away, when
$H_0d$ is much larger than 1000~km/s.  Furthermore, we can often
estimate what a galaxy's peculiar velocity will be by looking at the
nearby structures that will be pulling on it.\\

{\bf EXERCISE 6.3}~Suppose we observer two galaxies, one at a distance
of 35~Mly with a radial velocity of 580~km/s, and another at a
distance of $1,100~{\rm Mly}$ with a radial velocity of $25,400~{\rm km/s}$.
{\it (i)}~Calculate the Hubble constant for each of these two observations.
{\it (ii)}~Which of the two calculations would you consider to be more trustworthy? Why?
{\it (iii)}~Estimate the peculiar velocity of the closer galaxy.
{\it (iv)}~If the more distant galaxy had this same peculiar velocity, how would that change your calculated value of the Hubble constant?

We would expect intuitively that at any given time the universe ought
to look the same to observers in all typical galaxies, and in whatever
direction they look. (Hereafter we will use the label ``typical'' to
indicate galaxies that do not have any large peculiar motion of their
own, but are simply carried along with the general cosmic flow of
galaxies.) This hypothesis is so natural (at least since Copernicus)
that it has been called the {\it cosmological principle} by Milne~\cite{Milne}.

\begin{figure}
 \postscript{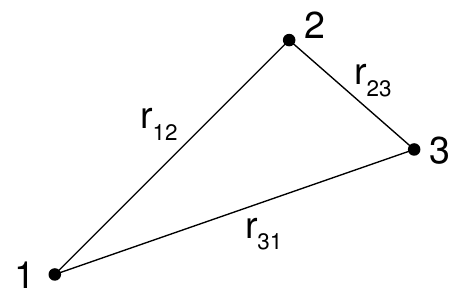}{0.8}
 \caption{A triangle defined by three galaxies in a uniformly expanding universe~\cite{Ryden:2003yy}.}
\label{barbara-triangle}
\end{figure}

As applied to the galaxies themselves, the cosmological principle
requires that an observer in a typical galaxy should see all the other
galaxies moving with the same pattern of velocities, whatever typical
galaxy the observer happens to be riding in. It is a direct
mathematical consequence of this principle that the relative speed of
any two galaxies must be proportional to the distance between them,
just as found by Hubble. To see this consider three typical galaxies
at positions $\vec r_1$, $\vec r_2$, and $\vec r_3$. They define the
triangle shown in Fig.~\ref{barbara-triangle}, with sides of length
\begin{eqnarray}
r_{12} & \equiv & |\vec r_1 - \vec r_2| \nonumber \\
r_{23} & \equiv & |\vec r_2 - \vec r_3| \nonumber \\
r_{31} & \equiv & |\vec r_3 - \vec r_1| \, .
\end{eqnarray}
In a homogeneous and uniform expanding universe the shape of the triangle
is preserved as the galaxies move away from each other. Maintaining
the correct relative lengths for the sides of the triangle requires an
expansion law of the form
\begin{eqnarray}
r_{12} (t) & = & a(t) \ r_{12} (t_0) \nonumber \\
r_{23} (t) & = & a(t) \ r_{23} (t_0) \nonumber \\
r_{31} (t) & = & a (t) \ r_{31} (t_0) \, ,
\end{eqnarray}
where $a(t)$ is a scale factor,  which is totally independent of location
or direction. The scale factor $a(t)$ tells us how the expansion (or
possibly contraction) of the universe depends on time. At any time $t$,
an observer in galaxy \#1 will see the other galaxies receding with a
speed
\begin{eqnarray}
v_{12} (t) & = & \frac{dr_{12}}{dt} = \dot a \, r_{12} (t_0) = \frac{\dot
    a}{a} r_{12} (t) \nonumber \\
v_{31} (t) & = & \frac{dr_{31}}{dt} = \dot a r_{31} (t_0) = \frac{\dot
    a}{a} r_{31} (t) \, .
\end{eqnarray}
You can easily demonstrate that an observer in galaxy \#2 or galaxy \#3
will find the same linear relation between observed recession speed
and distance, with $\dot a/a$  playing the role of the Hubble constant. Since
this argument can be applied to any trio of galaxies, it implies that
in any universe where the distribution of galaxies is undergoing
homogeneous, isotropic expansion, the velocity-distance relation
takes the linear form $v = Hr$, with $H = \dot a/a$.

If galaxies are currently moving away from each other, this implies
they were closer together in the past. Consider a pair of galaxies
currently separated by a distance $r$, with a velocity $v = H_0r$
relative to each other. If there are no forces acting to accelerate or
decelerate their relative motion, then their velocity is constant, and
the time that has elapsed since they were in contact is
\begin{equation}
t_{\rm H}= \frac{r}{v} = H_0^{-1} \,,
\end{equation}
independent of the current separation $r$ between galaxies. The time $H_0^{-1}$ is generally
referred to as the Hubble time. For \mbox{$H \approx 70~{\rm  km} \, {\rm
  s}^{-1} \, {\rm  Mpc}^{-1}$}, the
Hubble time is $H_0^{-1} \approx 14~{\rm Gyr}$. If the relative
velocities of galaxies have been constant in the past, then one Hubble
time ago, all the galaxies in the universe were crammed together into
a small volume. 

The observation of galaxy redshifts points naturally to a {\it big
  bang} description for the evolution of the universe. A big
  bang model could be broadly defined as a model in which the
universe expands from an initially highly dense state to its current
low-density state.  The Hubble time of $\sim 14~{\rm Gyr}$ is
comparable to the ages computed for the oldest known stars in the
universe. This rough equivalence is reassuring.  However, the age of
the universe (i.e the time elapsed since its original highly dense
state) is not necessarily exactly equal to $t_{\rm H}$.  On the one
hand, if gravity working on matter is the only force at work on large
scales, then the attractive force of gravity will act to slow down the
expansion. If this were the case, the universe was expanding more
rapidly in the past than it is now, and the universe is younger than
$H_0^{-1}$. On the other hand, if the energy density of the universe
is dominated by a cosmological constant $\Lambda$ (more on this
later), then the dominant gravitational force is repulsive, and the
universe may be older than $H_0^{-1}$.

The horizon distance is defined as the
greatest distance a photon can travel during the age of the universe.
The Hubble distance, ${\cal R}_{\rm H} = c/H_0 \approx 4.3~{\rm Gpc}$, provides a
natural distance scale. However, just as the age of the universe is roughly
equal to $H_0^{-1}$ in most big bang models, with the exact value depending
on the expansion history of the universe, one horizon
is roughly equal to $c/H_0$, with the exact value, again, depending on
the expansion history. 

Before proceeding any further,  two qualifications have to be
attached to the cosmological principle. First, it is obviously not
true on small scales -- we are in a Galaxy which belongs to a small
local group of other galaxies, which in turn
lies near the enormous cluster of galaxies in Virgo. In fact, of the
33 galaxies in Messier's catalogue, almost half are in one small part
of the sky, the constellation of Virgo. The cosmological principle, if
at all valid, comes into play only when we view the universe on a scale
at least as large as the distance between clusters of galaxies, or
about 100 million light years. Second, in using the cosmological
principle to derive the relation of proportionality between galactic
velocities and distances, we suppose the usual rule for adding $v \ll
c$. This, of course, was not a problem for Hubble in 1929, as none of
the galaxies he studied then had a speed anywhere near the speed of
light. Nevertheless, it is important to stress that when one thinks
about really large distances characteristic of the universe, as a
whole, one must work in a theoretical framework capable of dealing
with velocities approaching the speed of light.

Note how Hubble's law ties in with Olbers'  paradox. If the universe is
of finite age, $t_{\rm H} \sim H_0^{-1}$, then the night sky can be dark, even if the
universe is infinitely large, because light from distant galaxies has
not yet had time to reach us. Galaxy surveys tell us that the
luminosity density of galaxies in the local universe is
\begin{equation}
nL \approx 2 \times 10^8 L_\odot~{\rm Mpc}^{-3} \, .
\end{equation}
By terrestrial standards, the universe is not a well-lit place; this
luminosity density is equivalent to a single 40 watt light bulb within
a sphere 1~AU in radius. If the horizon distance is ${\cal R}_{\rm H} \approx c/H_0$, then
the total flux of light we receive from all the stars from all the
galaxies within the horizon will be
\begin{eqnarray}
F_{\rm gal} &\approx &  nL \int_0^{\cal R_{\rm H}} dr \sim nL
\frac{c}{H_0}  \sim  9 \times 10^{11} L_\odot~{\rm Mpc}^{-2} \nonumber \\
& \sim & 2 \times
10^{-11} L_\odot~{\rm AU}^{-2} \, .
\end{eqnarray}
By the cosmological principle, this is the total flux of starlight you
would expect at any randomly located spot in the universe. Comparing
this to the flux we receive from the Sun,
\begin{equation}
F_\odot = \frac{L_\odot}{4 \pi~{\rm AU}^2} \approx 0.08 L_\odot~{\rm
  AU}^{-2} \,,
\end{equation}
we find that $F_{\rm gal}/F_\odot \sim 3 \times 10^{-10}$. Thus, the
total flux of starlight at a randomly selected location in the
universe is less than a billionth the flux of light we receive from
the Sun here on Earth. For the entire universe to be as well-lit as
the Earth, it would have to be over a billion times older than it is;
and you would have to keep the stars shining during all that time.

\subsection{Friedmann-Robertson-Walker cosmologies}
\label{Sec-FRW}

In 1917 Einstein presented a model of the universe based on his theory
of general relativity~\cite{Einstein:1917ce}. It describes a
geometrically symmetric (spherical) space with finite volume but no
boundary. In accordance with the cosmological principle, the model is
homogeneous and isotropic. It is also static: the volume of the space
does not change. In order to obtain a static model, Einstein 
introduced a new repulsive force in his equations. The size of this
cosmological term is given by the cosmological constant
$\Lambda$. Einstein presented his model before the redshifts of the
galaxies were known, and taking the universe to be static was then
a reasonable assumption. When the expansion of the universe was discovered, this
argument in favor of a cosmological constant vanished. Einstein
himself later called it the biggest blunder of his life. Nevertheless,
the most recent observations seem to indicate that a non-zero
cosmological constant has to be present.

In 1922, Friedmann~\cite{Friedmann:1,Friedmann:2}
studied the cosmological solutions of Einstein equations. If $\Lambda
= 0$, only evolving, expanding or contracting models of the universe
are possible. The general relativistic derivation of the law of
expansion for the Friedmann models will not be given here. It is
interesting that the existence of three types of models and their law
of expansion can be derived from purely Newtonian considerations, with
results in complete agreement with the relativistic treatment.
Moreover, the essential character of the motion can be obtained from a
simple energy argument, which we discuss next.

\begin{figure}
 \postscript{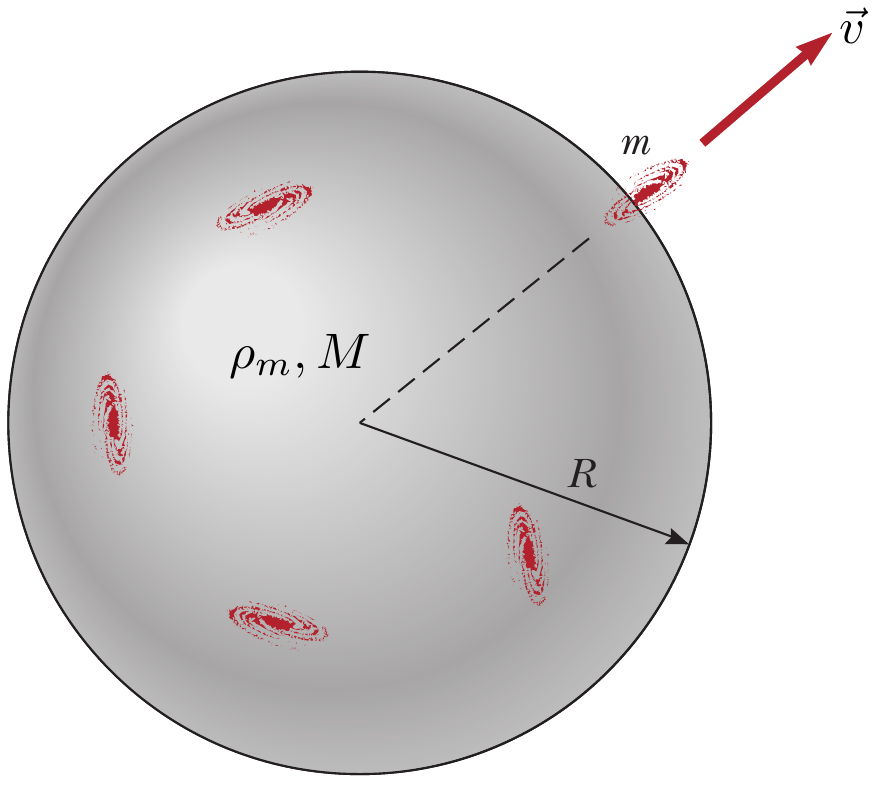}{0.6}
 \caption{Spherical region of galaxies with a larger radius than the
distance between clusters of galaxies, but smaller radius than any distance
characterizing the universe as a whole.}
\label{23}
\end{figure}

Consider a spherical region of galaxies of radius $R$. (For the
purposes of this calculation we must take $R$ to be larger than the
distance between clusters of galaxies, but smaller than any distance
characterizing the universe as a whole, as shown in Fig.~\ref{23}. 
We also assume $\Lambda =0$.) 
The mass of this sphere is
its volume times the cosmic mass density,
\begin{equation}
M = \frac{4\,\pi\,R^3}{3} \, \rho_m \, .
\end{equation}
We can now consider the motion of a galaxy of mass $m$ at the edge of
the spherical region. According to Hubble's law, the velocity of the
galaxy is
$v = HR$, and its corresponding kinetic energy
\begin{equation}
K = \frac{1}{2} m v^2 = \frac{1}{2} m H^2 R^2 \,\, .
\end{equation}
In a spherical distribution of matter, the gravitational force on a
given spherical shell depends only on the mass inside the shell. The
potential energy at the edge of the sphere is
\begin{equation}
U = - \frac{G Mm}{R} = - \frac{4 \pi m R^2 \rho_m G}{3}\, .
\end{equation}
Hence, the total energy is
\begin{eqnarray}
E = K + U & = & \frac{1}{2} m H^2 R^2 - G m \frac{4 \pi}{3} R^2 \rho_m
\, .
\label{context}
\end{eqnarray}
which has to remain constant as the universe expands. 
Likewise,
\begin{equation}
\frac{2E}{mR^2} = H^2 - \frac{8 \pi}{3} G \rho_m \, .
\label{worfri}
\end{equation}
Since we assume that the universe is homogeneous, $H$ and $\rho_m$ cannot be functions of $R$.
Thus, the left-hand-side of (\ref{worfri}) cannot depend on the chosen distance $R$ to the coordinate
center. However, the value of $2E/(mR^2)$ is time-dependent, because
the distance between us and the galaxy will change as the universe
expands. Since the mass $m$ of our test galaxy is arbitrary, we can
choose it such that $|2E/(mc^2)| = 1$ holds at an arbitrary moment as
long as $E \neq 0$. For different times, the left-hand-side scales as
$R^{-2}$ and thus we can rewrite (\ref{worfri}) as
\begin{equation}
\left(\frac{\dot a}{a}\right)^2 = \frac{8 \pi}{3} G \rho_m - \frac{k c^2}{a^2 R_0^2} \, .
\label{Friedmanntt}
\end{equation}
Note that because $E$ is constant, $k$ is constant too. Actually, $k = 0, \pm 1$ is
generally known as the curvature constant. Throughout the subscripted ``0''s indicate
that quantities (which in general evolve with time) are to be
evaluated at present epoch. Finally, we account for the equivalence of
mass and energy by including not only the mass but also the
energy density, $\rho = \rho_m c^2 + \cdots$ and so (\ref{Friedmanntt}) becomes
\begin{equation}
H^2 \equiv \left(\frac{\dot a}{a} \right)^2 = \frac{8 \pi}{3} G
\frac{\rho}{c^2} - \frac{k c^2}{a^2 R_0^2} \, .
\label{FriedmannGR}
\end{equation}
which is {\it Friedmann equation} (without cosmological constant) in
the Newtonian limit.  (\ref{FriedmannGR}) agrees exactly with the
equation derived from general relativity~\cite{Weinberg:1972kfs}. For
$k=0$, the value of $H$ fixes the so-called {\it critical density} as
\begin{equation}
\rho (k=0) \equiv \rho_c = \frac{3 H^2 c^2}{8 \pi G} \, .
\label{densityc}
\end{equation} 
Since we know the current value of the Hubble parameter to within
10\%, we can compute the current value of the critical density to
within 20\%. We usually
{\it hide} this uncertainty by introducing $h$,
\begin{equation}
H_0 = 100 \, h \, ~{\rm km} \, {\rm s}^{-1} \, {\rm Mpc}^{-1} \, ,
\end{equation}
such that 
\begin{eqnarray}
\rho_{c,0} & = & 2.77 \times 10^{11} h^2 M_\odot/{\rm Mpc}^3 \nonumber \\
           & = & 1.88 \times 10^{-29} h^2 {\rm g/cm}^3 \nonumber \\
& = & 1.05 \times 10^{-5} h^2~{\rm GeV/cm}^3 \, .
\end{eqnarray}
Note that since $h \approx 0.70^{+0.05}_{-0.03}$ a flat universe requires an energy
density of $\sim 10$ protons per cubic meter. 

The expansion of the universe can be compared to the motion of a mass
launched vertically from the surface of a celestial body. The form of
the orbit depends on the initial energy. In order to compute the
complete orbit, the mass of the main body and the initial velocity
have to be known. In cosmology, the corresponding parameters are the
mean density and the Hubble constant. On the one hand, if the density exceeds the critial density, the expansion of any
spherical region will turn to a contraction and it will collapse to a
point. This corresponds to the closed Friedmann model. On the other hand, if
$\rho_m < \rho_c$, the ever-expanding hyperbolic model is obtained.
These three models of the universe are called the standard models.
They are the simplest relativistic cosmological models for $\Lambda =
0$. Models with $\Lambda \neq 0$ are mathematically more complicated,
but show the same behaviour. The simple Newtonian treatment of the
expansion problem is possible because Newtonian mechanics is
approximately valid in small regions of the universe. However,
although the resulting equations are formally similar, the
interpretation of the quantities involved is not the same as in the
relativistic context. The global geometry of Friedmann models can only
be understood within the general theory of relativity~\cite{Weinberg:1972kfs}.

Next, we define the abundance $\Omega_i$ of the different players in
cosmology as their energy density relative to $\rho_c$.  For example,
the dimensionless mass density parameter is found to be
\begin{equation}
\Omega_m = \frac{\rho_m c^2}{\rho_c} = \frac{8 \pi G }{3 H^2} \, \rho_m
\, .
\label{omega-m}
\end{equation}
For simplicity, for the moment we will keep considering scenarios with $\Lambda
= 0$, but we advance the reader that
\begin{equation}
\Omega_\Lambda = \frac{\Lambda c^2 }{3
  H^2} \, .
\end{equation}

Now, what about our universe? On a large scale what is the overall
curvature of the universe? Does it have positive curvature, negative
curvature, or is it flat?  By solving Einstein equations,
Robertson~\cite{Robertson:1,Robertson:2} and Walker~\cite{Walker},
showed that the three hypersurfaces of constant curvature (the
hyper-sphere, the hyper-plane, and the hyper-pseudosphere) are indeed
possible geometries for a homegeneous and isotropic universe
undergoing expansion. The metric they derived, independently of each
other, is called the Friedmann-Robertson-Walker (FRW) metric. The line
element is most generally written in the form
\begin{equation}
ds^2 = c^2 dt^2 - a^2(t) \left[\frac{d\varrho^2}{1- k \varrho^2/R^2} +
  \varrho^2 d  \Omega^2\right] \,, 
\label{FRWmetric}
\end{equation}
where $d\Omega^2 = d\theta^2 + \sin^2 \theta d\phi^2$.  It is easily
seen that the spatial component of the FRW
metric consists of the spatial metric for a uniformly curved space of
radius $R$, scaled by the square of the scale factor $a(t)$.  If the
universe had a positive curvature $k=1$, then the universe would be closed,
or finite in volume. This would not mean that the stars and galaxies
extended out to a certain boundary, beyond which there is empty
space. There is no boundary or edge in such a universe. If a particle
were to move in a straight line in a particular direction, it would
eventually return to the starting point -- perhaps eons of time
later. On the other hand, if the curvature of the space was zero $k=0$
or negative $k=-1$, the universe would be open. It could just go on
forever.

Using the substitution 
\begin{equation}
\varrho = S_k (r) = \left\{\begin{array}{ll} 
R \sin (r/R) & ~~~~~ {\rm for} \ k = +1 \\
r & ~~~~~{\rm for} \ k=0 \\
R \sinh (r/R) & ~~~~~{\rm for} \ k=-1 
\end{array} \right. ;
\label{FRWsubs}
\end{equation}
the FRW line element can  be rewritten as
\begin{equation}
ds^2 =  c^2 dt^2 - a^2(t) \left[dr^2 + S_k^2(r) \, d\Omega^2 \right] \;;
\label{FRWsmetric}
\end{equation}
see Appendix~\ref{appD} for details.

The time variable $t$ in the FRW metric is the cosmological proper
time, called the cosmic time for short, and is the time measured by an
observer who sees the universe expanding uniformly around him. The
spatial variables $(\varrho, \theta, \phi)$ or $(r, \theta, \phi$) are
called the comoving coordinates of a point in space. If the expansion
of the universe is perfectly homogeneous and isotropic, the comoving
coordinates of any point remain constant with time.

To describe the time evolution of the scale
factor $a(t)$ we need an additional equation
describing how the energy content of the universe $\rho$ is affected
by expansion. The first law of thermodynamics,
\begin{equation}
dU = T dS - P dV, 
\label{1stLAW}
\end{equation}
with $dQ=0$ (no
heat exchange to the outside, since no outside exists)  becomes
\begin{equation}
dU = - P dV \Rightarrow \frac{dU}{dt} + P \frac{dV}{dt} = 0 \, .
\label{meijide1}
\end{equation}
There is a caveat to the statement that the expansion of a homogeneous
universe is {\it adiabatic}: when particles annihilate, such as electrons
and positrons, this adds heat and makes the expansion temporarily
non-adiabatic. This matters at some specific epochs in the very early
universe. 

For a sphere of comoving radius $R_0$,
\begin{equation}
V = \frac{4}{3} \pi \, R_0^3 \, a^3(t) \,,
\end{equation}
and so
\begin{equation}
\dot V = 4 \pi \, R_0^3  \, a^2 \, \dot a = 3 \frac{\dot a}{a} V \, .
\label{meijide2}
\end{equation}
Since $U = \rho V$,
\begin{equation}
\dot U = \dot \rho V + \rho \dot V = V \left(\dot \rho + 3 \frac{\dot
    a}{a} \rho \right) \, .
\label{meijide3}
\end{equation}
 Substituting (\ref{meijide2}) and (\ref{meijide3}) into (\ref{meijide1}) we have
\begin{equation}
V \left(\dot \rho + 3 \frac{\dot a}{a} \rho + 3 \frac{\dot a}{a} P
\right) = 0
\end{equation}
and thus
\begin{equation}
\dot \rho =- 3 \left(\rho + P \right) \frac{\dot a}{a} \,.
\end{equation}
This {\it fluid equation} describes the evolution of energy density in
an expanding universe. It tells us that the expansion decreases the
energy density both by dilution and by the work required to expand a
gas with pressure $P \geq 0$.

To solve this equation, we need an additional equation of state relating $P$ and $\rho$. Suppose we write this in the form
\begin{equation}
P = w \rho \, .
\end{equation}
In principle, $w$ could change with time, but we will assume that any
time derivatives of $w$ are negligible compared to time derivatives of
$\rho$. This is reasonable if the equation of state is determined by
``microphysics'' that is not directly tied to the expansion of the
universe. The fluid equation then implies
\begin{equation}
\frac{\dot \rho}{\rho} = - 3 (1 + w) \frac{\dot a}{a} \,,
\end{equation}
with solution
\begin{equation}
\frac{\rho}{\rho_0} = \left(\frac{a}{a_0} \right)^{-3 ( 1 + w)} \,
  .
\end{equation}
The pressure in a gas is determined by the thermal motion of its
constituents. For non-relativistic matter (a.k.a. cosmological dust),
\begin{equation}
w = \frac{P}{\rho} \sim \frac{m v^2}{m c^2} \sim
\frac{v}{c^2} \ll 1 \,, 
\end{equation}
where $v$ is the thermal velocity of particles with mass $m$.  To a
near-perfect approximation $w = 0$, implying $\rho_{ m} \propto
a^{-3}$. Light, or more generally any highly relativistic particle,
has an associated pressure (radiation pressure). Pressure is defined
as the momentum transfer onto a perfectly reflecting wall per unit
time and per unit area. Consider an isotropic distribution of photons
(or another kind of particle) moving with the speed of light. The
momentum of a photon is given in terms of its energy as $p = E/c=
h\nu/c$. Consider now an area element $dA$ of the wall; the momentum
transferred to it per unit time is given by the momentum transfer per
photon, times the number of photons hitting the area $dA$ per unit
time. We will assume for the moment that all photons have the same
frequency. If $\theta$ denotes the direction of a photon relative to
the normal of the wall, the momentum component perpendicular to the
wall before scattering is $p_\perp = p \cos \theta$, and after
scattering $p_\perp = - p \cos \theta$; the two other momentum
components are unchanged by the reflection. Thus, the momentum
transfer per photon scattering is $\Delta p = 2 p \cos \theta$. The number of
photons scattering per unit time within the area $dA$ is given by the
number density of photons, $n$ times the area element $dA$, times the
thickness of the layer from which photons arrive at the wall per unit
time. The latter is given by $c \cos \theta$, since only the perpendicular
velocity component brings them closer to the wall. Putting these terms
together, we find for the momentum transfer to the wall per unit time
per unit area the expression
\begin{equation}
P (\theta) = 2 h \nu \ n \ \cos^2 \theta \, .
\end{equation}
Averaging this expression over a half-sphere (only photons moving
towards the wall can hit it) then yields 
\begin{equation}
P = \frac{1}{3} h \nu n = \frac{1}{3} \rho \, .
\end{equation}
Then for radition, $w = 1/3$,
implying $\rho_{\rm rad} \propto a^{-4}$. This behavior also follows
from a simple argument: the number density of photons falls as $n
\propto a^{-3}$, and the energy per photon falls as $h\nu \propto
a^{-1}$ because of cosmological redshift (more on this below).

Next, we obtain an expression for the acceleration of the universe. If
we multiply our standard version of the Friedmann equation by $a^2$, we get
\begin{equation}
\dot a^2 = \frac{8 \pi G}{3 c^2} \rho a^2 - \frac{k c^2}{R_0^2} \, .
\label{cuchiufo}
\end{equation} 
Take the time derivative of (\ref{cuchiufo})
\begin{equation}
2 \dot a \ddot a = \frac{8 \pi G}{3 c^2} \left(\dot \rho a^2 + 2\rho  
  a \dot a \right) \, .
\end{equation}
divide by $2 \dot a a$
\begin{equation}
\frac{\ddot a}{a} = \frac{4 \pi G}{3c^2} \left(\dot \rho \frac{a}{\dot
  a} + 2 \rho \right) \,,
\end{equation}
and substitutte from the fluid equation
\begin{equation}
\dot \rho \frac{a}{\dot a}  = - 3 (\rho +P) 
\end{equation}
 to obtain the {\it acceleration equation}
\begin{equation}
\frac{\ddot a}{a} = - \frac{4 \pi G}{3 c^2}  (\rho + 3 P) \, .
\label{acceeq}
\end{equation}
We see that if $\rho$ and $P$ are positive, the expansion of the
universe decelerates.  Higher $P$ produces stronger deceleration for
given $\rho$, e.g., a radiation-dominated universe decelerates faster than
a matter-dominated universe.

In the remainder of this section, we consider a flat universe, i.e.,
$k=0$. It is easily seen that for non-relativistic matter,  the
solution to Friedmann equation (\ref{FriedmannGR}) is given by
\begin{equation}
a(t) = \left(\frac{t}{t_0} \right)^{2/3} \quad {\rm and} \quad \rho(t) = \frac{\rho_0}{a^3} = \frac{\rho_0 t_0^2}{t^2} \,,
\label{gear1}
\end{equation}
with
\begin{equation}
t_0 = \frac{2}{3} \frac{1}{H_0} \, ,
\label{gear2}
\end{equation}
where we have used (\ref{densityc}). Following the same steps for a
bizarre universe, which is dominated today by radiation pressure, yields the solution
\begin{equation}
a(t) = \left(\frac{t}{t_0} \right)^{1/2} \quad {\rm and} \quad
\rho(t) = \frac{\rho_0}{a^4} = \frac{\rho_0 t_0^2}{t^2} \, .
\label{gear3}
\end{equation}
From this simple exercise we can picture the the time evolution of the
universe as follows. In the early universe all matter is
relativistic and radiation pressure dominates: $a(t) \propto t^{1/2}$,
$\rho_{\rm rad} \propto t^{-2}$, and $\rho_m \propto a^{-3}
\propto t^{-3/2}$. The density of radiation then falls more quickly
than that of dust. On the other hand, when dust dominates: $a(t)
\propto t^{2/3}$, $\rho_m \propto t^{-2}$, and $\rho_{\rm
  rad} \propto a^{-4} \propto t^{8/3}$, hence dust domination
increases. \\

{\bf EXERCISE 6.4} Using the Hubble flow $v = H_0r$ show that the
expansion of the universe changes the particle number density
according to $\dot n = - 3 H_0 n$.\\

In closing, we discuss how to measure distances in the FRW
spacetime. Consider a galaxy which is far away from us, sufficiently
far away that we may ignore the small scale perturbations of spacetime
and adopt the FRW line element. In an expanding universe, the distance
between two objects is increasing with time. Thus, if we want to
assign a spatial distance between two objects, we must specify the
time $t$ at which the distance is the correct one. Suppose that you
are at the origin, and that the galaxy which you are observing is at a
comoving coordinate position $(r, \theta, \phi)$. We define a proper
distance, as the distance between two events $A$ and $B$ in a
reference frame for which they occur simultaneously ($t_A = t_B$). In
other words, the proper distance $d_{\rm p}(t)$ between two points in
spacetime is equal to the length of the spatial geodesic between them
when the scale factor is fixed at the value $a(t)$. The proper
distance between the observer and galaxy can be found using the FRW
metric at a fixed time $t$,
\begin{equation}
ds^2 = a^2 (t) \ \left[dr^2 + S_k^2 (r) \ d \Omega^2 \right] \, .
\end{equation}
Along the spatial geodesic between the observer
and galaxy, the angle $(\theta, \phi$) is constant, and thus
\begin{equation}
ds = a(t) \ dr \, .
\end{equation}
Likewise, using spatial variables $(\varrho, \theta, \phi)$ we have
\begin{equation}
ds = a(t)  [1 - k (\varrho/R)^2]^{-1/2} \, dr
\end{equation}
The proper distance $d_{\rm p}$ is found by integrating over the radial comoving
coordinate $r$
\begin{equation}
d_{\rm p} = a(t) \int_0^r dr = a(t) \ r \, ,
\label{eqdp}
\end{equation}
or using (\ref{FRWsubs})
\begin{equation}
 d_{\rm p} = a (t) \! \left\{\begin{array}{ll} k^{-1/2} \sin^{-1}
      (\sqrt{k} \varrho/R) & {\rm for} \, k=+1 \\
\varrho & {\rm for} \, k =0 \\
|k|^{-1/2} \sinh^{-1} (\sqrt{|k|} \varrho/R)  & {\rm for}
 \, k =-1 \\ 
\end{array} \right. \!\! \!\! \! .
\end{equation}
In a flat universe, the proper distance to an object is just its
coordinate distance, $d_{\rm p} (t) = a(t) \varrho$. Because
$\sin^{-1} (x) > x$ and $\sinh^{-1} (x) < x$, in a closed universe ($k
> 0$) the proper distance to an object is greater than its coordinate
distance, while in an open universe ($k < 0$) the proper distance to
an object is less than its coordinate distance.\\

{\bf EXERCISE 6.5}~A civilization that wants to conquer the universe,
which is homogeneous and isotropic, and hence is described by the
FRW metric, is getting ready to send out
soldiers in all directions to invade all the universe out to a proper
distance $d_{\rm p}$. Every soldier leaves the galaxy where the
civilization was born, and travels through the universe with its
spaceship along a geodesic, out to a distance $d_{\rm p}$ from the
original galaxy. At the end of the invasion, which occurs at a fixed
time $t$, all the soldiers stand on a spherical surface at a proper
distance $d_p$ from their original galaxy. The total volume that has
been invaded is the volume inside this spherical surface. What is the
total volume invaded? Answer this question for the following three
cases: {\it (i)}~A flat metric ($k=0$). {\it (ii)}~A closed metric ($k=+1$) with
radius of curvature $R$ at the cosmic time $t$ when the invaded volume
and the proper distance $d_{\rm p}$ are measured.  {\it (iii)}~An open
  metric ($k=-1$) with
radius of curvature $R$ at the cosmic time $t$ when the invaded volume
and the proper distance $d_{\rm p}$ are measured. \\

{\bf EXERCISE 6.6}~Consider a positively curved universe ($k =1$), in
which the sole contribution to the energy density comes from
non-relativistic matter. In this case the energy density has the
dependence $\rho_m = \rho_{m,0}/a^3$. {\it (i)}~Write down Friedmann equation for this
universe and show that the parametric solution,
\begin{eqnarray}
a(\theta) & = & \frac{4 \pi G \rho_{m,0} R_0^2}{3 c^4} (1 - \cos \theta)
\,, \nonumber \\
t(\theta) & = & \frac{4 \pi G \rho_{m,0} R_0^3}{3 c^5} (\theta - \sin \theta)
\,,
\label{ciento66}
\end{eqnarray}
satisfies the Friedmann equation. Here $\theta$ is a dimensionless
parameter that runs from 0 to $2\pi$, and $R_0$ is the present radius
of curvature if we have normalized the scale factor at present to
$a(t_0) = 1$. {\it (ii)}~What is $a_{\rm max}$, the maximum possible
scale factor for this universe? {\it (iii)}~What is the maximum value
that the physical radius of curvature ($aR_0$) reaches? {\it
  (iv)}~What is the age of the universe when this maximum radius is
reached? {\it (v})~What is $t_{\rm crunch}$, the time at which the
universe undergoes a {\it big crunch} (that is a recollapse to $a=0$)?
[{\it Hint:} Recall that $\dot a = da/dt = da/d\theta \, d\theta/
dt$.] \\

{\bf EXERCISE 6.7}~Consider a positively curved universe ($k =-1$), in
which the sole contribution to the energy density comes from
non-relativistic matter, and so the energy density has the
dependence $\rho_m = \rho_{m,0}/a^3$. {\it (ii)}~Write down Friedmann equation for this
universe and show that the parametric solution,
\begin{eqnarray}
a(\theta) & = & \frac{4 \pi G \rho_{m,0} R_0^2}{3 c^4} (\cosh \theta -1)
\,, \nonumber \\
t(\theta) & = & \frac{4 \pi G \rho_{m,0} R_0^3}{3 c^5} (\sinh \theta -
\theta)
\,,
\label{ciento67}
\end{eqnarray}
satisfies the Friedmann equation. {\it (ii)}~Compare the time
dependence of the scale factor for open, closed and critical
matter-dominated cosmological models in a log-log plot.\\

\subsection{Age and size of the Universe}

\begin{figure}
 \postscript{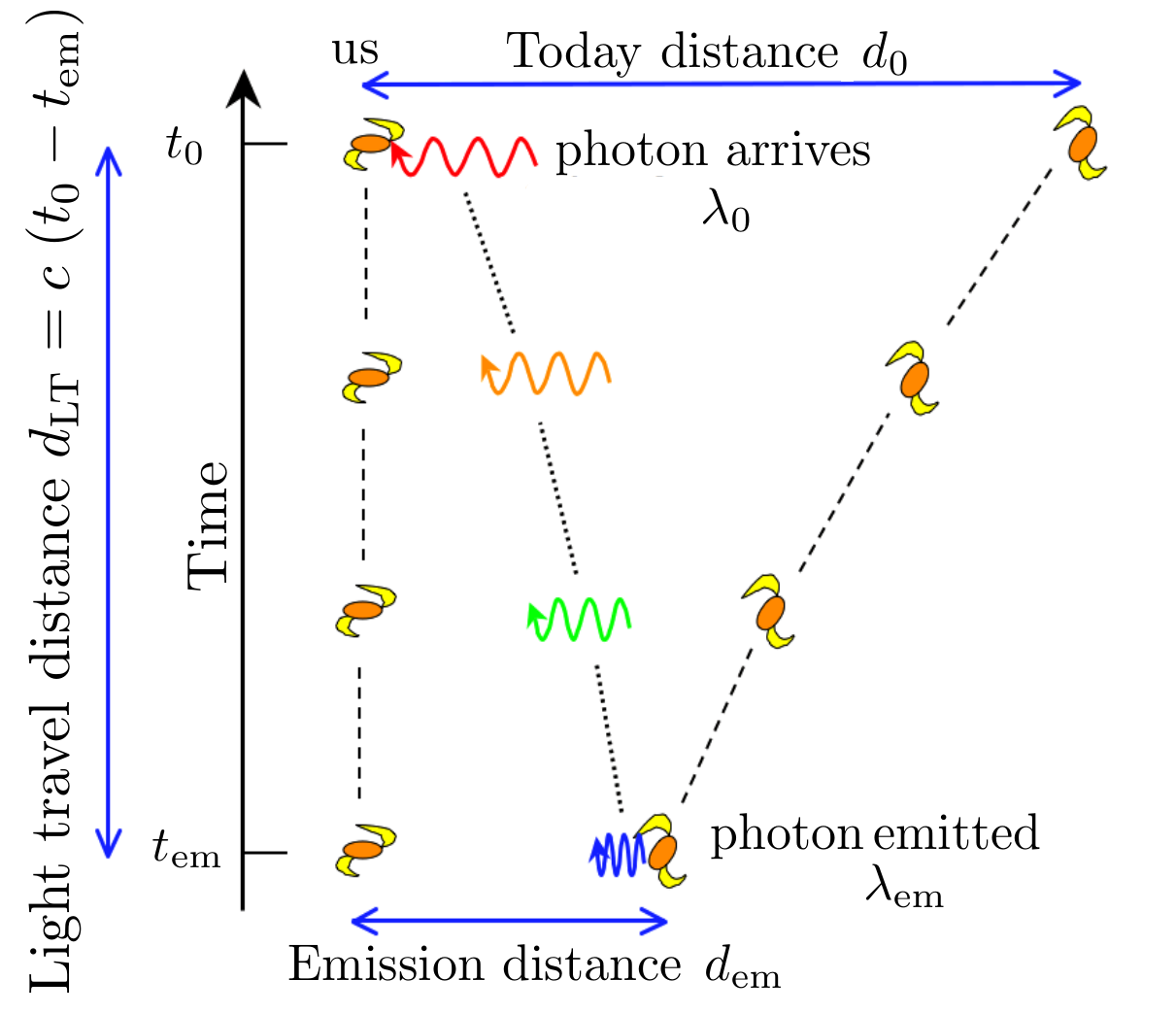}{0.99}
 \caption{Cosmological redshift.}
\label{fig-zt}
\end{figure}

In special (and general) relativity the propagation of light is along a null
geodesic ($ds = 0$).  If we place the observer at the origin ($\varrho = 0$),
and we choose a radial null geodesic ($d\theta = d\phi = 0$), we have
\begin{equation}
\frac{c dt}{a(t)} = \pm \frac{d\varrho}{[1 - k (\varrho/R)^2]^{1/2}} \,,
\end{equation}
where $+$ is for the emitted light ray and the $-$ is for a
received one. Imagine now that one crest of the light wave was emitted
at time $t_{\rm em}$ at distance $\varrho_{\rm em}$, and received at
the origin $\varrho_0 = 0$ at $t_0$, and that the next wave crest was
emitted at $t_{\rm em}+ \Delta t_{\rm em}$ and received at $t_0+\Delta
t_0$; see Fig.~\ref{fig-zt}. The two waves satisfy the relations:
\begin{equation}
\int_{t_{\rm em}}^{t_0} \frac{dt}{a(t)} = - \frac{1}{c}
\int_{\varrho_{\rm em}}^{\varrho_0} \frac{d\varrho}{\sqrt{1- k(\varrho/R)^2}}
\label{subsa}
\end{equation}
and 
\begin{equation}
\int_{t_{\rm em} + \Delta t_{\rm em}}^{t_0 +\Delta t_0} \frac{dt}{a(t)} = - \frac{1}{c}
\int_{\varrho_{\rm em}}^{\varrho_0} \frac{d\varrho}{\sqrt{1- k(\varrho/R)^2}} \, .
\label{subsb}
\end{equation}
Now, substract (\ref{subsa}) from (\ref{subsb})
\begin{equation}
\int_{t_{\rm em} + \Delta t_{\rm em}}^{t_0 +\Delta t_0}
\frac{dt}{a(t)} - \int_{t_{\rm em}}^{t_0} \frac{dt}{a(t)} =0
\end{equation}
and expand
\begin{eqnarray}
\int_{t_{\rm em} + \Delta t_{\rm em}}^{t_0 + \Delta t_0} \frac{dt}{a(t)}
& = & \int_{t_{\rm em}}^{t_0} \frac{dt}{a(t)} + \int_{t_0}^{t_0 + \Delta
  t_0} \frac{dt}{a(t)} \nonumber \\
& - & \int_{t_{\rm em}}^{t_{\rm em} + \Delta t_{\rm em} }\frac{dt}{a(t)}
\end{eqnarray}
to obtain
\begin{equation}
\int_{t_0}^{t_0 + \Delta t_0} \frac{dt}{a(t) } = \int_{t_{\rm
    em}}^{t_{\rm em} + \Delta t_{\rm em}} \frac{dt}{a(t)} \, .
\end{equation}
Any change in $a(t)$ during the time intervals between successive wave crests can be safely neglected, so that $a(t)$ is a constant with respect to the time integration. Consequently,
\begin{equation}
\frac{\Delta t_{\rm em}}{a(t_{\rm em})} = \frac{\Delta t_0}{a(t_0)} \,,
\end{equation}
or equivalently
\begin{equation}
\frac{\Delta t_{\rm em}}{\Delta t_0} =\frac{a(t_{\rm em})}{a(t_0)} \,
.
\end{equation}
The time interval between successive wave crests is the inverse of the
frequency of the light wave, related to its wavelength by the relation
$c = \lambda \nu$. Hence, from (\ref{zredshit}) the redshift is
\begin{equation}
z = \frac{\lambda_0}{\lambda_{\rm em}} -1 =\frac{a_0}{a (t_{\rm em})} - 1 \,\,;
\label{MUKA1}
\end{equation}
i.e., the redshift of a galaxy expresses how much the scale factor has
changed since the light was emitted.

The light detected today was emitted at some time $t_{\rm em}$ and,
according to (\ref{MUKA1}), there is a one-to-one correspondence between $z$ and
$t_{\rm em}$. Therefore, the redshift $z$ can be used instead of time
$t$ to parametrize the history of the universe. A given $z$
corresponds to a time when our universe was $1+z$ times smaller than
now.

Generally, the expressions for $a(t)$ are rather complicated and one
cannot directly invert (\ref{MUKA1}) to express the cosmic time $t \equiv t_{\rm
  em}$ in terms of the redshift parameter $z$. It is useful,
therefore, to derive a general integral expression for
$t(z)$. Differentiating (\ref{MUKA1}) we obtain
\begin{equation}
dz = - \frac{a_0}{a^2(t)} \dot a(t) dt = -(1 + z) H(t) dt \,,
\end{equation}
from which follows that
\begin{equation}
t = \int_z^\infty  \frac{dz}{H(z) (1+z)} \, .
\label{noire}
\end{equation}
A constant of integration has been chosen here so that $z \to \infty$
corresponds to the initial moment of $t=0$.

To obtain the expression for the Hubble parameter $H$ in terms of $z$
and the present values of $H_0$ and $\Omega_{m,0}$, it is convenient
to write the Friedmann equation (\ref{FriedmannGR}) in the form
\begin{equation}
H^2(z) + \frac{kc^2}{a_0^2 R_0^2} (1 + z)^2 = \Omega_{m,0} H_0^2
  \frac{\rho_m(z)}{\rho_{m,0}} \,,
\end{equation}
where the definitions in (\ref{omega-m}) and (\ref{MUKA1}) have been used. At $z=0$, this
equation reduces to
\begin{equation}
\frac{kc^2}{a_0^2 R_0^2} = (\Omega_{m,0} -1 ) H_0^2 \,, 
\end{equation}
allowing us to express the current value of  $a_0 R_0$ in
a spatially curved universe ($k \neq 0$) in terms of $H_0$ and
$\Omega_{m,0}$. Taking this into account, we obtain
\begin{eqnarray}
H(z) \!\! & = &  \!\! H_0 \sqrt{(1-\Omega_{m,0}) (1+z)^2 +
  \Omega_{m,0} \,
\rho_m(z)/\rho_{m,0}}  \nonumber \\
& = & \!\! H_0 \sqrt{(1 - \Omega_{m,0}) (1+z)^2 + \Omega_{m,0} (1+z)^3} \, .
\label{hachedez}
\end{eqnarray}

We can now complete our program by  finding an expression for the
comoving radial distance coordinate $r$ as a function of the reshift
$z$. Since photons travel on null geodesics of zero proper time, we
see directly from the metric (\ref{FRWsmetric}) that 
\begin{equation}
r = - \int \frac{c dt}{a(t)} = - \int c \frac{dt}{dz} (1 + z) dz = c \int
\frac{dz}{H (z)}
\, ,
\label{RrrrrrrrrR}
\end{equation}
with $H(z)$ given by (\ref{hachedez}).\\

\begin{figure}
 \postscript{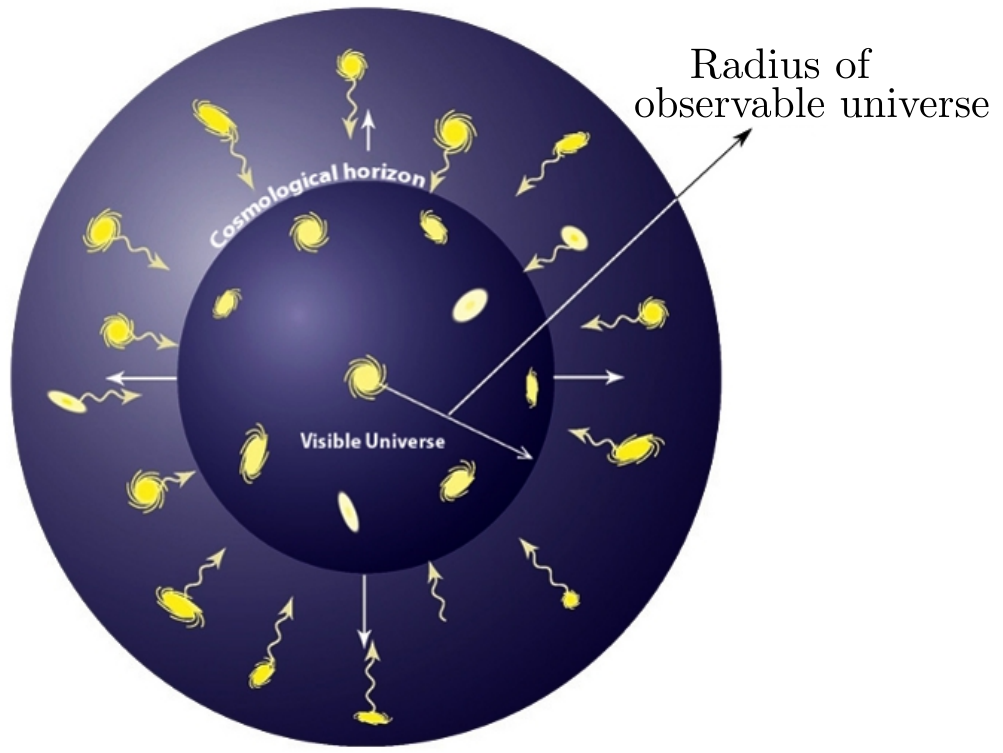}{0.99}
 \caption{Cosmological horizon.}
\label{fig-horizon}
\end{figure}

As the universe expands and ages, an observer at any point is able to
see increasingly distant objects as the light from them has time to
arrive, see Fig.~\ref{fig-horizon}. This means that, as time
progresses, increasingly larger regions of the universe come into
causal contact with the observer. The proper distance to the furthest
observable point (the particle horizon) at time $t$ is the ``horizon
distance'', $d_{\rm h}(t)$.

Again we return to the FRW metric, placing an observer at the origin
($\varrho = 0$) and letting the particle horizon for this observer at
time $t$ be located at radial coordinate distance $\varrho_{\rm
  h}$. This means that a photon emitted at $t = 0$ at $\varrho_{\rm
  h}$ will reach the observer at the origin at time $t$. 
Recalling photons move along null geodesics ($ds = 0$) and  considering only
radially traveling photons ($d\theta = d\phi = 0$), we find
\begin{equation}
\int_0^t \frac{dt'}{a(t')} = \frac{1}{c} \int_0^{\varrho_{\rm h}} \frac{d
    \varrho}{[1 - k (\varrho/R)^2]^{1/2} }\,,
\label{gear4}
\end{equation}
yielding
\begin{equation}
\varrho_{\rm h} = \left\{ \begin{array}{l l}
\sin\left[c \int_0^t dt'/a(t') \right] & ~~~{\rm for} \ k = +1 \\
c \int_0^t dt'/a(t') & ~~~{\rm for} \ k = 0 \\
\sinh \left[c \int_0^t dt'/a(t') \right] & ~~~{\rm for} \ k = -1 
\end{array} \right.  \, .
\label{poliuy}
\end{equation}
If the scale factor evolves with time as $a(t) = t^\alpha$, with
$\alpha > 1$, we can see that the time integral in (\ref{poliuy}) diverges as we
approach $t = 0$. This
would imply that the whole universe is in causal contact. However,
$\alpha =1/2$ and 2/3 in the radiation and matter-dominated eras, so there
is a horizon.

The proper distance from the origin to $\varrho_{\rm h}$ is given by
\begin{eqnarray}
d_{\rm h} (t) & = & a(t) \int_0^{\varrho_{\rm h}} \frac{d
    \varrho}{[1 - k (\varrho/R)^2]^{1/2} } \nonumber \\
& = & a(t) \int_0^t \frac{c
    dt'}{a(t')} \, .
\end{eqnarray}
For $k=0$, using (\ref{gear1}) and (\ref{gear3})  we obtain $d_{\rm h} = 2 ct$ in the
radiation-dominated era, and $d_{\rm h} (t)=3ct$ in the
matter-dominated era. Now, substituting (\ref{gear2}) into (\ref{gear1}) we have
\begin{equation}
a(t) = \left( \frac{3}{2} \, H_0 \, t \right)^{2/3}
\end{equation}
and so from (\ref{MUKA1}) it follows that
\begin{equation}
t = \frac{2}{3} \frac{1}{H_0 \, (1+z)^{3/2}} \, .
\end{equation}
For the matter-dominated era, the proper horizon distance is
\begin{equation}
d_{\rm h} = \frac{2 c}{H_0 \, (1+z)^{3/2}} \, .
\label{cientoochentayocho}
\end{equation}
For a flat universe with $\Omega_{m,0} =1$, we find that at present time,
\begin{equation}
d_{\rm h,0} = 2c/H_0 = 1.85 \times 10^{28} h^{-1}~{\rm cm} = 6 \,
h^{-1}~{\rm Gpc} \,.
\end{equation}
Note that because $a_0 =1$, we have
$\varrho_{{\rm h},0}  = d_{{\rm h},0}$.\\ 

{\bf EXERCISE 6.8}~Consider a flat model containing only matter, with
$\Omega_{m,0} = 1$, and present Hubble constant $H_0$. {\it (i)}~What
is the comoving distance to the horizon ($z = \infty$)? {\it
  (ii)}~What is the redshift at which the comoving distance is half
that to the horizon? {\it (iii)}~What is the ratio of the age of the
universe at that redshift, to its present age? {\it (iv)}~At which
redshift did the universe have half its present age?\\

In closing, we show that Hubble's law is indeed an approximation for small
redshift by using a Taylor expansion of $a(t)$,
\begin{eqnarray}
a(t) & = & a(t_0) + (t-t_0) \dot a(t_0) + \frac{1}{2} ( t - t_0)^2 \ddot
a(t_0) + \cdots \nonumber \\
& = & a(t_0) \left[ 1 + (t - t_0) H_0 - \frac{1}{2} (t - t_0)^2 q_0
  H_0^2 + \cdots \right] \,, \nonumber
\label{aTaylor}
\end{eqnarray}
where $q_0 \equiv - \ddot a(t_0) a(t_0)/ \dot a^2(t_0)$ is the
deceleration parameter (it is named ``deceleration'' because
historically, an accelerating universe was considered unlikely). If
the expansion is slowing down, $\ddot a<0$ and $q_0 > 0$.  For not too
large time-differences, we can use the Taylor expansion of $a(t)$ and
write
\begin{equation}
1-z \approx \frac{1}{1+z} = \frac{a(t)}{a(t_0)} \approx 1 + (t - t_0)
H_0 \, .
\end{equation}
 Hence Hubble's law, $z = (t_0- t)H_0 = d/cH_0$, is valid as long as
 $z \ll H_0 (t_0 - t) \ll 1$. Deviations from its linear form arises for
 $z \agt 1$ and can be used to determine $q_0$.

\subsection{Angular diameter and luminosity distances}
 
\begin{figure}
 \postscript{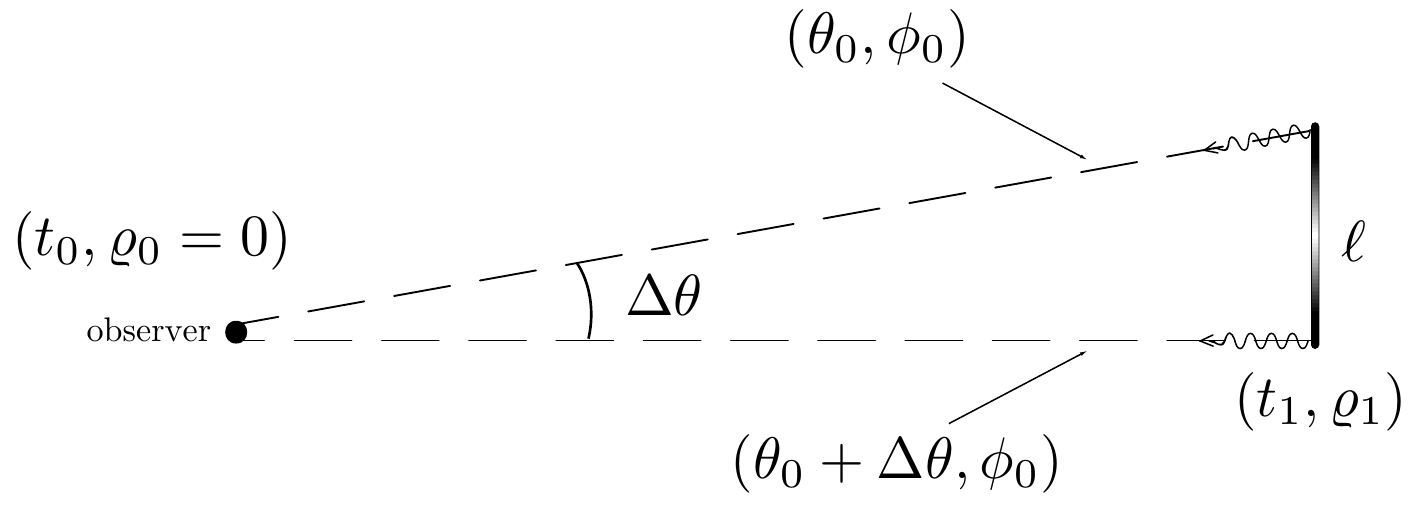}{0.8}
 \caption{Extended object of given transverse size $\ell$ at comoving
   distance $\varrho_1$ from the observer~\cite{Mukhanov:2005sc}.}
\label{fig-dA}
\end{figure}

\begin{figure}
 \postscript{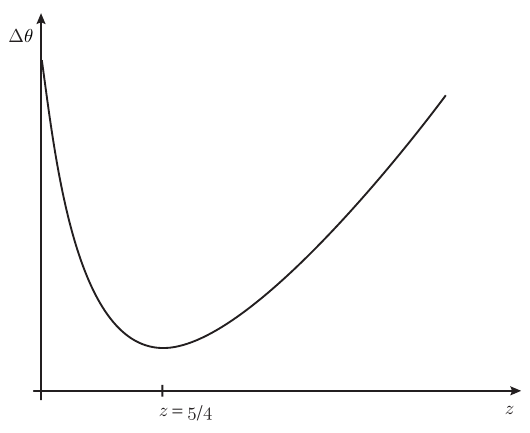}{0.8}
\caption{For a flat univrese filled with dust $d_A(z)$ has a maximum
  at $z = 5/4$, corresponding to the redshift at which objects of a
  given proper size $\ell$ will subtend the minimum angle $\Delta
  \theta$ on the sky. At redshifts $z > 5/4$ objects of a given proper
  size $\ell$ will appear bigger on the sky with increasing
  $z$~\cite{Mukhanov:2005sc}.}
\label{fig-dAz}
\end{figure}

The angular diameter distance to an object is defined in terms of the
object's actual size, $\ell$, and $\theta$ the angular size of the
object as viewed from earth. Consider a light source of size $\ell$ at
$\varrho = \varrho_1$ and $t = t_1$ subtending an angle $\Delta
\theta$ at the origin ($\varrho = 0,\, t = t_0$) as shown in
Fig.~\ref{fig-dA}. The proper distance $\ell$ between the two ends of
the object is related to $\Delta \theta$ by,
\begin{equation}
\Delta \theta = \frac{\ell}{a(t_1) \varrho_1} \,.
\end{equation}
We now define the angular diameter distance
\begin{equation}
d_A = \frac{\ell}{\Delta \theta}
\label{patoc1}
\end{equation}
so that 
\begin{equation}
d_A = a(t_1) \varrho_1 = \frac{\varrho_1}{1+z} \, .
\end{equation}
In analogy with (\ref{gear4}) we write
\begin{equation}
\int_0^{t_1} \frac{dt}{a(t)} = \frac{1}{c} \int_0^{\varrho_1} \frac{d
    \varrho}{[1 - k (\varrho/R)^2]^{1/2} }\, ,
\label{gear5}
\end{equation}
From an examination point of view, only proficiency in the $k = 0$
case will be expected. Hence,
\begin{equation}
\varrho_1 = c \int_0^{t_1} \frac{dt}{a(t)} = c \int_0^z \frac{dz}{H(z)} \,,
\end{equation}
where in the last equality we used (\ref{RrrrrrrrrR}). Then, for a flat universe
filled with dust, the angular diameter as a function of $z$ is
\begin{equation}
\Delta \theta (z) = \frac{\ell H_0}{2 c} \frac{(1+z)^{3/2}}{(1+z)^{1/2}
  -1 } \, .
\label{patocUNO}
\end{equation}
At low redshifts ($z \ll 1$), the angular diameter decreases in
inverse proportion to $z$, reaches a minimum at $z = 5/4$, and then
scales as $z$ for $z \gg 1$; see Fig.~\ref{fig-dAz}

Perhaps the most important relation for observational cosmology is
that between the monochromatic flux density and luminosity. Start by
assuming isotropic emission, so that the photons emitted by the source
pass with a uniform flux density through any sphere surrounding the
source. We can now make a shift of the origin, and consider the FRW metric
as being centred on the source. However, because of homogeneity, the
comoving distance between the source and the observer $\varrho_1$ is the
same as we would calculate when we place the origin at our
location. The photons from the source are therefore passing through a
sphere, on which we sit, of proper surface area $4\pi a_0^2 \varrho_1^2$.
However, the redshift still affects the flux density in four further ways: {\it
  (i)}~photon energies are redshifted, reducing the flux density by a
factor $1 + z$; {\it (ii)}~photon arrival rates are time dilated,
reducing the flux density by a further factor $1 + z$; 
{\it (iii)}~opposing this, the
bandwidth $d\nu$ is reduced by a factor $1 + z$, which increases the energy
flux per unit bandwidth by one power of $1 + z$;  {\it (iv)} finally, the
observed photons at frequency $\nu_0$ were emitted at frequency $(1 + z) \nu_0$.
Overall, the flux density is the luminosity at frequency $(1+z) \nu_0$, divided by the total area, divided by $(1 + z)$:
\begin{eqnarray}
{\cal F}_\nu (\nu_0) & = & \frac{L_\nu ([1+z] \nu_0)}{4 \pi a_0^2 \varrho_1^2(r) (1 +
  z)} \nonumber \\
& = & \frac{L_\nu (\nu_0)}{4 \pi a_0^2 \varrho_1^2 ( 1 + z)^{1 + \alpha}} \,,
\end{eqnarray}
where the second expression assumes a power-law spectrum $L \propto
\nu^{-\alpha}$. We can integrate over $\nu_0$ to obtain the corresponding
total or bolometric formulae
\begin{equation}
{\cal F} = \frac{L}{4 \pi a_0^2 \varrho_1^2 (1+z)^2} \, .
\end{equation}
The luminosity distance $d_L$ is
defined to satisfy the relation (\ref{L}).  Thus, 
\begin{equation}
d_L =  ( 1+ z) \varrho_1 = ( 1 + z)^2 d_A \,  ,
\label{pavilion}
\end{equation}
where we have taken $a_0=1$.

\section{The force awakens}

Independent cosmological observations have unmasked the presence of
some unknown form of energy density, related to otherwise empty space,
which appears to dominate the recent gravitational dynamics of the
universe and yields a stage of cosmic acceleration. We still have no
solid clues as to the nature of such dark energy (or perhaps more
accurately dark pressure). The cosmological constant is the simplest
possible form of dark energy because it is constant in both space and
time, and provides a good fit to the experimental data as of today. In
this section we will discuss the many observations that probes the dark
energy and we will describe the generalities of the concordance
model of cosmology with $\Lambda \neq 0$.

\subsection{Supernova Cosmology}

The expansion history of the cosmos can be determined
using as a ``standard candle'' any distinguishable class of
astronomical objects of known intrinsic brightness that can be
identified over a wide distance range. As the light from such beacons
travels to Earth through an expanding universe, the cosmic expansion
stretches not only the distances between galaxy clusters, but also the
very wavelengths of the photons en route. The recorded redshift and
brightness of each these candles thus provide a measurement of the
total integrated exansion of the universe since the time the light was
emitted. A collection of such measurements, over a sufficient range of
distances, would yield an entire historical record of the universe's
expansion.

Type Ia supernovae (SNe Ia) are the best cosmological yard sticks in
the market. They are precise distance indicators because they have a
uniform intrinsic brightness due to the similarity of the triggering
white dwarf mass (i.e., $M_{\rm Ch} = M_\odot$) and consequently the
amount of nuclear fuel available to burn. This makes SNe Ia the best
(or at least most practical) example of ``standardizable candles'' in
the distant universe.

Before proceeding, we pause to present some notation.  The apparent
magnitude ($m$) of a celestial object is a number that is a measure of
its apparent brightness as seen by an observer on Earth. The smaller
the number, the brighter a star appears. The scale used to indicate
magnitude originates in the Hellenistic practice of dividing stars
visible to the naked eye into six magnitudes. The brightest stars in
the night sky were said to be of first magnitude ($m = 1$), whereas
the faintest were of sixth magnitude ($m = 6$), which is the limit of
human visual perception (without the aid of a telescope).  In 1856,
Pogson formalized the system by defining a first magnitude star as a
star that is 100 times as bright as a sixth-magnitude star, thereby
establishing a logarithmic scale still in use today~\cite{Pogson:1856wh}. This implies
that a star of magnitude $m$ is $100^{1/5} \simeq 2.512$ times as bright
as a star of magnitude $m+1$. The apparent magnitude, $m$, in the
band, $x$, is defined as
\begin{equation}
m_x - m_{x,0} = -2.5 \log_{10} ({\cal F}_x/{\cal F}_{x,0}) \,,
\end{equation}
where ${\cal F}_x$ is the observed flux in the band $x$, whereas $m_{x,0}$
and ${\cal F}_{x,0}$ are a reference magnitude, and reference flux in
the same band $x$, respectively.  A difference in magnitudes, $\Delta m = m_1 - m_2$,
can then be converted to a relative brightness as  $I_2/I_1 \approx 2.512^{\Delta m}$.

\begin{figure}[tbp] \postscript{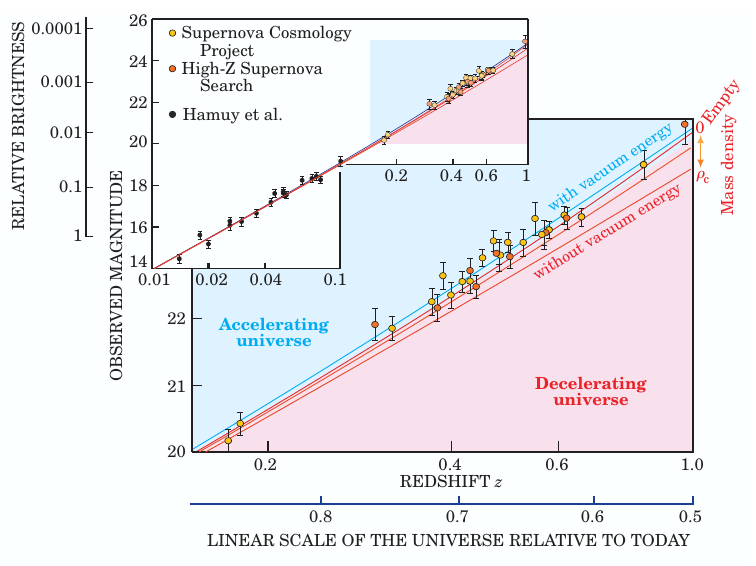}{0.9} \caption{Observed
    magnitude (and relative brightness) versus redshift is plotted for
    well-measured distant~\cite{Riess:1998cb,Perlmutter:1998np} and
    (in the inset) nearby~\cite{Hamuy:1993,Hamuy:1995in} SNe Ia. For
    clarity, measurements at the same redshift are combined. At
    redshifts beyond $z = 0.1$ (distances greater than about 109~ly),
    the cosmological predictions (indicated by the curves) begin to
    diverge, depending on the assumed cosmic densities of mass and
    vacuum energy. The red curves represent models with zero vacuum
    energy and mass densities ranging from 
    $\rho_c$ down to zero (an empty cosmos). The best fit (blue line)
    assumes a mass density of about $\rho_c/3$ plus a vacuum energy
    density twice that large, implying an accelerating cosmic
    expansion~\cite{Perlmutter:2003,Bahcall:1999xn}.}
\label{fig:SN}
\end{figure}

In Fig.~\ref{fig:SN} we show the observed magnitude (and relative 
brightness) versus redshift for well-measured distant and (in the
inset) nearby SNe Ia. The faintness (or distance) of the high-redshift
supernovae in Fig.~\ref{fig:SN} comes as a dramatic surprise. In the
(simplest) standard cosmological models described in
Sec.~\ref{Sec-FRW}, the expansion history of the cosmos is determined
entirely by its mass density. The greater the density, the more the
expansion is slowed by gravity. Thus, in the past, a high-mass-density
universe would have been expanding much faster than it does today. So
one should not have to look far back in time to especially distant
(faint) supernovae to find a given integrated expansion
(redshift). Conversely, in a low-mass-density universe one would have
to look farther back. But there is a limit to how low the mean mass
density could be. After all, we are here, and the stars and galaxies
are here. All that mass surely puts a lower limit on how far-that is,
to what level of faintness we must look to find a given redshift.
However, the high-redshift supernovae in Fig.~\ref{fig:SN} are fainter
than would be expected even for an empty cosmos.

\begin{figure}[tbp] \postscript{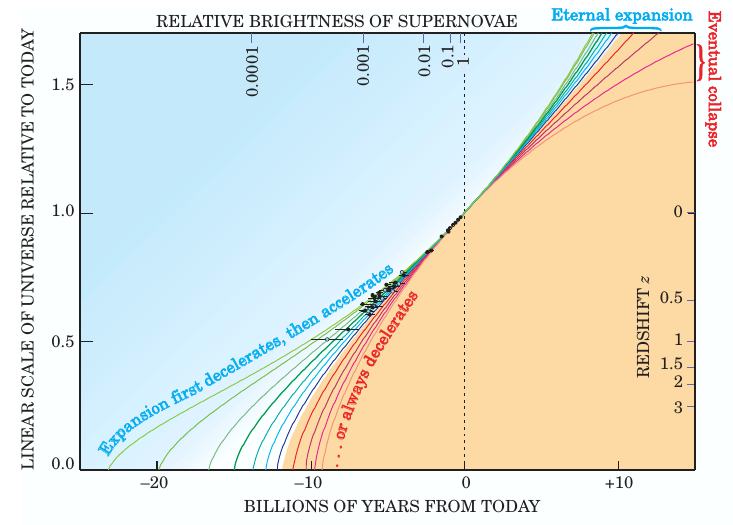}{0.9} \caption{The
    history of cosmic expansion, as measured by the high-redshift
    supernovae (the black data points), assuming flat cosmic
    geometry. The scale factor $a$ of the universe is taken to be 1 at
    present, so it equals $1/(1 + z)$. The curves in the blue shaded
    region represent cosmological models in which the accelerating
    effect of vacuum energy eventually overcomes the decelerating
    effect of the mass density. These curves assume vacuum energy
    densities ranging from $0.95~\rho_c$ (top curve) down to
    $0.4~\rho_c$. In the yellow shaded region, the curves represent
    models in which the cosmic expansion is always decelerating due to
    high mass density. They assume mass densities ranging (left to
    right) from $0.8~\rho_c$ up to $1.4~\rho_c$. In fact, for the last
    two curves, the expansion eventually halts and reverses into a
    cosmic collapse~\cite{Perlmutter:2003}.}
\label{fig:SN2}
\end{figure}

If these data are correct, the obvious implication is that the three
simplest models of cosmology introduced in
Sec.~\ref{Sec-FRW} must be too simple. The next to simplest
model includes an expansionary term in the equation of motion driven
by the cosmological constant $\Lambda$, which competes against
gravitational collapse.  The best fit to the 1998 supernova data shown
in Figs.~\ref{fig:SN} and \ref{fig:SN2} implies that, in the present
epoch, the vacuum energy density $\rho_\Lambda$ is larger than the
energy density attributable to mass $\rho_m$. Therefore, the cosmic
expansion is now accelerating.

To accommodate SNe Ia data we must add an additional term 
into the Friedmann equation (\ref{FriedmannGR}),
\begin{equation}
H^2 = \frac{8 \pi}{3} G
\frac{\rho}{c^2} - \frac{k c^2}{a^2 R_0^2} + \frac{\Lambda c^2}{3}\, .
\label{FriedmannLambda}
\end{equation}
The $\Lambda$ term also modifies the acceleration equation
(\ref{acceeq}), which becomes
\begin{equation}
\frac{\ddot a}{a} = \frac{\Lambda c^2}{3} - \frac{4 \pi G}{3 c^2}  (\rho + 3 P) \, ,
\label{acceeqLambda}
\end{equation}
and $H(z)$ in (\ref{hachedez}) is now given by
\begin{eqnarray}
H(z) & = & H_0 \left\{\Omega_{m,0} (1+z)^3  + \Omega_{{\rm rad},0}
  (1+z)^4 + \Omega_\Lambda 
\right.  \nonumber \\ 
& + & \left.
(1- \Omega_0) (1+z)^2  \right\}^{1/2} \, , 
\end{eqnarray}
where
\begin{equation}
\Omega = \Omega_m + \Omega_{\rm rad} + \Omega_\Lambda \,,
\label{energy-Omega}
\end{equation}
and $\Omega_{\rm rad}$ is the density fraction of relativistic matter
(radiation). We might note in passing that the quantity $k c^2/(a^2
R_0^2H_0^2)$ is sometimes referred to as $\Omega_k$.  This usage is
unfortunate, because it encourages us to think of curvature as a contribution
to the energy density of the universe, which is incorrect.\\

{\bf EXERCISE 7.1}~Imagine a class of astronomical objects that are
both standard candles and standard yardsticks. In other words, we know
both their luminosities $L$ and their physical sizes $\ell$. Recall
that the {\it apparent} brightness $I$ of an object is its flux on
Earth divided by its angular area, or solid angle on the sky, i.e. $I
= {\cal F}/\theta^2$, where $\theta$ the angular size. How does the
{\it apparent}  brightness depend on redshift for a general
cosmological model, for these objects with fixed $L$ and $\ell$?

\subsection{Cosmic Microwave Background}

\begin{figure}[tbp] \postscript{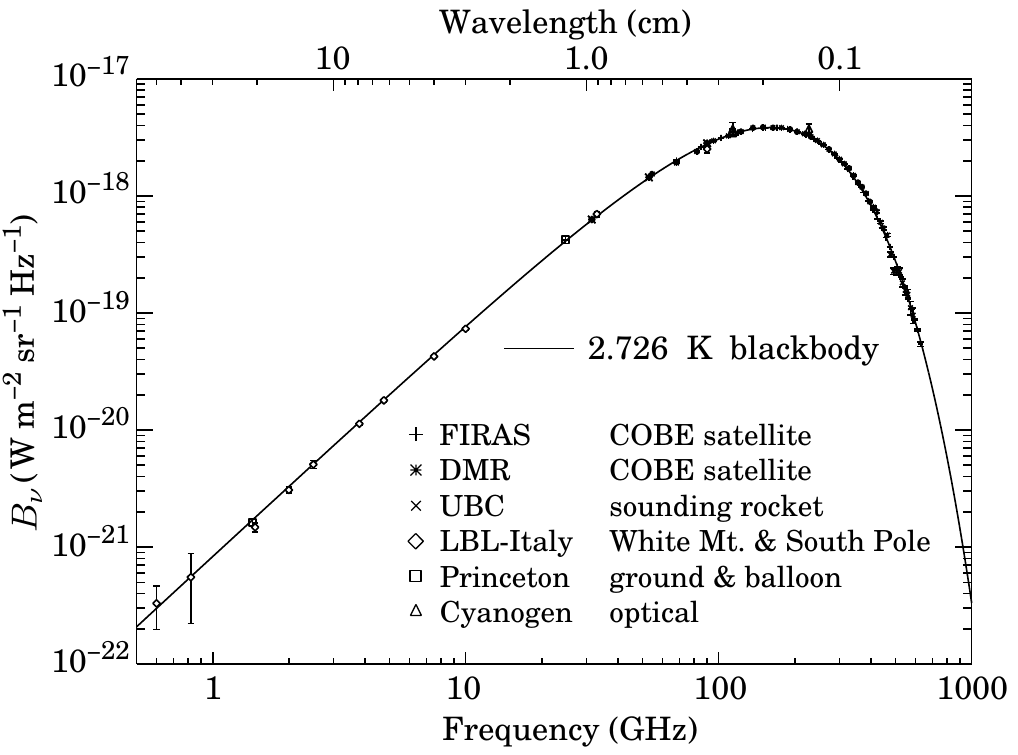}{0.9} \caption{The CMB blackbody spectrum as confirmed by measurements over a broad range of wavelengths~\cite{Smoot:1997xt}.}
\label{fig:CMB-blackbody}
\end{figure}

The cosmic microwave background (CMB) radiation was discovered in 1964
by Penzias and Wilson, using an antenna built for satellite
communication~\cite{Penzias:1965wn}.  The radiation was acting as a
source of excess noise (or ``static'') in the radio
receiver. Eventually, it became obvious that the source of noise was
actually a signal that was coming from outside the Galaxy. Precise
measurements were made at wavelength $\lambda = 7.35~{\rm cm}$. The
intensity of this radiation was found not to vary by day or night or
time of the year, nor to depend on the direction to a precision of
better than 1\%.  Almost immediately after its detection it was
concluded that this radiation comes from the universe as a whole: a
blackbody emission of hot, dense gas (temperature $T \sim 3000~{\rm
  K}$, peak wavelength $\lambda_{\rm max} \sim 1000~{\rm nm}$)
redshifted by a factor of 1000 to $\lambda_{\rm
  max} \sim 1~{\rm mm}$ and $T \sim 3~{\rm K}$~\cite{Dicke:1965zz}.  A
compilation of experimental measurements in the range $0.03~{\rm cm}
\alt \lambda \alt 75~{\rm cm}$ revealed an accurate blackbody
spectrum, see Fig.~\ref{fig:CMB-blackbody}. Actually, according to the
FIRAS (Far InfraRed Absolute Spectrometer) instrument aboard the COBE
({\it Cosmic Background Explorer}) satellite, which measured a
temperature of $T_0 = 2.726 \pm 0.010~{\rm K}$, the CMB is the most
perfect blackbody ever seen~\cite{Mather:1993ij}.

The CMB photons we see today interacted with matter for the last time
some $380~{\rm kyr}$ after the bang. Photon decoupling occurs when the
temperature has dropped to a point where there are no longer enough
high energy photons to keep hydrogen ionized: $  ^1{\rm H} \, \gamma\,
\slashed \leftrightharpoons \, e^- p^+$.  This era is known as recombination, even
though the atomic constituents had never been combined prior. The
ionization potential of hydrogen is 13.6~eV (i.e., $T \sim 10^5~{\rm
  K}$), but recombination occurs at $T_{\rm rec} \sim 3000~{\rm
  K}$. This is because the low baryon to photon ratio, $\eta \approx 5
\times 10^{-10}$,  allows the high
energy tail of the Planck distribution to keep the comparatively small
number of hydrogen atoms ionized until this much lower temperature.\\

{\bf EXERCISE 7.2}~{\it (i)}~For blackbody radiation, the energy density per
unit frequency is given by
\begin{equation}
u_\nu \ d\nu = \frac{8 \pi h \nu^3 d\nu}{c^3[\exp(h\nu/kT) -1]} \, .
\end{equation}
Since the energy of one photon is $h\nu$, the number density of
photons is given by the same expression above divided by $h\nu$. 
Calculate the present density of photons in the universe,
knowing that the CMB temperature is $T_0 \simeq 2.726~{\rm K}$. [{\it Hint:} you will
find it useful to know that $\int x^2 dx/(e^x -1 ) \simeq 2.404$.]
{\it (ii)}~If deuterium measurements
require a baryon to photon ratio of $\eta = 5.5 \times 10^{-10}$, what must the
current density of baryons be? {\it (iii)}~Assuming that the Hubble constant
is $H_0 = 70~{\rm km \ s^{-1} \ Mpc^{-1}}$, calculate what $\Omega_b$
  is. \\

\begin{figure}[tbp] 
\postscript{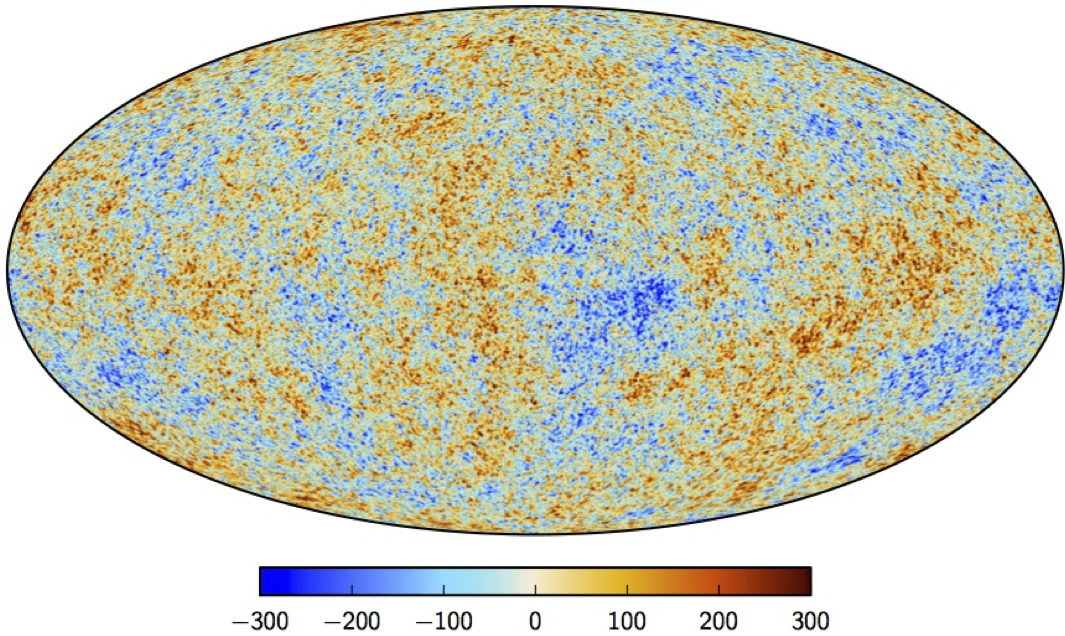}{0.85} 
\caption{
The CMB over the entire sky, color-coded to represent differences in temperature from the average 2.726~K: the color scale ranges from $+300~\mu{\rm K}$ (red) to $300~\mu{\rm K}$ (dark blue), representing slightly hotter and colder spots (and also variations in density.) Results are from the WMAP satellite~\cite{Bennett:2012zja} and the Planck mission~\cite{Adam:2015rua}.}
\label{wmap-planck}
\end{figure}

Before the recombination epoch the universe was an opaque ``fog'' of
free electrons and became transparent to photons
afterwards. Therefore, when we look at the sky in any direction, we
can expect to see photons that originated in the ``last-scattering
surface.'' This hypothesis has been tested very precisely by the
observed distribution of the CMB; see Fig.~\ref{wmap-planck}.  The
large photon-to-nucleon ratio implies that it is very unlikely for the
CMB to be produced in astrophysical processes such as the absorption
and re-emission of starlight by cold dust, or the absorption or
emission by plasmas.  Before the recombination epoch, Compton
scattering tightly coupled photons to electrons, which in turn coupled
to protons via electomagnetic interactions. As a consequence, photons
and nucleons in the early universe behaved as a single
``photon-nucleon fluid'' in a gravitational potential well created by
primeval variations in the density of matter. Outward pressure from
photons, acting against the inward force of gravity, set up acoustic
oscillations that propagated through the photon-nucleon fluid, exactly
like sound waves in air. The frequencies of these oscillations are now
seen imprinted on the CMB temperature fluctuations. Gravity caused the
primordial density perturbations across the universe to grow with
time. The temperature anisotropies in the CMB are interpreted as a
snapshot of the early stages of this growth, which eventually resulted
in the formation of galaxies~\cite{Weinberg:2008zzc,Dodelson:2003ft}.

The full sky CMB temperature anisotropy map,  as measured by the  Wilkinson
Microwave Anisotropy Probe (WMAP)~\cite{Bennett:2012zja} and the Planck mission~\cite{Adam:2015rua}, is shown
in Fig. ~\ref{wmap-planck}.  It is convenient to expand the difference
 $\Delta T (\hat n)$
between the CMB temperature observed in a direction given by the unit
vector $\hat n = (\theta, \phi)$ and the present mean value $T_0$ of the
temperature in spherical harmonics
\begin{equation}
\Delta T (\hat n) \equiv T(\hat n) - T_0 = \sum_{l =0}^\infty \
\sum_{|m| \leq l} \ a_{lm} Y_{l m} \,,  
\end{equation}
where 
\begin{equation}
T_0 = \frac{1}{4 \pi} \ \int d^2 \hat n \ T (\hat n) \, ,
\label{cmb-normalization}
\end{equation}
\begin{equation}
a_{lm} = \int \Delta T(\hat n) \, Y_{lm} (\hat n) \, d\Omega \,,
\end{equation}
and where $\Omega$ denotes the solid angle parametrized by the pair
$(\theta,\phi)$. The set $\{\ylm\}$ is complete and orthonormal, obeying
\begin{equation}
\int d\Omega\;Y_{{l_1}
  {m_1}}(\Omega)\,Y_{{l_2}{m_2}}(\Omega)=\delta_{{l_1}{l_2}}\,\delta_{{m_1}{m_2}}\, .
\label{eq:ortho}
\end{equation}
Since $\Delta T(\hat n)$ is real, 
we are interested in the real-valued, orthonormal $\ylm$'s, defined by
\begin{equation}
Y_{l m}(\theta,\phi)=
N(l,m)
\begin{cases}
P^l_m(x)(\sqrt{2}\cos(m\phi))&m>0\\
P_l(x)&m=0\\
P^l_m(x)(\sqrt{2}\sin(m\phi))\quad&m<0
\end{cases}\,,
\end{equation}
where 
\begin{equation}
N(l,m) = \sqrt{\frac{(2l+1)(l-m)!}{4\pi\,(l+m)!}} 
\end{equation}
is a normalization-factor, 
\begin{equation}
P_l^m (x) = \frac{(1-x^2)^{m/2}}{2^l l!} \frac{d^{m+l}}{dx^{m+l}} (x^2
- 1)^l 
\end{equation}
is the associated Legendre polynomial, $P_l=P^l_{m=0}$ is the Legendre
polynomial, and $x\equiv\cos\theta$; for further details see
e.g.~\cite{Anchordoqui:2013}.

\begin{figure*}[t]
\centering
\newcommand\nodalwidth{0.197\textwidth}
\includegraphics[width=\nodalwidth]{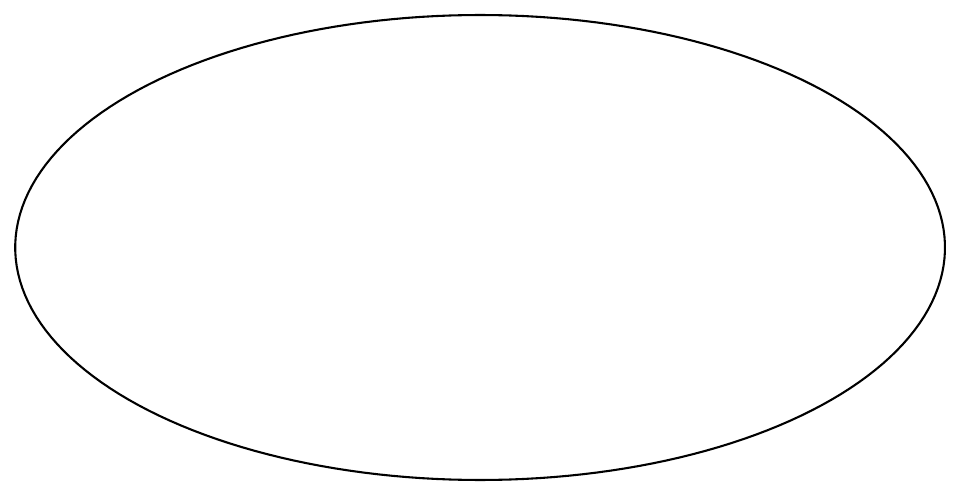}
\includegraphics[width=\nodalwidth]{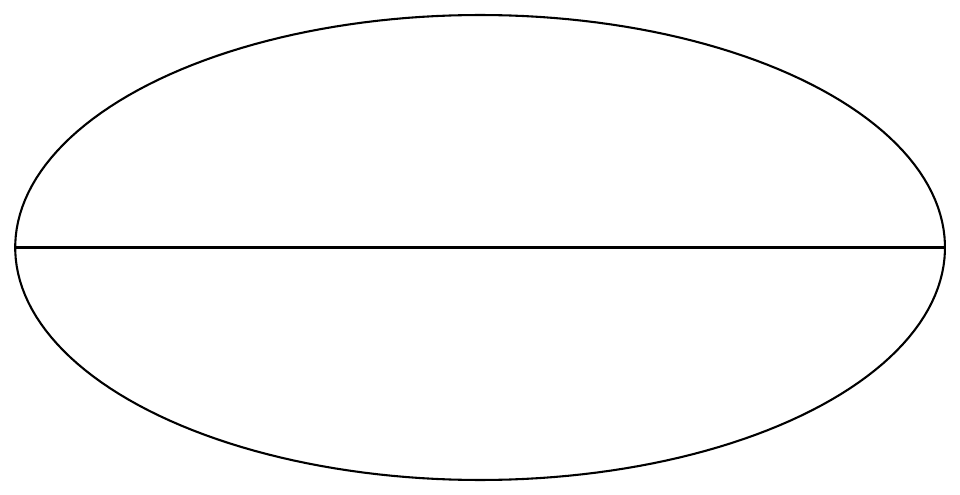}
\includegraphics[width=\nodalwidth]{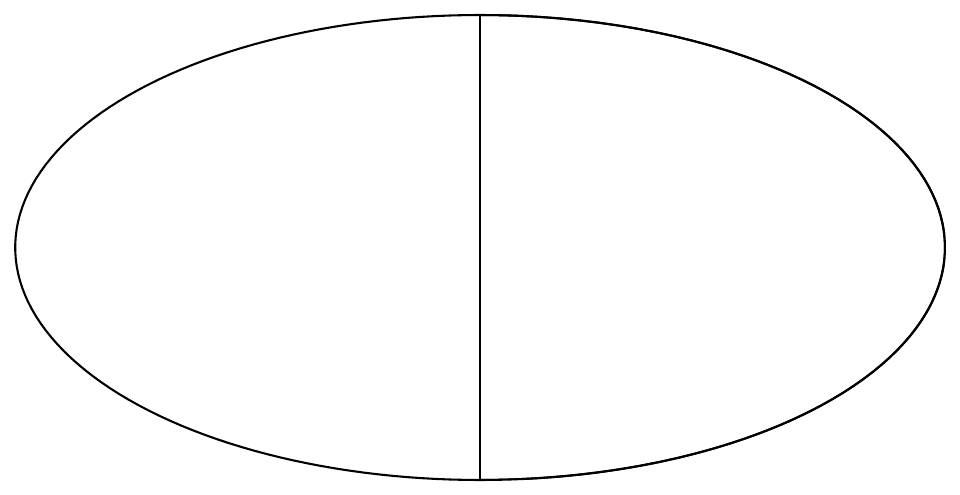}\\
\includegraphics[width=\nodalwidth]{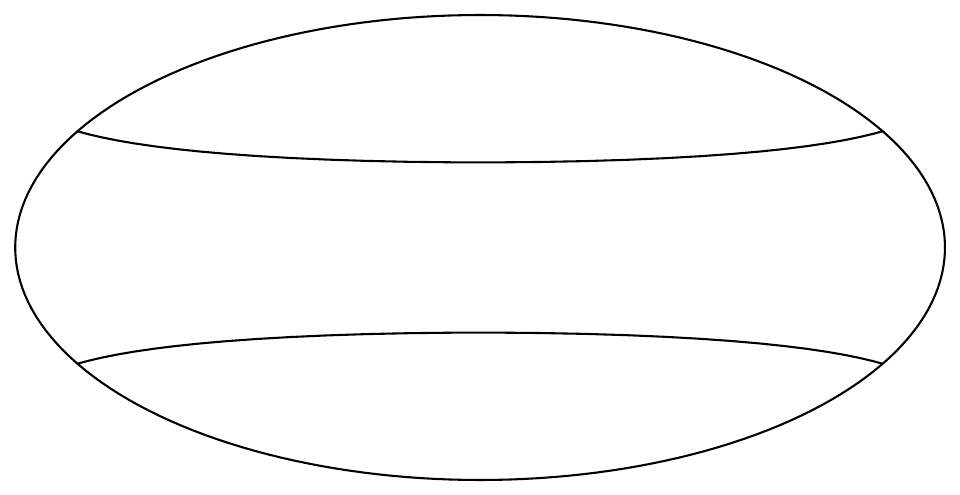}
\includegraphics[width=\nodalwidth]{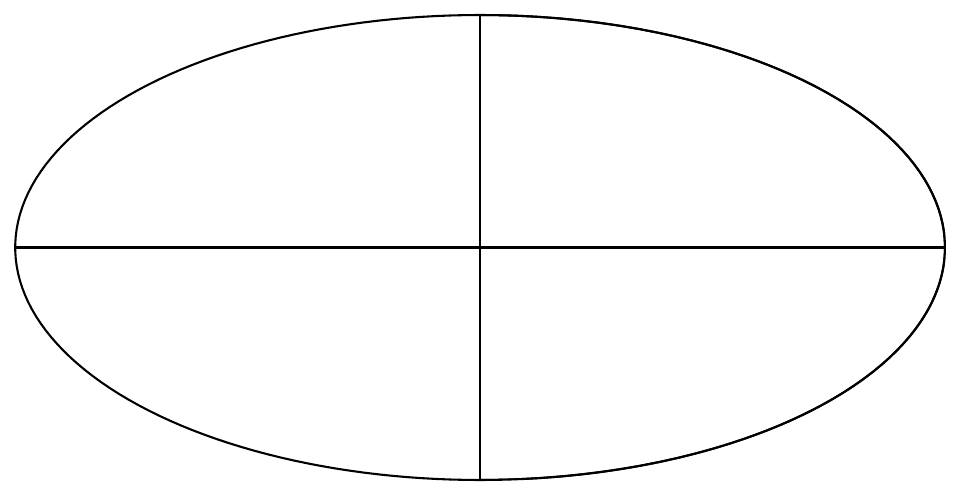}
\includegraphics[width=\nodalwidth]{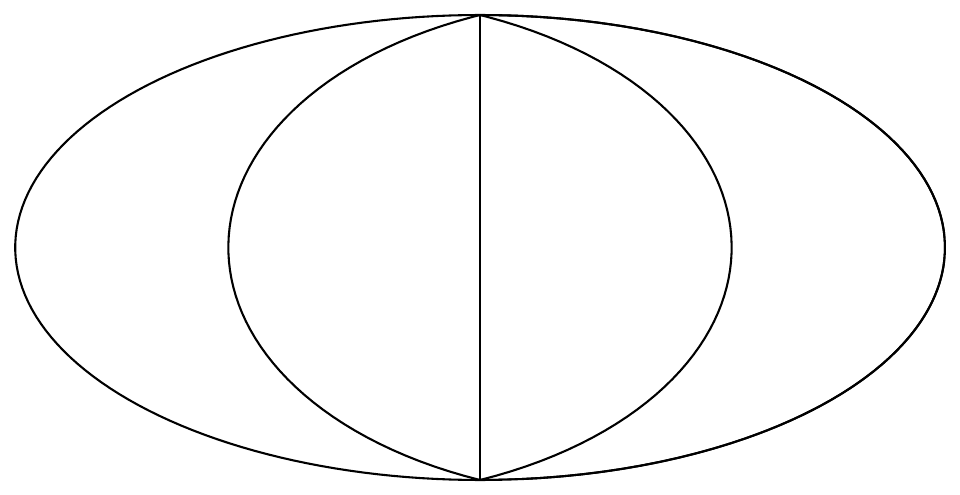}\\
\includegraphics[width=\nodalwidth]{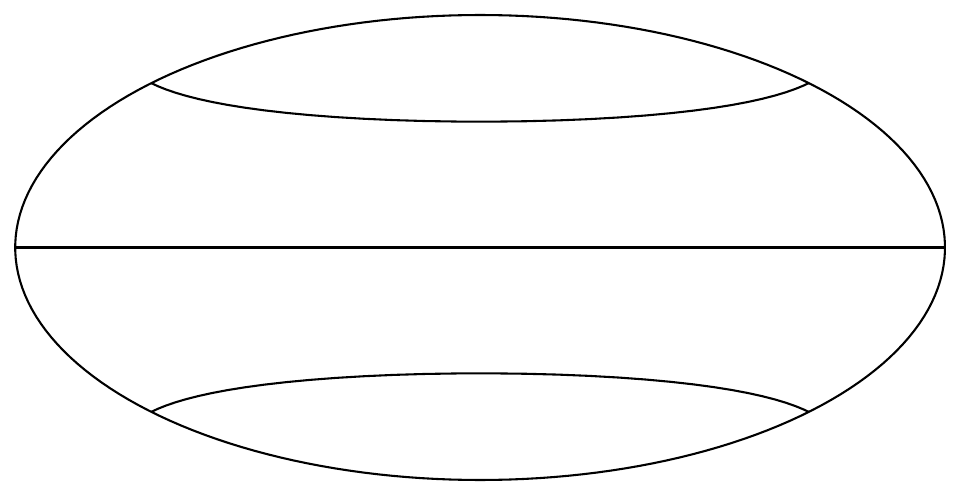}
\includegraphics[width=\nodalwidth]{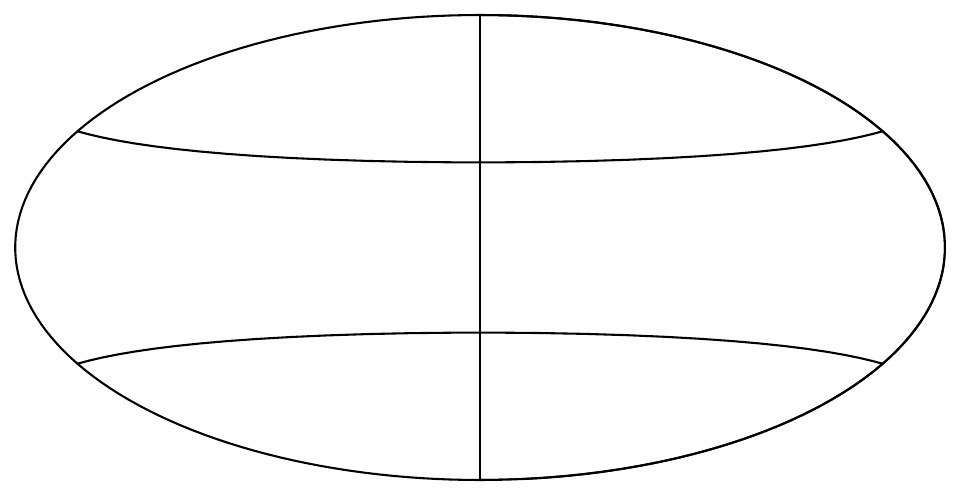} 
\includegraphics[width=\nodalwidth]{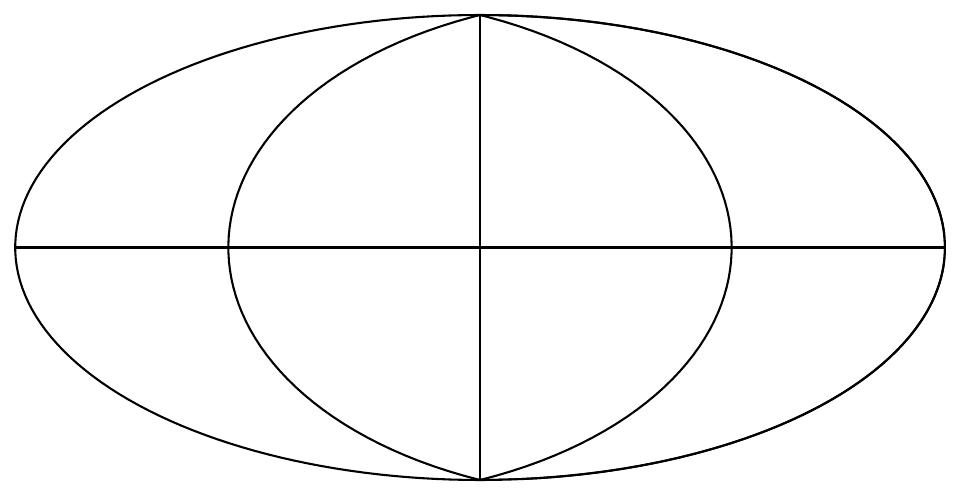}
\includegraphics[width=\nodalwidth]{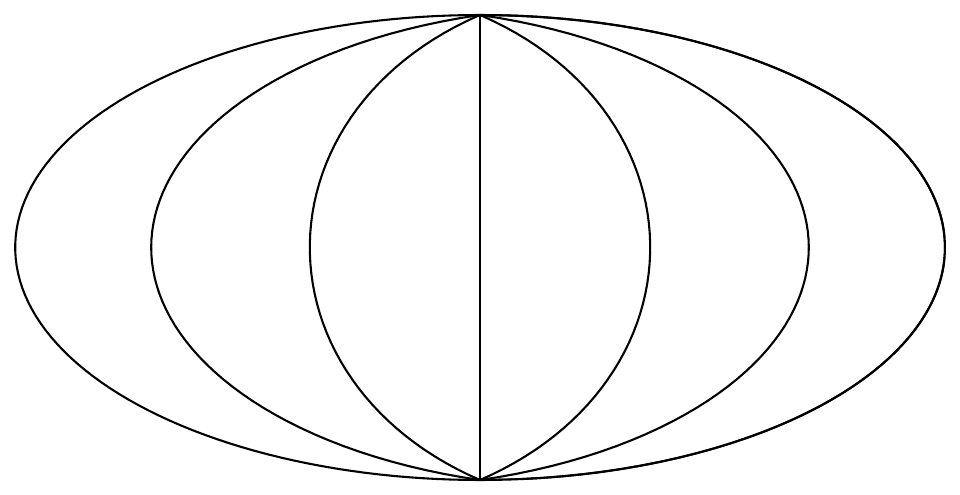}
\caption{Nodal lines separating excess and deficit regions of sky for various $(l, m)$ pairs.
The top row shows the $(0, 0)$ monopole, and the partition of the sky into two dipoles, $(1, 0)$ and $(1, 1)$.
The middle row shows the quadrupoles $(2, 0)$, $(2, 1)$, and $(2, 2)$.
The bottom row shows the $l=3$ partitions, $(3, 0)$, $(3, 1)$, $(3, 2)$, and $(3, 3)$~\cite{Denton:2014nga}.}
\label{fig:nodal}
\end{figure*}

The lowest multipole is the $l=0$ monopole, equal to the average
full-sky flux and is fixed by normalization (\ref{cmb-normalization}).
The higher multipoles ($l\ge 1$) and their amplitudes $a_{l m}$
correspond to anisotropies.  A nonzero $m$ corresponds to $2\,|m|$ longitudinal ``slices'' ($|m|$ nodal meridians).
There are $l+1-|m|$ latitudinal ``zones'' ($l-|m|$ nodal latitudes).
In Fig.~\ref{fig:nodal}
we show the partitioning described by some low multipole moments.\\

{\bf EXERCISE 7.3}~At every point in the sky, one observes a blackbody
spectrum, with temperature $T(\theta)$. The largest anisotropy is in
the $l = 1$ (dipole) first spherical harmonic, with amplitude $3.355
\pm 0.008~{\rm mK}$~\cite{Hinshaw:2008kr}. The dipole is interpreted to be the result of the
Doppler shift caused by the solar system motion relative to the nearly
isotropic blackbody field, as broadly confirmed by measurements of the
radial velocities of local galaxies. Show that the motion of an
observer with velocity $\beta = v/c$ relative to an isotropic
Planckian radiation field of temperature $T_0$ produces a
Doppler-shifted temperature pattern
\begin{equation} 
T(\theta) \approx T_0 \left[ 1 + \beta \cos \theta + \frac{\beta^2}{2} \cos
  (2 \theta) + {\cal O} (\beta^3) \right] \, .
\label{thetadeTq}
\end{equation}

\vspace{0.2in}

It is easily seen that the $a_{lm}$ coefficients are frame-dependent.
Note that a simple rotation in the $\phi$ coordinate will change the
$\sin\phi,\cos\phi$ part of the spherical harmonic for $m\neq0$ and a
rotation in the $\theta$ coordinate will change the associated
Legendre polynomial part for $l\neq0$.  So only the $\ell=m=0$
monopole coefficient is coordinate independent.  To combat this
problem, we use the power spectrum defined by
\begin{equation}
C_l\equiv\frac1{2l+1}\sum_{m=-l}^l a_{l m}^2\, .
\label{eq:power spectrum}
\end{equation}
A brief $C_{l}$ initiation is provided in Fig.~\ref{maps-ps}.\\

{\bf EXERCISE 7.4}~Show that the power spectrum $C_l$ is invariant
under  rotations.\\

\begin{figure}
 \postscript{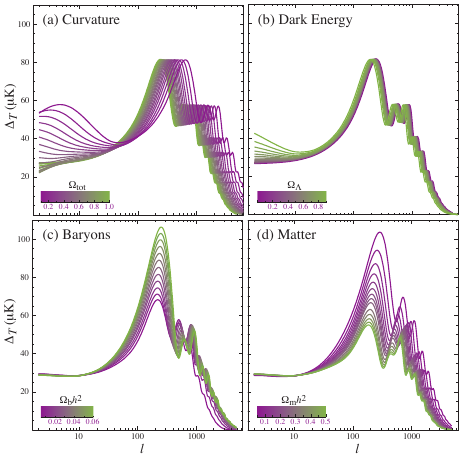}{0.99}
 \caption{Influence of several cosmological parameters on the angular
   power spectrum of the CMB~\cite{Kachelriess}.}
\label{ps}
\end{figure}

\begin{figure*}
 \postscript{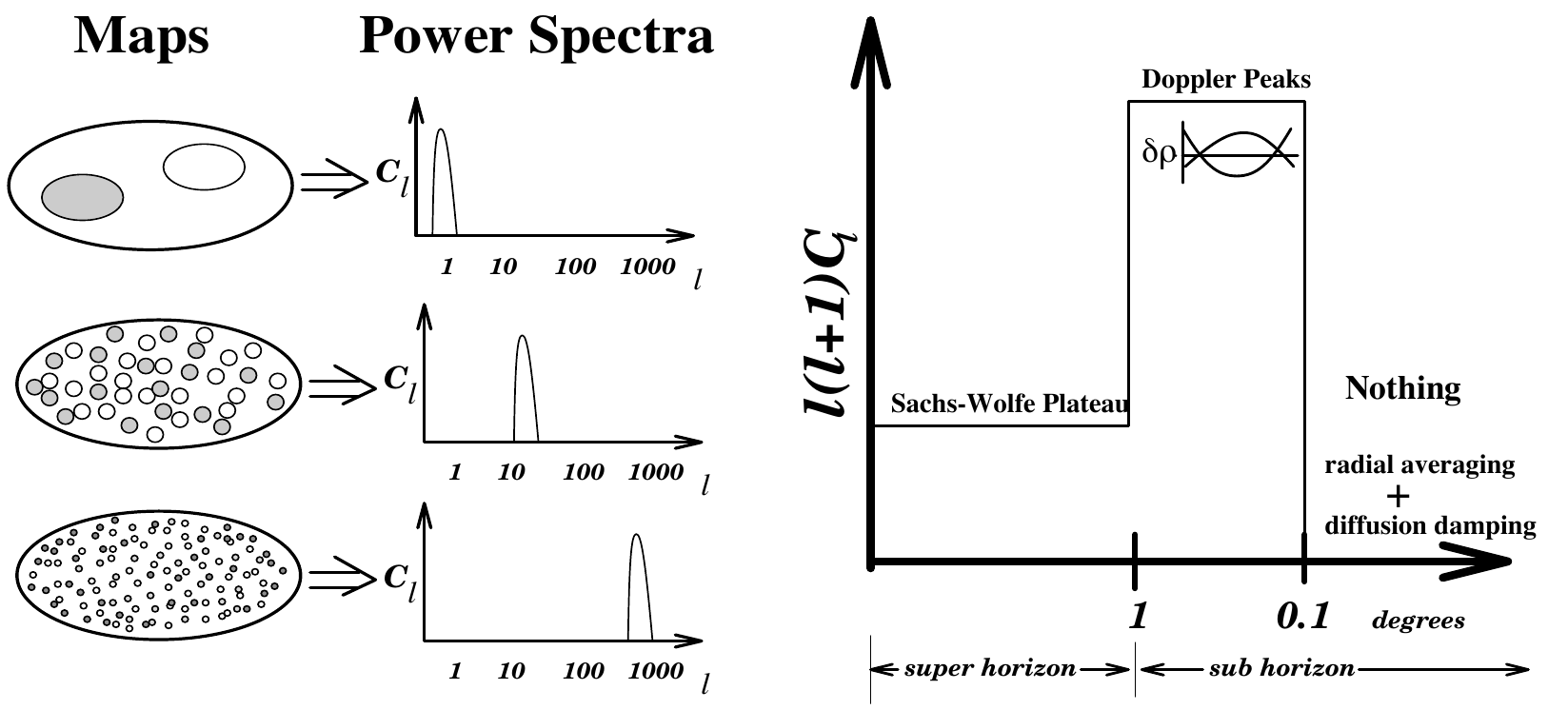}{0.99}
 \caption{{\bf Left panel.}   Illustrative
   sky maps and their angular power spectra. If a full-sky CMB map has
   only a dipole (top), its power spectrum is a delta function at $l =
   1$. If a map has only temperature fluctuations on an angular scale
   of $\sim 7^\circ$ (middle) then all of the power is at $l \sim
   10$. If all the hot and cold spots are even smaller (bottom) then
   the power is at high $l$. {\bf Right panel} Simplified CMB power
   spectrum. The CMB power spectrum can be crudely divided into three
   regions. The Sachs-Wolfe Plateau caused by the scale independence
   of gravitational potential fluctuations which dominate the spectrum
   at large super-horizon scales. The horizon is the angular scale
   corresponding to $ct_{\rm ls}$. The Doppler peaks on
   scales slightly smaller than the horizon are due to resonant
   acoustic oscillations. At smaller scales there is nothing because the
   finite thickness of the surface of last scattering averages small
   scale fluctuations along the line of sight. Diffusion damping
   (photons diffusing out of small scale fluctuations) also suppresses
   power on these scales~\cite{Lineweaver:1996tk}.}
\label{maps-ps}
\end{figure*}

To get a rough understanding of the power spectrum we can divide up
the multipole representation into super-horizon and sub-horizon
regions as shown in Fig.~\ref{maps-ps}. The angular scale
corresponding to the particle horizon size is the boundary between
super- and sub-horizon scales. The size of a causally connected region
on the surface of last scattering is important because it determines
the size over which astrophysical processes can occur. Normal physical
processes can act coherently only over sizes smaller than the particle
horizon. The relative size of peaks and locations of the power spectrum 
gives information about cosmological
parameters~\cite{Lineweaver:1996tk}. In Fig.~\ref{ps} we show the
influence of several cosmological parameters on the power
spectrum. For historical reasons, the quantity usually used in the
multipole representation is 
\begin{equation}
\Delta_T \equiv \left[ \frac{l (l+1)}{2 \pi}  C_l \right]^{1/2} \, .
\end{equation}

As an illustration, we sketch how to use the power spectrum to
determine the curvature of space.
At recombination the universe is already matter-dominated, so we can
substitute $z_{\rm ls} \simeq 1100$ into (\ref{cientoochentayocho})
to give an estimate of the horizon distance at the CMB epoch
\begin{equation}
d_{\rm h, ls} = \frac{2c}{H_0 (1 + z)^{3/2}} \approx 0.23~{\rm Mpc} \, .
\label{chaisp}
\end{equation}
This is the linear diameter of the largest causally connected region
observed for the CMB, $\ell_{\rm ls}$. Therefore, substituting
(\ref{chaisp}) into (\ref{patocUNO}) we find today's angular diameter
of this region in the sky
\begin{equation}
\theta = \frac{1}{(1+z)^{1/2}  - 1} = 0.03 \approx 1.8^\circ \, .
\label{angfound}
\end{equation}
The reason for this ``causality problem'' is that the universe expands
slower than light travels. Namely, as we have seen, when the age of the universe increases the
part observable to us increases linearly, $\propto ct$, while the scale factor
increases only with $t^{2/3}$ (or $t^{1/2}$). Thus we see more and more regions
that were never in causal contact for a radiation or matter-dominated
universe. We note that the sound horizon has approximately the same angular size,
because of $v_s \sim c/\sqrt{3}$. The sound horizon serves as a ruler at fixed
redshift $z_{\rm ls}$ to measure the geometry of spacetime. Moreover, the
fluid of photons and nucleons performs acoustic oscillations with
its fundamental frequency connected to the sound horizon plus higher
harmonics. The relative size of peaks and locations then gives information
about cosmological parameters.  The first panel of Fig.~\ref{ps} shows that, for a flat
universe ($\Omega_{\rm tot} \approx 1$, the first peak sits at $\theta \approx 1^\circ$ as we have
found in our simple estimate (\ref{angfound}). In Fig.~\ref{ps} we display a compilation of measurements of the CMB
angular power spectrum. The data agree with
high significance with models when they input dark energy as providing
$\approx 70\%$ of the energy in the universe, and when the total energy density
$\rho$ equals the critical density. The data also indicate that the
amount of normal baryonic matter in the universe $\Omega_b$ is only 4\% of the
critical density. What is the other 96\%?

\begin{figure}
 \postscript{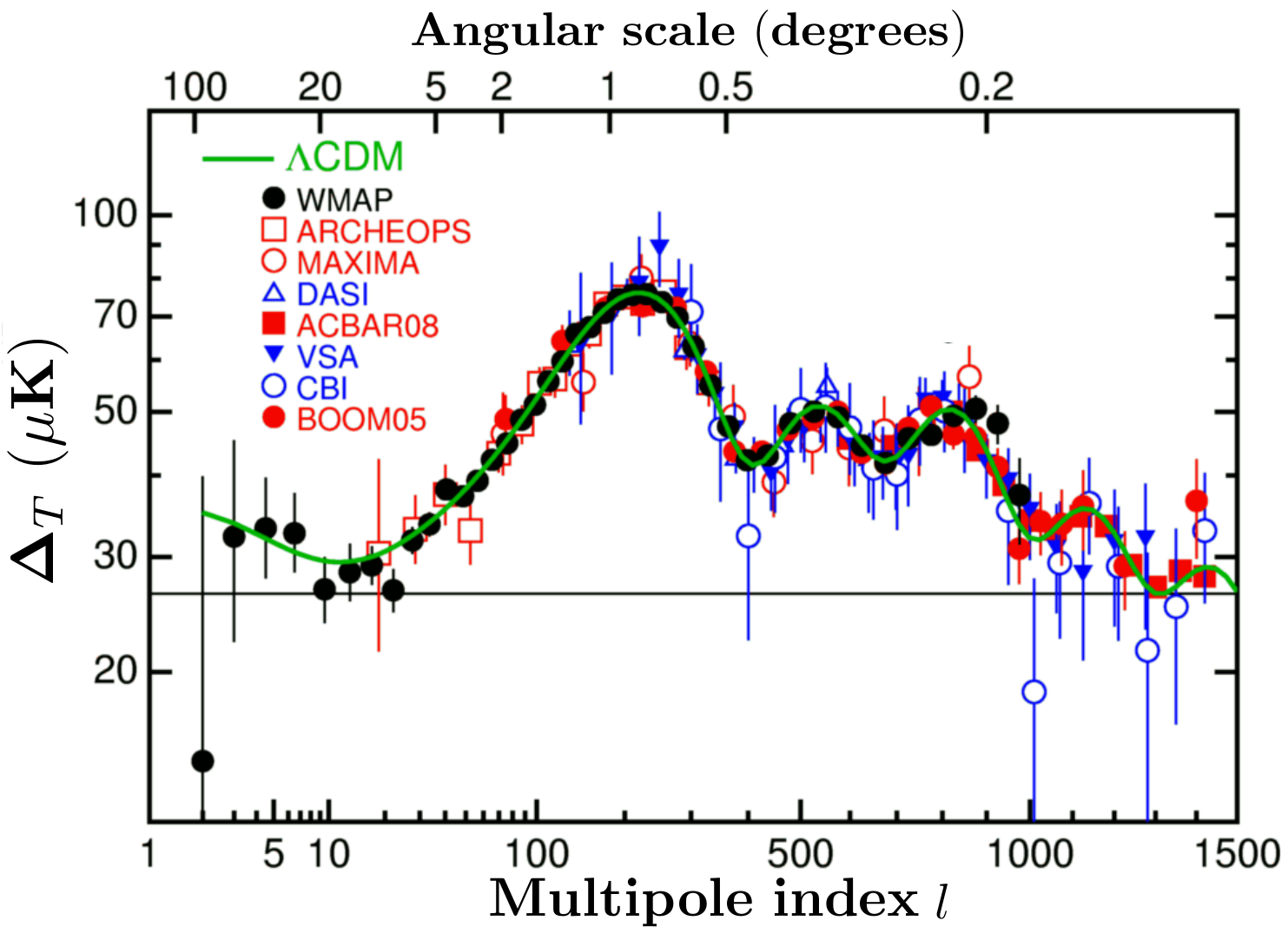}{0.99}
 \caption{A compilation of measurements of the CMB angular power
   spectrum spanning the region $2\alt l \alt 1500$. The best fit
   of the $\Lambda$CDM model is also shown.}
\label{ps-data}
\end{figure}

\begin{figure*}[tbp] \postscript{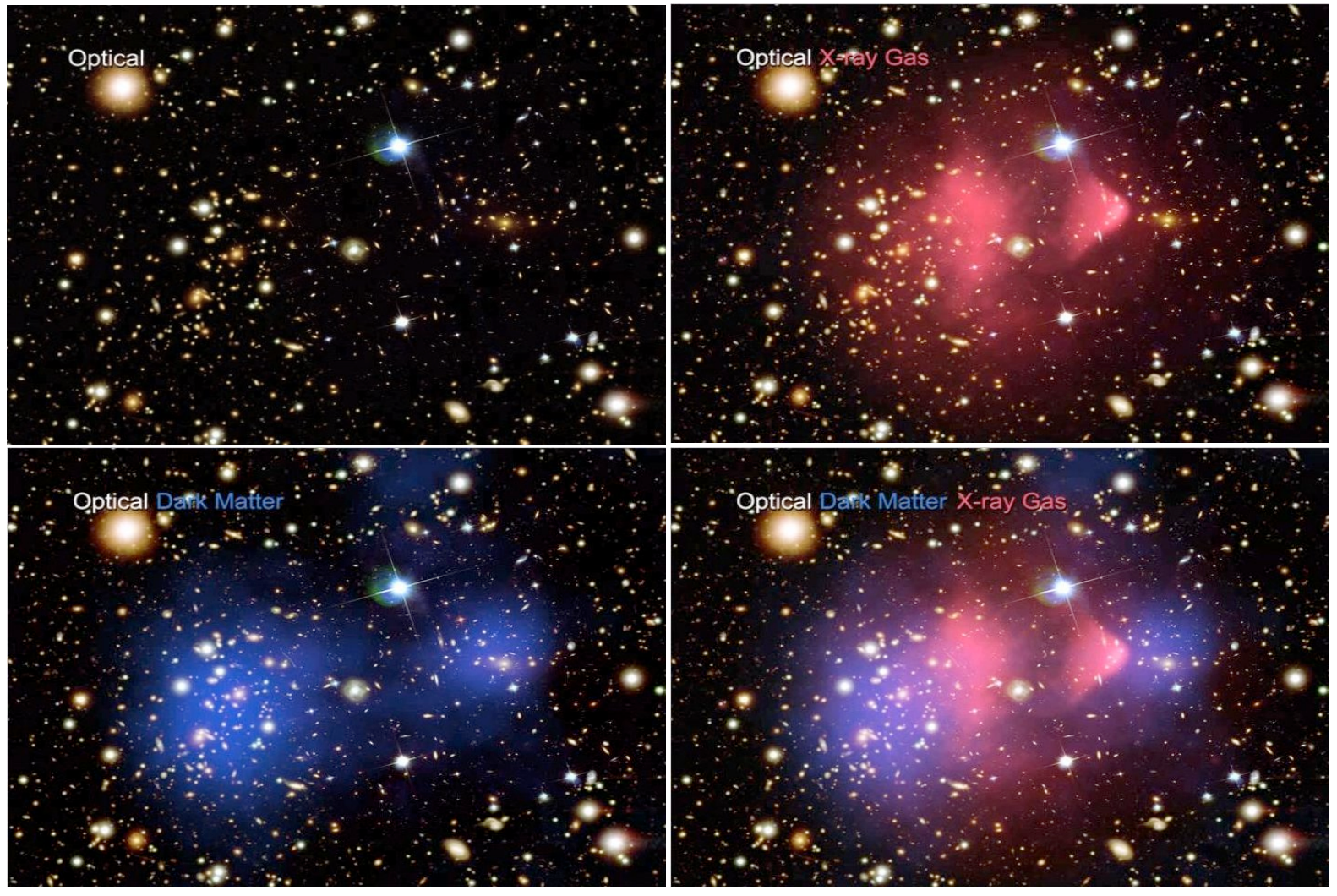}{0.9} \caption{The Bullet Cluster.}
\label{fig:BC}
\end{figure*}

There is a strong astrophysical evidence for a significant amount of
nonluminous matter in the universe referred to as cold dark matter
(CDM). For example, observations of the rotation of galaxies suggest
that they rotate as they had considerably more mass than we can
see~\cite{Rubin:1970zza,Rubin:1980zd,Rubin:1985ze}.  Similarly,
observations of the motions of galaxies within clusters also suggest
they have considerably more mass than can be
seen~\cite{Zwicky:1933gu}.  The most compelling evidence for CDM is
that observed at the Bullet Cluster~\cite{Clowe:2006eq}. In
Fig.~\ref{fig:BC} we show a composite image of the Bullet Cluster (1E
0657-558) that shows the X-ray light detected by Chandra in purple,
(an image from Magellan and the Hubble space telescope of) the optical
light in white and orange, and the CDM map (drwan up using data on
gravitational lensing from Magellan and European Space Observatory
telescopes at Paranal) in blue.  Galaxy clusters contain not only the
galaxies ($\sim 2\%$ of the mass), but also intergalactic plasma
($\sim 10\%$ of the mass), and (assuming the null hypothesis) CDM
($\sim 88\%$ of the mass). Over time, the gravitational attraction of
all these parts naturally push all the parts to be spatially
coincident. If two galaxy clusters were to collide/merge, we will
observe each part of the cluster to behave differently. Galaxies will
behave as collisionless particles but the plasma will experience ram
pressure. Throughout the collision of two clusters, the galaxies will
then become separated from the plasma. This is seen clearly in the
Bullet Cluster, which is undergoing a high-velocity (around 4500 km/s)
merger, evident from the spatial distribution of the hot, X-ray
emitting gas.  The galaxies of both concentrations are spatially
separated from the (purple) X-ray emitting plasma. The CDM clump
(blue), revealed by the weak-lensing map, is coincident with the
collisionless galaxies, but lies ahead of the collisional gas.  As the
two clusters cross, the intergalactic plasma in each cluster interacts
with the plasma in the other cluster and slows down. However, the dark
matter in each cluster does not interact at all, passing right through
without disruption. This difference in interaction causes the CDM to
sail ahead of the hot plasma, separating each cluster into two
components: CDM (and colissionless galaxies) in the lead and the hot
interstellar plasma lagging behind.  What might this nonluminous
matter in the universe be? We do not know yet. It cannot be made of
ordinary (baryonic)
matter, so it must consist of some other sort of elementary particle~\cite{Feng:2010gw} .\\

 {\bf EXERCISE 7.5}~We will examine galaxy rotation curves and show that
  they imply the existence of dark matter.  {\it (i)}~Recall that the
  orbital period ${\cal T}$ is given by ${\cal T}^2 = 4\pi^2a^3/GM$. Write down an
  expression that relates the orbital period and the orbital velocity
  for a circular orbit, and then write down an expression that relates
  the orbital velocity with the mass enclosed within $R$. {\it
    (ii)}~The Sun is 8~kpc from the center of the Milky Way, and
  its orbital velocity is $220~{\rm km/s}$. Use your expression from
  (i) to determine roughly how much mass is contained in a sphere
  around the center of the Milky Way with a radius equal to 8~kpc?
  {\it (iii)}~Assume that the Milky Way is made up of only luminous
  matter (stars) and that the Sun is at the edge of the galaxy (not
  quite true, but close). What would you predict the orbital velocity
  to be for a star 30~kpc  from the center? and for 100~kpc?
{\it (iv)}~Observations show that galaxy rotation curves are flat:
stars move at the same orbital velocity no matter how far they are
from the center. How much mass is actually contained within a sphere
of radius 30~kpc? 100~kpc? Take the orbital velocity at these
radii to be the same as the orbital velocity of the Sun. {\it
  (v)}~What do you conclude from all of this about the contents of our
galaxy? 

\subsection{$\bm{\Lambda}$CDM}

\begin{figure}
 \postscript{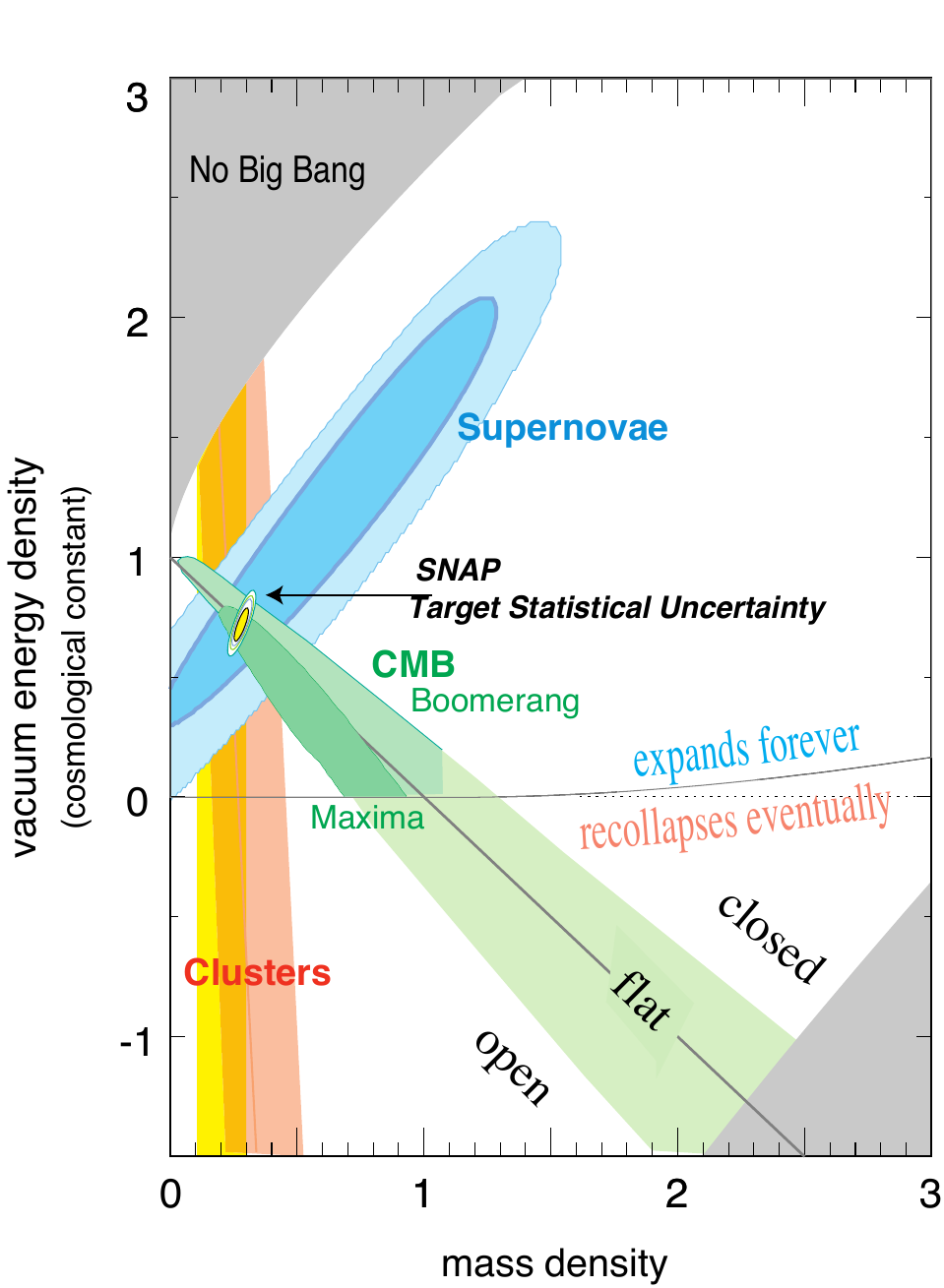}{0.99}
 \caption{Shown are three independent measurements of the cosmological
   parameters $(\Omega_\Lambda, \Omega_m)$. The high-redshift
   supernovae~\cite{Knop:2003iy}, galaxy cluster
   abundance~\cite{Allen:2002sr} and the CMB~\cite{Lange:2000iq,Balbi:2000tg}
   converge nicely near $\Omega_\Lambda = 0.7$ and $\Omega_m = 0.3$,
   as shown by the 68.3\%, 95.4\%, and 99.7\% confidence regions. The
   upper-left shaded region, labeled ``no {\it big bang},'' indicates
   bouncing cosmologies for which the universe has a turning point in
   its past~\cite{Carroll:1991mt}. The lower right shaded region
   corresponds to a universe which is younger than long-lived
   radioactive isotopes~\cite{Meyer:1986pw}, for any value of $H_0
   \geq 50~{\rm km}\ {\rm s}^{-1}\ {\rm Mpc}^{-1}.$ Also shown is the
   expected confidence region allowed by the future SuperNova /
   Acceleration Probe (SNAP) mission~\cite{Aldering:2002dp}.}
\label{28old}
\end{figure}

The concordance model of cosmology predicts the evolution of a
spatially flat expanding Universe filled with dark energy, dark
matter, baryons, photons, and three flavors of left- handed (that is, one
helicity state $\nu_L$) neutrinos (along with their right-handed
antineutrinos $\overline \nu_R$. The best fit to the most recent data
from the Planck satellite yields the following parameters:
$\Omega_{m,0} = 0.308 \pm 0.013$, $\Omega_{b,0} h^2 = 0.02234 \pm
0.00023$, $\Omega_{{\rm CDM},0} h^2 =0.1189 \pm 0.0022$, $h = 0.678
\pm 0.009$, and $1 - \Omega_0 < 0.005$~\cite{Ade:2015xua}. Note, however,
that the data only measure accurately the acoustic scale, and the
relation to underlying expansion parameters (e.g., via the
angular-diameter distance) depends on the assumed cosmology, including
the shape of the primordial fluctuation spectrum. Even small changes
in model assumptions can change $h$ noticeably. Unexpectedly, the
$H_0$ inference from Planck data deviates by more than $2\sigma$ from
the previous result from the maser-cepheid-supernovae distance ladder
$h = 0.738 \pm 0.024$~\cite{Riess:2011yx}. In what follows we will
take as benchmark: $\Omega_\Lambda \simeq 0.7$, $\Omega_{m,0} \simeq
0.3$, and $1 - \Omega_0 \simeq 0$, and $h \simeq 0.7$. As shown in
Fig.~\ref{28old}, this set of parameters is in good agreement with
cosmological and astrophysical observations. \\

  {\bf EXERCISE 7.6}~The Sun is moving around the center of the Milky
  Way galaxy along a roughly circular orbit at radius $R = 8~{\rm
    kpc}$, with a velocity $v = 220~{\rm km \, s}^{-1}$. Let us
  approximate the mass distribution and gravitational potential of the
  galaxy as spherically symmetric.  {\it (i)}~What is the total mass
  inside $R$?  {\it (ii)}~If the mass density varies with radius as
  $\rho \propto r^{-2}$, then what is the density at radius $R$?
  Express it in units of proton masses per ${\rm cm}^3$ (the proton mass is
  $m_p = 1.672 \times 10^{-24}~{\rm g}$).  {\it (iii)}~For the benchmark
  cosmological model, with $\Omega_\Lambda \simeq 0.7$ and $H_0 \simeq 70~{\rm
    km \ s^{-1} \ Mpc^{-1}}$, what
  is the density of the cosmological constant (or dark energy
  accounting for it)? Express it in the same units as in the previous
  question.  {\it (iv)}~The dark energy is spread out uniformly in the
  universe and it causes a gravitational repulsion which accelerates
  the expansion of the universe. Do you think the cosmological
  constant may be strongly affecting the dynamics of stars in our
  galaxy?\\

  {\bf EXERCISE 7.7}~Submillimeter Galaxies (SMGs), are extremely
  dusty starburst galaxies that were discovered at high redshifts ($z \sim
  1$  to 3). Assume the dust emission from SMGs are well characterized
  by blackbodies at a single dust temperature. If the observed
  spectrum of a SMG peaks at $180~\mu {\rm m}$, what would be its dust
  temperature if it is at a redshift of $z = 2$?\\

\begin{figure}
 \postscript{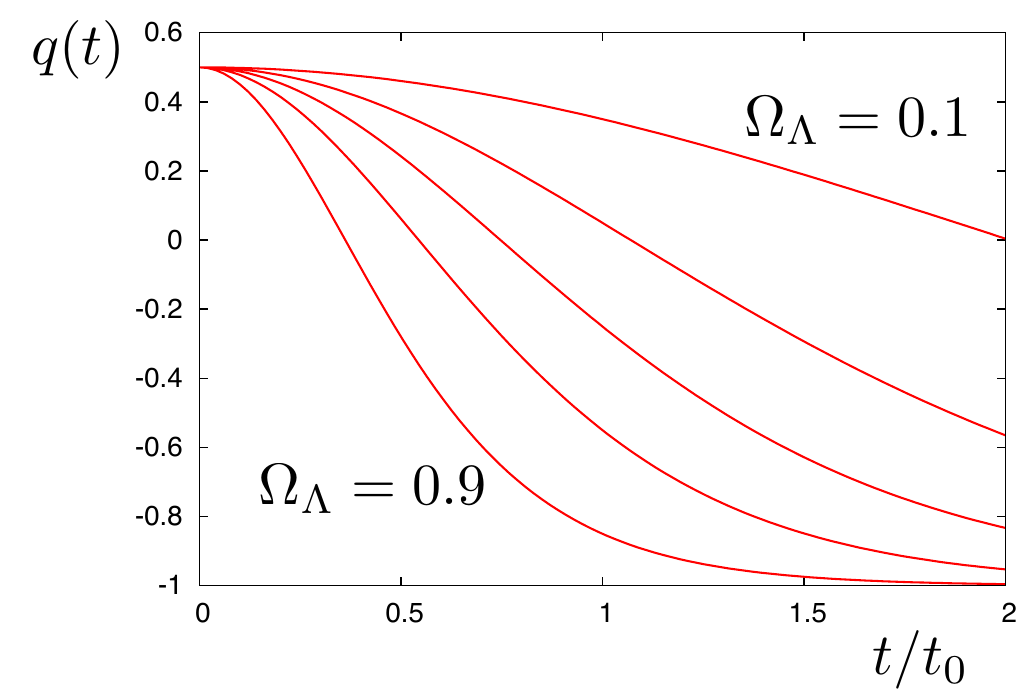}{0.99}
 \caption{The deceleration parameter $q$ as a function of $t/t_0$ for
   the $\Lambda$CDM model for various values of $\Omega_\Lambda = 0.1,\,
   0.3,\,  0.5,\, 0.7,\, 0.9$ from top to bottom~\cite{Kachelriess}.}
\label{deceleration}
\end{figure}

We now consider the benchmark model containing as its only two
components pressure-less matter and a cosmological constant,
$\Omega_{m,0} + \Omega_\Lambda =1$. Hence, the curvature term in the
Friedmann equation and the pressure term in the acceleration equation
play no role. Multiplying the acceleration equation (\ref{acceeqLambda}) by 2 and
adding it to the Friedmann equation (\ref{FriedmannLambda}), we eliminate $\rho_m$,
\begin{equation}
2 \frac{\ddot a}{a} + \left(\frac{\dot a}{a} \right)^2 = \Lambda c^2\, .
\end{equation}
Next, we rewrite first the left-hand-side and then the right-hand-side
as total time derivatives. Using
\begin{equation}
\frac{d}{dt} (a \dot a^2) = \dot a^3 + 2 a \dot a \ddot a = \dot a a^2
\left[\left(\frac{\dot a}{a} \right)^2 + 2 \frac{\ddot a}{a} \right]
\,,
\end{equation}
it follows that
\begin{equation}
\frac{d}{dt} ( a \dot a^2) = \dot a a^2 \Lambda c^2 = \frac{\Lambda c^2}{3}
\frac{d}{dt} (a^3)  \, .
\end{equation}
Integration is now trivial,
\begin{equation}
a \dot a^2 = \frac{\Lambda c^2}{3} a^3 + {\cal C} \, .
\label{terce}
\end{equation}
The integration constant,  ${\cal C} = 8 \pi G
\rho_{m,0}/3$, can be determined most easily by setting
$a(t_0) = 1$ and comparing (\ref{terce}) to the Friedmann equation
(\ref{FriedmannLambda}), with  $t = t_0$. Now, we introduce the new variable $x = a^{3/2}$ such that
\begin{equation}
\frac{da}{dt} = \frac{dx}{dt} \frac{da}{dx} = \frac{dx}{dt} \frac{2
  x^{-1/3}}{3} \, ,
\end{equation}
and (\ref{terce})  becomes
\begin{equation}
\dot x^2 - \frac{3}{4} \Lambda c^2 x^2 + \frac{9}{4} {\cal C} = 0 \, .
\end{equation}
Using an educated guess,
\begin{equation}
x(t) = A \sinh(\sqrt{3 \Lambda c}   t/2) \,,
\end{equation}
we fix $A = \sqrt{3 {\cal C}/\Lambda} c$. The  scale
factor is then
\begin{equation}
a(t) = A^{2/3} \sinh^{2/3} (\sqrt{3\Lambda c^2} t/2) \, .
\end{equation}
The time-scale of expansion is driven by $t_\Lambda = 2/\sqrt{3\Lambda
  c^2}$. The present age of the universe $t_0$ follows from the
normalization condition $a(t_0) = 1$ and is given by
\begin{equation}
t_0 = t_\Lambda \tanh^{-1} (\sqrt{\Omega_\Lambda}) \, .
\end{equation}
The deceleration,
\begin{equation}
q = - \frac{\ddot a}{a H^2},
\end{equation}
 is a key parameter for observational tests of the $\Lambda$CDM model. We
calculate first the Hubble parameter
\begin{equation}
H(t) = \frac{\dot a}{a} = \frac{2 }{3 t_\Lambda} \coth (t/t_\Lambda) \,,
\end{equation}
and after that
\begin{equation}
q(t) = \frac{1}{2} \left[ 1 - 3 \tanh^2 (t/t_\Lambda)\right] \, .
\end{equation}
Note that, as expected, for $t \to 0$ we have $q =1/2$, and for  $t \to
\infty$ we have $q=-1$.  Perhaps more interesting is the
transition region from a decelerating to an accelerating universe. As
shown in Fig.~\ref{deceleration} for $\Omega_\Lambda = 0.7$, this
transition takes place at $t \approx
0.55 \,  t_0$. This can be easily converetd to a redshift: $z_* =
a(t_0)/a(t_*) - 1 \approx 0.7$. Interestingly, $z_*$ can be directly probed by SNe Ia observations.\\

{\bf EXERCISE 7.8}~Consider the benchmark model, with $\Omega_{m,0} \simeq
0.3$, $\Omega_{\Lambda} \simeq 0.7$, with flat space geometry. What was
  the redshift at which the universe had half its present age?

\section{Hot thermal Universe}

Though we can {\it see} only as far as the surface of last scattering,
in recent decades a convincing theory of the origin and evolution of
the {\it early universe} has been developed. Most of this theory is
based on recent theoretical and experimental advances in elementary
particle physics. Hence, before continuing our look back through time,
we make a detour to overview the generalities of the standard
  model of particle physics.\footnote{You can find a more extensive
    but still qualitative discussion
    in~\cite{Anchordoqui:2015uww}. For a more rigorous treatment  see e.g.~\cite{Halzen:1984mc,Barger:1987nn,Quigg:2013ufa,Anchordoqui:2009eg}.}

\subsection{$\bm{SU(3)_C \otimes SU(2)_L \otimes U(1)_Y}$}

The {\it standard model} (SM) is our most modern attempt to answer
two simple questions that have been perplexing (wo)mankind
throughout the epochs: What is the Universe made of? Why is our world
the way it is?

\begin{table*}
\caption{The three generations of quarks and leptons in the Standard Model. \label{table:ql}}
\begin{tabular}{ccccccc}
\hline
\hline 
~~~~~~~~~~~~~~~~~~~~~~~~~~~~~~~~~& ~~~~Fermion~~~~ & ~~~~Short-hand~~~~ & ~~~~Generation~~~~ & ~~~~Charge~~~~ & ~~~~Mass~~~~ & ~~~~Spin~~~~ \\
\multirow{7}{4em}{Quarks} & up & $u$ & I & & $2.3^{+0.7}_{-0.5}~{\rm MeV}$ & \\
& charm & $c$ & II & $+\frac{2}{3}$ & $1.275 \pm 0.025~{\rm GeV}$ & $\frac{1}{2}$ \\
& top & $t$ & III & & $173.21\pm 0.51~{\rm GeV}$ & \\
\cline{4-7}
& down & $d$ & I &  & $4.8^{+0.5}_{-0.3}~{\rm MeV}$ & \\
& strange & $s$& II & $-\frac{1}{3}$ & $95^\pm5~{\rm MeV}$ & $\frac{1}{2}$ \\
& bottom & $b$& III &  & $4.18\pm 0.03~{\rm GeV}$ & \\
\hline
\multirow{7}{4em}{Leptons} & electron neutrino & $\nu_e$ & I & & $<2~{\rm eV}$ 95\% CL & \\
& muon neutrino & $\nu_\mu$ & II & 0 & $<0.19~{\rm MeV}$ 90\% CL & $\frac{1}{2}$ \\
& tau neutrino & $\nu_\tau$ & III & & $< 18.2~{\rm MeV}$ 95\%CL \\
\cline{4-7}
& electron &$e$ & I & & 0.511~{\rm MeV} & \\
& muon & $\mu$ & II & $-1$ & $105.7~{\rm MeV}$ & $\frac{1}{2}$ \\
& tau & $\tau$ & III & & 1.777~{\rm GeV} \\
\hline
\hline
\end{tabular}
\end{table*}

\begin{table*}
\caption{The four force carriers. \label{tablegb}}
\begin{tabular}{cccccc}
\hline \hline
~~~~~~~~~Force~~~~~~~~~ & ~~~~~~~~~Boson~~~~~~~~~ & ~~~~~~~~~Short-hand~~~~~~~~~ & ~~~~~~~~~Charge~~~~~~~~~ & ~~~~~~~~~Mass~~~~~~~~~ & ~~~~~~~~~Spin~~~~~~~~~ \\
\hline
Electromagnetic & photon & $\gamma$ & 0 & 0 & 1\\
Weak & $W$ & $W^\pm$ & $\pm 1$ & $80.385 \pm 0.015~{\rm GeV}$ & 1\\
Weak & $Z$ & $Z^0$ & 0 & $91.1876 \pm 0.0021~{\rm GeV}$ & 1 \\
Strong & gluon & $g$ & 0 & 0 & 1 \\
Gravitation & graviton & $G$ & 0 & 0 & 2 \\
\hline
\hline
\end{tabular}
\end{table*}

The elementary-particle model accepted today views quarks and leptons
as the basic (pointlike) constituents of ordinary matter.  By
pointlike, we understand that quarks and leptons show no evidence of
internal structure at the current limit of our resolution. Presently,
the world's largest microscope is the Large Hadron Collider (or LHC),
a machine that collides beams of protons at a cenetr-of-mass energy
$\sqrt{s} = 13~{\rm TeV}$. Remarkably, 70\% of the energy carried into
the collision by the protons emerges perpendicular to the incident
beams. At a given transverse energy $E_\perp$, we may roughly estimate
the LHC resolution as
\begin{eqnarray}
\ell_{\rm
  LHC}  &\approx & \hslash c/E_\perp \approx 2 \times 10^{-19}~{\rm TeV} \, {\rm
  m}/E_\perp \nonumber \\
& \approx & 2 \times 10^{-20}~{\rm m} \, .
\end{eqnarray}
There are six quarks and six leptons, together with their
antiparticles. These twelve elementary particles are all
spin-$\frac{1}{2}$ and fall naturally into three families or
generations.  Each generation consists of two leptons with electric
charges $Q = 0$ and $Q = -1$ and two quarks with $Q = + 2/3$ and $Q =
-1/3$. The masses of the particles increase significantly with each
generation, with the possible exception of the neutrinos~\cite{Agashe:2014kda}. The
properties of quarks and leptons are summarized in
Table~\ref{table:ql}. 

Now, an understanding of how the world is put together requires a
theory of how quarks and leptons interact with one
another. Equivalently, it requires a theory of the fundamental forces of
nature. Four such forces have been identified. They can be
characterized on the basis of the following four criteria: the types
of particles that experience the force, the relative strength of the
force, the range over which the force is effective, and the nature of
the particles that mediate the force.  Two of the forces, gravitation
and electromagnetism, have an unlimited range; largely for this reason
they are familiar to everyone. The remaining forces, which are called simply the
weak force and the strong force, cannot be perceived directly because
their influence extends only over a short range, no larger than the
radius of an atomic nucleus. The electromagnetic force is
carried by the photon, the strong force is mediated by gluons, the $W$
and $Z$ bosons transmit the weak force, and the quantum of the
gravitational force is called the graviton.  The main properties of the force carriers are summarized in Table~\ref{tablegb}.
A comparison of the (approximate) relative force strengths for two protons inside a nucleus is given in Table~\ref{fstrength}. Though gravity is the most obvious force in daily life, on a nuclear scale it is the weakest of the four forces and its effect at the particle level can nearly always be ignored.

In the SM quarks and  leptons are allotted several additive quantum
numbers: electric charge $Q$, lepton number $L= L_e + L_\mu + L_\tau$,
baryon number $B$, strangeness $s$, charmness $c$, bottomness $b$, and
topness $t$. For each particle additive quantum number $N$, the
corresponding antiparticle has the additive quantum number $-N$. 

The additive quantum numbers $Q$ and $B$ are assumed to be conserved
in strong, electromagnetic, and weak interactions. The lepton numbers
are not involved in strong interactions, but are strictly conserved in
both electromagnetic and weak interactions. The remainder, $s$, $c$,
$b$ and $t$ are strictly conserved only in strong and electromagnetic
interactions, but can undergo a change of one unit in weak
interactions.

The quarks have an additional {\it charge} which enables them to
interact strongly with one another. This {\it charge} is a three-fold
degree of freedom which has come to be known as
color~\cite{Fritzsch:1973pi}, and so the gauge theory
describing the strong interaction has taken on the name of quantum
chromodynamics (QCD).  Each quark
flavor can have three colors usually designated red, green, and
blue. The antiquarks are colored antired, antigreen, and
antiblue. Each quark or antiquark carries a single unit of color or
anticolor charge, respectively. The quanta of the color fields are
called gluons (as they glue the quarks together). There are eight
independent kinds of gluons in $SU(3)_C$, each of which carries a combination of a
color charge and an anticolor charge (e.g. red-antigreen). The strong
interactions between color charges are such that in nature the quarks
(antiquarks) are grouped into composites collectively called
hadrons~\cite{GellMann:1961ky,Ne'eman:1961cd,GellMann:1964nj}:
\begin{widetext}
\begin{equation}
\left\{ 
\begin{tabular}{llll} 
 $q \bar q$  &~~~~(quark + antiquark)~~~~&~~~~mesons~~~~& 
{\rm integral}\ {\rm spin} $\to$ {\rm Bose-Eisntein}\ {\rm statistics}~\cite{Bose:1924s,Einstein:1925s} \\
 $qqq$  &~~~~(three quarks)~~~~&~~~~baryons~~~~& 
   half-integral spin $\to$ Fermi-Dirac statistics~\cite{Fermi:1926s,Dirac:1926s} 
\end{tabular}
\right. . 
\label{GellM}
\end{equation}
\end{widetext}
In QCD each baryon, antibaryon, or meson
is colorless. However, these colorless particles may
interact strongly via residual strong interactions arising from their
composition of colored quarks and/or antiquarks. On the other hand the
colorless leptons are assumed to be structureless in the SM and
consequently do not participate in strong interactions. 

One may wonder what would happen if we try to see a single quark with
color by reaching deep inside a hadron. Quarks are so tightly bound to
other quarks that extracting one would require a tremendous amount of
energy, so much that it would be sufficient to create more quarks.
Indeed, such experiments are done at modern particle colliders and all
we get is not an isolated quark, but more hadrons (quark-antiquark
pairs or triplets). This property of quarks, that they are always
bound in groups that are colorless, is called confinement.  Moreover,
the color force has the interesting property that, as two quarks
approach each other very closely (or equivalently have high energy),
the force between them becomes small. This aspect is referred to as
asymptotic freedom~\cite{Gross:1973id,Politzer:1973fx}.

\begin{table}[t]
\caption{Relative force  strength for protons in  a nucleus.} 
\begin{tabular}{cc}
\hline
\hline
~~~~~~~~~~~Force~~~~~~~~~~~~&~~~~~~~~~~~~Relative Strength~~~~~~~~~~~~\\
\hline
Strong  & 1  \\
Electromagnetic & $10^{-2}$  \\
Weak  & $10^{-6}$ \\
Gravitational & $10^{-38}$  \\
\hline
\hline
\end{tabular}
\label{fstrength}
\end{table}

Before
proceeding we note that aside from
binding together quarks inside the hadrons, the strong force indirectly also binds
protons and neutrons into atomic nuclei. Such a nuclear force is
mediated by pions: spin-0 mesons  
with masses $m_{\pi^0} = 135.0~{\rm MeV}$ and $m_{\pi^\pm} = 139.6~{\rm MeV}$.

Electromagnetic processes between electrically charged particles are
mediated by massless neutral spin-1 photons. The interaction can be
described by a local $U(1)_{\rm EM}$ gauge theory called quantum
electrodynamics (QED). The symmetry properties of QED are unquestionably appealing~\cite{Schwinger:1948yk,Schwinger:1948yj,Tomonaga:1946zz,Feynman:1948ur,Dyson:1949bp,Dyson:1949ha,Feynman:1950ir}. Moreover,
QED has yielded results that are in agreement with experiment to an accuracy of about one part in a billion~\cite{Schwinger:1948iu}, which makes the theory the most accurate physical theory ever devised. It is the model for theories of the other fundamental forces and the standard by which such theories are judged.

Every quark and lepton of the SM interact weakly. The weak
interaction, mediated by the massive $W^+$, $W^-$ and $Z^0$ vector
bosons, fall into two classes: {\it (i)} charge-current (CC) weak
interactions involving the $W^+$ and $W^-$ bosons and {\it
  (ii)}~neutral current (NC) weak interactions involving the $Z^0$ boson. The CC
 interactions, acting exclusively on left-handed particles and
right-handed antiparticles, are described by a chiral $SU(2)_L$ local gauge
theory, where the subscript $L$ refers to left-handed particles
only.\footnote{A phenomenon is said to be chiral if it is not
  identical to its mirror image. The spin of a particle may be used to
  define a handedness for that particle.  The chirality of a particle
  is right-handed if the direction of its spin is the same as the
  direction of its motion. It is left-handed if the directions of spin
  and motion are opposite.}  On the other hand, the NC interactions act on both
left-handed and right-handed particles, similar to the electromagnetic
interactions. In fact the SM assumes that both the $Z^0$ and the
photon  arise from a mixing of two bosons, $W^0$ and $B^0$, via the
electroweak mixing angle $\theta_W$:
\begin{eqnarray}
\gamma & = & B^0 \cos \theta_W + W^0 \sin \theta_W \,, \nonumber \\
Z^0 & = & -B^0 \sin \theta_W + W^0 \cos \theta_W \, .
\end{eqnarray}
The electroweak interaction is described by a local gauge theory: $SU(2)_L \otimes U(1)_Y$,
where the hypercharge $U(1)_Y$ symmetry involves both left-handed and right-handed
particles~\cite{Glashow:1961tr,Weinberg:1967tq,Salam:1968rm}.  Experiment requires the masses of the weak gauge bosons $W$
and $Z$ to be heavy so that  weak interactions are very
short-ranged. The $W$ and $Z$ gauge bosons acquire masses through spontaneous
symmetry breaking $SU(2)_L \times U(1)_Y \to U(1)_{\rm EM}$. The
breaking of the symmetry triggers the Higgs mechanism~\cite{Higgs:1964pj,Englert:1964et}, which gives the relative masses of the $W$ and $Z$ bosons in terms of the
electroweak mixing angle,
\begin{equation}
M_W = M_Z \cos \theta_W \, ,
\end{equation}
while the photon remains massless.
In addition, by coupling originally massless fermions to the scalar
Higgs field, it is possible to produce the observed physical fermion
masses without violating the gauge invariance. 

The conspicuously well-known accomplishments of the $SU(3)_C \otimes
SU(2)_L \otimes U(1)_Y$ SM of strong and electroweak forces can be
considered as the apotheosis of the gauge symmetry principle to
describe particle interactions. Most spectacularly, the recent
discovery~\cite{ATLAS:2012ae,Chatrchyan:2012tx} of a new boson with
scalar quantum numbers and couplings compatible with those of a SM
Higgs has possibly plugged the final remaining experimental hole in
the SM, cementing the theory further.

In summary, the fundamental particles can be classified into spin-1/2
fermions (6 leptons and 6 quarks), and spin-1 gauge bosons ($\gamma$,
$W^\pm,$ $Z^0$, and $g$).  The leptons have 18 degrees of freedom:
each of the 3 charged leptons has 2 possible chiralities and its
associated anti-particle, whereas the 3 neutrinos and antineutrinos
have only one chirality (neutrinos are left-handed and antineutrinos
are right-handed). The quarks have 72 degrees of freedom: each of the
6 quarks, has the associated antiparticle, three different color
states, and 2 chiralities.  The gauge bosons have 27 degrees of
freedom: a photon has two possible polarization states, each massive
gauge boson has 3, and each of the eight independent types of gluon in
QCD has 2. The scalar spin-0 Higgs boson, with a mass $m_H \simeq
126~{\rm GeV}$, has 1 degree of freedom.\footnote{Recently,
  ATLAS~\cite{ATLAS} and CMS~\cite{CMS:2015dxe} announced the
  observation of a peak in the diphoton mass distribution around
  750~GeV, using (respectively) $3.2~{\rm fb}^{-1}$ and $2.6~{\rm
    fb}^{-1}$ of data recorded at a c.m. energy $\sqrt{s} =
  13~{\rm TeV}$.  The diphoton excesses could be interpreted as the
  decay products of a new massive particle $\digamma$, with spin 0, 2,
  or higher~\cite{Strumia:2016wys}.  Assuming a narrow width
  approximation ATLAS gives a local significance of $3.6\sigma$, or
  else a global significance of $2.0\sigma$ when the
  look-elsewhere-effect in the mass range $M_\digamma/{\rm GeV} \in [200 -
  2000]$ is accounted for. Signal-plus-background fits were also
  implemented for a broad signal component with a large decay
  width. The largest deviation from the background-only hypothesis
  corresponds to $M_\digamma \sim 750~{\rm GeV}$ with a total width
  \mbox{$\Gamma_{\rm total} \sim 45~{\rm GeV}$.} The local and global
  significances evaluated for the broad resonance fit are roughly 0.3
  higher than that for the fit using the narrow width approximation,
  corresponding to $3.9\sigma$ and $2.3\sigma$, respectively. The CMS
  data gives a local significance of $2.6\sigma$ and a global
  significance smaller than $1.2\sigma$. More recently, ATLAS and CMS
  updated their diphoton resonance
  searches~\cite{Delmastro,Musella,CMS:2016owr}. ATLAS reanalyzed the
  $3.2~{\rm fb}^{-1}$ of data, targeting separately spin-0 and spin-2
  resonances. For spin-0, the most significant deviation from the
  background-only hypothesis corresponds to $M_\digamma \sim 750~{\rm GeV}$
  and $\Gamma_{\rm total} \sim 45~{\rm GeV}$. The local significance
  is now increased to $3.9\sigma$ but the global significance remains
  at the $2 \sigma$ level. For the spin-2 resonance, both the local
  and global significances are reduced down to $3.6\sigma$ and
  $1.8\sigma$, respectively. The new CMS analysis includes additional
  data (recorded in 2015 while the magnet was not operated) for a
  total of 3.3~${\rm fb}^{-1}$. The largest excess is observed for
  $M_\digamma = 760~{\rm GeV}$ and $\Gamma_{\rm total} \approx 11~{\rm GeV}$,
  and has a local significance of $2.8\sigma$ for spin-0 and
  $2.9\sigma$ spin-2 hypothesis. After taking into account the effect
  of searching for several signal hypotheses, the significance of the
  excess is reduced to $< 1 \sigma$. CMS also communicated a combined
  search with data recorded at $\sqrt{s} = 13~{\rm TeV}$ and $\sqrt{s}
  = 8~{\rm TeV}$. For the combined analysis, the largest excess is
  observed at $M_\digamma = 750~{\rm GeV}$ and $\Gamma_{\rm total} = 0.1~{\rm
    GeV}$. The local and global significances are $\approx 3.4\sigma$
  and $1.6\sigma$, respectively. This could be the first observation of
  physics beyond the SM at the LHC.}

\subsection{Equilibrium thermodynamics}

The Universe we observe had its beginning in the {\it big bang}, the
cosmic firewall.  Because the early universe was to a good
approximation in thermal equilibrium, particle reactions can be
modeled using the tools of thermodynamics and statistical mechanics.
It will be helpfull then to take a second detour and revise some
concepts of statistical thermodynamics.

Consider a cubic box of volume $V$, and expand the fields inside into
periodic waves with harmonic boundary conditions. The density of
states in $k$-space is
\begin{equation}
dN = g \frac{V}{(2 \pi)^3} d^3k \,,
\end{equation}
where $g$ is a degeneracy factor and $k$ is the Fourier transform
wavenumber. The equilibrium phase space distribution (or occupancy)
function for a quantum state of energy $E$ is given by the familiar
Fermi-Dirac or Bose-Einstein distrubutions,
\begin{equation}
f = \frac{1}{e^{(E - \mu)/(kT)} \pm 1} \,,
\label{BEFD}
\end{equation}
where $T$ is the equilibrium temperature, $k$ is the Boltzmann
constant, $\mu$ is the chemical potential (if present), and $\pm$
corresponds to either Fermi or Bose statistics. Throughout we will
consider the case $|\mu| \ll T$ and neglect all chemical potentials
when computing total thermodynamic quantities. All evidence
indicates that this is a good approximation to describe particle
interactions in the super-hot primeval plasma~\cite{Kolb:1990vq} . 

The number density of a dilute weakly-interacting gas of particles in
thermal equilibrium with $g$ internal degrees of freedom is then
\begin{eqnarray}
n & = & \frac{1}{V} \int f  \, dN \nonumber \\
   & = & g \frac{1}{(2 \pi \hslash)^3} \int_0^\infty \frac{4 \pi p^2 dp}{e^{E/(kT)} \pm 1} \nonumber\\
& = & g \frac{1}{2 \pi^2 \hslash^3c^3} \int_{m c^2}^\infty \frac{(E^2 -
  m^2 c^3)^{1/2}}{e^{E/(kT)} \pm 1} E \,  dE \,,
\label{Ashley3}
\end{eqnarray}  
where in the second line we have changed to momentum space, $\vec p =
\hslash \vec k$, and in the third line we used the relativistic relation $E
= m^2c^4 + p^2c^2$.  The analogous expression for the energy density
is easily obtain since it is only necessary to multiply the integrand
in (\ref{Ashley3}) by a factor of $E$ for the energy of each mode,
\begin{equation}
\rho   =  \frac{g}{2 \pi^2 \hslash^3 c^3} \int_{m_i}^\infty  \frac{(E^2 -
  m^2c^4)^{1/2}}{e^{E/(kT)} \pm 1} \, E^2 \, dE  \, .
\end{equation}
Recalling that the pressure is the average value of the momentum
transfer $\langle p \rangle^2 c^2/E$ in a given direction, we have
\begin{equation}
P = \frac{g}{6 \pi^2 \hslash^3 c^3} \int_{m_i}^\infty  \frac{(E^2 - m^2c^4)^{3/2}}{e^{E/(kT)} \pm 1} \,  dE  \,,
\end{equation}
with the factor of $1/3$ associated with the assumed isotropy of the
momentum distribution. 

Let us now compute the above expressions in two asymptotic limits:
relativistic and non-relativistic particles, which will be sufficient
for our discussion of how the different particle species evolve in the
primeval plasma. For $kT \gg mc^2$, the particles behave as if they
were massless and the Bose-Einstein and Fermi-Dirac
distributions reduce to
\begin{equation}
f(y) = \frac{1}{e^y \pm 1} \,,
\end{equation}
where we have defined $y = |\vec p \, |/(kT)$. Using
\begin{equation}
\int_{0}^\infty \frac{z^{n-1}}{e^z - 1}\, dz = \Gamma(n)\, \zeta (n)
\label{usando1}
\end{equation}
and
\begin{equation}
\int_{0}^\infty \frac{z^{n-1}}{e^z + 1}\, dz = \frac{1}{2^n}\, (2^n -2)\,
\Gamma(n)\, \zeta (n)\,\,,
\label{usando2}
\end{equation}
we obtain 
\begin{eqnarray}
n & = & \left(\frac{kT}{c} \right)^3 \frac{4 \pi g}{(2 \pi \hslash)^3}
  \int_0^\infty \frac{y^2 dy}{e^y \pm 1}  \nonumber \\ & = & 
{\cal A}_\pm  \frac{\zeta(3)}{\pi^2} g \left(\frac{kT}{\hslash c}\right)^3 \,,
  \nonumber \\
\rho &  = & {\cal B}_\pm \frac{\pi^2 g}{30 (\hslash c)^3} (kT)^4 \,,
\nonumber \\
P & = & \frac{1}{3} \rho \,,
\label{gingseng}
\end{eqnarray}
where $\zeta (3) \approx 1.2$ and ${\cal A}_- = 1$ for bosons, and
${\cal A}_+ = 3/4$ for fermions, ${\cal B}_- = 1$ for bosons and
${\cal B}_+ = 7/8$ for fermions.\footnote{The Gamma function
  is an extension of the factorial function for non-integer and
  complex numbers. If $s$ is a positive integer, then $\Gamma(s) =
  (s-1)!.$ The Riemann zeta function of a real variable $s,$ defined
  by the infinite series $\zeta(s) = \sum_{n=1}^\infty 1/n^s,$
  converges $\forall s>1$. Note that (\ref{usando2}) follows from
  (\ref{usando1}) using the relation $\frac{1}{e^x +1 } =
  \frac{1}{e^x-1 }- \frac{2}{e^{2x} -1 }$.} 

For $kT \ll m c^2$, the exponential factor dominates the denominator in both
the Bose-Einstein and Fermi-Dirac distributions in (\ref{BEFD}), so that
the bosonic or fermionic nature of the particles becomes
indistinguishable. Furthermore, we have
\begin{eqnarray}
E & = & (p^2 c^2 + m^2 c^4)^{1/2} = m c^2 \left(1 + \frac{p^2}{m^2c^2}
\right)^{1/2} \nonumber \\ & \simeq & m c^2 + \frac{p^2}{2 m} \, . 
\label{maravillosa1}
\end{eqnarray}
Defining $x = |\vec p \,|/\sqrt{2m kT}$, for the number density we obtain the
Boltzmann distribution
\begin{eqnarray}
n & = & e^{-mc^2/(kT)} (2 m kT)^{3/2} \frac{4 \pi g}{(2 \pi \hslash)^3 }
\int_0^\infty e^{-x^2} x^2 dx \nonumber \\
 & = & \frac{g}{\hslash^3} \left(\frac{mkT}{2 \pi} \right)^{3/2}
 e^{-mc^2/(kT)} \, ,
\end{eqnarray}
where we have used 
\begin{equation}
\int_0^\infty x^n e^{-x^2} dx = \frac{1}{2} \Gamma\left(\frac{1+n}{2}
\right) \,,
\end{equation}
with $n=2$ and $\Gamma (3/2) = \sqrt{\pi}/2$. From (\ref{maravillosa1}) it is easily
seen that to leading order $\rho = m c^2 n$ in this case.  

To obtain the associated pressure, note that to leading order $|p^2
c^2|/E \simeq |p|^2/m$, so that
\begin{eqnarray}
P & \simeq & e^{-mc^2/(kT)} (2 m kT)^{5/2} \frac{4 \pi g}{(2 \pi
  \hslash)^3 } \frac{1}{3m} \int_0^\infty x^4 e^{-x^2} dx \nonumber \\
& = & e^{-mc^2/(kT)} (2 m kT)^{5/2} \frac{4 \pi g}{(2 \pi
  \hslash)^3 } \frac{1}{3m} \frac{3 \sqrt{\pi}}{8} \nonumber \\
& = & \frac{g}{\hslash^3} \left(\frac{mkT}{2 \pi} \right)^{3/2}
 e^{-mc^2/(kT)}  \, kT \nonumber \\
& = & n kT \,,
\label{maravillosa2}
\end{eqnarray}
where we have used $\Gamma(5/2) = 3 \sqrt{\pi}/4$. Note that
(\ref{maravillosa2}) is just but the
familar result for a non-relativistic perfect gas, $P = nkT$. Since $kT \ll
mc^2$, we have $P \ll \rho$ and the pressure may be neglected for a gas of
non-relativistic particles, as we had anticipated.

For a gas of non-degenerate, relativistic species, the average energy per particle is
\begin{eqnarray}
\langle E \rangle = \frac{\rho}{n} =  \left\{ \begin{array}{ll}
    \frac{\pi^4 k}{30 \zeta (3)} \, T  & \simeq 2.701 \,  T~{\rm for \, bosons} \\
 \frac{7\pi^4 k}{ 180 \zeta (3)} \, T & \simeq 3.151 \, T~{\rm for \, fermions}  
\end{array} \right. \,,
\label{plasticH}
\end{eqnarray}
whereas for a non-relativistic species
\begin{equation}
\langle E \rangle = m c^2 +\frac{3}{2} kT \, .
\end{equation}

The internal energy $U$ can be considered to be a function of two
thermodynamic variables among $P$, $V$ , and $T$. (These variables are
related by the equation of state.) Let us choose $V$ and $T$ to be the
fundamental variables. The internal energy can then be written as
$U(V,T)$. Let us differentiate this function:
\begin{equation}
dU = \left(\frac{\partial U}{\partial V} \right)_T dV + \left(
  \frac{\partial U}{\partial T}\right)_V d T \, .
\end{equation}
This equation can be combined with the first law (\ref{1stLAW}) to give
\begin{equation}
T dS = \left[\left(\frac{\partial U}{\partial V} \right)_T + P \right]
  dV + \left(\frac{\partial U}{\partial T} \right)_V dT \, .
\label{Ashley1}
\end{equation}
Now, since the  internal energy is a function of $T$ and $V$ we may therefore choose to view $S$ as a function of $T$ and $V$, and this gives rise to the differential
relation 
\begin{equation}
dS = \left(\frac{\partial S}{\partial T} \right)_V \, dT +
\left(\frac{\partial S}{\partial V} \right)_T \, dV \ .
\label{Ashley2}
\end{equation}
Substituting (\ref{Ashley2}) into (\ref{Ashley1}) and equating the $dV$ and $dT$ parts gives
the familiar
\begin{equation}
\frac{\partial U}{\partial T} = T \frac{\partial S}{\partial T} 
\end{equation}
and
\begin{equation}
S = \frac{U + PV}{T} \,,
\end{equation}
where we have used the relation for extensive quatities ($\partial
S/\partial V = S/V$ and $\partial U/ \partial V$).\footnote{Recall
  that an extensive property is any property that depends on the size
  (or extent) of the system under consideration. Take two identical
  samples with all properties identical and combine them into a single
  sample. Properties that double (e.g., energy, volume, entropy) are extensive. Properties that remain
  the same (e.g., temperature and pressure) are intensive.}
It is useful to define the entropy density $s = S/V$, which is thus given by
\begin{equation}
s=\rho+P \, .
\end{equation}

For photons, we can compute all of the thermodynamic quantities rather easily
\begin{eqnarray}
n_\gamma & = & \frac{2 \zeta(3)}{\pi^2} \left(\frac{kT_\gamma}{\hslash
    c^3} \right) = 60.42 \left(\frac{kT}{hc}\right)^3 \nonumber \\
& = & 20.28
\left(\frac{T}{K}\right)^3~{\rm photons\ cm}^{-3}\, , \nonumber \\
\rho_\gamma & = & \frac{\pi^2}{15} \frac{(kT_\gamma)^4}{(\hslash c)^3} = 0.66 \frac{(kT_\gamma)^4}{(\hslash c)^3} \,,
\nonumber \\
\langle E_\gamma \rangle & = & \frac{\rho}{n} = 3.73 \times 10^{16}
\left(\frac{T}{{\rm K}} \right)~{\rm erg} \,, \nonumber \\
 P_\gamma & = & \frac{1}{3} \rho_\gamma \,, \nonumber \\
 s_\gamma & = &  \frac{4}{3} \frac{\rho_\gamma}{T_\gamma} \, .
 \label{rhogama}
\end{eqnarray}
In the limit $kT\gg m_i c^2$, the total energy density can be conveniently expressed by
\begin{eqnarray}
\rho_{\rm rad} & = & \left(\sum_B g_B + \frac{7}{8} \sum_F g_F \right)
\frac{1}{(c\hslash)^3} \frac{\pi^2}{30} (kT)^4  \nonumber \\
 &= & \frac{1}{( \hslash c)^3} \frac{\pi^2}{30} \, g_\rho(T) \, (kT)^4 \,,
\label{NT}
\end{eqnarray}
where $g_{B(F)}$ is the total number of boson (fermion) degrees of
freedom and the sum runs over all boson (fermion) states with $m_i
c^2\ll k T$. The factor of $7/8$ is due to the difference between the
Fermi and Bose integrals. (\ref{NT}) defines the effective number of
degrees of freedom, $g_\rho(T)$, by taking into account new particle
degrees of freedom as the temperature is raised.  The change in
$g_\rho(T)$ (ignoring mass effects) is given in
Table~\ref{tab:NT}~\cite{Olive:2010mh}. At higher temperatures,
$g_\rho(T)$ will be model dependent.\\

\begin{table}
\caption{Effective numbers of degrees of freedom in  SM.}
\begin{center}
\begin{tabular}{llc}
\hline \hline
{\bf Temperature} & {\bf New particles} \qquad
&\boldmath$4g_\rho(T)$ \\
\hline\rule{0pt}{12pt}
$T < m_{ e}$   &     $\gamma$'s +   $\nu$'s & 29 \\
$m_{ e} <   T  < m_\mu$ &    $e^{\pm}$ & 43 \\
$m_\mu <  T  < m_\pi$  &   $\mu {}^{\pm}$ & 57 \\
$m_\pi <  T < T_{ c}^{*}$  & $\pi$'s & 69 \\
$T_{ c} <  T  < m_{\rm charm}$ \qquad &
  -  $\pi$'s + $  u,{\bar u},d,{\bar d},s,{\bar s}$ + gluons &  247 \\
$m_{ c} <  T < m_\tau$ &  $c,{\bar c}$ & 289 \\
$m_\tau < T < m_{\textrm{bottom}}$ & $\tau {}^{\pm}$ & 303 \\
$m_{ b} < T < m_{ W,Z}$ & $b,{\bar b}$ & 345 \\
$m_{ W,Z} <  T < m_{\textrm{Higgs}}$ & $W^{\pm}, Z$ & 381 \\
$ m_H< T < m_{\textrm{top}}$ & $H^0$ & 385 \\
$m_t< T $ & $t,{\bar t}$  & 427 \\
\hline
\hline
\end{tabular}
\end{center}
\vspace{-.2cm}
{\small *$T_{ c}$ 
corresponds to the confinement--deconfinement transition between
 quarks and hadrons.}
 \label{tab:NT}
\end{table}

{\bf EXERCISE 8.1}~If in the next $10^{10}~{\rm yr}$ the volume of the
universe increases by a factor of two, what then will be the
temperature of the blackbody radiation?

\subsection{The first millisecond}

The history of the universe from $10^{-10}$ seconds to today is based
on observational facts: the fundamental laws of high energy physics
are well-established up to the energies reached by the LHC. Before
$10^{-10}$ seconds, the energy of the universe exceeds 13~TeV and we
lose the comfort of direct experimental guidance. The physics of that
era is therefore as speculative as it is fascinating. Herein we will
go back to the earliest of times - as close as possible to the {\it
  big bang} - and follow the evolution of the Universe.

It is cear that as $a \to 0$ the temperature increases
without limit $T \to \infty$, but there comes a point at which the
extrapolation of classical physics breaks down. This is the realm of
quantum black holes, where the thermal energy of typical particles of
mass $m$ is such that their de Broglie wavelength is smaller than
their Schwarzschild radius.  Equating $h/mc$ to $2Gm/c^2$ yields a
characteristic mass for quantum gravity known as the Planck mass
$M_{\rm Pl}$.\footnote{Strictly speaking this is not quite the Planck
  mass. It is a factor of $\sqrt{\pi}$ larger. However, this heuristic
  derivation gives the right order of magnitude.} This mass scale,
together with the corresponding length $\hslash /(M_{\rm Pl} c)$ and
time $\hslash/(M_{\rm Pl} c^2)$ define the system of Planck units:
\begin{eqnarray}
M_{\rm Pl} & \equiv & \sqrt{\frac{\hslash c}{G}} \simeq 10^{19}~{\rm
  GeV} \,, \nonumber \\
\ell_{\rm Pl} & \equiv & \sqrt{\frac{\hslash G}{c^3} } \simeq
10^{-35}~{\rm m} \,, \nonumber \\
t_{\rm Pl} & \equiv & \sqrt{\frac{\hslash G}{c^5} }\simeq 10^{-43}~{\rm s} \, . 
\end{eqnarray}
The Planck time therefore sets the origin of time for the {\it
  classical big bang} era. It is inaccurate to extend the classical
solution of Friedmann equation to $a = 0$ and conclude that the
universe began in a singularity of infinite density. 

At $t \sim 10^{-43}$~s, a kind of {\it phase transition} is thought to
have occured during which the gravitational force {\it condensed out}
as a separate force. The symmetry of the four forces was broken, but
the strong, weak, and electromagnetic forces were still unified, and
there were no distinctions between quarks and leptons. This is an
unimaginably short time, and predictions can be only speculative. The
temperature would have been about $10^{32}~{\rm K},$ corresponding to
{\it particles} moving about every which way with an average kinetic
energy of
\begin{equation}
kT \approx \frac{1.4 \times 10^{-23}~{\rm J/K}\,\,\, 
10^{32}~{\rm K}}{1.6 \times 10^{-10}~{\rm J/GeV}} 
  \approx 10^{19}~{\rm GeV} \,,
\label{teuKR}
\end{equation}
where we have ignored the factor 2/3 in our order of magnitude
calculation. Very shortly thereafter, as the temperature had dropped to about
$10^{28}~{\rm K}$, there was another phase transition and the strong
force condensed out at about $10^{-35}~{\rm s}$ after the bang. Now
the universe was filled with a {\it soup} of  quarks and leptons. About
this time, the universe underwent an incredible exponential expansion,
increasing in size by a factor of $\agt 10^{26}$ in a tiny
fraction of a second, perhaps $\sim 10^{-34}~{\rm s}$.

\begin{figure}
 \postscript{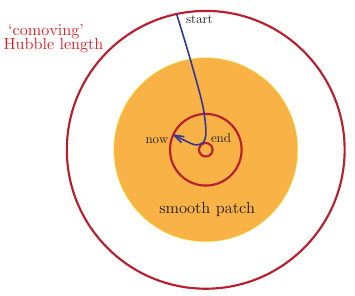}{0.8}
 \caption{Evolution of the comoving Hubble radius, $c/(aH)$, in the
   inflationary universe. The comoving Hubble sphere shrinks during
   inflation and expands after inflation. Inflation is therefore a
   mechanism to {\it zoom-in} on a smooth sub-horizon patch~\cite{Baumann:2009ds}.}
\label{inflation}
\end{figure}

As a matter of fact, the favored $\Lambda$CDM model implicitly
includes the hypothesis of a very early period in which the scale
factor of the universe expands exponentially: $a(t) \propto e^{Ht}$.
If the interval of exponential expansion satisfies \mbox{$\Delta t
  \agt 60/H$}, a small casually connected region can grow sufficiently
to accommodate the observed homogeneity and
isotropy~\cite{Guth:1980zm}. To properly understand why this is so,
we express the comoving horizon (\ref{gear4}) as an integral of the
comoving Hubble radius,
\begin{equation}
\varrho_{\rm h} \equiv c \int_0^t \frac{dt'}{a(t')} = c \int_0^a
\frac{da}{Ha^2} = c \int_0^a \frac{1}{aH} \ d \ln a \, .
\label{tacuarenocho}
\end{equation}
At this stage it is important to emphasize a subtle distinction
between the comoving horizon $\varrho_{\rm h}$ and the comoving Hubble
radius $c/(aH)$. If particles are separated by distances greater than
$\varrho_{\rm h}$, they never could have communicated with one
another; if they are separated by distances greater than $c/(aH)$,
they cannot talk to each other now. This distinction is crucial for
the solution to the horizon problem which relies on the following: It
is possible that $\varrho_{\rm h}$ is much larger than $c/(aH)$ now,
so that particles cannot communicate today but were in causal contact
early on. From (\ref{tacuarenocho}) we see that this might happen if
the comoving Hubble radius in the early universe was much larger than
it is now so that $\varrho_{\rm h}$ got most of its contribution from
early times. Hence, we require a phase of decreasing Hubble
radius, as illustrated in Fig.~\ref{inflation}.  The shrinking Hubble sphere is defined by $ d(aH)^{-1}/dt <0$.  From
the relation $ d(aH)^{-1}/dt = - \ddot a/(aH)^2$ we see immediately
that a shrinking comoving Hubble radius implies accelerated expansion
$\ddot a >0$. This explains why inflation is often defined as a period
of accelerated expansion. The second time derivative of the scale
factor may of course be related to the first time derivative of the
Hubble parameter according to
\begin{equation}
\frac{\ddot a}{a} = H^2 (1 - \epsilon) \,,
\end{equation}
where $\epsilon \equiv - \dot H/H^2$. Acceleration therefore
corresponds to $\epsilon <1$. All in all, $H$ is approximately
constant during inflation whereas $a$ grows exponentially, and so this
implies that the comoving Hubble radius decreases just as
advertised. Now, consulting (\ref{acceeq}) we infer that $\ddot a>0$
requires a negative pressure: $P < - \rho/3$. To see how this can be
realized in various physics models see e.g.~\cite{Baumann:2009ds,Riotto:2002yw}.\\

{\bf EXERCISE 8.2}~(\ref{FriedmannLambda})  can be rearranged to give
\begin{equation}
 \frac{8 \pi G \rho}{3 c^2 H^2} - \frac{k c^2}{H^2 a^2R_0^2} +
 \frac{\Lambda c^2}{3} = 1 \,
\label{flatness1}
\end{equation}
and so using (\ref{energy-Omega}) we rewrite (\ref{flatness1}) as 
\begin{equation}
\Omega -1 = \frac{k c^2}{a^2 R_0^2 H^2} \, .
\label{flatness2}
\end{equation}  
Now let us make a tremendous approximation and assume that Friedmann equation
is valid until the Planck era. From (\ref{flatness2}) we read that if
the universe is perfectly flat, then $\Omega = 1$ at all
times. However, if there is even a small curvature term, the time
dependence of $\Omega-1$ is quite different. In particular, for the
radiation dominated era we have, $H^2 \propto \rho_{\rm rad} \propto
a^{-4}$ and $\Omega -1 \propto a^2$, whereas during matter domination,
$\rho_m \propto a^{-3}$ and $\Omega -1 \propto a$. In both cases
$\Omega -1$ decreases going backwards in time. Since we know that
$\Omega_0 -1$ is of order unity at present, we can deduce its value at
$t_{\rm PL}$,
\begin{equation}
\frac{|\Omega - 1 |_{ T = T_{\rm Pl}}}{|\Omega -1 |_{T = T_0} }\approx
    \left(\frac{a_{\rm Pl}^2}{a_0^2} \right) \approx
    \left(\frac{T_0^2}{T_{\rm Pl}^2} \right) \approx {\cal O} (10^{-64}) \, .
\label{flatness3}
\end{equation}
This means that to get the correct value of $\Omega_0 - 1 \sim 1$
today, the value of $\Omega -1$ at early times has to be fine-tuned
to values amazingly close to zero, {\it but without being exactly
zero}. This has been dubbed the {\it flatness problem}.\footnote{A
didactic explanation of the flatness fine-tuning problem is given~\cite{Lineweaver:2003ie}.} Show that the
inflationary hypothesis elegantly solve the flatenss fine-tuning problem.\\ 

After the very brief inflationary period, the universe would have
settled back into its more regular expansion.  For
$10^{-34}~{\rm s} \alt t \alt 10^5~{\rm yr}$, the universe is thought
to have been dominated by radiation. This corresponds to $10^3~{\rm K}
\alt T \alt 10^{27}~{\rm K}$. We have seen that the equation of
state can be given by $w =1/3$.  If we neglect the contributions to
$H$ from $\Lambda$ (this is always a good approximation for small
enough $a$) then we find that $a \sim t^{1/2}$ and $\rho_{\rm rad}
\sim a^{-4}$. Substituting (\ref{NT}) into (\ref{FriedmannGR}) we can
rewrite the expansion rate as a function of the temperature in the
plasma
\begin{eqnarray}
H = \left( \frac{8 \pi G \rho_{\rm rad}}{3} \right)^{1/2} & = & \left(\frac{8 \pi^3}{90} g_\rho(T) \right)^{1/2} \, T^2/M_{\rm Pl} \nonumber \\
& \sim & 1.66 \sqrt{g_\rho(T)} \  T^2/M_{\rm Pl} \, ,  
\label{expansionT}
\end{eqnarray}
where we have adopted natural units ($\hslash = c = k =1$).
Neglecting the $T$-dependence of $g_\rho$ (i.e. away from mass
thresholds and phase transitions), integration of (\ref{expansionT})
yields (\ref{gear3}) and the useful commonly used approximation
\begin{equation}
t \simeq \left(\frac{3 M_{\rm Pl}^2}{32 \pi \rho_{\rm rad}}\right)^{1/2} \simeq 2.42 \frac{1}{\sqrt{g_\rho}} \, \left(\frac{T}{\rm MeV}\right)^{-2}~{\rm s} \, .
\label{alanparsons_time}
\end{equation}

At about $10^{-10}~{\rm s}$ the Higgs field spontaneously acquires a
vacuum expectation value, which breaks the electroweak gauge
symmetry. As a consequence, the weak force and electromagnetic force
manifest with different ranges. In addition, quarks and charged leptons interacting with the Higgs field
become massive. The fundamental interactions have by then taken their
present forms.

By the time the universe was about a microsecond old, quarks began to
condense into mesons and baryons. To see why, let us focus on the most
familiar hadrons: nucleons and their antiparticles. When the average
kinetic energy of particles was somewhat higher than 1~GeV, protons,
neutrons, and their antiparticles were continually being created out
of the energies of collisions involving photons and other
particles. But just as quickly, particle and antiparticles would
annihilate. Hence the process of creation and annihilation of nucleons
was in equilibrium. The numbers of nucleons and antinucleons were
high: roughly as many as there were electrons, positrons, or
photons. But as the universe expanded and cooled, and the average
kinetic energy of particles dropped below about 1~GeV, which is the
minimum energy needed in a typical collision to create nucleons and
antinucleons (940~MeV each), the process of nucleon creation could not
continue. However, the process of annihilation could continue with
antinucleons annihilating nucleons, until there were almost no
nucleons left; but not quite zero!

Manned and unmanned exploration of the solar system tell us that
it is made up of the same stuff as the Earth: {\it
  baryons}. Observational evidence from radio-astronomy and cosmic ray
detection indicate that the Milky Way, as well as interstellar space,
and distant galaxies are also made of baryons. Therefore, we can
cautiously conclude that the baryon number of the observable universe
is $B>0$. This requires that the early $q \bar q$ plasma contained a
tiny surplus of quarks. After all anti-matter annihilated with matter,
only the small surplus of matter remained
\begin{equation}
\eta = \frac{n_B - n_{\bar B}}{n_\gamma} = 5 \times 10^{-10}
\ \frac{{\rm excess \ baryons}}{{\rm photons}} \, .
\end{equation}
The tiny surplus can be explained by interactions in the
early universe that were not completely symmetric with respect to an
exchange of matter-antimatter, the so-called ``baryogenesis''~\cite{Sakharov:1967dj}.

\begin{figure*}[!t]
  \postscript{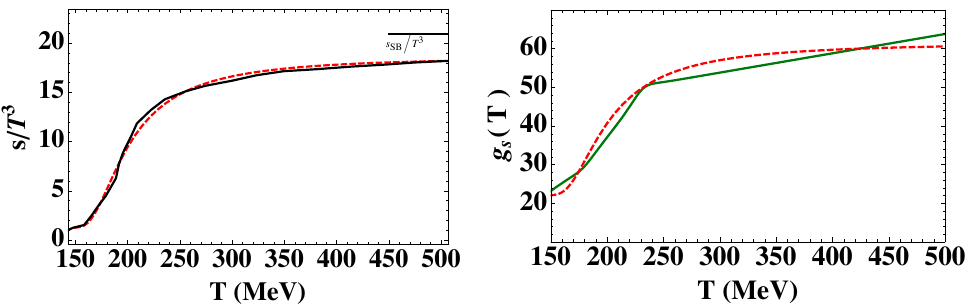}{0.9}
  \caption{{\bf Left.} The parametrization of the entropy density
    given in Eq.~(\ref{soverT})  (dashed line) superposed on the
    result from high statistics lattice
    simulations~\cite{Bazavov:2009zn} (solid line). {\bf Right.}~Comparison of $g_s (T)$ obtained using (\ref{gdet}) (dashed
    line) and the phenomenological estimate of~\cite{Laine:2006cp,Steigman:2012nb} (solid line)~\cite{Anchordoqui:2013wwa}.}
\label{fig:dos}
\end{figure*}

At this stage, it is worthwhile to point out that if some relativistic
particles have decoupled from the photons, it is necessary to
distinguish between two kinds of relativistic degrees of freedom
(r.d.o.f.): those associated with the total energy density $g_\rho$,
and those associated with the total entropy density $g_s$. At energies
above the deconfinement transition towards the quark gluon plasma,
quarks and gluons are the relevant fields for the QCD sector, such
that the effective number of interacting (thermally coupled)
r.d.o.f. is $g_s(T) = 61.75$.  As the universe cools down below the
confinement scale $\Lambda_{\rm QCD} \sim 200~{\rm MeV}$, the SM
plasma transitions to a regime where mesons and baryons are the
pertinent degrees of freedom. Precisely, the relevant hadrons present
in this energy regime are pions and charged kaons, such that $g_s(T) =
19.25$~\cite{Brust:2013xpv}. This significant reduction in the degrees
of freedom results from the rapid annihilation or decay of more
massive hadrons which may have formed during the transition. The
quark-hadron crossover transition therefore corresponds to a large
redistribution of entropy into the remaining degrees of freedom. To
connect the temperature to an effective number of r.d.o.f. we make use
of some high statistics lattice simulations of a QCD plasma in the hot
phase, especially the behavior of the entropy during the
changeover~\cite{Bazavov:2009zn}. Concretely, the effective number of
interacting r.d.o.f. in the plasma at temperature $T$ is given by
\begin{equation}
g_s (T) \simeq r (T) \left(g_B+ \frac{7}{8} g_{\rm F} \right) \,,
\end{equation}
where the coefficient $r (T)$ is unity for leptons, two for photon
contributions, and is the ratio $s(T )/s_{\rm SB}$ for the quark-gluon
plasma~\cite{Anchordoqui:2011nh}. Here, $s(T)$ and $s_{\rm SB}$ are
the entropy density and the ideal
Stefan-Bolzmann limit shown in Fig~\ref{fig:dos}. The entropy rise
during the confinement-deconfinement changeover can be parametrized,
for $150~{\rm MeV} < T < 500~{\rm MeV}$, by
\begin{equation}
\frac{s}{T^3} \simeq  \frac{42.82}{\sqrt{392 \pi}} \, e^{-C_1} + 18.62
\, \frac{ C_2^2}{ \left[
    e^{C_2} - 1\right]^{2}} \, \ e^{C_2} ,
\label{soverT}
\end{equation}
where
$C_1 = \left( T_{\rm MeV} - 151 \right)^2/392$
and 
$C_2 = 195.1/(T_{\rm MeV} - 134)$.
For the same energy range, we obtain
\begin{equation}
g_s(T) \simeq 47.5 \ r(T) + 19.25 \, .
\label{gdet}
\end{equation}
In Fig.~\ref{fig:dos} we show $g_s(T)$ as given by (\ref{gdet}). The
parametrization is in very good agreement with phenomenological
estimates~\cite{Laine:2006cp,Steigman:2012nb}.  

The entropy density is dominated by the contribution of relativistic
particles, so to a very good approximation
\begin{equation}
s = \frac{2 \pi^2}{45} g_s(T) \, T^3 \, .
\end{equation}
Conservation of $S = sV$ leads to
\begin{equation}
\frac{d}{dt} (s a^3) = 0 
\end{equation}
and therefore that $g_s(T) T^3 a^3$ remains constant as the universe
expands.  As one would expect, a non-evolving system would stay at
constant number or entropy density in comoving coordinates even though
the number or entropy density is in fact decreasing due to the
expansion of the universe.  Since the quark-gluon energy density in
the plasma has a similar $T$ dependence to that of the entropy (see
e.g. Fig.~7 in~\cite{Bazavov:2009zn}), hereafter we simplify the
discussion by taking $g= g_\rho = g_s$.

After the first millisecond has elapsed, when the majority of hadrons
and anti-hadrons annihilated each other, we entered the lepton era.

\subsection{Neutrino decoupling and BBN}

After the first tenth of a second, when the temperature was about $3
\times 10^{10}$~K, the universe was filled with a plasma of protons,
neutrons, electrons, positrons, photons, neutrinos, and antineutrinos
($p$, $n$, $\gamma$, $e^-$, $e^+$, $\nu$, and $\overline \nu$). The
baryons are of course nonrelativistic while all the other particles
are relativistic.  These particles are kept in thermal equilibrium by
various electromagnetic and weak processes of the sort $\bar \nu \nu
\rightleftharpoons e^+ e^-$, $\nu e^- \rightleftharpoons \nu e^-$, $n
\nu_e \rightleftharpoons p e^-$, $\gamma \gamma \rightleftharpoons e^+
e^-$, $\gamma p \rightleftharpoons \gamma p$, etc.  In complying with
the precision demanded of our phenomenological approach it would be
sufficient to consider that the cross section of reactions involving
left-handed neutrinos, right-handed antineutrinos, and electrons
is $\sigma_{\rm weak} \sim G^2_F E^2$, where $G_F = 1.16 \times 10^{-5}~{\rm
  GeV}^{-2}$ is the Fermi constant. If we approximate the energy $E$
of all particle species by their temperature $T$, their velocity by
$c$, and their density by $n \sim T^3$, then the interaction rate of
is ~\cite{Kolb:1990vq}
\begin{equation}
\Gamma_{{\rm int},\nu} (T) \approx \langle v \sigma \rangle \ n_\nu \ \approx G_F^2 T^5 \, .
\label{year34}
\end{equation}
Comparing (\ref{year34}) with the expansion rate (\ref{expansionT}),
calculated for $g(T) = 10.75$, we see that when the temperature drops
below some characteristic temperature $T_{\nu_L}^{\rm dec}$ neutrinos
{\it decouple}, i.e. they lose thermal contact with
electrons~\cite{Alpher:1953zz,Zeldovich:65,Zeldovich:67}.  The
condition
\begin{equation}
\Gamma_{{\rm int},\nu} (T_{\nu_L}^{\rm dec}) = H (T_{\nu_L}^{\rm
   dec})
\end{equation}
sets the decoupling temperature for left handed neutrinos:
$T_{\nu_L}^{\rm dec} \sim 1~{\rm MeV}$.

The much stronger electromagnetic interaction continues to keep the
protons, neutrons, electrons, positrons, and photons in equilibrium.
The reaction rate per nucleon, $\Gamma_{{\rm int},N} \sim T^3 \alpha^2/m_N^2$,  is
larger than the expansion rate as long as 
\begin{equation}
T> \frac{m_N^2}{\alpha^2 M_{\rm  Pl}} \sim {\rm a \ very \ low \ temperature} \,,
\end{equation} 
where the non-relativistic form of the electromagnetic cross
section, $\sigma \sim \alpha^2/m_N^2$, has been obtained by dimensional
analysis, with $\alpha$ the fine structure constant. The nucleons are thus mantianed in kinetic equilibrium. The
average kinetic energy per nucleon is $3T/2$. One must be careful to
distinguish between kinetic equilibrium and chemical
equilibrium. Reactions like $\gamma \gamma \to p \bar p$ have long
been suppressed, as there are essentially no anti-nucleons around.

For $T > m_e \sim 0.5~{\rm MeV} \sim 5 \times 10^9~{\rm K}$, the number of electrons, positrons, and photons are comparable, $n_{e^-} \sim n_{e^+} \sim n_\gamma$. The exact ratios are of course easily supplied by inserting the appropriate ``$g$-factors.'' Because the universe is electrically neutral, $n_{e^-} -  n_{e^+} = n_p$ and so there is a slight excess of electrons over positrons.
When $T$ drops below $m_e$, the process \mbox{$\gamma \gamma \to e^+ e^-$} is
severely suppressed by the Boltzmann factor $e^{-m_e/T}$, as only
very energetic photons in the ``tail-end'' of the Bose distribution
can participate. Thus positrons and electrons annihilate rapidly via
$e^+ e^- \to \gamma \gamma$ and are not replenished (leaving a small
number of electrons $n_{e^-} \sim n_p \sim 5 \times 10^{-10} n_\gamma$). As long
as thermal equilibrium was preserved, the total entropy remained
fixed.  We have seen that $sa^3 \propto g(T) T^3 a^3 =$ constant. For $T
\gtrsim m_e$, the particles in thermal equilibrium with the  
photons include the photon ($g_\gamma = 2$) and $e^\pm$ pairs
($g_{e^\pm} = 4$). The effective total number of particle species
before annihilation is $g_{\rm before} = 11/2.$ On the other hand,
after the annihilation of electrons and positrons, the only remaining
abundant particles in equilibrium are photons. Hence the effective
number of particle species is $g_{\rm after} = 2.$ It follows from
the conservation of entropy that
\begin{equation}
\left. \left. \frac{11}{2}\, (T_\gamma a)^3 \right|_{\rm before} = 
2\, (T_\gamma a)^3 \right|_{\rm after} \, .
\end{equation}
That is, the heat produced by the annihilation of electrons and positrons increases the quantity $T_\gamma a$ by a factor of
\begin{equation}
\frac{(T_\gamma a)|_{\rm after}}{(T_\gamma a)|_{\rm before}} = \left(\frac{11}{4} \right)^{1/3} 
\simeq 1.4 \,.
\end{equation}
Before the annihilation of electrons and positrons, the neutrino
temperature $T_{\nu}$ is the same as the photon temperature $T_\gamma$. But
from then on, $T_{\nu}$ simply dropped like $a^{-1},$ so for all
subsequent times, $T_{\nu} a$ equals the value before annihilation,
\begin{equation}
(T_{\nu} a)|_{\rm after} = (T_{\nu} a)|_{\rm before} = (T_\gamma a)|_{\rm before} \,\,.
\end{equation}
We conclude therefore that after the annihilation process is over, the photon temperature is higher than the neutrino temperature by a factor of
\begin{equation}
  \left. \left(\frac{T_\gamma}{T_{\nu}} \right)\right|_{\rm after}
 = \frac{(T_\gamma a)|_{\rm after}}{(T_{\nu} a)|_{\rm after}} \simeq 1.4 \,.
\end{equation}
Therefore, even though out of thermal equilibrium, the neutrinos and antineutrinos make an important contribution to the energy density.\\

{\bf EXERCISE 8.3}~By assuming that neutrinos saturate the dark matter
density derive an upper bound on the neutrino mass~\cite{Lee:1977ua}.

\begin{widetext}

The energy density stored in relativistic species is customarily given
in terms of the so-called {\it effective number of neutrino species},
$N_{\rm eff}$, through the relation \begin{equation} \rho_{\rm rad} =
  \left[1 + \frac{7}{8} \, \left( \frac{4}{11} \right)^{4/3} \, N_{\rm
      eff} \right] \rho_\gamma \, ,
\end{equation}
and so
\begin{equation}
  N_{\rm eff}  \equiv  \left(\frac{\rho_{\rm rad} - \rho_\gamma}{\rho_{\nu}}\right) 
       \simeq  \frac{8}{7} {\sum_B}' \frac{g_B}{2} \left(\frac{T_B}{T_{\nu}}\right)^4 + {\sum_F}' \frac{g_F}{2} \left(\frac{T_F}{T_{\nu}} \right)^4 \, ,
\end{equation} 
where $\rho_{\nu}$ denotes the energy density of a single species of
massless neutrinos, $T_{B(F)}$ is the effective temperature of boson
(fermion) species, and the primes indicate that electrons and photons
are excluded from the sums~\cite{Steigman:1977kc,Steigman:1986nh}. The
normalization of $N_{\rm eff}$ is such that it gives $N_{\rm eff} = 3$
for three families of massless left-handed standard model
neutrinos. For most practical purposes, it is accurate enough to
consider that neutrinos freeze-out completely at about $1~{\rm MeV}.$
However, as the temperature dropped below this value, neutrinos were
still interacting with the electromagnetic plasma and hence received a
tiny portion of the entropy from pair annihilations.  The
non-instantaneous neutrino decoupling gives a correction to the
normalization $ N_{\rm eff} =
3.046$~\cite{Dicus:1982bz,Dodelson:1992km,Mangano:2001iu,Mangano:2005cc}.

Near 1~MeV, the CC weak interactions,
\begin{equation}
n  \nu_e \rightleftharpoons p  e^{-} , \quad 
 n  e^{+} \rightleftharpoons p  \bar{\nu_{e}} , \quad 
 n \rightleftharpoons p  e^{-}  \bar \nu_e 
\end{equation}
guarantee neutron-proton chemical equilibrium. 
Defining $\lambda_{np}$ as the summed rate of the reactions which
convert neutrons to protons,
\begin{equation}
 \lambda_{np}  =  \lambda (n \nu_{e} \to p e^-) 
 + \lambda (n e^{+} \to p \bar \nu_{e})  +  \lambda (n \to p e^{-} \bar\nu_{e}) \ , 
\end{equation} 
the rate $\lambda_{pn}$ for the reverse reactions which convert
protons to neutrons is given by detailed balance:
\begin{equation}
 \lambda_{pn}  = \lambda_{np} \ e^{-\Delta m/T(t)} , 
 \label{detbal}
\end{equation} 
where $\Delta m \equiv m_{n} - m_{p} = 1.293~{\rm MeV}$.  The
evolution of the fractional neutron abundance $X_{n/N}\equiv n_n/n_N$
is described by the balance equation
\begin{equation} 
 \frac{d X_{n/N} (t)}{d t} = \lambda_{p n} (t) [1 - X_{n/N} (t)] -
  \lambda_{np} (t) X_{ n/N}(t)\ , 
\end{equation} 
where $n_{N}$ is the total nucleon density at this time,
$n_{N}=n_{n}+n_{p}$.
The 
equilibrium solution is obtained by setting $dX_{n/N}(t)/dt=0$:
\begin{equation} 
\label{nbypeqm}
 X_{n/N}^{\rm eq} (t) = \frac{\lambda_{pn} (t)}{\lambda_{pn} (t)+ \lambda_{np}(t)} = 
  \left[1 + e^{\Delta m/T(t)}\right]^{-1} \ . 
\end{equation}
The neutron abundance tracks its value in equilibrium until the inelastic neutron-proton scattering rate
 decreases sufficiently so as to become comparable to the
Hubble expansion rate. At this point the
neutrons freeze-out, that is they go out of {\em chemical} equilibrium. The
neutron abundance at the freeze-out temperature  $T_{n/N}^{\rm FO} =
0.75~{\rm MeV}$ can be approximated
by its equilibrium value (\ref{nbypeqm}),
\begin{equation}
\label{Xnfr}
 X_{n/N} (T_{n/N}^{\rm FO})  \simeq  X_{n/N}^{\rm eq} (T_{n/N}^{\rm
   FO})  =  \left[1 + e^{\Delta m / T_{n/N}^{\rm FO}}\right]^{-1} . 
\end{equation}
Since the ratio $\Delta m/T_{n/N}^{\rm FO}$ is of ${\cal O}(1)$, a substantial
fraction of neutrons survive when chemical equilibrium between
neutrons and protons is broken.

At this time, the photon temperature is already below the
deuterium binding energy $\Delta_{\rm D} \simeq 2.2~{\rm MeV},$ thus one would
expect sizable amounts of D to be formed via $n \, p \rightarrow
{\rm D}  \, \gamma$ process. However, the large photon-nucleon density
ratio $\eta^{-1}$ delays deuterium synthesis until the photo--dissociation process become
ineffective (deuterium {\it bottleneck}). Defining the onset of nucleosynthesis by the criterion
\begin{equation}
e^{\Delta_D/T_{\rm BBN}} \eta \sim 1 \,,
\label{onsetBBN}
\end{equation}
we obtain $T_{\rm BBN} \approx 89~{\rm keV}$.  Note that
(\ref{onsetBBN}) ensures that below $T_{\rm BBN}$ the high energy tail
in the photon distribution, with energy larger than $\Delta_D$, has
been sufficiently diluted by the expansion.  At this epoch, $N(T) =
3.36$, hence the time-temperature relationship
(\ref{alanparsons_time}) dictates that big bang nucleosynthesis (BBN) begins at
\begin{equation}
t_{\rm BBN} \simeq 167~{\rm s} \approx 180~{\rm s} \,,
\end{equation}
as widely popularized by Weinberg~\cite{Weinberg:1977ji}. 

Once D starts forming, a whole nuclear process network sets
in~\cite{Sarkar:1995dd,Olive:1999ij}. When the temperature dropped below  $\sim 80~{\rm keV},$ the universe has  
cooled sufficiently that the cosmic nuclear reactor can begin in earnest,  
building the lightest nuclides through the following sequence of
two-body reactions
\begin{equation}
\begin{array}{lll}
  p  \, n \to \gamma \, {\rm D},  & &  \\ 
  p \, {\rm D} \to ^{3}\!\!{\rm He} \, \gamma, \quad \quad & {\rm D} \, {\rm D} \to ^3\!\!{\rm He} \, n,
   \quad \quad &  {\rm D} \,  {\rm D} \to  p \, {\rm T}, \\
 {\rm T}  {\rm D} \to  ^4\!\!{\rm He}\, n, \ & ^4{\rm He} \,  {\rm T} \to  ^7\!\!{\rm Li} \, \gamma , &  \\
  ^3{\rm He} \, n \to  p \, {\rm T}, \quad & ^3{\rm He}  \, {\rm D} \to ^4\!\!{\rm He} \, p,
   \ & ^3{\rm He} \,  ^4{\rm He} \to ^7\!\!{\rm Be} \, \gamma,  \\
  ^7{\rm Li}  \, p \to ^4\!\!{\rm He} \, ^4{\rm He}, \quad  &
   ^7{\rm Be} \, n \to ^7\!\!{\rm Li} \, p,  &  \\
  ~~~~~~~~\vdots & &
\end{array} \, .
\label{reac}
\end{equation}
By this time the neutron abundance surviving at freeze-out has been depleted by $\beta$-decay to
\begin{equation}
X_{n/N}(T_{\rm BBN}) \simeq X_{n/N}(T_{n/N}^{\rm FO}) \, e^{-t_{\rm BBN}/\tau_n} \,,
\end{equation}
where $\tau_n \simeq 887~{\rm s}$ is the neutron lifetime. 
Nearly {\em all} of these surviving neutrons are captured in $^4$He
because of its large binding energy ($\Delta_{^4{\rm He}}=28.3~{\rm MeV}$) via
the reactions listed in (\ref{reac}). Heavier nuclei do not form in any
significant quantity both because of the absence of stable nuclei with
$A$=5 or 8, which impedes nucleosynthesis via
$n$ $^4$He, $p$ $^4$He or $^4$He $^4$He
reactions, and because of the large Coulomb barrier for reactions such as
$^4$He  T $\to$ $^7$Li  $\gamma$ and
$^3$He $^4$He $\to$ $^7$Be $\gamma$. 
By the time the temperature has dropped below  
$\sim 30$~keV, a time comparable to the neutron lifetime, the average  
thermal energy of the nuclides and nucleons is too small to overcome  
the Coulomb barriers; any remaining free neutrons decay, and BBN ceases.   
The resulting {\em mass} fraction of helium, conventionally referred to $Y_{\rm p}$, is simply given by
\begin{equation}
Y_{\rm p} \simeq 2 X_{n/N} (t_{\rm BBN}) = 0.251 \,,
\label{yprimordia}
\end{equation}
where the subscript p denotes primordial. The above calculation
demonstrates how the synthesized helium abundance depends on the
physical parameters. After a bit of algebra, (\ref{yprimordia}) can be
rweritten as~\cite{Sarkar:1995dd}
\begin{equation}
Y_{\rm p}  \simeq  0.251 + 0.014 \, \Delta N_\nu^{\rm eff} + 0.0002
\Delta \tau_n  +  0.009 \ln \left(\frac{\eta}{5 \times 10^{-10}} \right) \, .
\end{equation}
\end{widetext}

In summary, primordial nucleosynthesis has a single adjustable
parameter: the baryon density. Observations that led to the
determination of primordial abundance of $D$, $^3$He and $^7$Li can
determine $\eta$. The internal consistency of BBN can then be checked
by comparing the abundances of the other nuclides, predicted using
this same value of $\eta$, with observed abundances. Interestingly, in
contrast to the other light nuclides, the BBN-predicted primordial
abundance of $^4$He is very insensitive to the baryon density
parameter. Rather, the $^4$He mass fraction depends on the
neutron-to-proton ratio at BBN because virtually all neutrons
available at that time are incorporated into $^4$He. Therefore, while
D, $^3$He, and $^7$Li are potential baryometers, $^4$He provides a
potential chronometer.\\

{\bf EXERCISE 8.4}~Suppose that the difference in rest energy of the
neutron and proton were $0.1293~{\rm
  MeV}$, instead of $1.293~{\rm MeV}$, with all other physical parameters
unchanged. Estimate the maximum possible mass fraction
in $^4$He, assuming that all available neutrons are incorporated into
$^4$He nuclei.\\

{\bf EXERCISE 8.5}~A fascinating bit of cosmological history is that
of Gamow's prediction of the CMB in the late
1940s~\cite{Gamow:1946eb,Alpher:1948ve,Gamow:1949zz}. Unfortunately,
his prediction was premature; by the time the CMB was actually
discovered, his prediction had fallen into obscurity. This problem
reproduces Gamow's line of argument.  Gamow knew that nucleosynthesis
must have taken place at a temperature $T_{\rm BBN} \approx 10^9~{\rm
  K}$. He also knew that the universe must currently be $t_0 \sim
10^{10}$ years old. He then assumed that the universe was flat and
radiation dominated, even at the present time.  {\it (i)}~With these
assumptions, what was the energy density of the universe at the time
of nucleosynthesis?  {\it (ii)}~What was the Hubble parameter at the
time of nucleosynthesis?  (c) What was the age of the universe at BBN?
{\it (iv)}~Given the present age, what should the present temperature
of the CMB be? {\it (v)}~If we then assume that the universe changed
from being radiation dominated to matter dominated at a redshift
$z_{\rm eq} > 0$, will this increase or decrease the CMB temperature,
for fixed values of $T_{\rm BBN}$ and $t_0$?\\

The observationally-inferred primordial fractions of baryonic mass in
$^{4}$He ($Y_{\rm p} = 0.2472 \pm 0.0012$, $Y_{\rm p} = 0.2516 \pm
0.0011$, $Y_{\rm p} = 0.2477 \pm 0.0029$, and $Y_{\rm p} = 0.240 \pm
0.006$)~\cite{Izotov:2007ed,Peimbert:2007vm,Steigman:2007xt} have been
constantly favoring $N_\nu^{\rm eff} \lesssim
3$~\cite{Simha:2008zj}. Unexpectedly, two recent independent studies
yield $Y_{\rm p}$ values somewhat higher than previous estimates:
$Y_{\rm p} = 0.2565 \pm 0.001 ({\rm stat}) \pm 0.005 ({\rm syst})$ and
$Y_{\rm p} = 0.2561 \pm
0.011$~\cite{Izotov:2010ca,Aver:2010wd,Aver:2010wq}.  For $\tau_n =
885.4 \pm 0.9~{\rm s}$ and $\tau_n = 878.5 \pm 0.8~{\rm s}$, the
updated effective number of light neutrino species is reported as
$N_{\rm eff} = 3.68^{+0.80}_{-0.70}$ ($2\sigma$) and $N_{\rm
  eff} = 3.80^{+0.80}_{-0.70}$ ($2\sigma$), respectively.  The most
recent estimate of $Y_{\rm p}$ yields $ N_{\rm eff} = 3.58 \pm 0.25
(68\% {\rm CL}), \pm 0.40 (95.4\% {\rm CL}), \pm (99\% {\rm CL})$.
This entails that a non-standard value of $N_{\rm eff}$ is preferred
at the 99\% CL, implying the possible existence of additional types of
neutrino species~\cite{Izotov:2014fga}.\\

{\bf EXERCISE 8.6}~We have seen that he best multi-parameter fit of
Planck data yields a Hubble constant which deviates by more than
$2\sigma$ from the value obtained with the HST. The impact of the
Planck $h$ estimate is particularly important in the determination of
$N_{\rm eff}$. Combining observations of CMB data the Planck
Collaboration reported $N_{\rm eff} = 3.15 \pm
0.23$~\cite{Ade:2015xua}. However, if the value of $h$ is not allowed
to float in the fit, but instead is frozen to the value determined
from the maser-cepheid-supernovae distance ladder, the Planck CMB data
then gives $N_{\rm eff} = 3.62 \pm 0.25$, which suggests new
neutrino-like physics (at around the $2.3\sigma$
level)~\cite{Ade:2013zuv}. The hints of extra relativistic degrees of
freedom at BBN and CMB epochs can be explained, e.g., by means of the
right-handed partners of the three, left-handed, SM neutrinos. In
particular, milli-weak interactions of these Dirac states may allow
the $\nu_R$'s to decouple much earlier, at a higher temperature, than
their left-handed counterparts~\cite{Anchordoqui:2012qu}. Determine
the minimum decoupling temperature of the right-handed neutrinos which
is consistent with Planck data at the $1\sigma$ level.

\subsection{Quantum black holes}

As we have seen, black holes are the evolutionary endpoints of massive
stars that undergo a supernova explosion leaving behind a fairly
massive burned out stellar remnant. With no outward forces to oppose
gravitational forces, the remnant will collapse in on itself.

The density to which the matter must be squeezed
scales as the inverse square of the mass. For example, the Sun would
have to be compressed to a radius of 3~km (about four millionths 
its present size) to become a black hole. For the Earth to meet the
same fate, one would need to squeeze it into a radius of 9~mm, about a
billionth its present size. Actually, the density of a solar mass
black hole ($\sim 10^{19}$~kg/m$^3$) is about the highest that can be
created through gravitational collapse. A body lighter than the Sun
resists collapse because it becomes stabilized by repulsive quantum forces
between subatomic particles.

However, stellar collapse is not the only way to form black holes. The
known laws of physics allow matter densities up to the so-called
Planck value $10^{97}$~kg/m$^3,$ the density at which the force of
gravity becomes so strong that quantum mechanical fluctuations can
break down the fabric of spacetime, creating a black hole with a
radius $\sim 10^{-35}$~m and a mass of $10^{-8}$~kg. This is the
lightest black hole that can be produced according to the conventional
description of gravity. It is more massive but much smaller
in size than a proton.

The high densities of the early universe were a prerequisite for the
formation of primordial black holes but did not guarantee it. For a
region to stop expanding and collapse to a black hole, it must have
been denser than average, so the density fluctuations were also
necessary. As we have seen, such fluctuations existed, at least on
large scales, or else structures such as galaxies and clusters of
galaxies would never have coalesced. For primordial black holes to
form, these fluctuations must have been stronger on smaller scales
than on large ones, which is possible though not inevitable. Even in
the absence of fluctuations, holes might have formed spontaneously at
various cosmological phase transitions -- for example, when the
universe ended its early period of accelerated expansion, known as
inflation, or at the nuclear density epoch, when particles such as
protons condensed out of the soup of their constituent quarks.

The realization that black holes could be so small prompted Stephen
Hawking to consider quantum effects, and in 1974 his studies lead to
the famous conclusion that black holes not only swallow particles but
also spit them out~\cite{Hawking:rv,Hawking:sw}.  The strong gravitational fields
around the black hole induce spontaneous creation of pairs near the
event horizon. While the particle with positive energy can escape to
infinity, the one with negative energy has to tunnel through the
horizon into the black hole where there are particle states with
negative energy with respect to infinity.\footnote{One can alternatively
  think of the emitted particles as coming from the singularity inside
  the black hole, tunneling out through the event horizon to
  infinity~\cite{Hartle:tp}.} As the black holes radiate, they lose
mass and so will eventually evaporate completely and disappear. The
evaporation is generally regarded as being thermal in
character,\footnote{Indeed both the average number~\cite{Hawking:rv,Hawking:sw}
  and the probability distribution of the number~\cite{Parker:1975jm,Wald:1975kc,Hawking:1976ra}
  of outgoing particles in each mode obey a thermal spectrum.} with a
temperature inversely proportional to its mass $M_{\rm BH}$,
\begin{equation}
T_{\rm BH} = \frac{1}{8\pi G \mbh} = \frac{1}{4\,\, \pi\,\,r_s}\,,
\end{equation}
and an entropy $S = 2\, \pi\,\mbh\,r_s,$ where $r_s$ is the
Schwarzschild radius and we have set $c=1$.  
Note that for a solar mass black hole, the temperature is around $10^{-6}$~K,
which is completely negligible in today's universe. But for black holes of
$10^{12}$~kg the temperature is about $10^{12}$~K hot enough to emit
both massless particles, such as $\gamma$-rays, and massive ones, such as
electrons and positrons.  

The black hole, however, produces an effective potential barrier in
the neighborhood of the horizon that backscatters part of the outgoing
radiation, modifing the blackbody spectrum. The black hole absorption
cross section, $\sigma_s$ (a.k.a. the greybody factor), depends upon
the spin of the emitted particles $s$, their energy $Q$, and the mass
of the black hole~\cite{Page:df}. At high frequencies ($ Q r_s \gg 1$)
the greybody factor for each kind of particle must approach the
geometrical optics limit. The integrated power emission is reasonably
well approximated taking such a high energy limit. Thus, for
illustrative simplicity, in what follows we adopt the geometric optics
approximation, where the black hole acts as a perfect absorber of a
slightly larger radius, with emitting area given by~\cite{Page:df}
\begin{equation}
A = 27 \pi r_s^2 \,\,.
\end{equation}
Within this framework, we can conveniently write the greybody factor
as a dimensionless constant normalized to the black hole surface area
seen by the SM fields $\Gamma_s = \sigma_s/A_4,$ such that $\Gamma_{s=0} = 1$,
$\Gamma_{s=1/2} \approx 2/3$, and $\Gamma_{s=1} \approx 1/4$.

All in all, a black hole emits
particles with initial total energy between $(Q, Q+dQ)$ at a rate
\begin{equation}
\frac{d\dot{N}_i}{dQ} = \frac{\sigma_s}{8 \,\pi^2}\,Q^2 \left[
\exp \left( \frac{Q}{T_{\rm BH}} \right) - (-1)^{2s} \right]^{-1}
\label{rate}
\end{equation}
per degree of particle freedom $i$. The change of variables $u=Q/T,$ 
brings
Eq.~({\ref{rate}) into a more familar form,
\begin{equation}
\dot{N}_i = \frac{ 27 \,\Gamma_s\, T_{\rm BH}}{128 \,\pi^3}\, 
\int \frac{u^2}
{e^u - (-1)^{2s}} \,du.
\label{rate3}
\end{equation}
This expression can be easily integrated using (\ref{usando1}) and
(\ref{usando2}), and yields
\begin{equation}
\dot{N_i} = {\cal A}_\pm \frac{27\,\Gamma_s}{128\,\pi^3}  \,\Gamma(3) \, 
\zeta(3) \,T_{\rm BH}\, .
\end{equation}
Therefore, the black hole emission
rate is found to be
\begin{equation}
\dot{N}_i \approx 7.8 \times 10^{20}\, 
\left(\frac{T_{\rm BH}}{{\rm GeV}}\right)\,\, {\rm s}^{-1} \,\,,
\end{equation}
\begin{equation}
\dot{N}_i \approx 3.8 \times 10^{20}\, 
\left(\frac{T_{\rm BH}}{{\rm GeV}}\right)\,\, {\rm s}^{-1} \,\,,
\end{equation}
\begin{equation}
\dot{N}_i \approx 1.9 \times 10^{20}\,
\left(\frac{T_{\rm BH}}{{\rm GeV}}\right)\,\, {\rm s}^{-1} \,\,,
\end{equation}
for particles with $s = 0,\, 1/2, \,1,$ respectively. 

At any given time, the rate of decrease in the black hole mass is just
the total power radiated
\begin{equation}
\frac{d\dot M_{\rm BH}}{dQ} = - \sum_i g_i \,\frac{\sigma_s}{8 \pi^2}\, 
\frac{Q^3}{e^{Q/T_{\rm BH}} - (-1)^{2s}} \,\,,
\end{equation}
where $g_i$ is the number of internal degrees of freedom of particle 
species $i.$ A straightforward calculation gives
\begin{equation}
\dot M_{\rm BH} = - \sum_i g_i\,\, {\cal B}_\pm \,\, \frac{27\, \Gamma_s}{128 \, 
\pi^3} \, \Gamma(4) \, \zeta (4) \,\,T_{\rm BH}^2 \, .
\label{dife}
\end{equation}
Assuming that the effective high energy theory contains approximately
the same number of modes as the SM (i.e., $g_{s=1/2} = 90,$ and
$g_{s=1} = 27$), we find
\begin{equation}
\frac{dM_{\rm BH}}{dt} = 8.3 \times 10^{73}~{\rm GeV}^4 \frac{1}{M_{\rm BH}^2} \,\,.
\label{amateur}
\end{equation}
Ignoring thresholds, i.e., assuming that the mass of the black hole evolves
according to (\ref{amateur}) during the entire process of evaporation, we
can obtain an estimate for the lifetime of the black hole,
\begin{equation}
\tau_{\rm BH} = 1.2 \times 10^{-74}~{\rm GeV}^{-4}\,\,\int M_{\rm BH}^2\,\, dM_{\rm BH} \,\,.
\label{entree}
\end{equation}           
Using $\hslash = 6.58 \times 10^{-25}~{\rm GeV\, s},$ (\ref{entree})
can then be re-written as
\begin{eqnarray}
\tau_{\rm BH} & \simeq & 2.6 \times 10^{-99} 
\,\, (M_{\rm BH}/{\rm GeV})^3~{\rm s} \nonumber \\
             & \simeq &  1.6 \times 10^{-26} \,\, 
(M_{\rm BH}/{\rm kg})^3~{\rm yr} \,\,.
\label{ultima}
\end{eqnarray}
This implies that for a solar mass black hole, the lifetime is
unobservably long $10^{64}$~yr, but for a $10^{12}$~kg one, it is
$\sim 1.5 \times 10^{10}$~yr, about the present age of the universe.
Therefore, any primordial black hole of this mass would be completing
its evaporation and exploding right now.

The questions raised by primordial black holes motivate an empirical
search for them. Most of the mass of these black holes would go into
gamma rays (quarks and gluons would hadronize mostly into pions which
in turn would decay to $\gamma$-rays and neutrinos), with an energy
spectrum that peaks around 100~MeV. In 1976, Hawking and Don Page
realized that $\gamma$-ray background observations place strong
upper limits on the number of such black holes~\cite{Page:1976wx}.
Specifically, by looking at the observed $\gamma$-ray spectrum, they
set an upper limit of $10^{4}/{\rm pc}^3$ on the density of these
black holes with masses near $5 \times 10^{11}~{\rm kg}.$ Even if
primordial black holes never actually formed, thinking about them has
led to remarkable physical insights because they linked three
previously disparate areas of physics: general relativity, quantum
theory, and thermodynamics~\cite{Hawking:1976de}.\\

{\bf EXERCISE 8.7}~Very recently, it has become evident that a
promising route towards reconciling the apparent mismatch of the
fundamental scales of particle physics and gravity is to modify the
short distance behavior of gravity at scales much larger than the
Planck length. Such modification can be most simply achieved by
introducing extra dimensions (generally thought to be curled-up) in
the sub-millimiter
range~\cite{ArkaniHamed:1998rs}. In the canonical
example, spacetime is a direct product of ordinary 4-dimensional
spacetime and a (flat) spatial $n$-torus with circumferences of length
$2 \pi r_i$ ($i = 1, . . . , n$), generally of common linear size $r_i
= r_c$. The SM fields cannot propagate freely in the extra dimensions
without conflict with observations.  This is avoided by trapping the
fields to a 3-dimensional {\it brane-world}. Applying Gauss' law at $r
\ll r_c$ and $r \gg r_c$, it is easily seen that the effective Planck scale 
 is related to the fundamental scale of gravity $M_*$ simply by a volume factor,
\begin{eqnarray}
r_c  & = &  \left(\frac{M_{\rm Pl}}{M_*} \right)^{2/n} \frac{1}{M_*}
\nonumber \\
& = &  2.0 \times 10^{-17} \left(\frac{\rm TeV}{M_*} \right)
\left(\frac{M_{\rm Pl}}{M_*} \right)^{2/n}~{\rm cm} \,,
\end{eqnarray}
so that $M_*$ can range from $\sim {\rm TeV}$ to $10^{19}~{\rm GeV}$,
for $r_c \leq 1~{\rm mm}$ and $n \geq 2$. If nature gracefully picked
a sufficiently low-scale gravity, the first evidence for it would
likely be the observation of microscopic black holes produced in
particle collisions~\cite{Banks:1999gd}. Although the black hole
production cross section, ${\cal O}(M_W^{-1})$, is about 5 orders of
magnitude smaller than QCD cross sections, ${\cal O}(\Lambda_{\rm
  QCD}^{-1}$), it was proposed that such black holes could be produced
copiously at the LHC~\cite{Dimopoulos:2001hw,Giddings:2001bu} and in
cosmic ray collisions~\cite{Feng:2001ib,Anchordoqui:2001cg}, and that
these spectacular events could be easily filtered out of the QCD
background. To a first approximation it is reasonable to assume that
the evaporation process is dominated by the large number of SM brane
modes~\cite{Emparan:2000rs,Anchordoqui:2003ug}. Therefore, the emission rate per
degree of particle freedom $i$ of particles of spin $s$ with initial
total energy between $(Q,Q+dQ)$ can be approximated by (\ref{rate}). The
characteristic temperature of a $4+n$-dimensional black hole is~\cite{Anchordoqui:2002cp}
\begin{equation}
T_{\rm BH} = \frac{n +1}{4\,\pi\,r_s}\, ,
\end{equation}
where  
\begin{equation}
\label{schwarz}
r_s =
\frac{1}{M_*}
\left[ \frac{\mbh}{M_*} \,\, \frac{2^n \pi^{(n-3)/2}\Gamma({n+3\over 2})}{n+2}
\right]^{1/(1+n)}\,,
\end{equation}
is the Schwarzschild radius~\cite{Myers:1986un}.  As in the
convetional 
4-dimensional case, we can conveniently rewrite the
greybody factor as a dimensionless constant, $\Gamma_s =
\sigma_s/A_{4 \subset 4 +n}$, normalized to the black hole surface area
\begin{equation}
A_{4 \subset 4+n} = 4 \pi \left(\frac{n+3}{2} \right)^{2/(n+1)} \, \frac{n+3}{n+1} \, r_s^2
\label{area}
\end{equation}
seen by the SM fields~\cite{Emparan:2000rs}. The upper
limit on the accretion rate for a $4+n$-dimensional black hole is
\begin{equation}
\left. \frac{dM}{dt}\right|_{\rm accr} \approx \pi\,\left( \frac{n+3}{2} \right)^{2/(n+1)}\, \frac{n+3}{n+1} \, 
r^2_s\,\, \epsilon \,\,, 
\end{equation}
where $\epsilon$ is the nearby quark-gluon (or parton)  energy
density~\cite{Chamblin:2003wg}. The highest earthly value of energy
density of partonic matter is the one created at the LHC,
$\epsilon_{\rm LHC} < 500~{\rm GeV/fm}^3$. Consider the case with
$n=6$, which is well motivated by string
theory~\cite{Antoniadis:1998ig}. {\it(i)}~Show that the black holes
that could be produced at the LHC (or in any forseeable accelerator built
on Earth) would evaporate much too quickly to swallow the partons
nearby. {\it (ii)}~Determine the black hole lifetime. [{\it Hint}: For
$n=6$, you can evaluate the numerical results of~\cite{Harris:2003eg} at
$\langle Q \rangle$ and normalize the cross sections results to the
capture area $A_{4 \subset 4+n}$ to obtain $\Gamma_{s=1/2} \approx
0.33$ and $\Gamma_{s=1} \approx 0.34$.]

\section{Multi-messenger Astronomy}

For biological reasons our perception of the Universe is based on the
observation of photons, most trivially by staring at the night-sky
with our bare eyes. Conventional astronomy covers many orders of
magnitude in photon wavelengths, from $10^4$~cm radio-waves to
$10^{-14}$~cm gamma rays of GeV energy.  This 60 octave span in photon
frequency allows for a dramatic expansion of our observational
capacity beyond the approximately one octave perceivable by the human
eye.  

The $\gamma$-ray sky has been monitored since 1968.  The pioneering 
observations by the third Orbiting Solar Observatory (OSO-3) provided
the first $\gamma$-ray sky map, with 621 events detected above 50
MeV~\cite{Kraushaar}. In addition, these observations revealed the
existence of an isotropic emission.  The presence of an isotropic
diffuse $\gamma$-ray background (IGRB) has been confirmed by the Small
Astronomy Satellite 2 (SAS-2)~\cite{Fichtel} and the the Energetic Gamma Ray
Experiment Telescope (EGRET) on board of the Compton Gamma Ray
Observatory (CGRO)~\cite{Sreekumar:1997un,Hartman:1999fc}. Very recently, the
Fermi-LAT has released a new measurement of the IGRB spectrum from
100~MeV to 820~GeV at Galactic latitude $|b| > 20^\circ$~\cite{Ackermann:2014usa}; see Fig.~\ref{fig:fermi+icecube}. The LAT
has also measured the extragalactic $\gamma$-ray background (EGB),
which is the sum of the IGRB and the flux from detected sources. For
the first time a deviation from a power-law shape in the high-energy
part of the EGB and IGRB has been observed as an exponential cut off
with a break energy of about $E_\gamma = 280~{\rm GeV}$. The origin of
the IGRB is not yet fully understood.  This leaves intringuing puzzles for the next
generation of GeV $\gamma$ ray instruments to uncover~\cite{Felix}.
What happens at higher energies?

\begin{figure}[tbp]
\postscript{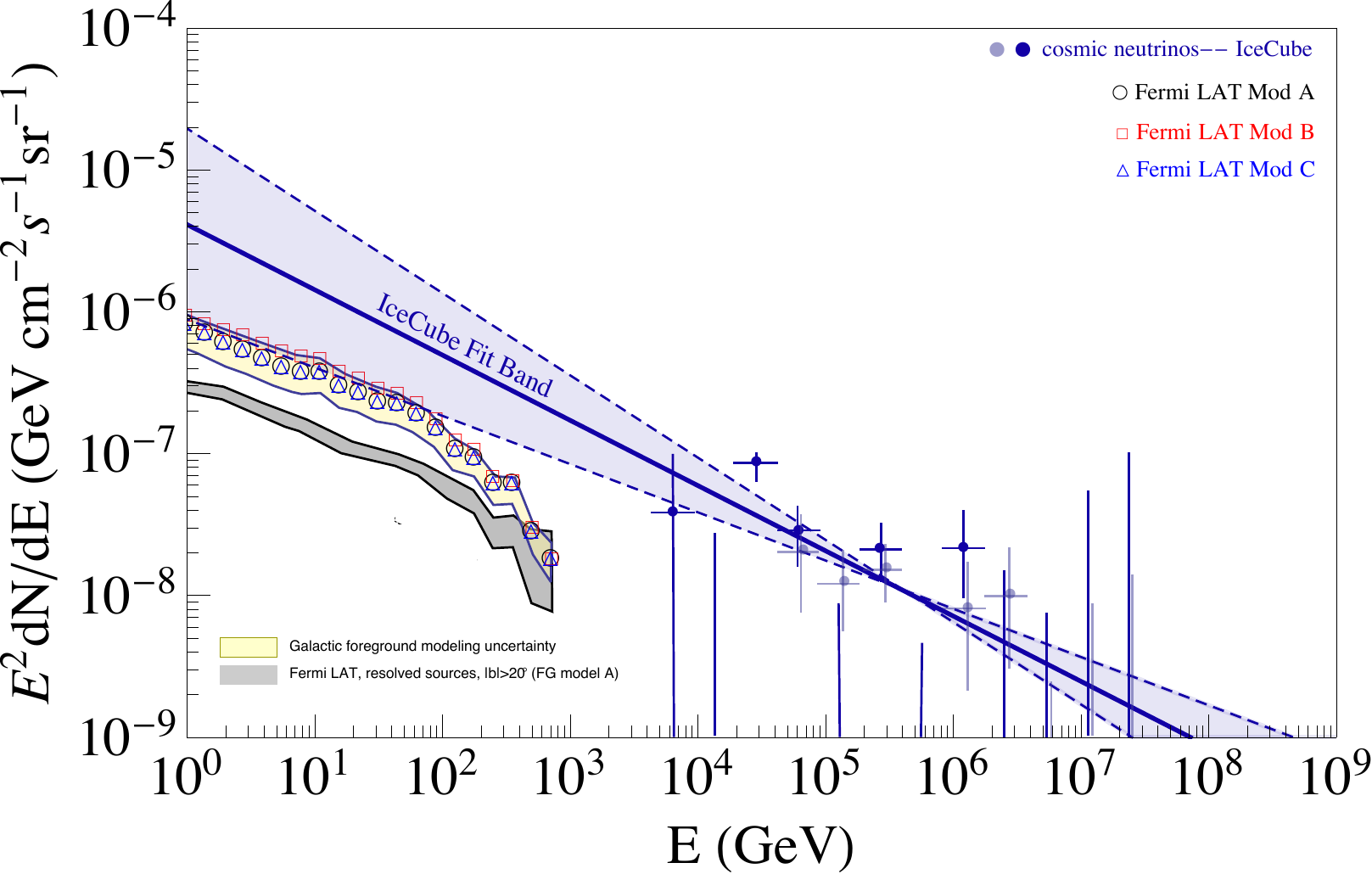}{0.99}
\caption{The open symbols represent the total extragalactic
  $\gamma$-ray background for different foreground (FG) models as
  reported by the Fermi Collaboration~\cite{Ackermann:2014usa}. For details on the modeling of 
the diffuse Galactic foreground emission in the benchmark FG models A,
B and C, see~\cite{Ackermann:2014usa}. The cumulative intensity from
  resolved Fermi-LAT sources at latitudes $|b| > 20^\circ$ is indicated by a (grey)
  band. The solid symbols indicate the neutrino flux reported by
  the IceCube Collaboration~\cite{Aartsen:2014muf}. The best fit to the data (extrapolated
  down to lower energies), $\Phi (E_\nu) = 2.06^{+0.4}_{-0.3} \times
  10^{-18} (E_\nu/10^5~{\rm GeV})^{-2.46 \pm 0.12}~{\rm GeV}^{-1}\ {\rm
      cm}^{-2} \ {\rm s}^{-1} \ {\rm sr}^{-1}$, is also shown
  for comparison~\cite{Anchordoqui:2014rca}. \label{fig:fermi+icecube}}
\end{figure}

Above a few~100 GeV the universe becomes opaque to the propagation of
$\gamma$ rays, because of $e^+ e^-$ production on the radiation fields
permeating the universe; see Fig.~\ref{37}. The pairs synchrotron
radiate on the extragalactic magnetic field before annihilation and so
the photon flux is significantly depleted.  Moreover, the charged
particles also suffer deflections on the $\vec B$-field camouflaging
the exact location of the sources. In other words, the injection
photon spectrum is significantly modified {\it en route} to
Earth. This modification becomes dramatic at around $10^6$~GeV where
interaction with the CMB dominates and
the photon mean free path is smaller than the Galactic radius.

\begin{figure}
 \postscript{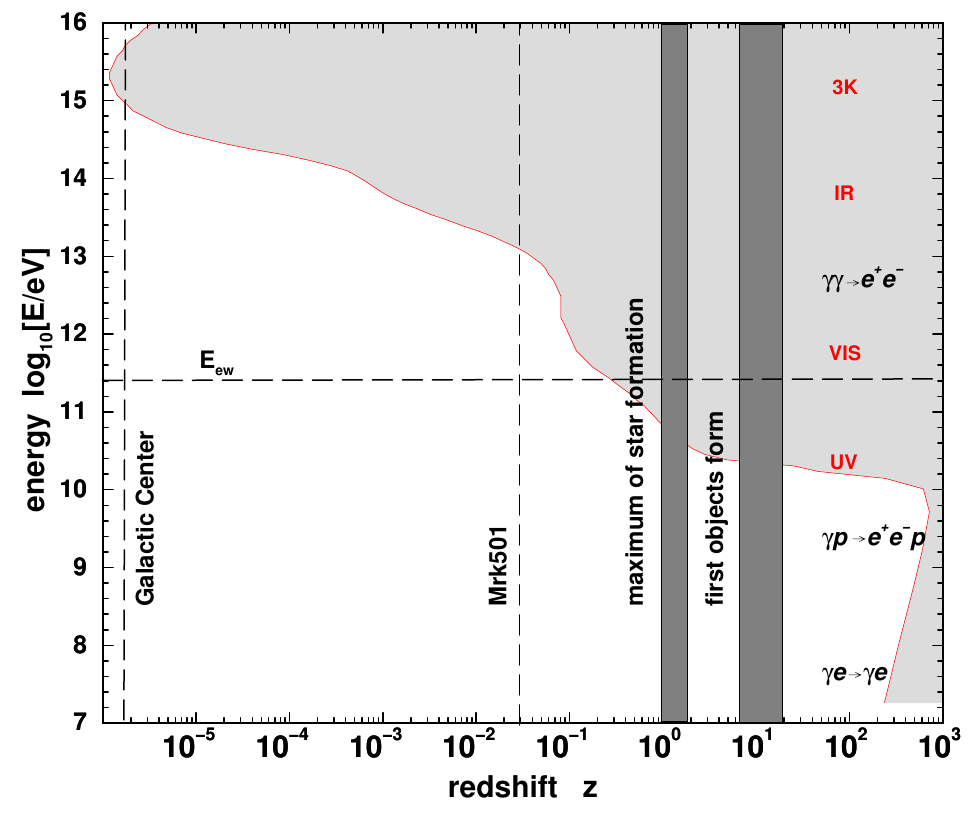}{0.9}
 \caption{Mean interaction length for photons on the
     ultraviolet (UV), visible (VIS), infrared (IR), and microwave (3K)
     backgrounds. The electroweak scale is indicated by a dashed line. 
The redshifts of the star formation epoch and the famous
$\gamma$-ray source Markarian 501 are also indicated~\cite{Learned:2000sw}.}
\label{37}
\end{figure}

Therefore, to study the high energy behavior of distance sources we
need new messengers. Nowadays the best candidates to probe the high
energy universe are cosmic rays, neutrinos, and gravitational waves. Of
course in doing multi-messenger astronomy one has to face new
challenges. It is this that we now turn to study. 

\subsection{Cosmic rays}

In 1912  Hess carried out a series of pioneering balloon flights
during which he measured the levels of ionizing radiation as high as
5~km above the Earth's surface~\cite{Hess}.  His discovery of
increased radiation at high altitude revealed that we are bombarded by
ionizing particles from above. These cosmic ray particles are now
known to consist primarily of protons, helium, carbon, nitrogen and
other heavy ions up to iron.

Below $10^{5}$~GeV the flux of particles is sufficiently large that
individual nuclei can be studied by detectors carried aloft in
balloons or satellites. From such direct experiments we know the
relative abundances and the energy spectra of a variety of atomic
nuclei, protons, electrons and positrons as well as the intensity,
energy and spatial distribution of $X$-rays and $\gamma$-rays.
Measurements of energy and isotropy showed conclusively that one
obvious source, the Sun, is not the main source. Only below 100~MeV
kinetic energy or so, where the solar wind shields protons coming from
outside the solar system, does the Sun dominate the observed proton
flux. Spacecraft missions far out into the solar system, well away
from the confusing effects of the Earth's atmosphere and
magnetosphere, confirm that the abundances around 1~GeV are strikingly
similar to those found in the ordinary material of the solar system.
Exceptions are the overabundance of elements like lithium, beryllium,
and boron, originating from the spallation of heavier nuclei in the
interstellar medium.\\

{\bf EXERCISE 9.1}~Consider a simple model of cosmic rays in the
Galaxy (height $H \ll$ radius $R$) in which the net diffusion of
cosmic rays is mainly perpendicular to the Galactic disk. In this case
the density of cosmic rays depends only on the vertical coordinate $z$
and follows the diffusion equation 
\begin{equation}
\frac{\partial n}{\partial t} = D \frac{\partial^2 n}{\partial z^2} +
Q(z,t) \,,
\end{equation}
where $D = \beta c \lambda/3$ is the diffusion coefficient, $\lambda$
is the mean free path, and the source term is given by $Q(z, t)$. Use
the approximation $Q(z, t) = Q_0 \delta (z)$ to describe a
time-independent concentration of stars close to $z = 0$, $\delta (z)$
is the Dirac delta function (see Appendix~\ref{appE}). {\it
  (i)}~Find the steady-state solution to the diffusion equation given
a vanishing cosmic ray density at the edges of the Galaxy, $n(z = +H)
= n(z = −H) = 0$. {\it (ii)}~Calculate the cosmic-ray column density 
\begin{equation}
N= \int_{-H}^{+H}  n(z) dz 
\end{equation}
and determine the average residence time $\tau_{\rm res}$ from $N =
Q_0 \tau_{\rm res}$. What is the mean free path for $H = 500~{\rm pc}$
and $\tau_{\rm res} = 10^7~{\rm yr}$?\\

Above $10^{5}$~GeV, the flux becomes so low that only ground-based
experiments with large apertures and long exposure times can hope to
acquire a significant number of events. Such experiments exploit the
atmosphere as a giant calorimeter. The incident cosmic radiation
interacts with the atomic nuclei of air molecules and produces
extensive air showers which spread out over large areas.  Already in
1938, Auger concluded from the size of extensive air showers
that the spectrum extends up to and perhaps beyond
$10^{6}$~GeV~\cite{Auger:1938,Auger:1939}. Nowadays substantial
progress has been made in measuring the extraordinarily low flux
($\sim 1$ event km$^{-2}$ yr$^{-1}$) above $10^{10}$~GeV. Continuously
running experiments using both arrays of particle detectors on the
ground and/or fluorescence detectors which track the cascade through the
atmosphere, have detected events with primary particle energies
somewhat above $10^{11}$~GeV~\cite{Bird:1994uy}.

\begin{figure}
\postscript{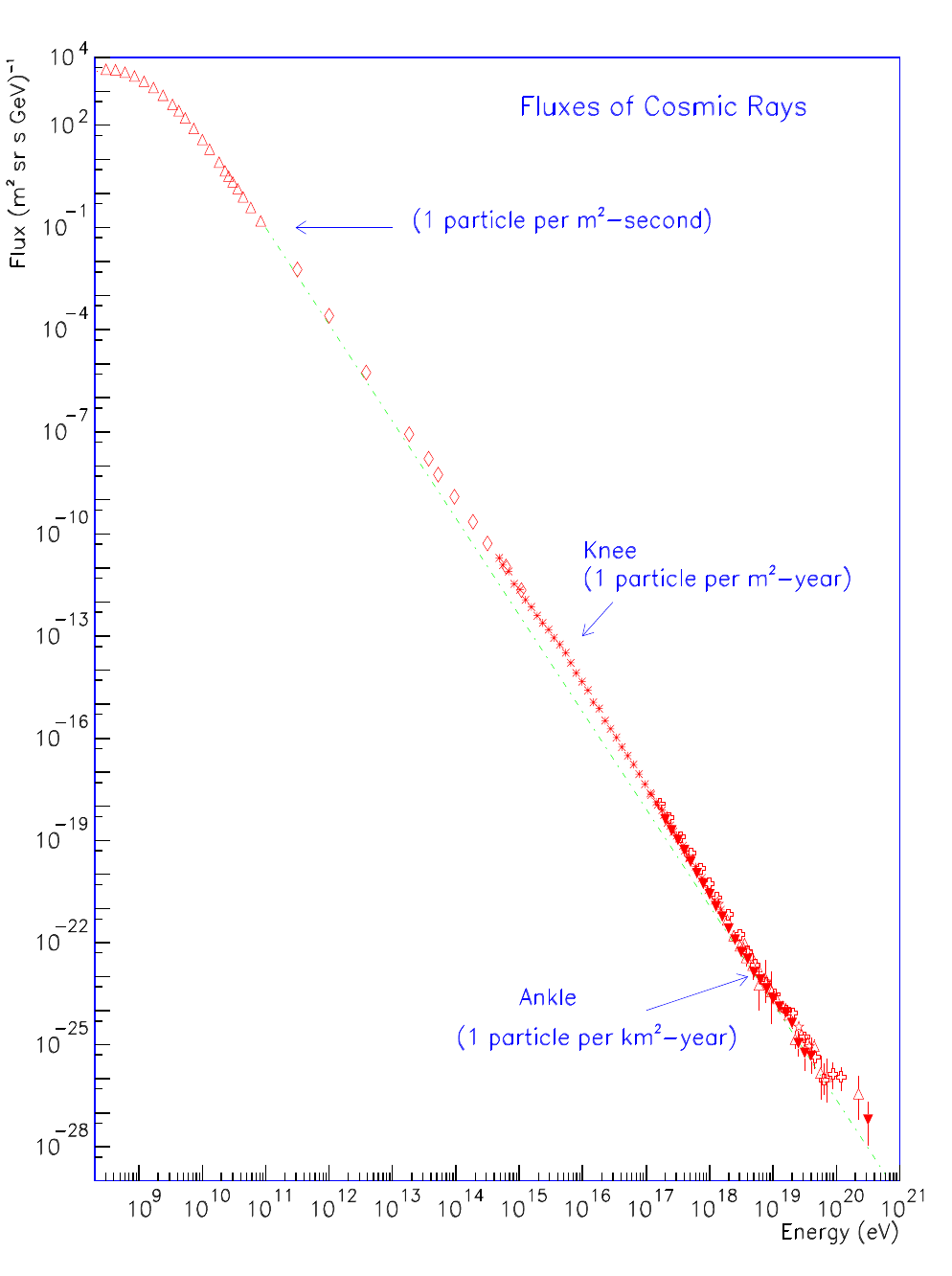}{0.95}
\caption{Compilation of measurements of the differential energy spectrum of 
cosmic rays. The dotted line shows an $E^{-3}$ power-law for comparison. 
Approximate integral fluxes (per steradian) are also shown.}
\label{38}
\end{figure}

The Pierre Auger Observatory employs the two detection
methods~\cite{ThePierreAuger:2015rma}. It consists of an array of
about $1,600$ water Cherenkov surface detectors (SD) deployed over a
triangular grid of 1.5~km spacing and covering an area of
$3,000$~km$^2$~\cite{Abraham:2010zz}. A SD event is formed when at
least 3 non-aligned stations selected by the local station trigger are
in spatial and temporal coincidence.  The ground array is overlooked
by 24 fluorescence telescopes, grouped in four sites, making up the
fluorescence detector (FD)~\cite{Abraham:2009pm}. The FD observes the
longitudinal development of the shower in the atmosphere by detecting
the fluorescence light emitted by excited nitrogen molecules and
Cherenkov light induced by shower particles in air. The two detection methods
have different strengths, and together allow for large statistics data
samples and unrivaled control over systematic uncertainties.

The FD provides a calorimetric measurement of the primary particle
energy, only weakly dependent on theoretical models. The most common
strategy to determine the nature of the primary cosmic ray is to study
the longitudinal shower profile of the electromagnetic component in
the atmosphere.  The slant depth is the amount of atmosphere
penetrated by a cosmic ray shower at a given point in its development,
and is customarily denoted by the symbol $X$. The value of $X$ is
calculated by integrating the density of air from the point of entry
of the air shower at the top of the atmosphere, along the trajectory
of the shower, to the point in question. The depth of the shower
maximum $X_{\rm max}$ is the position of the maximum of energy
deposition per atmospheric slant depth of an extensive air
shower. Lighter primaries penetrate the atmosphere deeper than heavier
primaries. In addition, due to the larger number of nucleons and the
larger cross section, the event-by-event fluctuations of $X_{\rm max}$
should be smaller for heavier nuclei. Therefore, the first two moments
of the $X_{\rm max}$ distribution, which are the mean $\langle X_{\rm
  max} \rangle$ and standard deviation $\sigma (X_{\rm max})$ provide
good discriminators between different primary cosmic rays; for details
see e.g.~\cite{Anchordoqui:2004xb}.

The mechanism(s) responsible for imparting an energy of more than one
Joule to a single elementary particle continues to present a major
enigma to high energy physics~\cite{Torres:2004hk}.  It is reasonable to assume that, in
order to accelerate a proton to energy $E$ in a magnetic field $B$,
the size $R$ of the accelerator must encompass the gyro radius of the
particle: $R > R_{\rm gyro} \sim E/B,$ i.e. the accelerating magnetic
field must contain the particle's orbit. By dimensional analysis, this
condition yields a maximum energy $E \sim \gamma BR.$ The $\gamma$-factor
has been included to allow for the possibility that we may not be at
rest in the frame of the cosmic accelerator, resulting in the
observation of boosted particle energies. Opportunity for particle
acceleration to the highest energies is limited to dense regions where
exceptional gravitational forces create relativistic particle flows.
All speculations involve collapsed objects and we can therefore
replace $R$ by the Schwarzschild radius $R \sim GM/c^2$ to obtain $E <
\gamma BM.$

At this point a reality check is in order. Such a dimensional analysis
applies to the Fermilab accelerator: 10 kilogauss fields over several
kilometers (covered with a repetition rate of $10^5$ revolutions per
second) yield 1~TeV. The argument holds because, with optimized design
and perfect alignment of magnets, the accelerator reaches efficiencies
matching the dimensional limit. It is highly questionable that nature
can achieve this feat.

\begin{figure*}[t!]
\centering \includegraphics[width=0.7\linewidth]{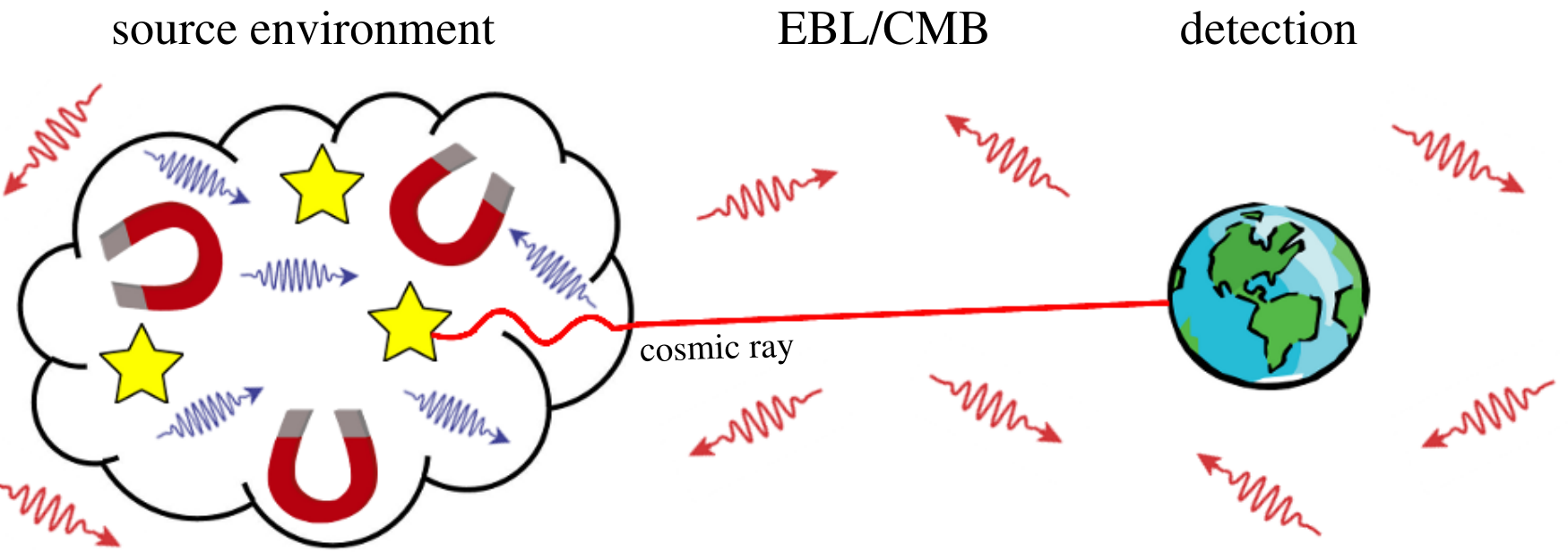}
\caption{Sources (yellow stars) inject cosmic ray nuclei with a power
  law in energy into a surrounding region of radiation and turbulent
  magnetic fields.  After propagation through this local environment
  and intergalactic space, these cosmic rays and their spallation
  products are detected at Earth.  The photon energies in the source
  environment are characteristically of much higher energy than in the
  extragalactic background light (EBL)~\cite{Unger:2015laa}.}
\label{fig:illustration}
\end{figure*}

Given the microgauss magnetic field of our galaxy, no structures are
large or massive enough to reach the energies of the highest energy
cosmic rays. Dimensional analysis therefore limits their sources to
extragalactic objects. A common speculation is that there may be
relatively nearby active galactic nuclei powered by a billion solar
mass black holes. With kilo-Gauss fields we reach $10^{11}$~GeV. The
jets (blazars) emitted by the central black hole could reach similar
energies in accelerating sub-structures boosted in our direction by a
$\gamma$-factor of 10, possibly higher. \\

{\bf EXERCISE~9.2} {\it (i)}~Derive the magnetic field strength needed
to hold a charge on a circular orbit of radius $R$ given its momentum
$p$. Assume that the magnetic field is uniform, that the motion of the
particle is perpendicular to the magnetic field and let $\beta \sim
1$. {\it (ii)}~Given the circumference ($\sim 26.659~{\rm km}$) of the
LHC, determine the uniform magnetic field strength needed to keep 7
TeV protons in orbit. {\it (iii)}~Using this magnetic field strength
find what would be the required size needed for an LHC-like
accelerator to launch particles to cosmic-ray energies $\sim
10^{11}~{\rm GeV}$. Compare this with the orbits in the solar system
and estimate the cost of the accelerator. {\it (iv)}~Which of the
following astrophysical objects are able to keep ultrahigh energy
cosmic rays in orbit? Neutrons stars ($R \sim 10^{-13}~{\rm pc},$ $B
\sim 10^{12}~{\rm G}$), AGN jets ($R \sim 1~{\rm kpc},$ $B \sim
10^{-5}~{\rm G}$), supernova remnants $(R = 1~{\rm pc}$ $B \sim
10^{-4}~{\rm G})$. Consider protons and iron nuclei of energy
$10^{11}~{\rm GeV}$ and that you are at rest in the frame of the
cosmic accelerator.\\

The almost structureless power law spectrum
spans many decades of energy,  $10^1~{\rm GeV} < E < 10^{11}~{\rm GeV}$. A close
examination of Fig.~\ref{38} reveals three major features: {\it
  (i)}~the steepening of the spectrum dubbed the {\it knee} centered
at $10^{6.6}~{\rm GeV}$ ~\cite{Antoni:2005wq}; {\it (ii)}~a pronounced
hardening of the spectrum at about $10^{9.6}~{\rm GeV}$, the so-called {\it
  ankle} feature~\cite{Bird:1993yi};
{\it (iii)}~a cutoff around $10^{10.6}~{\rm
  GeV}$~\cite{Abbasi:2007sv,Abraham:2008ru}.  Three additional more
subtle features have been recently spotted between the knee and the
ankle: a hardening of the spectrum at around $10^{7.3}~{\rm
  GeV}$~\cite{Apel:2012tda,Aartsen:2013wda} followed by two softenings at $\approx 10^{7.9}~{\rm
  GeV}$~\cite{Apel:2012tda, Aartsen:2013wda} and $\approx 10^{8.5}~{\rm
  GeV}$~\cite{AbuZayyad:2000ay, Bergman:2007kn}. The latter is traditionally
referred to as the {\it second knee}.

The variations of the spectral index reflect various aspects of cosmic
ray production, source distribution, and propagation.  The first and
second knee have unequivocal explanations, as reflecting the maximum
energy of Galactic magnetic confinement or acceleration capability of
the sources, both of which grow linearly in the charge $Z$ of the
nucleus; the first knee being where protons drop out and the second
knee where the highest-$Z$ Galactic cosmic rays drop out.  As the
energy increases above the second knee to the ankle, the nuclear composition
switches from heavy to light~\cite{Kampert:2012mx} whereas the cosmic
ray arrival directions are isotropic to high accuracy throughout the
entire range~\cite{Abreu:2011ve,Auger:2012an,ThePierreAuger:2014nja}.
Lastly, as the energy increases above the ankle, not only does the
spectrum harden significantly, but the composition gradually becomes
heavier (interpreting the data using conventional extrapolations of
accelerator-constrained particle physics
models)~\cite{Aab:2014kda,Aab:2014aea}.

The observed evolution in the extragalactic cosmic ray composition and
spectral index presents a complex puzzle.  A pure proton composition
might be compatible with the observed spectrum of extragalactic cosmic
rays~\cite{Berezinsky:2002nc} when allowance is made for experimental
uncertainties in the energy scale and the fact that the real local
source distribution is not homogeneous and
continuous~\cite{Ahlers:2012az} (although the sharpness of the ankle
is difficult to accommodate).  However, a pure proton composition is
incompatible with the $X_{\rm max}$ and $\sigma (X_{\rm max})$
distributions reproted by the Auger
Collaboration~\cite{Aab:2014kda,Aab:2014aea} unless current
extrapolations of particle physics are incorrect.  On the other hand,
models which fit the spectrum and composition at highest energies,
predict a deep gap between the end of the Galactic cosmic rays and the
onset of the extragalactic cosmic rays. Models can be devised to fill
this gap: fine-tuning is required to position this new population so
as to just fit and fill the
gap~\cite{Aloisio:2013hya,Giacinti:2015hva}, unless we consider
interactions in the region surrounding the accelerator as illustrated
in Fig.~\ref{fig:illustration}.

The discovery of a suppression above $10^{10.6}~{\rm GeV}$ 
was first reported by the HiRes and Auger
collaborations~\cite{Abbasi:2007sv,Abraham:2008ru} and later confirmed
by the Telescope Array Collaboration~\cite{AbuZayyad:2012ru}; by now
the significance is well in excess of $20\sigma$ compared to a
continuous power law extrapolation beyond the ankle
feature~\cite{Abraham:2010mj}. This suppression is consistent with the
Greisen-Zatsepin-Kuzmin (GZK) prediction that interactions with cosmic
background photons will rapidly degrade cosmic ray
energies~\cite{Greisen:1966jv,Zatsepin:1966jv}. Intriguingly, however,
there are also indications that the source of the suppression may be
more complex than originally anticipated. The trend toward heavier
composition above the ankle could reflect the endpoint of cosmic
acceleration, with heavier nuclei dominating the composition near the
end of the spectrum, which coincidentally falls off near the expected
GZK cutoff region~\cite{Aloisio:2009sj}. If this were the case, the
suppression would constitute an imprint of the accelerator
characteristics rather than energy loss in transit. It is also
possible that a mixed or heavy composition is emitted from the
sources, and photodisintegration of nuclei and other GZK energy losses
suppress the flux.

The main reason why this impressive set of data fails to reveal the
origin of the particles is undoubtedly that their directions have been
scrambled by the microgauss Galactic magnetic fields. However,
above $10^{10}$~GeV proton astronomy could still be possible because
the arrival directions of electrically charged cosmic rays are no
longer scrambled by the ambient magnetic field of our own Galaxy.
Protons point back to their sources with an accuracy determined by
their gyroradius in the intergalactic magnetic field $B$,
\begin{equation}
\theta \simeq \frac{d}{R_{\rm gyro}} = \frac{dB}{E} \,\,,
\end{equation}
where $d$ is the distance to the source. Scaled to units relevant to the 
problem,
\begin{equation}
\frac{\theta}{0.1^\circ} \simeq \frac{(d/{\rm Mpc})\, (B/{\rm nG})}{E/10^{11.5}~{\rm GeV}} \,\,.
\end{equation}
Speculations on the strength for the inter-galactic magnetic field
range from $10^{-7}$ to $10^{-9}$~G. For the distance to a nearby
galaxy at 100~Mpc, the resolution may therefore be anywhere from
sub-degree to nonexistent. Moreover, neutrons with energy $\agt
10^{9}$~GeV have a boosted $c\tau_n$ sufficiently large to serve as
Galactic messengers~\cite{Anchordoqui:2003vc}.\footnote{Neutron
  astronomy from the nearby radio galaxy Centaurus A may also be
  possible~\cite{Anchordoqui:2001nt}.} The decay mean free path of a
neutron is $c\,\gamma_n\,\overline\tau_n=9.15\,(E_n/10^9~{\rm
  GeV})$~kpc, the lifetime being boosted from its rest-frame value,
$\overline\tau_n=886$~s, to its lab value by $\gamma_n=E_n/m_n$.  It
is therefore reasonable to expect that the arrival directions of the
very highest energy cosmic rays may provide information on the
location of their sources.\footnote{For a more extensive discussion of
  this 
subject see e.g.~\cite{Anchordoqui:2011gy}.}

\subsection{Cosmic neutrinos}

For a deep, sharply focused examination of the universe a telescope is
needed which can observe a particle that is not much affected by the
gas, dust, and swirling magnetic fields it passes on its journey.
The neutrino is the best candidate. As we have seen, neutrinos constitute
much of the total number of elementary particles in the universe, and
these neutral, weakly-interacting particles come to us almost without
any disruption straight from their sources, traveling at very close to
the speed of light. A (low energy) neutrino in flight would not notice
a barrier of lead fifty light years thick. When we are able to see
outwards in neutrino light we will no doubt receive a wondrous new view
of the universe.\\

{\bf EXERCISE 9.3}~In 1987, the astronomical world was electrified
with the news of a supernova exploding in the Large Magellanic Cloud,
a dwarf galaxy companion to the Milky Way, at a distance of
150,000~ly. It was the nearest supernova to have gone off in 400~yr,
and was studied in great detail. Its luminosity was enormous; the
explosion released as much visible light energy in a few weeks as the
Sun will emit in its entire lifetime of $10^{10}~{\rm yr}$. It was
easily visible to the naked eye from the Southern hemisphere. However,
models of the mechanisms taking place in the supernovae predict that
the visible light represents only 1\% of the total energy of the
supernova; there is 100 times more energy emitted in the form of
neutrinos, in a blast lasting only a few seconds. {\it (i)}~Calculate
the total amount of energy emitted by the supernova in
neutrinos. Express your answer in Joules. {\it (ii)}~Each neutrino has
an energy of roughly $\langle E_\nu \rangle \sim 1.5 \times 10^{-12}~{\rm J}$. Calculate how many
neutrinos are emitted by the supernova. (This is an easy calculation,
but will give you a very large number). {\it (iii)}~Kamiokande is one
of the largest neutrino detectors. In 1987, it consisted of 2.140~kton
of water (it has since been expanded).  Calculate how many electron
neutrinos should have been detected by Kamiokande if the detection
efficiency at $\langle E_{\nu_e} \rangle$ ia bout 60\%~\cite{Schramm:1987ra}.\\

We have seen that MeV neutrinos are are produced by nuclear reaction
chains in the central core of stars. Moving up in energy, neutrinos
would also be inevitably produced in many of the most luminous and
energetic objects in the universe.  Whatever the source, the machinery
which accelerates cosmic rays will inevitably also produce neutrinos,
guaranteeing that high energy neutrinos surely arrive to us from the
cosmos.

Neutrino detectors must be generally placed deep underground, or in
water, in order to escape the backgrounds caused by the inescapable
rain of cosmic rays upon the atmosphere. These cosmic rays produce
many muons which penetrate deeply into the earth, in even the deepest
mines, but of course with ever-decreasing numbers with depth. Hence
the first attempts at high energy neutrino astronomy have been
initiated underwater and under ice~\cite{Anchordoqui:2009nf}.

The IceCube facility is located near the Amundsen-Scott station below
the surface of the Antarctic ice sheet at the geographic South
Pole~\cite{Achterberg:2006md}. The main part of the detector is the {\it InIce} array, which covers
a cubic kilometer of Antarctic glacial ice instrumented with digital
optical modules (DOMs) that detect Cherenkov ligh~\cite{Abbasi:2008aa}. The DOMs are
attached to km-long supply and read-out cables called strings.  Each
string carries 60 DOMs spaced evenly along 1 km. The full baseline
design of 86 strings was completed in December 2010. In addition to
the InIce array, IceCube also possesses an air shower array called
{\it IceTop} which comprises 80 stations, each of which consists of two
tanks of water-ice instrumented with 2 DOMs to detect Cherenkov light~\cite{IceCube:2012nn}. The hybrid observations of air showers in the InIce and IceTop
arrays have mutual benefits, namely significant air shower background
rejection (for neutrino studies) and an improved air shower muon
detection (for cosmic ray studies).

In 2012, the IceCube Collaboration famously announced an observation
of two $\sim$~1 PeV neutrinos discovered in a search for the nearly
guaranteed cosmogenic neutrinos (which are expected to produced as
secondaries in the GZK chain reaction~\cite{Beresinsky:1969qj})~\cite{Aartsen:2013bka}. The
search technique was later refined to extend the neutrino sensitivity
to lower energies~\cite{Schonert:2008is,Gaisser:2014bja}, resulting in
the discovery of an additional $26$ neutrino candidates with energies
between $50$~TeV and $2$~PeV, constituting a $4.1\sigma$ excess for
the combined 28 events compared to expectations from neutrino and muon
backgrounds generated in Earth's
atmosphere~\cite{Aartsen:2013jdh}. Interpretation of these results,
however, does not appear to be entirely straightforward. For instance,
if one makes the common assumption of an unbroken $E_\nu^{-2}$
neutrino energy spectrum, then one expects to observe about 8-9 events
with higher energies than the two highest energy events observed thus
far. The compatibility between IceCube observations and the hypothesis
of an unbroken power-law spectrum requires a rather steep spectrum,
$\Phi (E_\nu) \propto E^{-2.3}$~\cite{Anchordoqui:2013qsi}. Very
recently, the IceCube results have been
updated~\cite{Aartsen:2014gkd,Aartsen:2015knd,Aartsen:2015zva}.  At
the time of writing, 54 events have been reported in four years of
IceCube data taking (1347 days between 2010 -- 2014). The data are
consistent with expectations for equal fluxes of all three neutrino
flavors~\cite{Aartsen:2015ivb}. The best-fit power law is $E_\nu^2
\Phi(E_\nu) = 2.2 \pm 0.7 \times 10^{-8} (E_\nu/100~{\rm
  TeV})^{-0.58}~{\rm GeV} \, {\rm cm}^{-2} \, {\rm s}^{-1} {\rm
  sr}^{-1}$ and rejects a purely atmospheric explanation at more than
$5.7\sigma$. Splitting the data into two sets, one from the northern
sky and one from the souther sky, allows for a satisfactory power law
fit with a different spectral index for each hemisphere. The best-fit
spectral index in the northern sky is $\gamma_N = 2.0^{+0.3}_{-0.4}$,
whereas in the southern sky it is $\gamma_S = 2.56 \pm 0.12$~\cite{Aartsen:2015knd}. The
discrepancy with respect to a single power law corresponds to
$1.1\sigma$ and may indicate that the neutrino flux is anisotropic~\cite{Neronov:2015osa,Neronov:2016bnp}. The largest concentration of events is at or near the Galactic
center, within uncertainties of their reconstructed arrival
directions~\cite{Razzaque:2013uoa,Bai:2014kba,Anchordoqui:2016dcp}. There are
numerous proposed explanations for the origin of IceCube’s
events~\cite{Anchordoqui:2013dnh}. However, considerably more data are
yet required before the final verdict can be given.

\subsection{Gravitational waves}

\begin{figure}
 \postscript{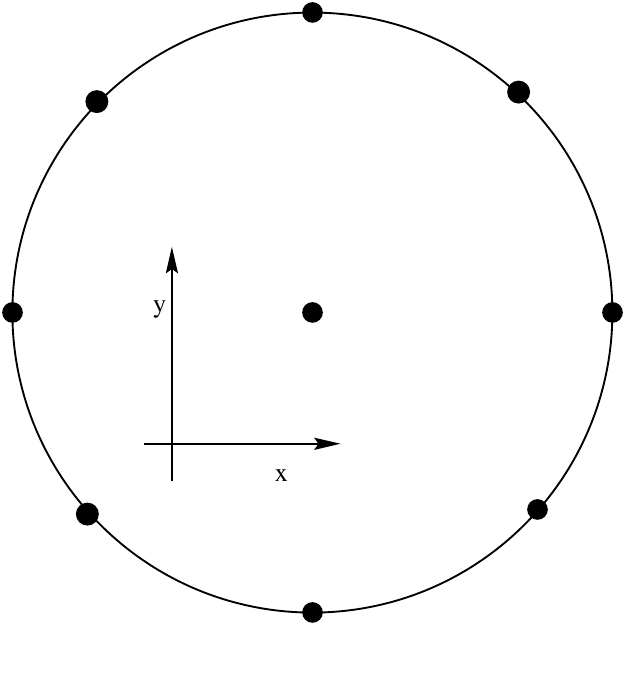}{0.55}
 \caption{Initial configuration of test particles on a circle of radius $L$ before a gravitational wave hits them.}
\label{40}
\end{figure}

\begin{figure}
 \postscript{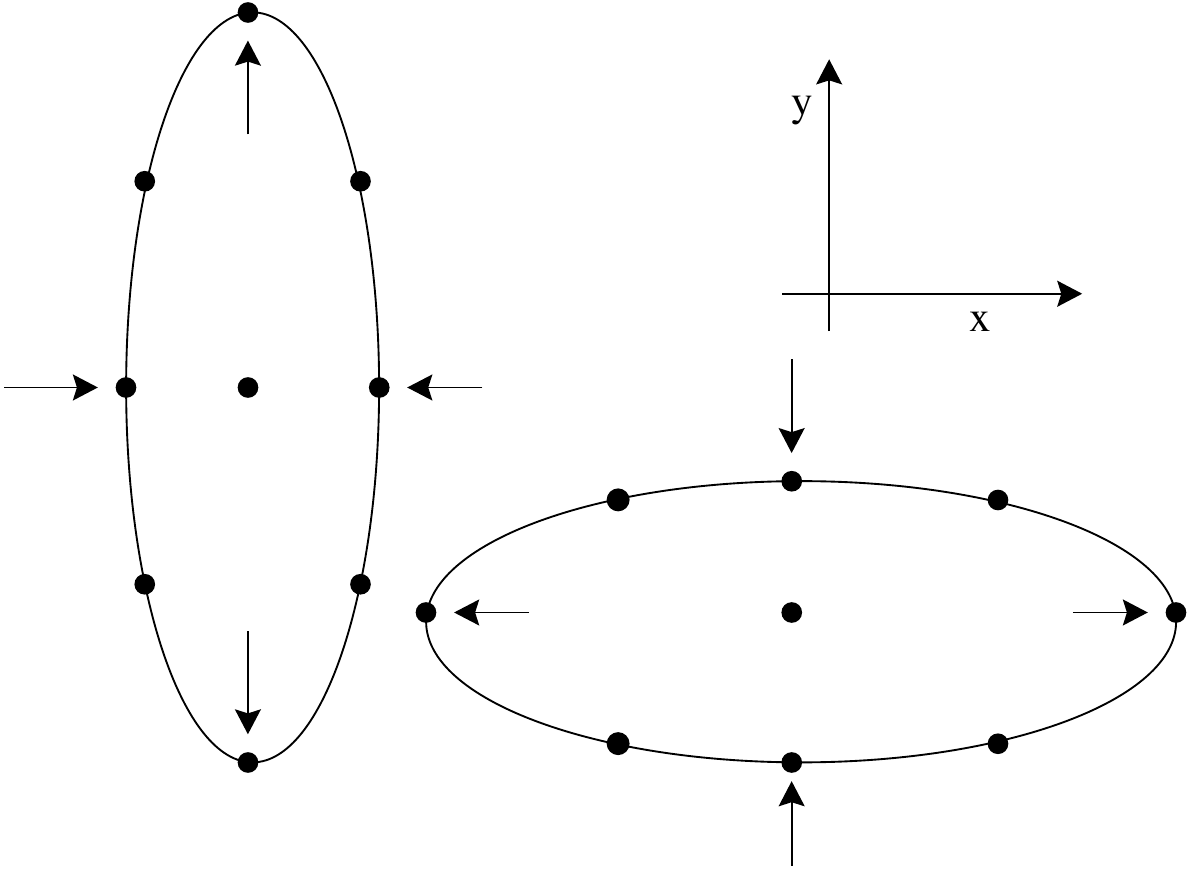}{0.9}
 \caption{The effect of a plus-polarized gravitational wave 
on a ring of particles. The amplitude shown in the figure is roughly 
$h = 0.5$. Gravitational waves passing through the Earth are many billion
  billion times weaker than this.}
\label{41}
\end{figure}

\begin{figure}
 \postscript{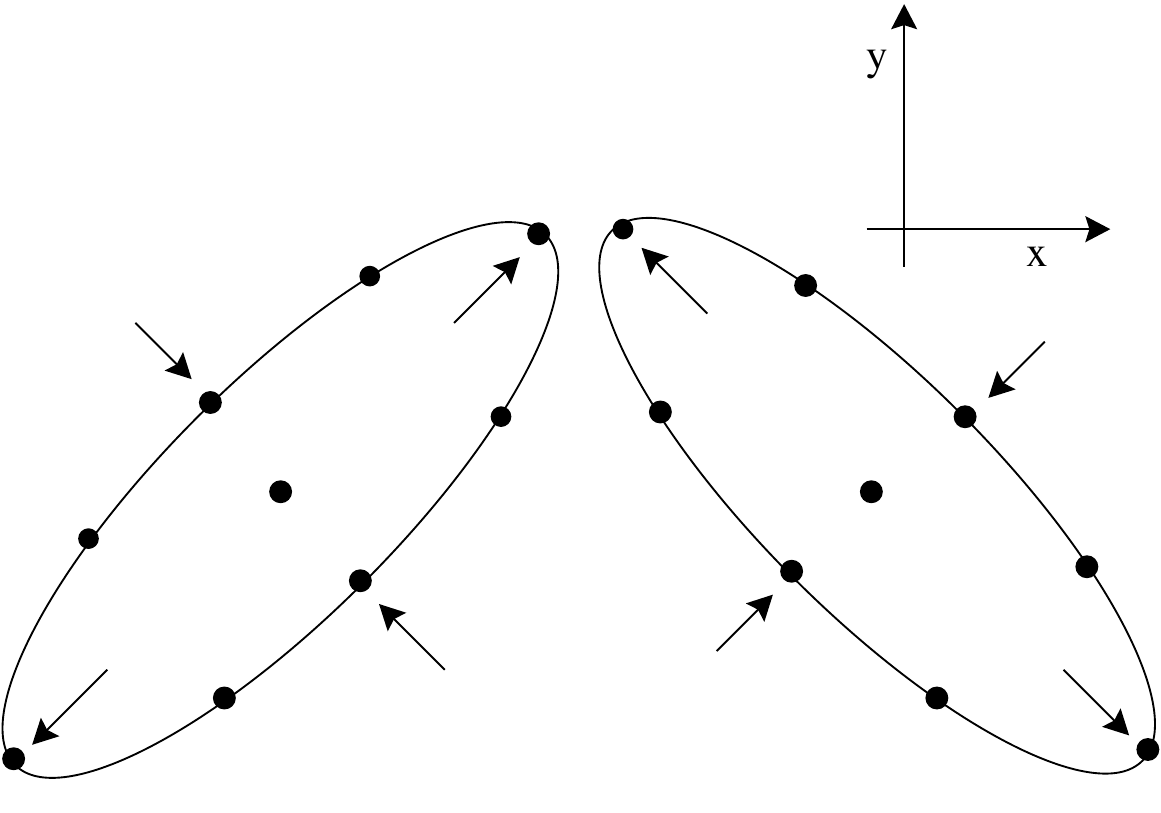}{0.9}
 \caption{The effect of cross-polarized gravitational waves 
on a ring of particles.}
\label{42}
\end{figure}

\begin{figure}
 \postscript{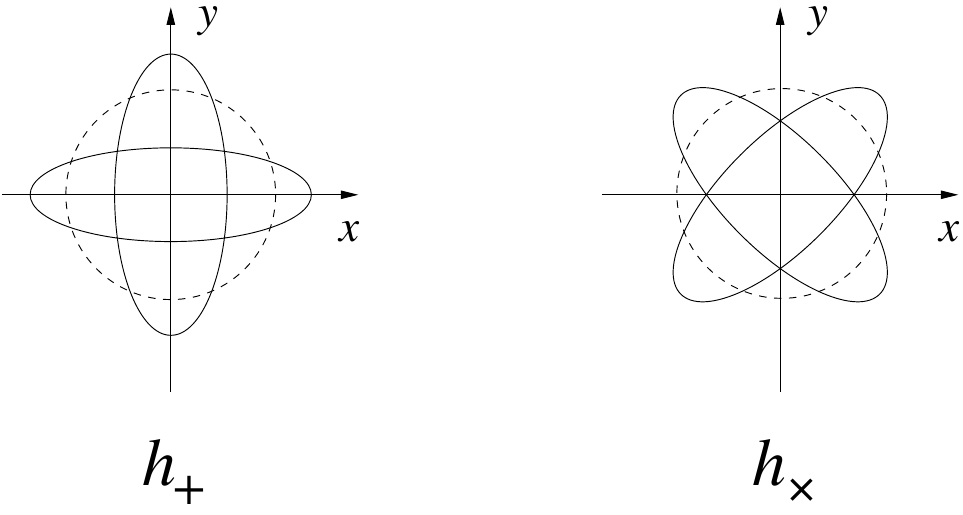}{0.9}
 \caption{Two linearly independent polarizations of a gravitational
   wave are illustrated by displaying their effect on a ring of free
   particles arrayed in a plane perpendicular to the direction of the
   wave. The figure shows the distortions in the original circle that
   the wave produces if it carries the plus-polarization or the
   cross-polarization. In general relativity there are only 2
   independent polarizations. The ones shown here are orthogonal to
   each other and the polarizations are transverse to the direction of
   the wave.}
\label{43}
\end{figure}

Ever since Newton in the XVII century, we have learned that
gravity is a force that acts immediately on an object. In Einstein
theory of general relativity, however, gravity is not a {\it force} at
all, but a curvature in space~\cite{Einstein:1916vd}. In other words,
the presence of a very massive body does not affect probed objects
directly; it warps the space around it first and then the objects move
in the curved space.  Inherit from such a redefinition of gravity is
the concept of gravitational waves: as massive bodies move around,
disturbances in the curvature of spacetime can spread outward, much
like a pebble tossed into a pond will cause waves to ripple outward
from the source.  Propagating at (or near) the speed of light, these
disturbances do not travel {\it through} spacetime as such -- the fabric
of spacetime itself is oscillating!

The simplest example of a strong source of gravitational waves is a
spinning neutron star with a small mountain on its surface. The
mountain's mass will cause curvature of the spacetime. Its movement
will {\it stir up} spacetime, much like a paddle stirring up water. The
waves will spread out through the universe at the speed of light,
never stopping or slowing down.

As these waves pass a distant observer, that observer will find
spacetime distorted in a very particular way: distances between
objects will increase and decrease rhythmically as the wave passes.
To visualize this effect, consider a perfectly flat region of
spacetime with a group of motionless test particles lying in a plane,
as shown in Fig.~\ref{40}.  When a weak gravitational wave arrives, passing
through the particles along a line perpendicular to the ring of radius
$L$, the test particles will oscillate in a {\it cruciform} manner, as
indicated in Figs.~\ref{41} and \ref{42}. The area enclosed by the
test particles does not change, and there is no motion along the
direction of propagation. The principal axes of the ellipse become $L
+\Delta L$ and $L-\Delta L$. The amplitude of the wave, which measures
the fraction of stretching or squeezing, is $h = \Delta L/L.$ Of 
course the size
of this effect will go down the farther the observer is from the
source. Namely, $h \propto d^{-1},$ where $d$ is the source distance. 
Any gravitational waves expected to be seen on Earth will be
quite small, $h \sim 10^{-20}.$ 

The frequency, wavelength, and speed of a gravitational wave are
related through $\lambda = c \nu.$ The polarization of a gravitational
wave is just like polarization of a light wave, except that the
polarizations of a gravitational wave are at $45^\circ$, as opposed to
$90^\circ$. In other words, the effect of  a {\it cross}-polarized gravitational
wave ($h_\times$) on test particles would be
basically the same as a wave with plus-polarization ($h_{+}$), 
but rotated by $45^\circ.$ The different polarizations are summarized 
in Fig.~\ref{43}.

In general terms, gravitational waves are radiated by very massive
objects whose motion involves acceleration, provided that the motion
is not perfectly spherically symmetric (like a spinning, expanding or
contracting sphere) or cylindrically symmetric (like a spinning disk).
For example, two objects orbiting each other in a quasi-Keplerian
planar orbit will radiate~\cite{Thorne:1980rt,Thorne:1980ru}. The power given off by
a binary system of masses $M_1$ and $M_2$ separated a distance $R$ is~\cite{Peters:1963ux}
\begin{equation}
P = -\frac{32}{\pi} \frac{G^4}{c^5} \frac{(M_1\, M_2)^2 (M_1 + M_2)}{R^5} \,\,.
\label{PGWs}
\end{equation}
For the Earth-Sun system $R$ is very large and $M_1$ and $M_2$ 
are relatively very small, yielding 
\begin{widetext}
\begin{equation}
 P = - \frac{32}{\pi}\, \frac{(6.7 \times 10^{-11}\, \frac{\mathrm{m}^3} 
{\mathrm{kg}\, \mathrm{s}^{2}})^4}{(3\times10^8\, \mathrm{m/s})^5}\, 
\frac{(6\times 10^{24}\,\mathrm{kg} \,\, 2\times 10^{30}\,\mathrm{kg})^2 (6\times 10^{24}\,\mathrm{kg} + 2\times 10^{30}\,\mathrm{kg})} 
{(1.5\times10^{11} \mathrm{m})^5} = 313~{\rm W} \,\,. 
\end{equation}
\end{widetext}
Thus, the total power radiated by the Earth-Sun system in the form of
gravitational waves is truly tiny compared to the total
electromagnetic radiation given off by the Sun, which is about $3.86 \times
10^{26}~{\rm W}.$ The energy of the gravitational waves comes out of
the kinetic energy of the Earth's orbit. This slow radiation from the
Earth-Sun system could, in principle, steal enough energy to drop the
Earth into the Sun. Note however that the kinetic energy of the Earth orbiting
the Sun is about $2.7 \times 10^{33}$~J. As the gravitational radiation is
given off, it takes about 300~J/s  away from the orbit.
At this rate, it would take many billion times more than the current
age of the universe for the Earth to fall into the Sun. 

Although the power radiated by the Earth-Sun system is minuscule, we
can point to other sources for which the radiation should be
substantial. One important example is the pair of stars (one of which
is a pulsar) discovered by Hulse and Taylor~\cite{Hulse:1974eb}. The
characteristics of the orbit of this binary system can be deduced from
the Doppler shifting of radio signals given off by the pulsar. Each of
the stars has a mass about 1.4 $M_\odot$. Also, their orbit is about
75 times smaller than the distance between the Earth and Sun, which
means the distance between the two stars is just a few times larger
than the diameter of our own Sun. This combination of greater masses
and smaller separation means that the energy given off by the
Hulse-Taylor binary will be far greater than the energy given off by
the Earth-Sun system, roughly $10^{22}$ times as much.

The information about the orbit can be used to predict just how much
energy (and angular momentum) should be given off in the form of
gravitational waves. As the energy is carried off, the orbit will
change; the stars will draw closer to each other. This effect of
drawing closer is called an inspiral, and it can be observed in the
pulsar's signals. The measurements on this system were carried out
over several decades, and it was shown that the changes predicted by
gravitational radiation in general relativity matched the observations
very well, providing the first experimental evidence for gravitational
waves.

Inspirals are very important sources of gravitational waves. Any time
two compact objects (white dwarfs, neutron stars, or black holes) come
close to each other, they send out intense gravitational waves. As the
objects come closer and closer to each other (that is, as $R$ becomes
smaller and smaller), the gravitational waves become more and more
intense. At some point these waves should become so intense that they
can be directly detected by their effect on objects on the Earth. This
direct detection is the goal of several large experiments around the
world.

The great challenge of this type of detection, though, is the
extraordinarily small effect the waves would produce on a detector.
The amplitude of any wave will fall off as the inverse of the distance
from the source. Thus, even waves from extreme systems like merging
binary black holes die out to very small amplitude by the time they
reach the Earth.  For example, the amplitude of waves given off by the
Hulse-Taylor binary as seen on Earth would be roughly $h \approx
10^{-26}.$ However, some gravitational waves passing the Earth could
have somewhat larger amplitudes, $h\approx
10^{-20}$~\cite{Thorne:1980rt,Thorne:1980ru}. For an object 1~m in
length, this means that its ends would move by $10^{-20}$~m relative
to each other. This distance is about a billionth of the width of a
typical atom.

A simple device to detect this motion is the laser interferometer,
with separate masses placed many hundreds of meters to several
kilometers apart acting as two ends of a bar. Ground-based
interferometers are now operating, and taking data. The most sensitive
is the Laser Interferometer Gravitational Wave Observatory
(LIGO)~\cite{Abramovici:1992ah}. This is actually a set of three
devices: one in Livingston, Louisiana; the other two (essentially on
top of each other) in Hanford, Washington. Each consists of two light
storage arms which are 2 to 4~km in length. These are at $90^\circ$
angles to each other, and consist of large vacuum tubes running the
entire 4 kilometers. A passing gravitational wave will then slightly
stretch one arm as it shortens the other. This is precisely the motion
to which an interferometer is most sensitive.

On September 14, 2015 at 09:50:45~UTC gravitational waves were
detected by both of the twin LIGO detectors~\cite{Abbott:2016blz}. The waves originated in
the collision and merger of two black holes (with 29 and 36 $M_\odot$)
approximately 400~Mpc from Earth.  About 3 times the mass of the sun
was converted into gravitational waves in a fraction of a second, with
a peak power output about 50 times that of the whole visible
universe. This detection inaugurates a new era of astronomy in which
gravitational waves are tools for studying the most mysterious and
exotic objects in the universe.\\

{\bf EXERCISE 9.4} {\it (i)}~Estimate the power radiated in gravitational waves by a
neutron star of $M_\star = 1.4 M_\odot$ orbiting a black hole of
$M_{\rm BH} = 20 M_\odot$, assuming the orbital radius is $R= 6GM_{\rm
  BH}/c^2.$ {\it (ii)}~If the kinetic energy of the neutron star orbiting
the black hole is about $7 \times 10^{47}~{\rm J},$ how much time 
will it take the neutron star to fall into the black hole?

\subsection{Looking ahead}

The recent observation of a diffuse astrophysical flux of high energy
neutrinos and the direct detection of gravitational waves represents
the {\it first light} in the nascent field of multimessenger
astronomy. The search for correlations in the different data sample
has already
started~\cite{Aartsen:2014mfp,Aartsen:2015dml,Fermi-LAT:2016qqr,Adrian-Martinez:2016xgn}. Thus
far, there are no excesses beyond randomly expected.

An in-depth exploration of the neutrino universe requires a
next-generation IceCube detector. IceCube-Gen2 is based upon
the robust design of the current detector~\cite{Aartsen:2014njl}. The goal for this  new
observatory is to deliver statistically significant samples of very
high energy astrophysical neutrinos, in the $10^6~{\rm GeV} \alt E_\nu
\alt 10^9~{\rm GeV}$ range, and yield hundreds of neutrinos across all
flavors at energies above 100~TeV. This will enable detailed spectral
studies, significant point source detections, and new
discoveries. Companion experiments in the deep Mediterranean are
moving into construction phase, and the space-based CHerenkov from Astrophysical
Neutrinos Telescope (CHANT)  is in the R\&D phase~\cite{Neronov:2016zou}.

Resolving the fundamental questions of UHECR composition and origins,
and investigating particle physics above accelerator energies, will
require both enhanced experimental techniques implemented at the
existing observatories, as well as a significant increase in exposure
to catch the exceedingly rare highest energy events. In the very near
future the upgrade of the Pierre Auger Observatory, named {\it Auger
  Prime}, will allow: {\it (i)}~a precise reconstruction of mass
dependent energy spectrum; {\it (ii)}~the identification of primaries,
event-by-event, up to the highest energies; {\it (iii)}~a systematic
study of arrival direction(s) of an enhanced proton data
sample~\cite{Aab:2016vlz}.

Even before we know the results from Auger Prime it seems clear that
still larger aperture observatories with much better energy and
$X_{\rm max}$ resolution will be called for in order to measure the
spectra and composition distribution of individual sources. It is
inspiring to note that some 5 million UHECRs above about $5.5 \times
10^{10}~{\rm GeV}$ strike the Earth's atmosphere each year, from which
we currently collect only about 50 or so with present
observatories. In this sense, there exists some 5 orders of magnitude
room for improvement!  It may well be that the best hope to make
inroads in this area is to take the search for UHECR sources into
space from which a huge volume of atmosphere can be viewed using the
fluorescence technique. To this end several path finder efforts are
underway to develop the requisiste technologies. For example, in 2017
a NASA/CNES supported mission to fly a {\it super-pressure}
stratospheric ballom with a fluorescence detector will take
place. Such ballons can fly for hundreds of days, and may observed the
first air showers from above. Eventually these technologies may lead
to a permanently orbiting satellite to detect UHECRs. An optimist
might even imagine an eventual constellation of satellites to tap the
remaining 5 orders of magnitude of UHECR ``luminosity'', accessing
naturally occuring particle beams at energies far in excess available
to terrestrial colliders with an event rate opening up a new window on
beyond-the-standard-model phenomena.

\acknowledgments{I'm thankful to Michael Unger for entertaining
  discussions and Walter Lewin for a very thorough reading of the
  notes and insightful comments. I'd also like to thank Heinz
  Andernach for helpful remarks. This work has been supported by
  U.S. National Science Foundation (NSF) CAREER Award PHY1053663 and
  by the National Aeronautics and Space Administration (NASA) Grant
  No. NNX13AH52G.}


\onecolumngrid

\section*{Answers and Comments on the Exercises}

1.1~This is a simple application of Kepler's third law. For $a =
30.066~{\rm AU}$, (\ref{Kepler-h}) gives 164.85~yr. The answer is given to five
significant figures, the same number as we have for the semi-major
axis. {\it (ii)}~The exact same calculation for Pluto gives 248.1~yr,
to four significant figures.  {\it (iii)}~The ratio of orbital times
is 248/164.8 = 1.505 to four significant figures. This is quite close
(within 0.3\%) to a ratio of $3:2$. That is, every time Pluto makes
two orbits around the Sun, Neptune makes three orbits. These
resonances are actually quite common in the Solar System. It turns out
that many of the gaps in Saturn’s rings are due to resonances with the
various moons of Saturn, and more complicated resonances explain some
of the stunning detailed features seen in those rings. {\it (iv)}~The
eccentricity of the orbit is very close to zero, so we are not
surprised that the aphelion distance, $a(1 + e) = 30.4~{\rm AU}$, is
very close to the semi-major axis. Here, the number of significant
figures is subtle. You might think that the eccentricity is known to
only a single significant figure, so that the aphelion should be given
to the same significance. In fact, what counts in this calculation is
the quantity $1 + e$, which has three significant figures. {\it
  (v)}~The perihelion is $a(1 - e) = 39.48  (1 − 0.250)~{\rm AU} =
29.61~{\rm AU}$,
and the aphelion, similarly, is $a(1 + e) = 49.35~{\rm AU}$. The perhelion
distance of Pluto is less than the aphelion distance of Neptune, so
indeed, Pluto is sometimes a bit closer to the Sun than is Neptune. It
only gets a little inside Neptune's orbit, and it turns out that it
was last inside Neptune's orbit from 1979 to 1999.\\

1.2~This is another application of Kepler's third law; we have a
period (24 hours) and want to find a radius of the orbit. However,
note that this is not an orbit around the Sun, and so Kepler's
third law in its original form is not valid. Rather, we can use
Newton's form of Kepler's third law:
\begin{equation}
a^3 = \frac{G M_\oplus {\cal T}^2}{4 \pi^2} \, .
\end{equation}
where $M_\oplus$ is the mass of the Earth and we will approximate as
${\cal T} \approx 90,000~{\rm s}$. Thus $a \sim 4.2 \times 10^7~{\rm
  m}$. Are we done? Well, we were asked for the distance from the
Earth's surface, whereas what we have calculated is from the Earth's
center. So we need to subtract from this the radius of the Earth,
$6371~{\rm km}$, leaving roughly $35,629~{\rm km}$. Besides, we are asked to express this number
in Earth radii; dividing by $6371~{\rm km}$ gives about 5.6 Earth
radii.\\

1.3~{\it (i)}~The radius of the orbit was $R \approx 6371~{\rm km} + 200~{\rm
  km} \approx 6600~{\rm km}$. The circumference of this circular orbit
is $2\pi R = 4.2 \times 10^4~{\rm km}$. So, the station has done $8.6
\times 10^4$ orbits. Now, how long does each orbit take? This is the
inverse of the problem we have done in exercise 1.2. We know the semi-major
axis of the orbit (the radius for a circular orbit), and can use
Kepler's law applied to the Earth to find the period:
\begin{equation}
{\cal T}_{\rm Mir} = \left(\frac{4 \pi^2}{GM_\oplus}\right)^{1/2} a^{3/2}
\approx 5,400~{\rm s} = 90~{\rm minutes}.
\end{equation}
So, if every orbit takes roughly 90 minutes, $8.6 \times 10^4$ orbits
take $4.6 \times 10^8~{\rm s}$, or 15~yr, rounding off to the needed
precision. {\it (ii)}~We found that the period of each orbit was close
to 90 minutes. Therefore, it made 16 turns per day. Now,
what would an orbit of 20 revolutions per day look like? Let's find
the radius of this orbit. The period of this orbit is shorter by 16/20
from 90 minutes for Mir (or 72 minutes). We can find its orbital
radius the hard way by using the full Newton's version of Kepler's law
again, or else, we can remember that Kepler's law holds for both orbits
around the Earth, so we can find the ratio of both
expressions. Namely:
\begin{equation}
\left(\frac{{\cal T}_{\rm new \, sat}}{{\cal T}_{\rm Mir}} \right)^2 =
\left( \frac{a_{\rm new \, sat}}{a_{\rm Mir}} \right)^3 \Rightarrow
a_{\rm new\, sat} = a_{\rm Mir} \left(\frac{{\cal T}_{\rm new \,
      sat}}{{\cal T}_{\rm Mir}} \right)^{2/3} = 0.86 a_{\rm Mir} = 5,700~{\rm km} \, .
\end{equation}
This would have been a nice circular orbit, except for the fact that it is 700~km underground.\\

2.1~The Earth radius is $R_\oplus \simeq 6378~{\rm km}$ and we
consider the two measurements separated 784~km in latitude. Then the
two cities are separated by $7^\circ$ in latitude. If we interpret the
$7^\circ$ separation in terms of stellar parallax then the distance to
the Sun would be $d \approx 784~{\rm km}/ \tan 7^\circ \sim 6,385~{\rm km}$. If
we interpret night and day as the Sun revolving around a flat disk,
then the Sun will  crash into the Earth.\\

2.2~{\it (i)}~The solar constant is ${\cal F}_\odot = 1.3 \times
10^3$~W/m$^2$. {\it (ii)}~The absolute luminosity is $L_\odot = 3.7 \times
10^{26}$~W. {\it (iii)}~The Sun-Mars distance is $D =1.524~{\rm
  AU}$. Then  
\begin{equation}
F = {\cal F}_\odot \left(\frac{d_{\rm Sun-Earth}}{d_{\rm Sun-Mars} } \right)^2
  = 590~{\rm W} \, {\rm m}^{-2} \, .
\end{equation}
{\it (iii)} The power is 
$P = F \ A_{\rm solar \, panels} \ \epsilon = 767~{\rm W}$,
where $A_{\rm solar \, panels } = 1.3~{\rm m^2}$ and $\epsilon = 0.2$
is the efficiency. \\

2.3~Let $D_1$ be the distance between Mercury and Saturn when they are
closest to each other and $D_2$ the distance when they are further
appart. {\it (i)}~Then the distance from Mercury to the Sun is $D_{\rm
  MS} = (D_2 - D_1)/2 = 3.2~{\rm lm} \simeq 0.385~{\rm AU}$. {\it
  (ii)}~The distance between Saturn and the Sun is $D_{\rm SS} = D_1 +
D_{\rm MS} = 79.5~{\rm  lm} \simeq 9.57~{\rm AU}$. NASA's MESSENGER
spacecraft slammed into the surface of Mercury on the 30 April 2015
bringing a groundbreaking mission to a dramatic end. The probe crashed
at 3:26~pm red-sox time (1926 GMT), gouging a new crater into
Mercury's heavily pockmarked surface. This violent demise was
inevitable for MESSENGER, which had been orbiting Mercury since March
2011 and had run out of fuel. The 10-foot-wide (3 meters) spacecraft
was traveling about 8,750~mph (14,080~km/h) at the time of impact, and
it likely created a smoking hole in the ground about 52 feet (16~m)
wide in Mercury's northern terrain. No observers or instruments
witnessed the crash, which occurred on the opposite side of Mercury
from Earth. Cassini is the fourth space probe to visit Saturn and the first to enter orbit, and its mission is ongoing as of 2016. It has studied the planet and its many natural satellites since arriving there in 2004. \\

2.4~The observed luminosity from the Sun when it is not eclipsed is
$\pi R_\odot^2 \sigma T_\odot^4$. When Jupiter passes in front of the
Sun, it blocks an area of size $\pi r_{\rm J}^2$, and the observed
luminosity decreases to $\pi (R_\odot^2 - r_{\rm J}^2) \sigma
T_\odot^4$, where $r_{\rm J} = 71,492~{\rm km}$. The fractional
decrease in the apparent surface brightness is
\begin{equation}
\frac{\Delta I}{I_\odot} = \frac{\pi (R_\odot^2 - r^2_{\rm J}) \sigma T_\odot ^4}{\pi R_\odot^2 \sigma
T_\odot^4}  - 1 = - \frac{r_{\rm J}^2}{R_\odot^2} \approx - 0.01 \, ,
\label{Jeclipse}
\end{equation}
The eclipse only reduces the brightness by about 1\%.\\

2.5~{\it (i)}~The range of distances consistent with the measured
parallax angle is $1/0.006~{\rm pc} < D < 1/0.004~{\rm pc}$, or
equivalently $ 167~{\rm pc} < D < 250~{\rm pc}$. {\it (ii)}~The
faintest stars that can be detected with the HST have apparent
brightnesses which are $4 \times 10^{21}$ fainter than the Sun. This
implies that the HST apparent brightness lower threshold is $I_{\rm
  th} = 2.5 \times 10^{-22} \, I_\odot$. For a star $\star$ like the
Sun, $L_\star = L_\odot$. Hence using (\ref{L}) and the ratio method
the distace to the star is found to be $d_L = d \,
\sqrt{I_\odot/I_{\rm th}} = 6.3 \times 10^{10}~{\rm AU} = 306~{\rm
  kpc} = 10^6~{\rm ly},$ where $d = 1~{\rm AU}$ is the Earth-Sun
distance and $1~{\rm pc} = 206265~{\rm AU} = 3.262~{\rm ly}$.  {\it
  (iii)}~For Cepheids, $L_{\rm C} = 2 \times 10^4L_\odot$. Because we
want to calculate the limiting distance for observing this object we
take $I_{\rm C} = I_{\rm th}$ and so $d_L = d \sqrt{(L_{\rm C}
  I_\odot)/(L_\odot I_{\rm th})} = 8.94 \times 10^{12}~{\rm AU} =
43.4~{\rm Mpc} = 1.41 \times
10^8~{\rm ly}$.\\

2.6~{\it (i)}~As the Earth moves, the direction to which we point to
Eris changes. In order to do calculations, we need to know how fast
the Earth travels around the Sun. This can be done simply by
remembering that the Earth travels the circumference of the Earth's
orbit in the time of one year, thus the speed is given by the distance
divided by the time,
\begin{equation}
 v \approx \frac{2 \pi \times 1.5 \times 10^8{\rm km}}{3 \times 10^7~{\rm s}}
   =30~{\rm km/s} ,
\end{equation}
to one significant figure (we made the standard approximation that
$\pi \approx 3$).  At this speed, how far does Earth move in 5 hours
(i.e. $18,000~{\rm s}$)? Rounding to $20,000~{\rm s}$, the answer is
$\ell = 600, 000~{\rm km}$. The parallax diagram is a long skinny
triangle, with the $600,000~{\rm km}$ of the Earth's path at the base,
the distance $d$ to Eris its length, and a $7.5''$ angle at its apex.
If we measure angles in radians and if the angles are small, then
there is a very simple relation between the long side $d$ and short
side $\ell$ of long skinny triangles: $\theta =\ell/d$. Now, since
$7.5'' \approx 4 \times 10^{-5}$ radians, the distance to Eris is $d=
\ell/\theta \approx 1.5 \times 10^{10}~{\rm km}$.  There are $1.5
\times 10^8~{\rm km}$ in an AU, so the distance to Eris is 100
AU. (For this problem, we can take the distance from the Sun to Eris,
and from the Earth to Eris, as essentially the same).  The semi-major
axis of Pluto’s orbit is 40~AU; Eris is quite a bit more
distant. Interestingly, Eris is on a highly elliptical orbit, $e =
0.44$ (much larger than any of the other planets in the Solar System)
and is currently very close to aphelion. Its perhelion is actually
within the orbit of Pluto, but it will not be there for another 280
years, given its 560-year orbital period. {\it (ii)}~The brightness of
the Sun as seen at Eris is given by the inverse square law: $I =
L_\odot/(4\pi d^2) $. This is the amount of energy per unit time per
unit area received at Eris. The cross sectional area of Eris is $\pi
r^2$, and thus the total energy per unit time falling on Eris is the
product of the apparent brightness and that area, namely: $L_\odot
r^2/(4 d^2)$. However, only a fraction $\mathfrak{a}$ of that light is
reflected, the rest is absorbed. So our final answer is: $r^2
\mathfrak{a} L_\odot/ (4d^2)$. {\it (iii)}~What we have calculated in
part {\it (ii)} is the luminosity (in reflected light, at least), of
Eris. We are observing it a distance $d$ away (again, we are taking
the approximation that the distance from the Sun to Eris, and from
Eris to the Earth, are essentially the same). So its brightness just
follows from the inverse square law, namely:
\begin{equation}
I_{\rm Eris} = \mathfrak{a} L_\odot \left(\frac{r}{2d} \right)^2 \frac{1}{4 \pi d^2} =
\mathfrak{a} L_\odot \frac{r^2}{16 \pi d^4} \, .
\label{Ieris}
\end{equation}
The brightness of a distant asteroid falls off as the inverse fourth
power of the distance. This is why these distant guys took so long to
be discovered; they are really faint. {\it (iv)}~Let's solve (\ref{Ieris}) for $r$
\begin{equation}
r = 4 \pi d^2 \sqrt{\frac{\pi b}{\mathfrak{a} L_\odot}} \sim 1.5
\times 10^5~{\rm m} \, .
\end{equation}
This is seriously large, larger than Pluto itself.  The discovery of
Eris (named after the goddess of strife and discord in Greek
mythology) set off a controversy in the astronomical community about
whether it should be called a planet, and what the definition of a
planet is. Reams have been written on this subject. The basic problem
is that the concept of a planet has evolved as we have learned more,
and we now realize that things that might conceivably be called
planets now fall into a variety of categories:
\begin{itemize}
\item {\it terrestrial planets}, relatively small rocky objects in the
  inner solar system, including Mercury, Venus, Earth and Mars;
\item {\it gas giants}, much much larger and more massive bodies in
  the outer part of the solar system, including Jupiter, Saturn,
  Uranus and Neptune (many have argued that Uranus and Neptune should
  be in their own category of {\it ice giants} as they are actually mostly
  frozen gas, not vaporous gas like Jupiter and Saturn);
\item {\it dwarf planets}, some found in the main asteroid belt
  between Mars and Jupiter, but a class to which Pluto, Eris, and
  other recent discoveries belong.
\end{itemize}
And this list does not yet include the massive moons of the Earth,
Jupiter, and Saturn (the largest of which are considerably larger than
Pluto), or the planets discovered around other stars. The term {\it
  planet} is now too broad to allow a single, all-encompassing clean
definition, and our field has become richer with the discovery of Eris
and its brethren. {\it (v)}~The diameter of Eris is $D_{\rm Eris}
\approx 3,000~{\rm km}$, and the distance is $d = 100~{\rm
  AU}$. Imagining the triangle covering the diameter of Eris on one
end, and Earth at the other vertex, the angular size of Eris is then
\begin{equation}
\theta = D_{\rm Eris}/d = 2 \times 10^{-7} \approx 40~{\rm
milliarcseconds} \,,
\end{equation}
where we used the conversion $1~{\rm radian} \approx 200, 000~{\rm
  arcseconds}$. This is at the very limit of the resolution of HST,
so you can resolve it (in fact, HST did). {\it (vi)}~This is a simple
application of Newton's form of Kepler's third law, which relates the
period and semi-major axis of an orbiting body to the mass of the
object it orbits. Here we are given the period and the semi-major
axis, and we need to calculate the mass. Solving for the mass gives:
\begin{equation}
M = \frac{4 \pi^2 a^3}{G {\cal T}^2} \, .
\label{mdeuz}
\end{equation}
However, do we have the semi-major axis? What we’re given is the angle
the semi-major axis subtends in our HST
images. Another opportunity to use the small-angle formula. Consider a
very long skinny triangle, with length given by the distance from the
Earth to Eris (100 AU) and interior angle $0.53''$; we want to find the
base of the triangle. It is: 
\begin{equation}
a = \theta d = 0.53~{\rm arcsec} \ \frac{1~{\rm radian}}{2 \times
    10^5~{\rm arcsec}}  1.5 \times 10^{10}~{\rm km} = 37,000~{\rm
      km} \, .
\end{equation}
If we plug this into (\ref{mdeuz}) we find:
\begin{equation}
M = \frac{4 \pi^2 \times (3.7 \times 10^7~{\rm m})^3}{6.674 \times
  10^{-11}~{\rm m}^3 \ {\rm s}^{-2} \ {\rm kg}^{-1} \times (15.8~{\rm days}
  \times 86,400~{\rm s}/{\rm day})^2} = 1.6 \times 10^{22}~{\rm kg} \,
.
\end{equation}
This is remarkable: Eris is a bit more massive than is Pluto.\\

2.7~{\it (i)}~The luminosity is not isotropic, because the solid
angle subtended by the blackbody from different directions will be
different, and the surface brightness should be constant, so each
observer will see a different flux depending on the direction. {\it
  (ii)}~The maximum flux will be seen along the $z$-axis, because that
is where the subtended solid angle is greatest. The solid angle is
$\Omega = \pi(a/d_L)^2$, because the blackbody appears as a circular
disk of angular radius $a/d_L \ll 1$. In addition, using $a \ll d_L$,
the flux is simply $F = \int I\, d\Omega$ (because we can approximate
$\cos\theta\simeq 1$, where $\theta$ is the angle between each point
and the center of the observed object, whenever the object observed
has an angular size much smaller than 1 radian). Using $I = B =
(\sigma T^4 / \pi)$ at the surface of the blackbody, $F = I \Omega =
\sigma T^4 (a/d_L)^2$. {\it (iii)}~The minimum flux will be seen along
any direction along the equator, where the subtended solid angle will
be smallest. Now, the projected image of the ellipsoid is an ellipse,
with solid angle $\Omega=\pi(ab/d_L^2)$, so just as before, $F = I
\Omega = \sigma T^4 (ab/d_L^2)$.  {\it (iv)}~Everyone sees the same
{\it apparent} surface brightness, $I=\sigma T^4/\pi$.  {\it (v)}~The total
luminosity is the area times $\sigma T^4$. All we need is to find the
area of the ellipsoid, which can be done for example by dividing the
ellipsoid into thin rings of radius $x$ parallel to the $x-y$
plane. For the area A we find:
\begin{equation}
A = 2\times 2\pi \int_0^a r\, dr\, \left[ 1+ \left( {dz\over dr}
\right)^2 \right]^{1/2} ~,
\end{equation}
where $z=b (1-r^2/a^2)^{1/2}$, and
\begin{equation}
L = (\sigma T^4)\, 4\pi\, \int_0^a dr\, r\,
\left[{ a^2 - x^2(1-q^2) \over a^2 - x^2 } \right]^{1/2} ~,
\end{equation}
where $q=b/a$. If you have the patience, the integral can actually be
solved.  {\it (vi)}~The galaxy is different because it is made of
individual stars, and each star is spherical (or even if they are
oblate because they are rotating, they should have their spin axes
randomly oriented and uncorrelated). The
condition $NR^2 \ll ab$ guarantees that stars do not block each other,
so the flux observed is the sum of the flux from each star and the luminosity is
isotropic. This is true also for any optically thin gas, where each
atom emits isotropically. Therefore, the flux is the same for all
observers along different directions:
\begin{equation}
F = { L\over 4\pi d_L^2} ~,
\end{equation}
where $L$ is the total luminosity of the galaxy, and the {\it
  apparent} surface
brightness is different. If all the luminosity is contained within
the oblate ellipsoid, the average apparent surface brightness within the
projected ellipsoid as seen by the observer on the $z$-axis is
\begin{equation}
I = F/\Omega = {L\over 4\pi^2 a^2 } \,,
\end{equation}
and for the observer on the equator,
\begin{equation}
I = F/\Omega = {L\over 4\pi^2 a b } \, .
\end{equation}
The apparent surface brightness is higher when the galaxy is seen edge-on because
the surface density of stars is greater, since the line of sight
crosses a greater pathlength through the galaxy. {\it (vii)}~The
answer would be modified because stars would block each other so some
stars would be occulted. This would be a really compact galaxy since
its mean surface brightness would be similar to that of the Sun, and
the stellar atmospheres would be actually heated by the other stars in
the galaxy significantly. The galaxy would not be stable because every
star would on average have a physicall collision with another star
once every orbit (a good recipe for making a big black hole and some
fireworks, more on this below).\\

2.8~Taking the $\log_{10}$ of (\ref{Luminosity}) we have
\begin{equation}
\log_{10} L = \log_{10} (4 \pi R^2 \sigma T^4) = \log_{10} (4 \pi \sigma) + \log_{10} R^2 +
\log_{10} T^4 = \log_{10} (4 \pi \sigma) + 2 \log_{10} R + 4 \log_{10} T \, .
\end{equation}
If $R$ is constant then $\log_{10} (4 \pi \sigma) + 2 \log_{10} R =$
constant. Then, for constant $R$, the slope of $\log_{10} L$ vs. $\log_{10} T$
plot is 4, but since in the HR diagram $T$ is plotted increasing to
the left the slope is
$-4$.\\

3.1~From the inverse Lorentz transformation (\ref{dop1}) we get
\begin{equation}
\tan \theta = \frac{\sin \theta_0}{\gamma (\beta - \cos \theta_0)} \, .
\end{equation}
Using the identity $1 + \tan^2 \theta = {\rm sec}^2 \theta$ it is
straightforward to obtain (\ref{dopplerang}).\\

3.2~The distribution in the $S$ system is given by
\begin{equation}
\frac{dN}{d\Omega} = \frac{dN}{d\Omega_0} \frac{d\Omega_0}{d\Omega} \,,
\end{equation}
so we need $d\cos \theta_0/d\cos \theta$. Invert (\ref{dopplerang}) to obtain
\begin{equation}
\cos \theta_0 = \frac{\beta + \cos \theta}{\beta \cos \theta + 1}
\quad {\rm and \ so} \frac{dN}{d \Omega} = \frac{\varkappa (1 -\beta^2)}{(1 +
  \beta \cos \theta)^2} \, .
\end{equation}
This is for a source moving away from $O$. To get the result for
motion towards $O$, just replace $\beta \to - \beta$.\\

3.3~To show that for $v\ll c$ the Doppler shift in wavelength is
approximately $v/c$ we use the binomial expansion:
\begin{equation}
\lambda'  =   \lambda \, (1 + v/c)^{1/2} \,\, (1-v/c)^{-1/2} 
          \approx  \lambda \left[ 1 + \frac{v}{2c} 
                     + {\cal O} \left(\frac{v^2}{c^2}\right)\right] \left[ 1 
- \left(- \frac{v}{2c} \right) + {\cal O} \left(\frac{v^2}{c^2}\right) \right] 
         \approx  \lambda [1 + v/c + {\cal O} (v^2/c^2)] \, ,
\end{equation}
and so $\Delta \lambda /\lambda \approx v/c$.\\

3.4~The wavelengths from single electron energy level transitions are
inversely proportional to the square of the atomic number of the
nucleus. Therefore, the lines from singly-ionized helium are usually
one fourth the wavelength of the corresponding hydrogen lines. Because
of their redshift, the lines have 4 times their usual wavelength
(i.e., $\lambda' = 4 \lambda$) and so
$4 \lambda = \lambda \sqrt{(1+v/c)/ (1 -v/c)} \Rightarrow v = 0.88~c$.\\

3.5~Let 
\begin{equation}
p^\mu = \left( \frac{h\nu}{c}, -\frac{h \nu}{c} \cos \theta , -
  \frac{h \nu}{c} \sin \theta, 0 \right)
\end{equation}
be the momentum 4-vector for the photon as seen in $S$ and 
\begin{equation}
{p'}^\mu = \left( \frac{h\nu'}{c}, -\frac{h \nu'}{c} \cos \theta' , -
  \frac{h \nu'}{c} \sin \theta', 0 \right)
\end{equation}
in $S'$. Do the direct Lorentz transformation from $S \to S'$ to get
$\nu' \cos \theta' = \nu \gamma (\cos \theta + \beta)$, $\nu' \sin
\theta' = \nu \sin \theta$, and $\nu' = \nu \gamma (1+ \beta \cos
\theta)$.  Use the last relation to eliminate $\nu$ and $\nu'$ giving
\begin{equation}
\cos \theta' = \left( \frac{\beta + \cos \theta}{1 + \beta \cos
    \theta} \right) \quad {\rm and} \quad \sin \theta' = \frac{\sin
  \theta}{\gamma (1 + \beta \cos \theta)} \, .
\end{equation}
For small $\beta$, this gives $\cos \theta' \approx \cos \theta +
\beta \sin^2 \theta$. Now, using $\theta' = \theta - \alpha$, since
$\alpha$ is small we have $\cos
(\theta - \alpha) = \cos \theta + \alpha \sin \theta$, and so $\alpha
\approx \beta \sin \theta$. This is in agreement with the data of
Bradley~\cite{Bradley:1727}, a result which caused problems for the
\ae ther theory of
electromagnetic waves (for details see
e.g.~\cite{Anchordoqui:2015xca}). \\

3.6~The universality of Newton's law of gravitation tells us
that all the equations and conclusions derived for the Sun and Earth
interaction also hold for any system consisting of a star and a single
planet orbiting around it. In particular, the period $T$ of the planet
is the same as the period of the star: the time it takes each of them
to complete one orbit around their common center-of-mass. The period
is the easiest parameter to determine, since precisely what is
detected is a periodic motion of the star. (At least this is true in
theory; in practice, determining the period from a finite set of
observations can prove tricky.) The other quantity that we are usually
able to estimate fairly accurately is the mass $M$ of the star, based
on the spectral type and luminosity of the star.  From (\ref{Kepler-h}) it follows
that the ratio $T^2/a^3$ is not exactly the same for all planets, but
is very close to $4 \pi^2/(GM)$, since the ratio $m/M$ is, almost by
definition of a planet, very close to zero. {\it (i)}~As a consequence, the
first fact about the unseen planet that we can infer immediately from
the knoweledge of $M$ and $T$ is its distance from the star, $a =
[(GMT^2)/(4 \pi^2)]^{1/3} = 0.046~{\rm AU}$. {\it (ii)}~In the simplest case of a nearly
circular orbit, the planet describes a circle of radius a at constant
velocity $v$, and the star describes a circle with constant velocity $V$,
both orbits having period $T$. Then $vT = 2 \pi a$, and since $T$ determines $a$,
by Kepler's third law, we also have the velocity $v$ of the
planet.  Then, from (\ref{Rc.m.}) we conclude that $vm = V M$  so that we could
determine the mass $m$ of the planet, if we knew the value of $V$. If the
plane of the orbit contained our line of sight, then $V$ would simply be
the maximal radial velocity. In general, if $i$ denotes the angle of
inclination between the normal to the plane of the orbit and our line
of sight to the star, then the maximal radial velocity would be $K = V
\sin i$, and hence we can deduce the quantity $m \sin i = K
M/v$. Using the measured value of $i$ we get $m = 0.63 M_{\rm J}$,
where $M_{\rm J} = 1.898 \times 10^{27}~{\rm kg}$ is the mass of
Jupiter. {\it (iii)}~From (\ref{Jeclipse}) it follows that $r = \sqrt{0.02 R^2}
\approx 1.5 r_{\rm J}$. \\

4.1~The flux density of neutrinos at Earth is
\begin{equation}
F_\nu = \frac{10^{38}~{\rm neutrinos/s}}{4 \pi d^2} = 3.5 \times
10^{10}~\frac{\rm neutrinos}{\rm cm^2 \, s} \, ,
\end{equation} 
where $d$ is the Sun-Earth distance. Thus, the flux of neutrinos
passing through the brain per second is
\begin{equation}
\mathscr F_\nu = F_\nu  \, A_{\rm brain} = 3.5 \times
10^{10}~\frac{\rm neutrinos}{\rm cm^2 \, s} \frac{\pi D_{\rm brain}^2}{4}
\approx 6.2 \times 10^{12}~\frac{\rm neutrinos}{\rm s} \,,
 \end{equation}
where we have assumed that the  diameter of the brain is
$D_{\rm brain} \simeq 15~{\rm cm}$.\\

4.2~The equation of hydrostatic support is (\ref{staruno}). For a
constant density, we can set $M(r) = 4\rho_0 \pi r^3/3$. Separation of
variables and integration from the center to the surface (where $P =
P_s = 0$) yields
\begin{equation}
\int_{P_c}^{P_s}  dP = 0 - P_c = -\int_0^R \frac{4}{3} \ \frac{G \rho_0 \pi r^3
  \rho_0}{r^2} \ dr = \frac{4}{3} \pi G \rho_0^2 \int_0^R r dr = -
\frac{4}{6} \pi G \rho_0^2 R^2 \, .
\end{equation}
Substituting $\rho_0$ by $M/V = M/(4 \pi R^3/3)$ we obtain the result
\begin{equation}
P_c = \frac{3}{8 \pi} G \frac{M^2}{R^4} \, .
\end{equation}
{\it (ii)}~The mass within a radius $r$ is
\begin{equation}
M(r) = \int_0^r \rho(x) 4 \pi x^2 \ dx = 4 \pi \rho_c
\left( \frac{1}{3} r^3 - \frac{1}{4} \frac{r^4}{R} \right) \, .
\end{equation}
For $r = R$ the total mass of the star is found to be $M = \pi \rho_c
R^3/3$ and therefore we can express the central density in terms of
$M$ and $R$
\begin{equation}
\rho_c = \frac{3}{\pi} \frac{M}{R^3}, \quad {\rm with \ which} \quad M(r) =
12 M \left(\frac{1}{3} \xi^3 - \frac{1}{4} \xi^4 \right),
\end{equation}
where $\xi = r/R$ is the scaled radius. The integral of the equation of hydrostatic support is then
\begin{equation}
P_c = 12 M \frac{3}{\pi} \frac{M^2}{R^4} G \int_0^1 \frac{(\frac{1}{3}
  \xi^3 - \frac{1}{4} \xi^4) (1 - \xi)}{\xi^2} d \xi = \frac{5}{4 \pi}
G \frac{M^2}{R^4} \, .
\end{equation}

\vspace{0.2in}

4.3~From (\ref{lagarde}) we have 
\begin{equation}
P = \frac{1}{3} n m v^2 = \frac{2}{3} n \langle E \rangle  \, ,
\end{equation}
where $n$ is the particle density. For a non-relativistic degenerate
electron gas we have 
\begin{equation}
P = \frac{2}{3} \frac{\rho}{\mu_e m_p} \langle E \rangle =
\frac{1}{20} \left(\frac{3}{\pi} \right)^{2/3} \frac{h^2}{m_e}
\left(\frac{\rho}{\mu_e m_p} \right)^{5/3} \,, 
\end{equation}
where $\rho$ is the mass density. Equating the two relations and
solving for $\langle E \rangle$ gives
\begin{equation}
\langle E \rangle = \frac{3}{40} \left(\frac{3}{\pi} \right)^{2/3}
\frac{h^2}{m_e} \left(\frac{\rho}{\mu_e m_p} \right)^{2/3} \, .
\end{equation}
Using the numerical value of the density of Sirius B we obtain:
$\langle E \rangle = 155.27~{\rm keV}$. This corresponds to a Lorentz
factor $\gamma = 1.30$ and $\beta = 0.64$. The electrons are thus
mildly relativistic. The non-relativistic approximation agrees with the full relativistic
result  to an accuracy of 20\% (note that the
derivation of the equation of state uses the electron momentum). For
larger densities the non-relativistic equation of state is surely not
appropiate and we need to use the relativistic one.\\

4.4~By setting $P_c = P$ we obtain
\begin{equation}
\alpha \frac{G}{\pi} \frac{M^2}{R^4} = \frac{1}{8}
\left(\frac{3}{\pi}\right)^{1/3} h c \left(\frac{\rho}{\mu_e m_p}
\right)^{4/3} = \frac{1}{8} \left(\frac{3}{\pi} \right)^{1/3} h c
\left(\frac{3}{4 \pi \mu_e m_p} \right)^{4/3} \frac{M^{4/3}}{R^4} \,,
\end{equation}
with $\alpha = 3/8$ and $5/4$ for constant and linear density,
respectively. Solving for $M$ yields
\begin{equation}
M_{\rm Ch} = \alpha^{-3/2} \frac{9}{256 \pi} \sqrt{\frac{3}{2}}
\left(\frac{h c}{G} \right)^{3/2} \left(\frac{1}{\mu_e m_p} \right)^2
\, .
\end{equation}
This evaluates numerically to
$M_{\rm Ch}^{\rm const} = 0.44 M_\odot$ and $M_{\rm Ch}^{\rm linear} =
0.07 M_\odot$. For a constant density, the value is about a factor of
3 smaller than the exact result.\\

4.5~A type Ia supernova explosion is 10 billion times more luminous
than the Sun (for a few days). Using the result of exercise 2.5 we write $D = d \sqrt{(L_{\rm Ia}
  I_\odot))/(L_\odot I_{\rm th})} = 6.32 \times 10^{15}~{\rm AU} =
3.07~{\rm Gpc} = 10^{11}~{\rm ly}$,
where $I_{\rm th}$ is the limiting brightness for detection with HST. \\

4.6~{\it (i)}~The radius of the blast wave can be read off the figures taking
into account the height of the tower. Note that the shock wave is not
at the border of the fireball, but at the end of the compression layer
that is growing with time (seen as a faint layer in e.g. the figure at
0.053~s). Using a ruler we estimate the numbers given in
Table~\ref{t-shock}, with an estimated precision of 6~m (corresponding
to the 1/16th inch sub-division of the ruler). {\it (ii)}~The Sedov-Taylor
expansion is described by~(\ref{lagarde-sedov-taylor}). In a log-log plot this corresponds to a line $\log_{10} r
= a + b \log_{10} t$, with $a = 1/5 \log_{10} (E/\rho_1)$ and $b = 2/5$. The fitted
slope, $b =0.42$, is indeed close to the value expected for the
Sedov-Taylor phase of $2/5 = 0.4$ (least-square fitting of a power law
with error bars yields $0.40 \pm 0.02$, i.e. perfect compatibility within
one standard deviation). The best value of a is the one minimizing the
squared distances to the data points:
\begin{equation}
\frac{d}{da} \left\{ \sum_{i=1}^N \left[ \log_{10}(r_i) - a - b \log_{10}(t_i)
\right] \right\}^2 = -2 \sum_{i=1}^N \left[\log_{10} (r_i) - a - b \log_{10} (t_i) \right]
\end{equation}
leading to
\begin{equation}
a = \frac{1}{N} \left[ \sum\log_{10} (r_i) - b \sum \log_{10} (t_i) \right] = 2.77
\end{equation}
The results of the fit are shown in Fig.~\ref{figtrinityradius}.
Solving a for $E$ yields $E = 10^{5a} \rho_1$. The density of air at
$1,100~{\rm m}$ above sea level is about $1.1~{\rm kg/m^3}$ and 1 ton
TNT equivalent is $4.184 \times 10^9~{\rm J}$. Therefore, $E = 7.78
\times 10^{13}~{\rm J} = 18,595~{\rm ton \, TNT}$. The official yield
estimate of Trinity is $16,800~{\rm ton \, TNT} < E < 23,700~{\rm ton
  \, TNT}$~\cite{Taylor:1950b}.  Note that the Sedov--Taylor
expression was derived on the basis of a spherical explosion, whereas
in this case the blastwave expands hemispherically. Following the
orignal reasoning by Taylor~\cite{Taylor:1950b}, we have implicitely
assumed ``\dots that it may be justifiable to assume that most of the
energy associated with the part of the blast wave which strikes the
ground is absorbed
there.''\\

\begin{table}
\caption{Expansion of the shock front as a function of time. \label{t-shock}}
\begin{tabular}{cc}
\hline \hline
 ~~~~~~~~~~~~~~~ time (s) ~~~~~~~~~~~~~~~&~~~~~~~~~~~~~~~ shock radius (m) ~~~~~~~~~~~~~~~\\
\hline
0.006 & 75 \\
0.016 & 108 \\
0.025 & 138 \\
0.053 & 200\\
0.062 & 206 \\
0.090 & 218 \\
\hline
\hline
\end{tabular}
\end{table}

\begin{figure}
 \postscript{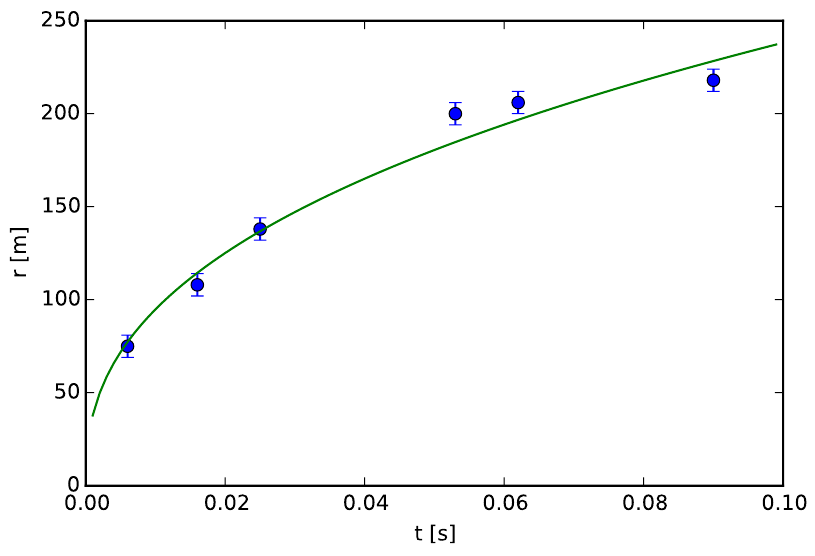}{0.8}
 \caption{Expansion of the shock front as a function of time~\cite{michael}.}
\label{figtrinityradius}
\end{figure}

 5.1~{\it (i)}~Differentiating $\vec \sigma(u,v)$ with respect to $u$ and $v$ yields
\begin{equation}
\frac{\partial \vec \sigma}{\partial u} (u,v) \equiv \vec \sigma_u (u,v) =
\left( \begin{array}{c} -\sin u \, \sin v\\ \cos u \, \sin v\\
    0 \end{array} \right) 
\quad \quad 
{\rm and} \quad \quad
\frac{\partial \vec \sigma}{\partial v} (u,v) \equiv \vec \sigma_v (u,v) = \left( \begin{array}{c} \cos u \,  \cos v\\ \sin u \, \cos v\\ -\sin v \end{array} \right) \, . 
\end{equation}
The coefficients of the first fundamental form may be found by taking the dot product of the partial derivatives
\begin{equation}
E =  \vec \sigma_u \, \cdot \, \vec \sigma_u  = \sin^2 v, \quad
F = \vec \sigma_u \, \cdot \, \vec \sigma_v  = 0, \quad
G =  \vec \sigma_v \, \cdot \, \vec \sigma_v  = 1 \, . 
\end{equation}
The line element  may be expressed in terms of the coefficients of the
first fundamental form as $ds^2 = \sin^2 v \, du^2 +   dv^2$.
{\it (ii)}~The surface area is given by
\begin{equation} 
A= \left. \int_0^\pi \int_0^{2\pi} \sqrt{EG -F^2} du \, dv  =  \int_0^\pi  \int_0^{2\pi} \sin v \, du \, dv 
 =   2 \pi (-\cos v) \right|_0^\pi =  4\pi \, .
\end{equation}
The coefficients of the second fundamental form are
\begin{equation}
e =  \vec \sigma_{uu} \, \cdot \, \hat n  = \sin^2 v, \quad 
f =  \vec \sigma_{uv} \, \cdot \, \hat n  = 0, \quad 
g =  \vec \sigma_{vv} \, \cdot \, \hat n =  1 \, .
\end{equation}
{\it (iii)}~The Gaussian curvature is
\begin{equation}
K  = \frac{{\rm det\,  II}}{{\rm det \, I}} = \frac{eg - f^2}{EG - F^2} =  1 \, .
\end{equation}

\vspace{0.2in}

5.2~{\it (i)}~The coefficients of the first fundamental form are
\begin{equation}
E  =  {\rm tanh}^2\;\! u, \quad
F =  0 , \quad
G  =  {\rm sech}^2 \;\! u  \, .
\end{equation}
The line element is $ds^2 =  {\rm tanh}^2\;\! u \, du^2 + {\rm sech}^2
\;\! u  \, dv^2$. {\it (ii)}~The surface area is
\begin{equation}
A =  2 \int_0^{2\pi} \int_0^\infty {\rm sech} \!\; u \, {\rm tanh}
\!\; u \, du \, dv 
 =  4\pi \,
\end{equation}
which is exactly that of the sphere. {\it (iii)}~The coefficients of the second fundamental form are
\begin{equation}
e  =  - {\rm sech}\;\! u \, {\rm tanh} \;\! u, \quad
f  =  0 , \quad
g  =  {\rm sech} \;\! u  \,  {\rm tanh} \;\! u \, .  
\end{equation}
The Gaussian curvature is $K = -1$.\\

5.3~We have
\begin{equation}
\frac{d}{dt} ||\dot \gamma||^2 = \frac{d}{dt} (\dot{\vec \gamma} \cdot \dot
{\vec \gamma}) = 2 \ddot {\vec \gamma} \cdot \dot {\vec \gamma} \,. 
\end{equation}
Since $\gamma$ is geodesic, $\ddot{\vec  \gamma}$ is perpendicular to the
tangent plane which contains $\dot{\vec  \gamma}$. Hence, $\ddot{\vec  \gamma} \cdot
\dot {\vec \gamma} = 0$. Subsequently, $d||\dot \gamma||^2/dt =0$. Therefore, the speed
$||\dot \gamma||$ is constant.\\

5.4~The tangent plane is spanned by $\vec \sigma_u$ and $\vec
\sigma_v$. By definition the curve $\gamma$ is a geodesic if $\ddot {\vec
\gamma} \cdot \vec \sigma_u = \ddot{\vec \gamma} \cdot \vec \sigma_v =
0$. Since $\dot {\vec \gamma} = \dot u \vec \sigma_u + \dot v \vec
\sigma_v$, it follows that $\ddot{\vec \gamma} \cdot \vec \sigma_u =
0$ becomes
\begin{equation}
\left[\frac{d}{dt} ( \dot u \vec \sigma_u + \dot v \vec \sigma_v) \right]
\cdot \vec \sigma_u = 0 \, .
\end{equation}
We rewrite the left hand side of the above equation:
\begin{eqnarray}
\left[\frac{d}{dt} (\dot u \vec \sigma_u + \dot v \vec \sigma_v)
\right] \cdot \vec \sigma_u & = & \frac{d}{dt} \left[ (\dot u \vec
  \sigma_u + \dot v \vec \sigma_v) \cdot \vec \sigma_u \right] - (\dot
u \vec \sigma_u + \dot v \vec \sigma_v) \cdot \frac{d \vec
  \sigma_u}{dt} \nonumber \\
 & = & \frac{d}{dt} (E \dot u + F \dot v) - (\dot u \vec \sigma_u +
\dot v \vec \sigma_v) \cdot ( \dot u \vec \sigma_{uu} + \dot v \vec
\sigma_{uv}) \nonumber \\
& = & \frac{d}{dt} (E \dot u + F \dot v) - \left[\dot u^2 (\vec
  \sigma_u \cdot \vec \sigma_{uu}) + \dot u \dot v (\vec \sigma_u
  \cdot \vec \sigma_{uv} + \vec \sigma_v \cdot \vec \sigma_{uu}) +
  \dot v^2 (\vec \sigma_v \cdot \vec \sigma_{uv} ) \right] \, .
\label{mutt-hawaii}
\end{eqnarray}
We have that
\begin{equation}
\vec \sigma_u \cdot \vec \sigma_{uu} = \frac{1}{2}
\frac{\partial}{\partial u} (\vec \sigma_u \cdot \vec \sigma_u) =
\frac{1}{2} E_u, \quad \quad \vec \sigma_v \cdot \vec \sigma_{uv} =
\frac{1}{2} G_u, \quad \quad \vec \sigma_u \cdot \vec \sigma_{uv} +
\vec \sigma_v \cdot \vec \sigma_{uu} = F_u \, .
\end{equation}
Substituting them into (\ref{mutt-hawaii}), we obtain
\begin{equation}
\left[\frac{d}{dt} (\dot u \vec \sigma_u + \dot v \vec \sigma_v)
\right] \cdot \vec \sigma_u = \frac{d}{dt}(E \dot u + F \dot v) - \frac{1}{2}
(E_u \dot u^2 + 2 F_u \dot u \dot v + G_u \dot v^2) \, .
\end{equation}
This establishes the first differential equation (\ref{geodesic1}). Similarly,  (\ref{geodesic2})
can be established from
\begin{equation}
\left[\frac{d}{dt}(\dot u \vec \sigma_u + \dot v \vec \sigma_v)
\right] \cdot \vec \sigma_v = 0 \, .
\end{equation}

\vspace{0.2in}

5.5~For the parametrization in (\ref{esfera}) the first fundamental form is found
to be $ds^2 = d\theta^2 + \cos^2 \theta d\phi^2$, with $E=1$, $F=0$,
and $G = \cos^2 \theta$. We restrict to unit-speed curves $\gamma (t)
= \vec \sigma(\theta(t), \phi(t))$, so that
\begin{equation}
E \dot \theta^2 + 2 F \dot \theta \dot \phi + G \dot \phi^2 = \dot
\theta^2 + \dot \phi^2 \cos^2 \theta = 1 \, .
\label{esfera4}
\end{equation}
If $\gamma$ is a geodesic, then  (\ref{geodesic2}) is satisfied. Here
(\ref{geodesic2}) reduces to
\begin{equation}
\frac{d}{dt} (\dot \phi \cos^2 \theta) = 0, \quad {\rm or \
equivalently} \quad \dot \phi \cos^2 \theta = \zeta,
\label{esfera6}
\end{equation}
where $\zeta$ is a constant.   There are two cases: {\it (i)}~$\zeta =
0$; then $\dot \phi = 0$. In this case, $\phi$ is constant and
$\gamma$ is part of a meridian. {\it (ii)}~$\zeta \neq
0$. Substituting (\ref{esfera6}) into the unit-speed condition (\ref{esfera4}), we
have
\begin{equation}
\dot \theta^2 = 1 - \frac{\zeta^2}{\cos^2 \theta} \, .
\end{equation}
Combining the above with (\ref{esfera6}), along the geodesic it holds that
\begin{equation}
\left(\frac{d\phi}{d \theta}\right)^2 = \frac{\dot \phi^2}{\dot
  \theta^2} = \frac{1}{\cos^2 \theta (\cos^2 \theta/\zeta^2 -1) }\, .
\end{equation}
Integrate the derivative $d\phi/d\theta$:
\begin{equation}
\phi - \phi_0 = \pm \int \frac{d\theta}{\cos \theta \sqrt{\cos^2
    \theta/\zeta^2 - 1}} \,,
\end{equation}
where $\phi_0$ is a constant. The substitution $u = \tan \theta$ yields
\begin{equation}
\phi - \phi_0 = \pm \int \frac{du}{\sqrt{\zeta^{-2} - 1 - u^2}} =
\sin^{-1} \left( \frac{u}{\sqrt{\zeta^{-2} - 1}} \right) \,,
\end{equation}
which leads to
\begin{equation}
\tan \theta = \pm \sin( \phi - \phi_0) \, \sqrt{\zeta^{-2} - 1} \, .
\end{equation}
Multiply both sides of the above equation by $\cos \theta$:
\begin{equation}
\sin \theta = \pm \sqrt{\zeta^{-2} -1} (\cos \phi_0 \cos \theta \sin
\phi - \sin \phi_0 \cos \theta \cos \phi) \, .
\end{equation}
Since $\vec \sigma(\theta,\phi) = (xy,z)$, we have
\begin{equation}
z = \mp (\sin \phi_0 \sqrt{\zeta^{-2} -1 }) \, x \pm ( \cos \phi_0
\sqrt{\zeta^{-2} -1}) \, y \, .
\end{equation}
Clearly, $z = 0$ when $x = y = 0$. Therefore, $\gamma$ is contained in
the intersection of $S^2$ with a plane through the center of the
sphere. Hence it is part of a great circle.\\

5.6~(\ref{Sch-escape}) gives the rate at which a clock at radius $r$ ticks relative to one infinitely far away. Here we are asked to compare the rate of a clock at radius $r$ relative to one at the radius of the Earth (i.e., at the Earth's surface). We can think about this by considering the rate of each of these clocks relative to a distant clock. That is, the clock on the Earth's surface has a rate a factor
\begin{equation}
\sqrt{1 - \frac{1}{c^2} \frac{2GM_\oplus}{R_\oplus}}
\label{aquafina1}
\end{equation}
slower than the distant clock, while the clock at radius $r$ has a rate
\begin{equation}
\sqrt{1 - \frac{1}{c^2} \frac{2GM_\oplus}{r}} \, .
\label{aquafina2}
\end{equation}
Note that both these expressions are less than one, but because $r > R_\oplus$, the stationary clock at radius $r$ ticks faster than that at the Earth's surface. Indeed, the relative rate of the two is just the ratio of these two expressions
\begin{equation}
\sqrt{\left(1 - \frac{1}{c^2} \frac{2GM_\oplus}{r} \right)  \left(1 -
    \frac{1}{c^2} \frac{2 GM_\oplus}{R_\oplus} \right)^{-1}} \, .
\label{aquafina3}
\end{equation}
Again, the expression in (\ref{aquafina3}) is greater than 1. {\it
  (ii)}~Circular motion at speed $v$ at
a radius $r$ gives rise to an acceleration $v^2/r$, which we know is
due to gravity. Thus if the astronaut has a mass $m$, Newton's second
law yields
\begin{equation}
\frac{GM_\oplus m}{r^2} = m \frac{v^2}{r}, \quad {\rm or\ solving \
  for} \ v \ {\rm gives} \quad  v = \sqrt{\frac{G M_\oplus}{r}} \, .
\end{equation}
The time dilation in special relativity is due to the by now familiar factor $(1
-v^2/c^2)^{1/2}$, which gives
\begin{equation}
\sqrt{1 - \frac{G M_\oplus}{r c^2}} \, .
\label{aquafina4}
\end{equation}
Note again how similar this looks to the expression above for time
dilation due to gravity. Again, this is the rate that an orbiting
clock at radius $r$ ticks relative to a stationary clock at the same
radius. {\it (iii)}~In part (\ref{aquafina3}), we calculated the ratio of rates of
stationary clocks at radius $r$ and $R_\oplus$ (due to general relativity),
while in part (\ref{aquafina4}), we calculated the ratio of the rates of an orbiting
clock at radius $r$ to a stationary clock at the same radius. Therefore, the
ratio of the rate of an orbiting clock at radius $r$ to a stationary one
on the ground is simply the product of these two results; namely,
\begin{equation}
\sqrt{\left(1-  \frac{2GM_\oplus}{rc^2} \right)   \left( 1-
   \frac{2 GM_\oplus}{R_\oplus c^2} \right)^{-1} \left( 1 -  \frac{GM_\oplus}{rc^2} \right)} \, .
\label{bigequation}
\end{equation}
 {\it  (iv)} We now simplify (\ref{bigequation}). We will do this in
 pieces, starting from (\ref{aquafina3}) we can write:
\begin{equation}
\sqrt{\left(1 - \frac{1}{c^2} \frac{2GM_\oplus}{r} \right)  \left(1 -
    \frac{1}{c^2} \frac{2 GM_\oplus}{R_\oplus} \right)^{-1}} \approx \sqrt{\left(1 - \frac{1}{c^2}
  \frac{2 GM_\oplus}{r} \right) \left( 1 + \frac{1}{c^2} \frac{2
    GM_\oplus}{R_\oplus} + \cdots \right) } \, .
\label{aquafina10}
\end{equation}
 We then use $(1 -x) (1-y) \approx 1 - (x+y)$ to re-write
 (\ref{aquafina10}) as
\begin{equation}
\sqrt{1 - \frac{GM_\oplus}{c^2} \left(\frac{2}{r} -
    \frac{2}{R_\oplus}\right)} \, .
\end{equation}
This then gets multiplied by  (\ref{aquafina4}), yielding
\begin{equation}
\sqrt{1 - \frac{GM_\oplus}{c^2} \left(\frac{3}{r} - \frac{2}{R_\oplus}
  \right)} \approx 1 - \frac{GM_\oplus}{2c^2} \left(\frac{3}{r} -
  \frac{2}{R_\oplus} \right) \, .
\label{aquafina-final}
\end{equation}
However, we do need to justify the use of the various approximations
we have made. We dealt with a variety of expressions of the form $1 -
x$; in every case $x$ is of the form $(GM_\oplus)/ (rc^2)$. The smallest $r$ we
considered, and therefore the largest the expression $(GM_\oplus)/(rc^2)$, is
$r=R_\oplus$. So we plug in numbers at $r=R_\oplus$ to obtain
\begin{equation}
\frac{GM_\oplus}{R_\oplus c^2} = \frac{2}{3} \frac{10^{-10}~{\rm m}^3
\ {\rm s}^{-2} \ {\rm kg}^{-1} \times 6 \times 10^{24}~{\rm kg}}{6.4
\times 10^6~{\rm m} \times (3 \times 10^8~{\rm m/s})^2} \approx 7
\times 10^{-9} \, ,
\end{equation}
which is indeed a number much much smaller than 1. {\it (v)}~We are
asked to find the radius at which (\ref{aquafina-final}) is equal to
unity. This clearly holds when $3/r - 2/R_\oplus = 0$, or $r = 1.5
R_\oplus$. Given the radius of the Earth is $6,400~{\rm km}$, the
critical radius $r = 1.5 R_\oplus$ is at
a distance of $9,600~{\rm km}$ from the Earth's center, or $3,200~{\rm
  km}$ above the Earth's surface. Now, (\ref{aquafina-final}) is less
than 1 for $r < 1.5 R_\oplus$, and so astronauts on the
space shuttle age less than those staying home.\\

5.7~{\it (i)}~The Schwarzschild radius of a black hole of mass $M$ is
$R_{\rm Sch} = 2 GM/c^2$. The volume of a sphere of this radius is
just the familiar $4 \pi R_{\rm Sch}^3$. The density is the mass
divided by the volume, giving:
\begin{equation}
\rho_{\rm BH} = M \times \left[ \frac{4}{3} \pi \left( \frac{2 G M}{c^2} \right)^3 \right]^{-1} = \frac{3 c^6}{32 \pi G^3 M^2} \, .
\label{hh1}
\end{equation}
The more massive the black hole, the smaller the density. Thus there
is a mass at which the black hole has the density of paper, which is
what we are trying to figure out. {\it (ii)}~The density is the mass
per unit volume. If we can figure out the volume of a square meter of
paper (whose mass we know, 75~g), we can calculate its
density. The volume of a piece of paper is its area times its
thickness. The thickness is 0.1~mm, or $10^{-4}~{\rm m}$, and so the
volume of a square meter of paper is $10^{-4}~{\rm m}^3$. Therefore,
the density of paper is
\begin{equation}
\rho  = \frac{7.5 \times 10^{-2}~{\rm kg}}{10^{-4}~{\rm m}^3} =
7.5 \times 10^2~{\rm kg}/{\rm m}^3 \, ,
\label{hh2}
\end{equation}
similar to (but slightly less than) the density of water (remember,
paper is made of wood, and wood floats in water). {\it (iii)}~Here we
equate (\ref{hh1})  with (\ref{hh2}) and solve for the mass
\begin{equation}
M = \sqrt{\frac{3 c^6}{32 \pi G^3 \rho}} \approx 3 \times 10^{38}~{\rm kg} \,,
\end{equation}
where we have made all the usual approximations of $\pi = 3$, 32 = 10,
and so on. We need to express this in solar masses, so we divide by
$M_\odot = 2 \times 10^{30}~{\rm kg}$ to obtain $M \approx 1.5 \times
10^8 M_\odot$. A black hole 150 million times the mass of the Sun has
the same density as a piece of paper. We know the Hitchhiker's Guide
to the Galaxy is science fiction, but do such incredibly massive black
holes actually exist? Indeed they do: the cores of massive galaxies
(including our own Milky Way) do contain such enormous black
holes. Actually, the most massive such black hole known to exist is in
the core of a particularly luminous galaxy known as M87, with a mass
of 3 billion solar masses. We still have to calculate the
Schwarzschild radius of a black hole. We could plug into the formula
for a Schwarzschild radius and calculate away, but here we outline a
simpler approach. We know the Schwarzschild radius is proportional to
the mass of a black hole, and we happen to remember that a $M_\odot$
mass black hole has a Schwarzschild radius of 3~km, (\ref{Sch3km}). So
a 150 million $M_\odot$ black hole has a Schwarzschild radius 150
million times larger, or $4.5 \times 10^8~{\rm km}$. We are asked to
express this in terms of AU; 1~AU $= 1.5 \times 10^8~{\rm km}$, so the
Schwarzschild radius of such a black hole is 3 AU. {\it (iv)}~We know
the entire mass of the black hole. If we can calculate the mass of a
single piece of paper, the ratio of the two gives the total number of
pages. So we now turn to calculate the mass of a single piece of paper. We know
that a square meter of paper has a mass of 75~g. How many square
meters is a standard-size sheet? One inch is $2.5~{\rm cm} = 2.5
\times 10^{-2}~{\rm m}$. So $8.5 \times 11~{\rm inch}^2 \approx
100~{\rm inch}^2 \approx 6 \times 10^{-2}~{\rm m}^{-2}$. Thus the mass
is
\begin{equation}
{\rm Mass \ of \ a \ piece \ of \ paper} = 7.5 \times 10^{-2}~{\rm kg/m^2}   \times 6 \times 10^{-2}~{\rm m}^2 \approx 5  \times 10^{-3}~{\rm kg} .
\end{equation}
That is, a piece of paper weighs about $5~{\rm g}$. We divide this into the mass we
calculated above:
\begin{equation}
{\rm Number \, of \, sheets \,  of \,  paper}= \frac{\rm Mass \ of \ rule \ book}{\rm Mass \ per \ page} = \frac{3 \times 10^{38}~{\rm kg}}{5 \times 10^{-3}~{\rm kg/page}} = 6 \times 10^{40}~{\rm pages} \, .
\label{hh3}
\end{equation}
Strictly speaking, if the rule book is printed on both sides of the
page, we should multiply the above result by a factor of two. That is
one seriously long set of rules!  Finally, note that because the
density of a more massive black hole is smaller (\ref{hh1}), the mass
and number of pages of the Brockian Ultra Cricket rule book as given
by (\ref{hh3}) is really just a lower
limit. That is, if the rule book were even larger than what we have just calculated, it would still collapse into a black hole.\\

5.8~{\it (i)}~The Schwarzschild radius of a $3 M_\odot$ black hole is
$R_{\rm Sch} = 2GM/c^2 = 9~{\rm km}$. If you remember that for $1~M_\odot$
black hole the Schwarzchild radius is 3~km, you can scale from there. {\it
  (ii)}~Using the Newton's law of gravitation, we can write the
difference in gravitational forces acting on two bodies of mass $m$
which are located at distances $r_1$ and $r_2$ from the massive body
of mass $M$
\begin{equation}
\delta F \equiv F_1 - F_2 = \frac{GmM}{r_1^2} - \frac{GmM}{r_2^2} =
GmM \left(\frac{1}{r_1^2} - \frac{1}{r_2^2} \right) \, .
\label{aquafina20}
\end{equation}
We are interested in the difference in gravitational forces in two
locations that are close to each other, since the height of the person
falling into the black hole is small compared to the Schwarzchild
radius. We take $r_2 = r_1 + \Delta$, where $\Delta  \ll r_1$. Now, we simplify
(\ref{aquafina20}); dropping the subscript 1 we have 
\begin{equation}
\delta F = GmM \left[\frac{1}{r^2} - \frac{1}{(r + \Delta)^2}
    \right] = G m M \frac{(r + \Delta)^2 - r^2}{r^2 (r + \Delta)^2} =
    GmM \frac{r^2 + 2 r \Delta + \Delta^2 -r^2}{r^2 (r + \Delta)^2}
      \approx GmM \frac{2 \Delta}{r^3} \, .
\label{aquafina21}
\end{equation}
In obtaining this expression, we used the approximations $r + \Delta =
r (1 + \Delta/r) \approx r$ and $2 r \Delta + \Delta ^2 = 2 r \Delta
[1 + \Delta/(2r)] \approx 2 r \Delta$, because $\Delta \ll r$. Next,
we use (\ref{aquafina21}) to find the distance $r_{\rm crit}$ from the
black hole where the relative stretching force between your head and
your legs is equal to some critical force
\begin{equation}
\delta F_{\rm crit} = GmM \frac{2 \Delta}{r_{\rm crit}^3} \quad {\rm
  and \ so} \quad r_{\rm crit} = \left( \frac{2 G mM \Delta}{\delta F_{\rm crit}}
\right)^{1/3} \, .
\label{aquafina22}
\end{equation} 
Finally, we can plug in the numbers. The mass of the black hole is $M
= 3M_\oplus = 6 \times 10^{30}~{\rm kg}$. The mass of the body is $m =
70~{\rm kg}$, $\delta F_{\rm crit} = 10~{\rm kN}$. The critical radius
is then, $r_{\rm crit} \approx 2000~{\rm km}$ or recalling that the
the Schwarzchild radius for $3 M_\odot$ black hole is $R_{\rm Sch}
\sim 10~{\rm km}$, we have $r_{\rm crit} \sim 200 R_{\rm Sch}$. Note,
that in this case a significant amount of stretching occurs already
relatively far from the black hole. {\it (iii)}~The force with which
the metal plate is pulling on you is given by the Newton's second law,
$F = m_{\rm sp} g$, where $m_{\rm sp}$ is the mass of the steel plate,
and $g \sim10~{\rm m/s^2}$ is the gravitational acceleration on the
Earth. Thus $m_{\rm sp} = 10~{\rm kN}/(10~{\rm m/s^2}) = 1,000~{\rm
  kg}$, or 1~ton. If you still have hard time imagining how much
weight 10~kN is, this is the weight of a typical car. So, imagine
attaching a car to your feet: not pleasant. Most likely this is enough
to kill or at the very least severely disable a person. {\it (iv)}~We
can apply the formula for $r_{\rm crit}$ from (\ref{aquafina22}). Now the mass
of the black hole is $1.3 \times 10^6$ times larger, so the radius
increases by $(1.3 \times 10^6)^{1/3} \approx 100$ times. The answer
is then $r_{\rm crit} = 2 \times 10^5~{\rm km}$. In terms of the
Schwarzchild radii, remember that $R_{\rm Sch}$ is linearly
proportional to the mass. For 4 million solar mass black hole, the
Schwarzchild radius is then $1.3 \times 10^6 \times 10~{\rm km}
\approx 107~{\rm km}$ and so $r_{\rm crit} \approx 2 \times
10^{-2}~R_{\rm Sch}$. Since $r_{\rm crit} < R_{\rm Sch}$, the
``spaghettification'' happens inside the Schwarzchild radius. {\it
  (v)}~As we saw in (iv), the radius at which the tidal force reaches
10~kN grows as the third root of the mass of the black hole, but the
Schwarzchild radius grows linearly with the mass of the hole. In part
(iii) for the $3 M_\odot$ hole the critical radius was outside $R_{\rm
  Sch}$, while in part (iv) for $4 \times 10^6 M_\odot$ hole the
critical radius was inside. Thus, there should be a minimum mass of
the hole, at which $r_{\rm crit} = R_{\rm Sch}$, i.e., we can just
barely pass through the horizon before getting fatally stretched. We
find this mass setting $r_{\rm crit} = R_{\rm Sch}$, which leads to
\begin{equation}
\left(\frac{2GmM_{\rm min} \Delta}{\delta F_{\rm crit}} \right)^{1/3} =
\frac{2GM_{\rm min}}{c^2} \, ,
\end{equation}
and so
\begin{equation}
M_{\rm min} = \left(\frac{c^3}{2G} \right) \left(\frac{m\Delta}{\delta
    F_{\rm crit}} \right)^{1/2} = 2 \times 10^{34}~{\rm kg} = 10^4
M_\odot \, .
\end{equation}
So, if you fall into a $10^4 M_\odot$ black hole, you will be killed
right as you go through the horizon. If the black hole is more
massive, then you can go through the horizon while still alive, and
enjoy the sights! Sadly, you will not have much time to enjoy the view
anyways, because you will be crushed by the singularity in 0.01~{\rm
  s} seconds for this $10^4 M_\odot$ black hole. This time is
proportional to mass of the hole.\\

5.9~ {\it (i)}~The Schwarzschild metric is given by (\ref{Sch-metric}). Firstly, we have $dt=0$, $dr=0$, $d\theta=0$, and $\theta =
\pi/2$, and so (\ref{Sch-metric}) simplifies to
\begin{equation}
ds^2 = r^2 \sin^2 \theta \ d\phi^2 = r^2 d\phi^2 \, .
\end{equation}
 Now, to find the circunference we can integrate this function from $0
 \leq \phi \leq 2\pi$,
\begin{equation}
C = \int_0^{2 \pi} R d\phi = 2 \pi R \, .
\label{SPR427}
\end{equation}
{\it (ii)}~Secondly, we have $dt =0$,
$d\theta =0$, and $d\phi=0$, and so (\ref{Sch-metric}) simplifies to
\begin{equation}
ds^2 = \left(1 - \frac{2GM}{rc^2}\right)^{-1} dr^2 \rightarrow ds =
\left(1 - \frac{2GM}{rc^2}\right)^{-1/2} dr \, .
\end{equation}
This can also be rewritten as 
\begin{equation}
R_{\rm phys} = \int_0^R \left( 1 - \frac{2GM}{rc^2} \right)^{-1/2} dr
\, .
\end{equation}
If we multiply both top and bottom by $rc^2$ and divide top and bottom
by $2GM$ we get an expression of the form
\begin{equation}
R_{\rm phys} = \int_0^R \sqrt{\frac{rc^2/(2GM)}{r c^2/(2 GM-1)}} \ dr
\, .
\label{SPRuiz}
\end{equation} 
Now, let 
\begin{equation}
\xi = \frac{rc^2}{2GM} \Rightarrow \frac{2GM}{c^2} d \xi = dr \, .
\end{equation}
(\ref{SPRuiz}) can be made to look like (\ref{SCHint1}) and
(\ref{SCHint2})  by multiplying the denominater by $-1$ and taking the
absolute value of this function; namely
\begin{equation}
R_{\rm phys} = \frac{2GM}{c^2} \int_0^\alpha \sqrt{\frac{\xi}{1-\xi}} d
  \xi = \frac{2 GM}{c^2} \left[\int_0^1 \sqrt{\frac{\xi}{1-\xi} } d
    \xi + \int_0^\alpha \sqrt{\frac{\xi}{1-\xi} } d
    \xi \right] \, .
\end{equation}
It follows that
\begin{equation}
R_{\rm phys} = \frac{2GM}{c^2} \left[\frac{\pi}{2}  + \ln
  \left(\sqrt{\alpha -1} + \sqrt{\alpha}\right) + \sqrt{\alpha - 1 }
  \sqrt{\alpha} \right] \,,
\end{equation}
where $\alpha = Rc^2/(2GM)$; equivalently,
\begin{equation}
R_{\rm phys} = \frac{2GM}{c^2} \left[\frac{\pi}{2} + \ln\left(
    \sqrt{\frac{Rc^2}{2GM} - 1} + \sqrt{\frac{Rc^2}{2GM}} \right) +
    \sqrt{\frac{Rc^2}{2GM} - 1}  \sqrt{\frac{Rc^2}{2GM}} \right] \, .
\end{equation}
{\it (iii)}~Finally, we use the answers of {\it (i)} and {\it (ii)} to
compute $\Pi$ where
\begin{equation}
C = 2 \Pi R_{\rm phys} \, .
\end{equation}
Using (\ref{SPR427}) we find
\begin{equation}
2 \pi R = 2 \Pi R_{\rm phys} = 2 \Pi \frac{2GM}{c^2} \left[\frac{\pi}{2} + \ln\left(
    \sqrt{\frac{Rc^2}{2GM} - 1} + \sqrt{\frac{Rc^2}{2GM}} \right) +
    \sqrt{\frac{Rc^2}{2GM} - 1}  \sqrt{\frac{Rc^2}{2GM}} \right] \,,
\end{equation}
and solving for $\Pi$ we have
\begin{equation}
\Pi = \pi \frac{Rc^2}{2GM} \left[\frac{\pi}{2} + \ln\left(
    \sqrt{\frac{Rc^2}{2GM} - 1} + \sqrt{\frac{Rc^2}{2GM}} \right) +
    \sqrt{\frac{Rc^2}{2GM} - 1}  \sqrt{\frac{Rc^2}{2GM}} \right]^{-1} \,,
\end{equation}
which can also be written as
\begin{equation}
\Pi = \pi \xi \left[\frac{\pi}{2} + \ln \left(\sqrt{\xi -1}
    +\sqrt{\xi} \right) + \sqrt{\xi-1} \sqrt{\xi} \right]^{-1} \, .
\end{equation}
{\it (iv)}~A plot of $\Pi$ versus $R/R_{\rm Sch}$ is shown in
Fig.~\ref{fig:Pi}. We can see that $\Pi$ approaches the value of $\pi$
measured in a flat space, as $\xi \to \infty$.\\

\begin{figure}
 \postscript{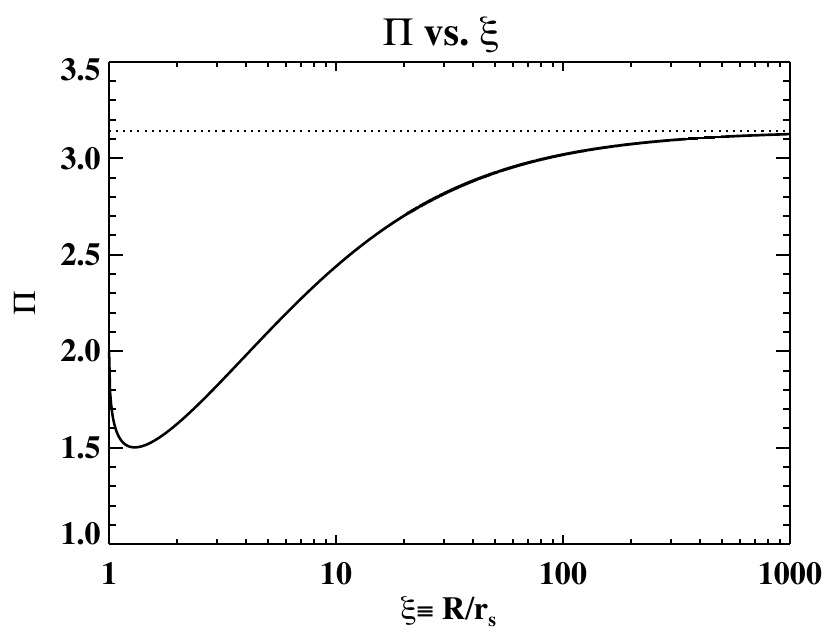}{0.95}
 \caption{$\Pi$ is not a constant~\cite{Ma}.}
\label{fig:Pi}
\end{figure}

5.10~{\it (i)}~From Fig.~\ref{quasar} estimate an initial position of
about 32~ly at a time 2002.12 and 45~ly at 2002.73. The apparent
velocity is therefore $v_{\rm app} \approx 13~{\rm ly}/0.61~{\rm yr} =
21~{\rm ly}/{\rm yr} = 21 c$, which is in agreement with the value
$v_{\rm app} = (25.6 \pm 4.4) c$ given in~\cite{Piner:2005px}
reporting this measurement. The apparent velocity of the blob is thus
highly superluminal. {\it (ii)}~The light emitted at point $A$ at time
$t_{i,1}$ will reach the observer located at a distance $d_1$ at time
$t_1 = d_1/c$; see Fig.~\ref{appv}. The blob travels with
``true'' velocity $v$ from $A$ to $B$ a distance $H$ which takes a time
\begin{equation}
\Delta t_{A \to B} = t_{i,2} - t_{i1} = \frac{1}{v} \frac{L}{\sin \theta} \, ,
\end{equation}
where the only hypothesis is that the signal travels at the speed of light $c$.
The remaining distance to the observer is $d_2 = d_1 - L/ \tan \theta$ and
therefore the light from position $B$ will arrive at
\begin{equation}
t_2 = \frac{1}{v} \frac{L}{\sin \theta} + \frac{1}{c} \left(d_1 -
  \frac{L}{\tan \theta} \right) \,  .
\end{equation}
The time difference between the signals from $A$ and $B$ is
\begin{equation}
\Delta t = t_2 - t_1 = \frac{1}{v} \frac{L}{\sin \theta} + \frac{1}{c}
\frac{L}{\tan \theta} = L \left( \frac{c - v \cos \theta}{vc \sin
    \theta}\right) = L \left( \frac{1 - \beta \cos
    \theta}{v \sin \theta} \right) \,,
\end{equation}
where $\beta = v/c$, and the apparent transverse velocity is therefore 
\begin{equation}
\beta_{\rm app} = \frac{1}{c} \frac{L}{\Delta t} = \frac{\beta \, \sin
  \theta}{1 - \beta \cos \theta} \, .
\label{dosseiscuatro}
\end{equation}
{\it (iii)}~Using $x = \tan\theta/2$,  (\ref{dosseiscuatro}) can be re-written with
standard trigonometry as $\beta_{\rm app} = 2\beta
x/[(1-\beta)+(1+\beta ) x^2]$ so that after a bit of algebra it
follows that
\begin{equation}
v_{\rm app} = kc \Leftrightarrow [(1 + \beta) x - \beta/k]^2 =
(\beta/k - \gamma^{-1}) \, (\beta/k + \gamma^{-1}) \,,
\end{equation}
where $\gamma = (1 - \beta^2)^{-1/2}$ is the Lorentz factor. The left
hand term is positive and the equation in $x$ admits at least one
solution if $\beta/k \geq \gamma^{-1}$. Superluminal motion $v_{\rm app} \geq c$ (i.e. $k \geq 1$) is
then possible as long as $\gamma \beta \geq \beta_{\rm app}$. A direct consequence of the
previous equation is $\gamma \geq \beta_{\rm app}$, which proves that, even for
moderate superluminal motions, the true velocity is relativistic. The angle that maximizes the apparent transverse velocity
can be found by differentiating $v_{\rm app}$ and solving for
$\theta_{\rm max}$, we have
\begin{equation}
\frac{d v_{\rm app}}{d \theta} = \frac{\beta \cos \theta}{1 - \beta
  \cos \theta} - \frac{\beta^2 \sin^2 \theta}{(1 - \beta \cos
  \theta)^2} = 0 \Rightarrow \cos \theta_{\rm max} = \beta \, .
\end{equation}
The maximum apparent transverse velocity is therefore
\begin{equation}
\beta_{\rm app}^{\rm max} = \frac{\beta \sin \theta_{\rm max}}{1 -
  \beta \cos \theta_{\rm max}} = \frac{\beta \sqrt{1 - \cos^2
    \theta_{\rm max}}}{1 - \beta \cos \theta_{\rm max}} = \frac{\beta
  \sqrt{1- \beta^2}}{1 - \beta^2} = \beta \gamma \, .
\end{equation}
Therefore,
\begin{equation}
\beta_{\rm app}^{\rm max} = \beta \gamma = \gamma \sqrt{1 -
  1/\gamma^2} \Rightarrow \gamma = \sqrt{1+ \beta^2_{\rm app}} \,,
\end{equation}
that is to say the plasma blob moves with a highly relativistic
Lorentz factor of at least $\sim 21$.\\

\begin{figure}
 \postscript{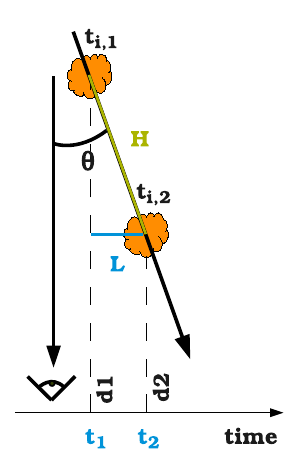}{0.3}
 \caption{The situation in exercise 5.10.}
\label{appv}
\end{figure}

6.1~{\it (i)}~Imagine a circle with radius $x$ around the observer. A fraction
$s(x),$ $0 \leq s(x) \leq 1$, is covered by trees. Then we'll move a
distance $dx$ outward, and draw another circle. There are $2 \pi n x
dx$ trees growing in the annulus limited by these two circles. They
hide a distance $2 \pi xn D dx,$ or a fraction $n D dx$ of the
perimeter of the circle. Since a fraction $s(x)$ was already hidden,
the contribution is only $[1-s(x)] n D dx$. We get
\begin{equation}
s(x+dx) = s(x) + [1 - s(x)] \, n\,D\, dx\,,
\end{equation}
which gives a differential equation for $s$:
\begin{equation}
\frac{ds(x)}{dx} = [1 - s(x) ] \, n\, D \, .
\end{equation}
This is a separable equation which can be integrated:
\begin{equation}
\int_0^s \frac{ds}{1-s} = \int_0^x n\, D\, dx \,\, .
\end{equation}
This yields the solution
\begin{equation}
s(x) = 1 - e^{-nDx} \,\,.
\end{equation}
This is the probability that in a random direction we can see at most
to a distance $x$. This function $x$ is a cumulative probability
distribution. It is as if we have compressed the 2-dimensional forest
into an imaginary 1-dimensional structure, with a characteristic mean
free path. The corresponding probability density is its derivative
$ds/dx.$ The mean free path $\lambda$ is the expectation of this
distribution
\begin{equation}
\lambda = \int_0^\infty x \, \, \frac{ds(x)}{dx} \, dx = \frac{1}{nD} \,\,.
\end{equation}
For example, if there are 2000 trees per hectare, and each trunk is
10~cm thick, we can see to a distance of 50~m, on average.
{\it (ii)}~The result can be easily generalized into 3 dimensions. Assume
there are $n$ stars per unit volume, and each has a diameter $D$ and a
surface $A= \pi D^2$ perpendicular to the line of sight. Then we have
\begin{equation}
s(x) = 1 - e^{-nAx}\,\,,
\end{equation}
where $\lambda = (nA)^{-1}$. For example, if there were one sun per
cubic parsec, the mean free path would be $1.6 \times 10^4$~pc. If the
universe were infinite old and infinite in size, the line of sight
would eventually meet a stellar surface in any direction, although we
could see very far indeed.\\

6.2~{\it (i)}~The relation between luminosity, distance, and brightness is given by the inverse-square law, namely
\begin{equation}
{\rm brightness} = \frac{{\rm luminosity}}{4 \pi \, {\rm distance}^2} \,
.
\end{equation}
Here we are given the luminosity of each galaxy (the four are the
same, namely $4 \times 10^{37}~{\rm  J/s}$), and the brightness, in units of Joules/meters$^2$/second. Solving for the distance gives
\begin{equation}
{\rm distance} = \left( \frac{{\rm luminosity}}{4 \pi \, {\rm
      brightness}} \right)^{1/2} \, ,
\end{equation}
and so we find: galaxy \#1, distance $= 6.5 \times 10^{24}~{\rm m} =
210~{\rm Mpc}$; galaxy \#2, distance $= 8.4 \times 10^{24}~{\rm m} =
270~{\rm Mpc}$; galaxy \#3, distance $= 1.1 \times 10^{25}~{\rm m} =
360~{\rm Mpc}$; galaxy \#4, distance $= 1.5 \times 10^{25}~{\rm m} =
490~{\rm Mpc}$. {\it (ii)}~The redshift is given by $z = (\lambda
-\lambda_0)/\lambda_0$, where $\lambda_0 = 3935~{\rm \AA}$ and
$3970~{\rm \AA}$ for the two calcium lines. The measured wavelengths
for each of the two lines in each of the galaxies, the corresponding
redshift from each of the lines, and the average redshift are given in
Table~\ref{table-Hubble1}. {\it (iii)}~ The redshift is equal to the
velocity of recession divided by the speed of light. So we can
calculate the velocity of recession as the redshift times the speed of
light.  The Hubble constant is given in Table~\ref{table-Hubble2}. The
four galaxies give consistent values of the Hubble constant, at about
60~km/s/Mpc.  Not identical to the modern value of 70~km/s/Mpc, but
close.  That seemed quite straightforward; so why is there such
controversy over the exact value of the Hubble constant? The difficult
point is getting an independent measurement of the luminosity of each
galaxy. The problem stated that each of the galaxies has the same
luminosity of the Milky Way. That is only approximately true; the
numbers were adjusted somewhat to make this come out with a reasonable
value for $H_0$.\\
 
\begin{table}
\caption{Redshifts of the four galaxies in exercise 6.2 \label{table-Hubble1}}
\begin{tabular}{cccccc}
\hline \hline
~~~~Galaxy~~~~ & $\lambda$ (\AA) & $\lambda$ (\AA) & $z$ & $z$  & $z$ \\
& ~~~~first line~~~~ & ~~~~second line~~~~ & ~~~~first line~~~~ & ~~~~second line~~~~ & ~~~~average~~~~ \\
\hline
1 & 4100 & 4135 & 0.042 & 0.041 & 0.042 \\
2 & 4145 & 4185 & 0.053 & 0.054 & 0.053 \\
3 & 4215 & 4255 & 0.071 & 0.072 & 0.071 \\
4 & 4318 & 4360 & 0.097 & 0.098 & 0.098 \\
\hline \hline
\end{tabular}
\end{table}

\begin{table}
\caption{Determination of the Hubble constant. \label{table-Hubble2}}
\begin{tabular}{ccccc}
\hline \hline 
~~~Galaxy~~~ & ~~~~Redshift~~~~ & ~~~~Velocity (km/s)~~~~ &
~~~~Distance (Mpc)~~~~ & ~~~$H_0$ 
(km/s/Mpc)~~~ \\
\hline
1 & 0.042 & 12600 & 210 & 60 \\
2 & 0.053 & 15900 & 270 & 59 \\
3 & 0.071 & 21300 & 360 & 59 \\
4 & 0.098 & 29400 & 490 & 59\\
\hline \hline
\end{tabular}
\end{table} 

6.3~{\it (i)}~In the ``local Universe'' approximation where we pretend
that cosmological redshifts are Doppler shifts and it is a good
approximation to pretend that galaxies which are getting more distant
from us due to the expansion of space are flying away from us at a
given velocity, we can use the Hubble's law. Then for the closer
galaxy, we get $H_0 = 580~{\rm km/s}/35~{\rm Mly} = 17~{\rm km} \, {\rm
  s}^{-1} \, {\rm Mly}^{-1}$. For the farther galaxy, we get $H_0 =
25,400~{\rm km/s}/1,100~{\rm Mly} = 23~{\rm km} \, {\rm s}^{-1} \, {\rm
  Mly}^{-1}$. {\it (ii)}~The calculation from the more distant
galaxy. Reason: peculiar velocities are always a few hundred
km/s. They are random, so they could be anywhere from minus a few
hundred to plus a few hundred. Potentially, this could be a large
fraction of the 580~km/s of the closer galaxy. However, it will be a
small fraction of the 25,400~km/s of the more distant galaxy.  It is
noteworthy that closer galaxies tend to be brighter and therefore the
measurements are less likely to suffer from observational
errors. While this is true, the exercise presents numbers to (at least
approximately) equivalent significant figures in both cases. It may
well have taken a lot more telescope time and effort to get the
numbers on the more distant galaxy, but you have them. The peculiar
velocity issue, however, is an intrinsic effect that perfect
observations cannot get around. We will always have to deal with
galaxies moving about in the universe even if we have amazing data.
{\it (iii)}~We use the value of $H_0$ derived from the distant galaxy
to calculate the receding $v$ for the closer galaxy, $v = H_0 d =
805~{\rm km/s}$, which implies the peculiar velocity of the nearby
galaxy is $v_{\rm pec} = -220~{\rm km/s}$, that is $220~{\rm km/s}$
toward us. {\it (iv)}~Assuming that 220~km/s of the 25,400~km/s we
observed for the more distant galaxy were due to peculiar velocity,
from (\ref{vpeculiar}) we
have $H_0 = 25,620~{\rm km/s}/1,100~{\rm Mly} = 23.3~{\rm
  km/s/Mly}$. Note that $1,100~{\rm Mly}$ only has two significant
figures and so the difference is smaller than the precision of our
measurement. This is a specific illustration of why, given that all
galaxies will tend to have peculiar velocities of a few hundred km/s,
more distant galaxies give you a more reliable estimate of the Hubble
constant.\\

6.4~The Hubble flow $v = H_0r$ induces the flux $vn$ through the surface
$4\pi r^2$ of a sphere with radius $r$, and thus $\dot N = 4 \pi r^2
vn$. These particles escape from the sphere containing the
total number of particles $N = V n$. Hence $\dot N = - 4 \pi r^2 v n =
4 \pi r^3 \dot n /3$, or $\dot n = - 3 vn/r = -3 H_0 n$.\\

6.5~At the final time of the invasion, $t$,  the invaders are at proper
distance $d_{\rm p}$, and therefore comoving distance $r = d_{\rm
  p}/a(t)$; see (\ref{eqdp}).  {\it (i)}~For a flat space, the proper volume is
obviously the usual one in Euclidean geometry,
\begin{equation}
V = \frac{4 \pi}{3} d_{\rm p}^3 \, .
\end{equation}
{\it (ii)}~In a closed universe, and if $R$ is the comoving radius of
curvature, the proper area of a sphere at comoving coordinate $r$ is
$4\pi \, a^2(t) \, R^2 \, \sin^2(r/R)$, and the proper distance between
two spheres at $r$ and $r + dr$ is just $a(t) dr$, as obtained from
the FRW metric. Therefore, the proper volume of
each spherical shell between $r$ and $r + dr$ is $4 \pi a^3(t)R^2
\sin^2(r/R) dr$, and the proper volume of the sphere is
\begin{equation}
V = 4 \pi a^3(t) R^2 \int_0^r \sin^2 (r/R) dr = 4 \pi a^3(t) R^3
\int_0^{r/R} \sin^2 (r/R) \, d(r/R) = 4 \pi a^3 R^3 \left[\frac{d_{\rm
      p}}{2aR} - \frac{\sin (2 d_{\rm p}/a/R)}{4} \right] \, .
\end{equation}
{\it (iii)}~In an open universe, the calculation is just like for the
closed universe but with the substitution $\sin (r/R)$ by  $\sinh(r/R)$,
\begin{equation}
V = 4 \pi a^3(t) R^3 \int_0^{r/R} \sinh^2 (r/R) \, d(r/R) = 4 \pi a^3
R^3 \left(\frac{\sinh(2 d_{\rm p}/a/R)}{4} - \frac{d_{\rm p}}{2aR}
\right) \, .
\end{equation}

\vspace{0.2in}

6.6~For the case when the universe contains only matter with negligible
pressure, the energy density changes as $\rho_m(t) = \rho_{m,0}
/a^3(t)$. Multiplying (\ref{FriedmannGR}) by $a^2(t)$, we have
\begin{equation}
(\dot a)^2 = \frac{8 \pi G \rho_{m,0}}{3 c^2 a} - \frac{c^2}{R_0^2} \,
  .
\label{buck2}
\end{equation}
Now, using $\dot a = (da/dt) = (da/d\theta) (d\theta/dt)$, we find the
left-hand-side of (\ref{buck2}) is
\begin{equation}
(\dot a)^2 = \frac{c^2}{R_0^2} \frac{\sin^2 \theta}{(1 - \cos \theta)^2}
  = \frac{c^2}{R_0^2} \frac{1 + \cos \theta}{1 - \cos \theta} \,,
\label{rafauno}
\end{equation}
where the last equality follows from $\sin^2 \theta = 1 - \cos^2
\theta  = (1 - \cos \theta)(1 + \cos \theta)$, and the right-hand-side
of (\ref{buck2}) is
\begin{equation}
\frac{8 \pi G \rho_{m,0}}{3 c^2 a} - \frac{c^2}{R_0^2} =
\frac{c^2}{R_0^2} \left( \frac{2}{1 - \cos \theta} -1 \right) =
\frac{c^2}{R_0^2} \frac{1 + \cos \theta}{1 - \cos \theta} \, .
\end{equation}
So the two sides of (\ref{buck2}) are indeed equal, confirming that
this parametric solution given as $a(\theta)$ and $t(\theta)$ is indeed a solution
of Friedmann's equation. {\it (ii)}~The maximum value of a occurs at $\theta = \pi$, and is 
\begin{equation}
a_{\rm max} = \frac{8 \pi G \rho_{m,0} R_0^2}{3 c^4} \, .
\end{equation}
{\it (iii)}~Correspondingly, the maximum value of the proper radius of curvature is
\begin{equation}
a_{\rm max} R_0 = \frac{8 \pi G \rho_{m,0} R_0^3}{3 c^4} \, .
\end{equation}
{\it (iv)}~The age of the universe at $\theta = \pi$ is
\begin{equation}
t_{\rm max} = \frac{4 \pi^2 G \rho_{m,0} R_0^3}{3 c^5} \, .
\end{equation}
{\it (v)}~The {\it big crunch} happens when $\theta = 2 \pi$, and we then have
\begin{equation}
t_{\rm crunch} = \frac{8 \pi G \rho_{m,0} R_0^3}{3 c^5} \, .
\end{equation}

\vspace{0.2in}

6.7~Multiplying (\ref{FriedmannGR}) by $a^2(t)$, we have
\begin{equation}
(\dot a)^2 = \frac{8 \pi G \rho_{m,0}}{3 c^2 a} + \frac{c^2}{R_0^2} \,  .
\label{buckhyp}
\end{equation}
Now, using the relations
\begin{equation}
\sin (ix) = \frac{e^{i(ix)} - e^{-i(ix)}}{2i} = \frac{e^{-x} -
  e^{x}}{2i} = - \frac{e^x - e^{-x}}{2i} = - \frac{1}{i} \sinh x = i
  \sinh x \,
\end{equation}
\begin{equation}
\cos(ix) = \frac{e^{i(ix)} + e^{-i(ix)}}{2} = \frac{e^{-x} + e^x}{2} = \cosh
x \,,
\end{equation} 
\begin{equation}
\cosh^2 x - \sinh^2 x = \cos^2 (ix) - [\sin(ix)/i]^2 = \cos^2(ix) +
\sin^2 (ix) = 1
\end{equation}
we can rewrite (\ref{rafauno}) as
\begin{equation}
(\dot a)^2 =  \frac{c^2}{R_0^2} \frac{\sinh^2 \theta}{(\cosh \theta-1)^2}
  = \frac{c^2}{R_0^2} \frac{\cosh \theta + 1}{\cosh \theta -1} \,,
\label{rafauno-dos}
\end{equation}
and the right-hand-side of (\ref{buckhyp}) is 
\begin{equation}
\frac{8 \pi G \rho_{m,0}}{3 c^2 a} + \frac{c^2}{R_0^2} =
\frac{c^2}{R_0^2} \left( \frac{2}{\cosh \theta-1} +1 \right) =
\frac{c^2}{R_0^2} \frac{\cosh \theta + 1}{\cosh \theta -1} \, .
\end{equation}
The two sides of (\ref{buck2}) are indeed equal, confirming that
this parametric solution given as $a(\theta)$ and $t(\theta)$ is indeed a solution
of Friedmann's equation. {\it (ii)}~A comparison of the scale factors
in (\ref{gear1}), (\ref{ciento66}), and (\ref{ciento67}) corresponding
to solutions with $k=0$, $k=1$, and 
$k=-1$, respectively is exhibited in Fig.~\ref{scale67}. \\

\begin{figure}
 \postscript{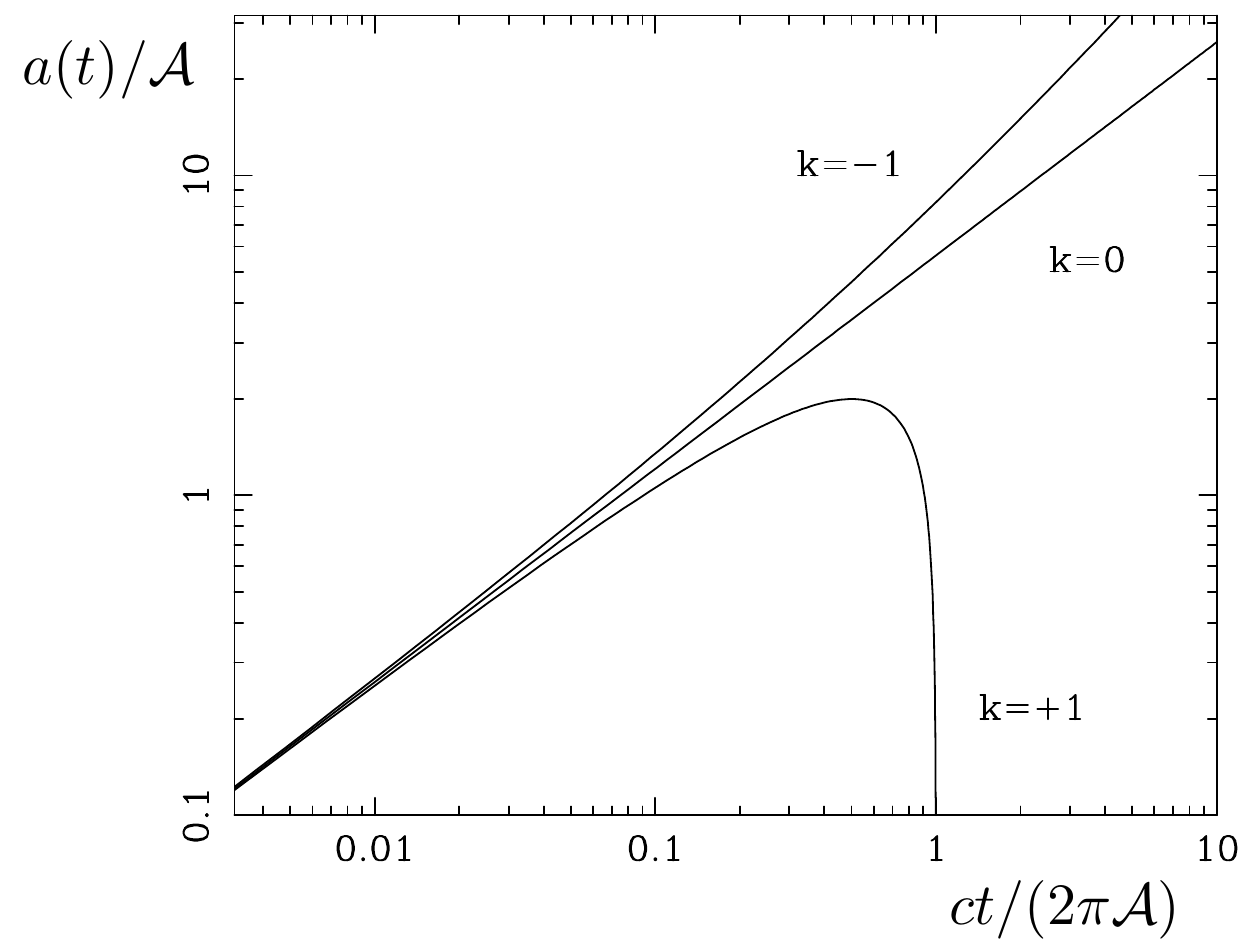}{0.7}
 \caption{The time dependence of the scale factor for open, closed and
   critical matter-dominated cosmological models. The upper line
   corresponds to $k = -1$, the middle line to the flat $k = 0$ model, and
   the lowest line to the recollapsing closed $k = +1$ universe. The log
   scale is designed to bring out the early-time behaviour, although
   it obscures the fact that the closed model is a symmetric cycloid
   on a linear plot of $a$ against $t$. We have set ${\cal A} = 4\pi G
   \rho_{m,0} R_0^3/(3 c^2)$~\cite{Peacock}.}
\label{scale67}
\end{figure}

6.8~{\it (i)}~For the model with $\Omega_{m,0} = 1$, the comoving distance is
\begin{equation}
r = c \int_0^z \frac{dz}{H(z)} = \frac{c}{H_0} \int_0^z \frac{dz}{(1 +
  z )^{3/2}} = \frac{2c}{H_0} \left(1 - \frac{1}{\sqrt{1 + z}} \right)
\, .
\end{equation}
The comoving distance to the horizon, at $a = 0$ or $z = \infty$, is
$r = 2c/H_0$. {\it (ii)} For this model, half the comoving distance to
the horizon is $r = c/H_0$, and the redshift at which the comoving
distance has this value is obtained as:
\begin{equation}
1 - \frac{1}{\sqrt{1 + z}} = \frac{1}{2} \Rightarrow z = 3 \, .
\end{equation}
{\it (iii)}~For this same model, and at $z = 3$, the age of the
universe is obtained from
\begin{equation}
t(z) = \int_z^\infty \frac{dz}{(1 + z) \ H(z)} = \frac{2}{3}
\frac{1}{H_0} \frac{1}{(1 + z)^{3/2}} \, .
\end{equation}
The present age of the universe is of course $t_0 = 2/(3H_0)$, and so the ratio of the age
at $z = 3$ to its present age is just
\begin{equation}
\frac{t(z=3)}{t_0} = \frac{1}{(1 + z)^{3/2}} = \frac{1}{8} \, .
\end{equation}
{\it (iv)}~From the same equation as above, we find
\begin{equation}
\frac{t(z)}{t_0} = \frac{1}{(1 +z)^{3/2} }= \frac{1}{2} \Rightarrow z =
  2^{2/3} - 1 = 0.5874 .
\label{snaple1}
\end{equation}
Note that all these equations are of course valid only for the specific model that is flat and contains only matter, with $\Omega_{m,0} = 1$.\\

7.1~The flux of one of these objects is ${\cal F} = L/(4\pi d_L^2)$, and its
angular size is $\theta = \ell/d_A$. Hence, the apparent surface brightness is 
$I \propto {\cal F}/\theta^2$, or
\begin{equation}
I = {\rm constant} \times \frac{{\cal F}}{\theta^2} = {\rm constant} \times
  \frac{d_A^2}{d_L^2} = {\rm constant} \times (1+z)^{-4}\, ,
\end{equation}
where in the last equality we have used (\ref{pavilion}). Note that $L$, $\ell$, and
$4\pi$ are constants and so can be absorbed in the constant of
proportionality. Therefore, the apparent surface brightness $I$ will always
decrease with redshift as $(1 + z)^{-4}$ compared to the {\it intrinsic} surface
brightness $B$, without any dependence on the cosmological model.\\

7.2~{\it (i)}~The number density of photons per unit frequency is equal to the energy density per unit frequency divided by $h\nu$, or
\begin{equation}
n_\nu d\nu = \frac{8 \pi \nu^2 d\nu}{c^3 \{\exp[h \nu/(kT)] - 1\}} \, .
\label{ndeT}
\end{equation}
The total number density is found by integrating over frequency, which
gives
\begin{equation}
n = \frac{8 \pi}{c^3} \int \frac{\nu^2 d\nu}{exp[h\nu/(kT)] -1} \, .
\end{equation}
Substituting $x = h \nu/(kT)$, we find
\begin{equation}
n = \frac{8 \pi}{c^3} \left(\frac{kT}{h} \right)^3 \int \frac{x^2
  dx}{e^x -1} \, .
\end{equation}
For $T_0 = 2.725$, we find $410.4~{\rm cm}^{-3}$. {\it (ii)}~The current
density of baryons must then be 
\begin{equation}
n_b =  5.5 \times 10^{-10}  \ 410.4~{\rm cm^{-3}} = 2.25 \times
10^{-7}~{\rm cm^{-3}} \, .
\end{equation}
{\it (iii)}~Every baryon weighs approximately like the mass of a
proton (this is not exact because, for example, the helium nuclei
weigh a little less than 4 protons because of the helium nucleus
binding energy, but the difference is rather small). So the density of
baryons is $n_bm_p = 3.78 \times 10^{-31} {\rm g/ cm^3}$. The critical
density is $3H_0^2/(8\pi G) = 9.2
\times 10^{-30}~{\rm g/cm^3}$, so $\Omega_b = 0.041$.\\

  7.3~To analyze the measurement of our own galaxy through the CMB, it
  is useful to consider the density $N_\gamma (\vec p \,)$ of photons
  in phase space, defined by specifying that there are $N_\gamma (\vec
  p \, ) d^3p$ photons of each polarization (right or left circularly
  polarized) per unit spatial volume in a momentum-space volume $d^3p$
  centered at $\vec p$. Since $|\vec p \,| = h\nu/c$ and $4\pi
  h^3\nu^2 d\nu/c^3$ is the momentum-space volume between frequencies
  $\nu$ and $d\nu$, (\ref{ndeT}) gives
\begin{equation}
N_\gamma (\vec p \,) = \frac{1}{2} \frac{n_T (c|\vec p \,|)/h)}{4 \pi h^3
  \nu^2/c^3} = \frac{1}{h^3} \frac{1}{\exp[|\vec \, p| c/(kT) -1] } \,,
\end{equation}
where $n_T$ is the number density of photons in equilibrium with
matter at temperature $T$ at photon frequency between $\nu$ and $\nu +
d \nu$, and the factor $1/2$ takes account of the fact that $n_T$ includes both
possible polarization states. This is of course the density that
would be measured by an observer at rest in the radiation
background. The phase space volume is Lorentz invariant, and the
number of photons is also Lorentz invariant, so $N_\gamma$
is a scalar, in the sense that a Lorentz transformation to a coordinate system
moving with respect to the radiation background that takes $\vec p$ to
${\vec p}^{\,\prime}$
also takes $N_\gamma$ to $N'_\gamma$ , where
\begin{equation}
N'_\gamma ({\vec p}^{\, \prime}) = N_\gamma (\vec p \,) \, .
\end{equation}
If the Earth is moving in the $x$-direction with a velocity (in units
of $c$) of $\beta$, and we take $\vec p$ to be the photon momentum in
the frame at rest in the CMB and ${\vec p}^{\, \prime}$ to be the photon momentum
measured on Earth, then from (\ref{dop1}) it follows that
\begin{equation}
|{\vec p}^{\, \prime}| = \gamma (1 - \beta \cos \theta) |\vec p\,|
\end{equation}
where $\theta$ is the angle between $p$ and the $x$-axis. Thus
\begin{equation}
N'_\gamma ({\vec p}^{\, \prime}) = \frac{1}{h^3} \ \frac{1}{\exp[
  |\vec p^{\, \prime}|
  c/(kT')] - 1} \,,
\end{equation}
where the temperature is a function of the angle between the direction
of the photon and the Earth's velocity
\begin{equation}
T = T' \gamma (1-\beta \cos \theta) \, .
\end{equation}
 This means that the temperature $T(\theta)$ observed in the direction $\theta$, is given in terms of the average temperature $T_0$ by
\begin{eqnarray}
T(\theta) & = &T_0 \frac{\sqrt{1 - \beta^2}}{1 -
  \beta \cos \theta} = T_0 \left(1 - \beta^2 \right)^{1/2} \left(1
  - \beta \cos \theta \right)^{-1} \approx T_0 \left(1 -
 \beta^2/2 + \cdots \right) \left( 1 + \beta \cos \theta
  + \beta^2 \cos^2 \theta + \cdots \right) \nonumber \\
& \approx & T_0 \left[ 1 + \beta \cos \theta + \beta^2 \left(\cos^2
\theta - 1/2 \right) + \cdots \right] \, .
\end{eqnarray}
Using the trigonometric relation $\cos^2x = (1 + \cos2 x)/2$ we obtain
(\ref{thetadeTq}).  The motion of the observer (us) gives rise to both
a dipole and other, higher order corrections. The observed dipole
anisotropy implies that~\cite{Weinberg:2008zzc} 
\begin{equation}
\vec v_\odot - \vec v_{\rm CMB} = 370 \pm 10~{\rm km/s} \quad {\rm
  towards} \quad \phi = 267.7 \pm 0.8^\circ, \quad \theta = 48.2 \pm
0.5^\circ \,,
\end{equation}
where $\theta$ is the colatitude (polar angle) and it is in the range
$0 \leq \theta \leq \pi$ and $\phi$ is the longitude (azimuth) and it
is in the range $0 \leq \phi \leq 2 \pi$.  Allowing for the Sun's
motion in the Galaxy and the motion of the Galaxy within the Local
Group, this implies that the Local Group is moving with 
\begin{equation}
\vec v_{\rm LG} - \vec v_{\rm CMB} \approx 600~{\rm km/s} \quad {\rm
  towards} \quad \phi = 268, \quad \theta = 27^\circ \, .
\end{equation}
This ``peculiar'' motion is
subtracted from the measured CMB radiation, after which the intrinsic
anisotropy is isolated (Fig.~\ref{wmap-planck}), and revealed to be about few parts
in $10^5$. Even though miniscule, these primordial perturbations provided
seeds for the structure of the Universe.\\

7.4~For a discrete set of directions in the sky, the normalized intensity function is $I(\Omega)=\frac{1}{N}\sum_{i=1}^N\delta(\vec{u}_i,\Omega)$.
The spherical harmonic coefficients acan be written as
\begin{equation}
\label{discrete alm}
\bar a_l^m=\frac1N\sum_{i=1}^NY_l^{m*}(\vec{u}_i)\,,
\end{equation}
where $\vec{u}_i$ is the unit vector to the $i$th direction, $1\le i \le N$.
Next, we construct an estimation of the corresponding power spectrum by squaring the $a_l^m$'s followed by a sum over $m$:
\begin{equation}
\bar C_l \equiv\frac{1}{2 l+1} \sum_{|m|\le l} |\bar a_l^m |^2
=\frac{1}{N^2(2 l+1)}\sum_{|m|\le l} \left| \sum_{i=1}^N Y_l^{m*}(\vec{u}_i)\right|^2\,.
\end{equation}
Because all the sums are finite they could be rearranged and expanded to
\begin{equation}
\bar C_l=
\frac1{N^2(2l+1)}\sum_{i=1}^N\sum_{|m|\le l}| Y_l^{m}(\vec{u}_i)|^2
+\frac2{N^2(2 l+1)}\sum_{i<j}\sum_{|m|\le l}Y_l^{m*}(\vec{u}_i) Y_l^m(\vec{u}_j)\,.
\label{eq:cl ylm big sum}
\end{equation}
The formula for addition of spherical harmonics is given by~\cite{Anchordoqui:2013}
\begin{equation}
P_l(\vec x\cdot\vec y)=\frac{4\pi}{2 l+1}\sum_{|m|\le l}Y_l^{m*}(\vec
x) Y_l^m(\vec y)\, .
\label{eq:m sum different}
\end{equation}
Now, since $P_l(1)=1$ we set the unit direction vectors $\vec x$ and
$\vec y$ to be equal in (\ref{eq:m sum different}) to obtain
\begin{equation}
\frac{2 l+1}{4\pi}=\sum_{|m|\le l}| Y_l^m(\vec x)|^2\,.
\label{eq:m sum same}
\end{equation}
Combining (\ref{eq:cl ylm big sum}), (\ref{eq:m sum different}), and
(\ref{eq:m sum same}) leads to
\begin{equation}
\bar C_\ell=\frac1{4\pi N}+\frac1{2\pi N^2}\sum_{i<j}P_l(\vec u_i\cdot\vec u_j)\,.
\end{equation}
Experimentally, only $\bar a_l^m$ and $\bar C_l$ could be measured, but these are estimates of their continuous counterparts
$a_l^m,C_l$ respectively. Therefore, since inner products are
invariant under rotations, it follows that the $C_l$ are also invariant under rotations~\cite{Denton:2014hfa}.\\

7.5~{\it (i)}~The circumference of a circle of radius $a$ is $2\pi a$, so the
orbital speed is the circumference divided by period:
\begin{equation}
v = \frac{2 \pi a}{{\cal T}} = \frac{2 \pi a}{\sqrt{4 \pi^2 a^3/(GM)}} =
\sqrt{\frac{GM}{a}} \, .
\end{equation}
{\it (ii)}~According to the Birkoff's theorem, the orbit about a mass
distributed within a sphere is the same as if the mass is all
concentrated in the center of the sphere. So, we can use the velocity
formula derived in (i), and invert it to obtain the mass enclosed by an
orbit:
\begin{equation}
M(a) = \frac{a}{G} v^2 \, .
\label{zzol}
\end{equation}
The enclosed mass at 8~kpc is then $M(8~{\rm kpc}) \approx 2 \times
10^{41}~{\rm kg} = 10^{11} M_\odot$. {\it (iii)}~If the mass enclosed
by the orbit stays at $10^{11} M_\odot$ as the radius increases, which
follows from the fact that the Sun is at the edge of the luminous
galaxy, then at different radii the velocity given by (\ref{zzol})
will decrease with square root of the distance. At 30~kpc, it is $v
\approx 110~{\rm km/s}$. At 100~kpc, we have $v \approx 60~{\rm
  km/s}$. {\it (iv)}~Let's look again at (\ref{zzol}), which says that
if the orbital velocity stays the same, the mass enclosed will
increase linearly with a as the radius of the orbit grows. We already
calculated the mass enclosed by 8~kpc orbit in part (ii). So, at
30~kpc, the mass enclosed will be 30~kpc/8~kpc times larger,
or $3.8\times 10^{11} M_\odot$. At 100~kpc, the mass enclosed will
100~kpc/8~kpc times larger, or $1.3 \times 10^{12}
M_\odot$. {\it (v)}~We see that the mass of the gravitating matter is
increasing linearly with radius, and exceeds by more than factor of 10
the mass of the luminous matter (e.g. stars and gas). We thus infer
that the outer halo of the galaxy is dominated by invisible dark
matter.\\

7.6{\it (i)}~The total mass inside $R$ is obtained from $GM(R)/R = v_c^2(R)$.
The answer can of course be found by substituting for the value of $G$
and everything else in your favorite system of units, and if you are
lucky not to make any mistake you may even get the right
answer. Often, it is faster and safer to work it out using
proportionality comparing to an example that you know and love. What
could this example be but the Earth moving around the Sun? For the
Earth, with $M_\odot$ and an orbit of 1~AU the velocity is $30~{\rm km
s}^{-1}$ (if you did not know how fast the Earth moves around the Sun,
this is a good number to remember). So, the mass inside radius $R$ is
\begin{equation}
M(R) = 8 \times 10^{10} M_\odot
\end{equation}
{\it (ii)}~If the density at $R$ is $\rho_0$, then the density at any
other radius $r$ is $\rho_0 (R/r)^2$, so the mass inside $R$ is
\begin{equation}
M(R) = 4 \pi \int_0^R dr r^2 \rho_0 \left(\frac{R}{r} \right)^2 = 4
\pi \rho_0 R^3 \, .
\end{equation}
Hence the density at $R$ is
\begin{equation}
\rho_0 = \frac{M(R)}{4 \pi R^3} = 0.51 \, m_p~{\rm cm}^{-3} \, .
\end{equation}
The result is most easily computed remembering that the solar mass
contains $1.19 \times 10^{57}$ proton masses (another useful number to
remember), and a parsec is $3.086 \times 10^{16}~{\rm m}$. {\it
  (iii)}~The density is
\begin{equation}
\rho_\Lambda = \Omega_\Lambda \frac{3 H_0^2}{8 \pi G} = 5.5 \times
10^{-6} \, \Omega_\Lambda \, m_p~{\rm cm}^{-3} = 4 \times 10^{-6} \,
m_p~{\rm cm}^{-3}  \, .
\end{equation}
{\it (iv)}~Because the dark energy is spread out uniformly, whereas
the dark matter and baryonic matter are highly concentrated in the
inner parts of galaxies, the density of dark energy is very small
compared to the density of matter inside the radius of the solar orbit
in the Milky Way. The dark energy therefore must have a tiny dynamical
effect.\\

7.7~The relation between the emitted $T_{\rm em}$ and the observed
$T_{\rm obs}$ is
\begin{equation}
T_{\rm em} = T_{\rm obs} (1 + z) = \frac{2.9 \times 10^{-3}~{\rm
    mK}}{\lambda_{\rm max}^{\rm obs}} (1 + z) \,,
\end{equation}
where in the last equality we used Wien's displacement
law~\cite{Wien:1894}. For $\lambda_{\rm max}^{\rm obs} = 180~\mu{\rm m}$ and $z=2$, we
have $T_{\rm em} \simeq 48~{\rm
  K}$. If we did not account for redshift, we would have thought the
galaxy was only at 16~K.\\

7.8~In the benchmark model, at the present moment, the ratio of the
vacum energy density  to the energy density in matter is
\begin{equation}
\frac{\rho_\Lambda}{\rho_{m,0}} = \frac{\Omega_\Lambda}{\Omega_{m,0}} \approx 2.3 \, .
\end{equation}
In the past, however, when the scale factor was smaller, the ratio of
densities was
\begin{equation}
\frac{\rho_\Lambda}{\rho_m (a)} = \frac{\rho_\Lambda}{\rho_{m,0}/a^3}
= \frac{\rho_\Lambda}{\rho_{m,0}} a^3 \, .
\end{equation}
If the universe has been expanding from an initial very dense state, at some moment in the past, the energy density of matter and $\Lambda$ must have been equal. This moment of matter-$\Lambda$ equality occurred when
\begin{equation}
a_{m\Lambda}^3 = \frac{\Omega_{m,0}}{\Omega_{\Lambda}} =
  \frac{\Omega_{m,0}}{1 - \Omega_{m,0}} \approx 0.43 \, .
\label{babiche1}
\end{equation}
where we have used the normalization $a_0 = 1$ for the present.
Next, we generalize (\ref{noire}) to write the age of the universe at
any redshift $z$, for a flat model with matter and a cosmological
constant,
\begin{equation}
t(z) = \frac{1}{H_0} \int_z^\infty \frac{dz}{(1+z) \sqrt{\Omega_{m,0}
    (1+z)^3 + \Omega_\Lambda}} = \frac{1}{H_0 \sqrt{\Omega_\Lambda}}
\int_z^\infty \frac{dz}{(1+z) \sqrt{1 + (\Omega_{m,0}/\Omega_\Lambda)
    (1 +z)^3}} \, .
\end{equation}
With the change of variables $y = \sqrt{1 + (\Omega_{m,0}/\Omega_\Lambda)
  (1+z)^3}$ we find $2 y dy = 3 (y^2 -1) dz/(1 +z)$ and so
\begin{equation}
t = \frac{2}{3 H_0 \sqrt{\Omega_\Lambda}} \int_y^\infty \frac{dy}{y^2
    -1 } \, .
\end{equation}
The integral can be solved analytically as:
\begin{equation}
- \int \frac{dy}{y^2 -1} = \frac{1}{2} \ln \left[\frac{y+1}{y-1}
\right] \,,
\end{equation}
which yields for $t$
\begin{eqnarray}
t & = & \frac{2}{3 H_0 \sqrt{\Omega_\Lambda}}
  \left.\ln\left(\frac{1}{\sqrt{y^2 -1}} + \frac{y}{\sqrt{y^2-1}}
    \right) \right|_{\infty}^{\sqrt{1 +\Omega_{m,0} (1+z)^3
      /\Omega_\Lambda}} \nonumber \\ & = & \frac{2}{3 H_0
    \sqrt{\Omega_\Lambda}} \ln 
      \left( \sqrt{\frac{\Omega_\Lambda}{\Omega_{m,0} (1+z)^3}} +
        \sqrt{1 + \frac{\Omega_\Lambda}{\Omega_{m,0} (1+z)^3}} \right) \, .
\label{babiche2}
\end{eqnarray}
Using (\ref{MUKA1}) and (\ref{babiche1}) we can rewrite (\ref{babiche2}) as
\begin{equation}
H_0 t = \frac{2}{3 \sqrt{1 - \Omega_{m,0}}} \ln
  \left[(a/a_{m\Lambda})^{3/2} + \sqrt{1 + (a/a_{m\Lambda})^3} \right] = \frac{2}{3 \sqrt{1-\Omega_{m,0}}} \ln A \,,
\end{equation}
and so
\begin{equation}
H_0 t_0 =  \frac{2}{3 \sqrt{1-\Omega_{m,0}}} \ln A_0 \,,
\end{equation}
where we have defined 
\begin{equation}
A_0 = a_{m\Lambda}^{-3/2} + \sqrt{1 + a_{m\Lambda}^{-3}} \simeq 3.35
\, .
\end{equation}
Now, we want to find the value of $a$ for which $t = t_0/2$. This
implies 
\begin{equation}
\ln A = \frac{1}{2} \ln A_0 \,,
\label{snaple2}
\end{equation}
where
\begin{equation}
A = x + \sqrt{1 +x^2} \quad {\rm with} \quad x =
\left(\frac{a}{a_{m\Lambda}} \right)^{3/2} \, .
\end{equation}
(\ref{snaple2}) implies $A = \sqrt{A_0}$ and so
\begin{equation}
x + \sqrt{1+x^2} = \sqrt{A_0} \quad \Rightarrow \quad 1 + x^2 = \left(\sqrt{A_0} -
  x \right)^2 \quad \Rightarrow \quad x = \frac{A_0 -1}{2 \sqrt{A_0}} \simeq 0.64   \,,
\end{equation}
yielding
\begin{equation}
a = a_{m\Lambda} x^{2/3} = 0.56 \quad \quad {\rm and} \quad \quad z =
\frac{1}{a} - 1 = 0.78 \, .
\end{equation}
We see that the redshift at which the age of the universe was half the
present age is larger in this benchmark model than in the model with
$\Omega_{m,0} = 1$, see (\ref{snaple1}). This is because in the benchmark model,
which contains vacuum energy, the universe has started to accelerate
recently, roughly since the epoch at $a = a_{m\Lambda}$. The universe
took a longer time to expand to $a = 0.56$ and then it picked up speed
again in its expansion up to the present $a_0 = 1$.\\

8.1~We have seen in (\ref{rhogama}) that $s_\gamma \propto T^3$, so
that $S_\gamma \propto VT^3$. For a reversible adiabatic expansion,
the entropy of the (non-interacting) CMB photons remains
unchanged. Hence, when $V$ doubles, $T$ will decrease by a factor
$(2)^{-1/3}$. So after $10^{10}~{\rm yr}$ the average tempreature of
the blackbody will become $\langle T \rangle =
2.2~{\rm K}$.\\

8.2~Since during inflation the Hubble rate is constant
\begin{equation}
\Omega -1 = \frac{kc^2}{a^2 H^2 R_0^2} \propto a^{-2} \, .
\end{equation}
On the other hand, (\ref{flatness3}) suggests that to reproduce today's observed
value $\Omega_0 -1 \sim 1$ the initial value at the beginning of the
radiation-dominated phase must be $|\Omega -1| \sim 10^{-54}$. Since
we identify the beginning of the radiation-dominated phase with the
end of inflation we require
\begin{equation}
|\Omega -1|_{t = t_{\rm f}} \sim 10^{-54} \, .
\end{equation}
During inflation
\begin{equation}
\frac{|\Omega -1|_{\rm t = t_{\rm f}}}{|\Omega - 1 |_{t = t_{\rm i}}}
= \left(\frac{a_{\rm i}}{a_{\rm f}} \right)^2 = e ^{-2 H \Delta t} \, .
\end{equation}
Taking $|\Omega - 1|_{t = t_{\rm i}}$ of order unity, it is enough to
require that $\Delta t \agt 60/H$ to solve the flatness problem. Thus,
inflation ameliorates the fine-tuning problem, by explaining a tiny
number ${\cal O} (10^{-54})$ with a number ${\cal O} (60)$. \\

8.3~The effective number of neutrinos and antineutrinos is $g_{\nu_L}
= 6$ and the temperature of the cosmic neutrino background is $T_\nu
= 0.7 \, T_\gamma \approx 1.9~{\rm K}$. Now, from (\ref{gingseng}) we have
\begin{equation}
n_\nu = \frac{3}{4} \frac{\zeta (3)}{\pi^2} g_{\nu_L}
\left(\frac{kT_\nu}{\hslash c}\right)^3 \approx 45.63
\left(\frac{T_\nu}{K} \right)^3 \approx 313~{\rm neutrinos}/{\rm cm}^3
\, .
\end{equation}
If neutrinos saturate the dark
matter density the upper bound on the neutrino mass is then
\begin{equation}
m_\nu  <  \frac{0.26 \,\,\rho_{\rm c}}{n_\nu} 
\approx \frac{2.6 \times 10^{-27}~{\rm kg/m}^3}{3.13 \times 10^{8}~{\rm neutrino/m}^3} \approx 10^{-35}~\frac{\rm kg}{\rm neutrino} \times \frac{9.38 \times 10^8 {\rm eV}/c^2}{1.67 \times 10^{-27}~{\rm kg}} \sim 5.6~{\rm eV}/c^2 \,\,.
\label{mnulimit}
\end{equation}
where we have used $m_p = 938~{\rm MeV}/c^2 = 1.67 \times
10^{-27}~{\rm kg}$ to obtain the result in natural units.\\

8.4~If we change the difference between the proton and neutron mass to
$\Delta m = 0.129~{\rm MeV}$ while all other parameters remain the
same, then the time of freeze-out of the neutron abundance occurs at
the same temperature $T_{n/N}^{\rm FO} = 0.75~{\rm MeV}$. Therefore,
the neutron abundance freezes out at $n_n/n_p =
  e^{-0.1293/0.75} = 0.84$. If there were no neutron decays and all
  neutrons combined to form helium, the maximum primordial $^4$He
  abundance would then be
\begin{equation}
Y_{\rm p}^{\rm max} =  \frac{2n_n}{n_n + n_p}  = 0.91 \, .
\end{equation}
Note that neutrons would in fact not decay, they would be stable
because the difference with the mass of the proton would be less than
the mass of the electron.  It would be rather unfortunate if the
neutron had a mass so close to the proton mass: almost all the matter
in the universe would have turned to helium in the beginning of the
universe, and main-sequence stars would not live very long with the
very small amount of hydrogen they would have left. The Sun would live
for less than 1 billion years and the planet Earth would not have had
enough time to sustain life on it for us to be here now.\\

8.5~{\it (i)}~From (\ref{rhogama}), the energy density of photons at the time of BBN was 
\begin{equation}
\rho_{\gamma, {\rm BBN}} =  0.66 \frac{(kT_{\rm BBN})^4}{(\hslash c)^3} = 7.56 \times
10^{20}~{\rm J/m^3}  \, ,
\end{equation}
where $T_{\rm BBN} \approx 10^9~{\rm K}$.
Note that in reality we should also account for the neutrinos, but
Gamow did not know much about the three families of neutrinos and
their interactions. A better estimate of the energy
density at BBN goes as follows. The effective number of neutrinos and
antineutrinos is 6, or 3 times the effective number of species of
photons. On the other hand, the fourth power of the neutrinos
temperature is less than the fourth power of the photon temperature by
a factor of $3^{-4/3}$. Thus the ratio of the energy density of
neutrinos and antineutrinos to that of photons is
\begin{equation}
\rho_\nu/ \rho_\gamma = 3^{-4/3}\,\,\, 3 = 0.7 \,\, . 
\label{bvo}
\end{equation}
Hence the total energy density after electron
positron annihilation is
\begin{equation}
\rho_{\rm BBN}  \simeq \rho_{\nu,{\rm BBN}} + \rho_{\gamma,{\rm BBN}}
= 1.7 \rho_{\gamma,{\rm BBN}} \simeq 1.3 \times 10^{21}~{\rm J/m}^3 \,\,.
\end{equation}
{\it (ii)}~Since the universe was radiation dominated, the
critical density at BBN had to be equal to this radiation density, so
\begin{equation}
\frac{3 c^2 H^2}{8 \pi G} = \rho_{\rm BBN}
\end{equation}
This gives a Hubble parameter at the time of BBN in Gamow's radiation dominated
universe of
$H = 2.17 \times 10^{-3}~{\rm s}^{-1}$.  {\it (iii)}~Readjusting
(\ref{gear3}), the time for BBN is found to be
\begin{equation}
t_{\rm BBN, G} = \frac{1}{2 H} = 231~{\rm  s} \, .
\end{equation}
{\it (iv)}~For a present age $t_0 \approx 10^{10}$~yr, the temperature is given
by 
\begin{equation}
\frac{3 c^2 H_0^2}{8\pi G} = \frac{3 c^2}{32 \pi G t_0^2} = 0.66
\frac{(kT_{0,{\rm G}})^4}{(\hslash c)^3}
\end{equation}
which gives $T_{0,{\rm G}}
= 27~{\rm K}$. Note that actually this temperature just depends on
$t_0$ and the assumption of a flat, radiation-dominated universe, but
it does not depend on $T_{\rm BBN}$.  {\it (v})~If the universe
changed from being radiation dominated to matter dominated at some
redshift $z_{\rm eq}$, then at the present time the matter density is
greater than the radiation density by a factor $1 + z_{\rm eq}$; so
$\rho_{\rm rad} = \rho_m c^2/(1 + z_{\rm eq})$. In a flat universe with
{\it only matter and radiation}, the total density has to be equal to
the critical density, therefore $\rho_{\rm rad} + \rho_m c^2 =
\rho_{\rm rad} (2 + z_{\rm eq}) = \rho_c$. So,
\begin{equation}
\rho_{\rm rad} = \frac{3 H_0^2 c^2}{8 \pi G} \frac{1}{2 + z_{\rm eq}} =
0.66 \frac{(kT_0)^4}{(\hslash c)^3} \,,
\end{equation}
and the radiation temperature is smaller by a factor $(2 + z_{\rm eq})^{-1/4}$.\\

8.6~The effective number of neutrino species contributing to r.d.o.f. can be written as
\begin{equation}
N_{\rm eff} = 3 \left[1 + \left(\frac{T_{\nu_R}}{T_{\nu_L}}\right)
  ^4\right] \, .
\end{equation}
Taking into account the isentropic heating of the rest of the plasma
between $\nu_R$ decoupling temperature $T^{\rm dec}_{\nu_R}$ and the end of
the reheating phase,
\begin{equation}
\delta N_\nu = 3 \left(\frac{g(T_{\nu_L}^{\rm dec})}{g(T_{\nu_R}^{\rm dec})} \right)^{4/3} \,,
\label{7}
\end{equation}
where  $T^{\rm dec}_{\nu_L}$ is the temperature at the end of the
reheating phase (when left-handed neutrinos decouple), and we have
taken $N_{\rm eff} = 3 + \delta N_\nu$. To be consistent with  Planck
data at $1\sigma$ we require  $N_{\rm eff} < 3.68$. We take $g(T_{\nu_L}^{\rm dec}) = 10.75$ reflecting
$(e_L^-+e_R^+ + e_R^- + e_L^+ \nu_{eL} + \bar \nu_{eR} + \nu_{\mu L} +
\bar \nu_{\mu R} + \nu_{\tau L} + \bar \nu_{\tau R} + \gamma_L +
\gamma_R)$. From (\ref{7}) the allowable range is $g(T_{\nu_R}^{\rm dec})
> 33$. This is achieved for  $r(T_{\nu_R}^{\rm dec}) > 0.29$. Using (\ref{soverT})
this can be translated into a decoupling temperature: $T_{\nu_R}^{\rm
  dec} > 185~{\rm MeV}$.\\

8.7~At a given time, the rate of decrease in the BH mass is just
the total power radiated
\begin{equation}
\frac{d\dot{M}_{\rm BH}}{dQ} = - \sum_{i} g_i\, \frac{\sigma_s}{8 \,\pi^2}\,\,Q^3 \left[
\exp \left( \frac{Q}{T_{\rm BH}} \right) - (-1)^{2s} \right]^{-1}\,\, .
\label{rate2}
\end{equation}
Integration of (\ref{rate2}) leads to
\begin{equation}
\dot{M}_{\rm BH} = - \sum_i g_i \,\,{\cal B}_\pm\,\, \frac{\Gamma_s}{8\,\pi^2} \,\, \,\Gamma(4) \,\,
\zeta(4)\, \,T^4_{\rm BH}\,A_{4 \subset 4 +n} \, .
\label{m}
\end{equation}
The net change of the BH mass is therefore
\begin{equation}
\frac{dM_{\rm BH}}{dt} = \left. \frac{dM_{\rm BH}}{dt}\right|_{\rm
  accr} + \left. \frac{dM_{\rm BH}}{dt}\right|_{\rm evap} \,\, .
\end{equation}
Substituting $\mbh \sim \sqrt{\hat s}$ into (\ref{m}), where
$\sqrt{\hat s}$ is the center-of-mass energy of the constituents of
the protons (quarks and gluons), a rather lengthy but
straightforward calculation shows that $dM/dt > 0 \Leftrightarrow
\epsilon > 10^{10}~{\rm GeV/fm}^3$. Note that the energy desnity of
partonic matter produce at the LHC is more than 7 orders of magnitude
smaller. {\it (ii)}~Since the ratio of
degrees of freedom for gauge bosons, quarks and leptons is 29:72:18
(the Higgs boson is not included), from (\ref{m}) we obtain a rough
estimate of the mean lifetime,
\begin{equation}
\tau_{_{\rm BH}} \approx  1.67 \times 10^{-27}
\left(\frac{\mbh}{M_*}\right)^{9/7} \left(\frac{\rm TeV}{M_*}\right)
\,{\rm s}\, .
\label{lifetime}
\end{equation}
then ({\ref{lifetime}) indicates that black holes that could
  be produced at the LHC would evaporate instantaneously into visible
  quanta. For further thoughts on this 
subject~\cite{Giddings:2008gr,Giddings:2008pi}. \\

9.1~{\it (i)} For a steady state, $\partial n/\partial t =0$ and
\begin{equation}
\frac{\partial^2 n}{\partial z^2} = - \frac{Q_0}{D} \delta (z) \, .
\end{equation}
Integration yields
\begin{equation}
\frac{\partial n}{\partial z} = A - \frac{Q_0}{D} \Theta (z) \,,
\end{equation}
where $A$ is an integration constant and $\Theta(z)$ the Heaviside
step function (see Appendix~\ref{appE}). A second integration leads to
\begin{equation}
n(z) = B + A z - \frac{Q_0}{D} z \Theta (z) \,,
\end{equation}
where $B$ is an integration constant. From $n(-H) = 0$ we can conclude
that $B = AH$ and so 
\begin{equation}
n(z) = AH + Az - \frac{Q_0}{D} z \Theta (z) \, . 
\end{equation}
On the other hand, $n(+H) = 0$ yields
\begin{equation}
2 AH - \frac{Q_0}{D} H = 0
\end{equation}
or
\begin{equation}
A = \frac{1}{2} \frac{Q_0}{D} \, .
\end{equation}
Then the particle density in the range $-H \leq z \leq H$ is 
\begin{equation}
n(z) = \frac{1}{2} \frac{Q_0}{D} (H+z) - \frac{Q_0}{D} z \Theta (z) \,,
\end{equation}
which can be rewritten as
\begin{equation}
n(z) = \frac{1}{2} \frac{Q_0}{D} (H - |z|)
\end{equation}
{\it (ii)} The column density is
\begin{equation}
N = \int_{-H}^{+H} n(z) \, dz = 2 \int_0^H \frac{1}{2} \frac{Q_0}{D}
(H -z) dz = \frac{Q_0 H^2}{2D} \, .
\end{equation}
Using $N = Q_0 \tau_{\rm res}$ we have
\begin{equation}
\tau_{\rm res} = \frac{H^2}{2D} \Rightarrow D = \frac{H^2}{2 \tau_{\rm res}}  \, .
\end{equation}
Using $D \beta c \lambda/3$ the mean free path is
\begin{equation}
\lambda = \frac{3}{2} \frac{H^2}{\beta c \tau_{\rm res}} \, .
\end{equation}
For $H = 500~{\rm pc}$, $\tau_{\rm res} = 10^7~{\rm yr}$ and $\beta \sim 1$ the
mean free path is about 0.1~pc. \\

9.2~{\it (i)}~The equation of motion is
\begin{equation}
\frac{d \vec p}{dt} = \vec F = \frac{d}{dt} (\gamma m \vec v) = Ze
(\vec v \times \vec B)
\end{equation}
where $e$ is the elementary charge and $Z$ is the charge number. The
acceleration in a magnetic field is always perpendicular to the
velocity, $\vec v \perp \dot{\vec v} = \vec a$ and hence $\dot \gamma= 0$. Therefore
\begin{equation}
\gamma m \dot{\vec v} = Ze (\vec v \times \vec B) \, .
\end{equation}
For $\vec v \perp \vec B$ we can write down the component-wise
differential equations which read as
\begin{equation}
\dot v_x = \frac{Ze}{\gamma m} v_y B \quad \quad {\rm and} \quad \quad
\dot v_y = - \frac{Z e}{\gamma m} v_x B \, .
\end{equation}
The solution is
\begin{equation}
v_x = v \sin \left( \frac{Z e B}{\gamma m } \, t \right) \quad \quad {\rm
  and} \quad \quad v_y = v \cos \left( \frac{Z eB}{\gamma m} \, t
\right) \, ,
\end{equation}
which leads to 
\begin{equation}
x = - \frac{v \gamma m}{ZeB} \cos \left(\frac{Z e B}{\gamma m} \, t
\right)  \quad \quad {\rm and} \quad \quad y =  \frac{v \gamma m}{ZeB} \sin \left(\frac{Z e B}{\gamma m} \, t
\right) \, .
\end{equation}
The radius is therefore
\begin{equation}
R = \sqrt{x^2 + y^2} = \frac{v \gamma m}{ZeB} \approx \frac{c \gamma
  m}{ZeB} \,,
\end{equation}
where in the last step we set $v\approx c$. {\it (ii)}~For a given radius $R$, the magnetic field strength can thus be expressed as
\begin{equation}
B = \frac{c \gamma m}{ZeR} \, .
\end{equation}
For $R \simeq 27~{\rm km}/ (2 \pi)$ and $c \gamma m_p = E/c$ we can
calculate the average magnetic field at the LHC, $B_{\rm LHC} \approx
5.43~{\rm T}$. Note that in reality the particles in a collider are
not in a uniform magnetic field, but the collider ring is composed of
alternating sections for bending, accelerating and focussing the
particles. Therefore the actual magnetic field strengths needed are
slightly larger than the ones calculated above. At the LHC, the
magnets produce a field of 8.7~Tesla. {\it (iii)}~Useful formulae for the radius
of a particle in a magnetic field can be be obtained by introducing $E
= \gamma m c^2$ and evaluating the numerical constants, which gives
the rule of thumb for particle physics detectors
\begin{equation}
R = 3.3~{\rm m} \, \frac{E/(\rm GeV)}{Z(B/T)} \,,
\end{equation}
and the rule of thumb for cosmic ray acceleration (sometimes called
the {\it Hillas criterion}~\cite{Hillas:1985is})
\begin{equation}
R = 1.1~{\rm kpc} \, \frac{(E/EeV)}{Z(B/\mu {\rm G})}
\end{equation}
The radius of a collider, with the average magnetic field of the
LHC, that is expected to launch particles to $10^{11}~{\rm GeV}$ would
be $6 \times 10^{10}~{\rm m}$. This radius is comparable to the
Sun-Mercury distance, which is $5.76 \times 10^{10}~{\rm m}$ (see
exercise 2.3). Hence, such a collider would be priceless! {\it
  (iv)}~The maximum attainable energies in the given astrophysical
objects are: for neutron stars, $E_{\rm max}^p \sim 10^{11}~{\rm GeV}$ and  $E_{\rm max}^{^{56}{\rm
    Fe}} \sim 2.6 \times 10^{12}~{\rm GeV}$; for AGN jets, $E_{\rm max}^p \sim 10^{10}~{\rm GeV}$ and $E_{\rm max}^{^{56}{\rm
    Fe}} \sim 2.6 \times 10^{11}~{\rm GeV}$; for supernova remnants
$E_{\rm max}^p \sim 10^{7}~{\rm GeV}$ and $E_{\rm max}^{^{56}{\rm
    Fe}} \sim 2.6 \times 10^{8}~{\rm GeV}$.\\

9.3~{\it (i)}~We are told that the energy emitted by the supernova in visible
light is equal to that emitted by the Sun in $10^{10}$~yr. We can look
up the luminosity of the Sun (energy emitted per second), and simply
multiply by the $10^{10}$~yr,
\begin{equation}
{\rm Total \ energy \ emitted \ in \ visible \ light} = 4 \times 10^{26}~{\rm J/s}
\times 10^{10}~{\rm yr} \times \frac{3 \times10^{7}~{\rm s}}{1~{\rm
    yr}} \approx 10^{44}~{\rm J} \, .
\end{equation} 
The energy associated with the neutrinos is 100 times larger still
than that, namely $10^{46}~{\rm J}$. {\it (ii)}~If each neutrino has
an energy of $\langle E_\nu \rangle \sim 1.5 \times 10^{-12}~{\rm J}$, the total number of
neutrinos emitted by the star is $\sim 7 \times 10^{57}$. {\it
  (iii)}~These neutrinos are emitted essentially all at once, and
thereafter, travelling at the speed of light, they expand into a huge
spherical shell of ever-increasing radius. Thus, by the time they
impinge on the Earth, they are spread out over a spherical shell of
radius $150,000~{\rm ly}$. The number density on the shell is:
\begin{equation}
\frac{\rm Number}{\rm Surface \ Area} = \frac{7 \times 10^{57}~{\rm
    neutrinos}} 4 \pi (1.5 \times 10^{5}~{\rm ly} \times 10^{16}~{\rm
  m/ly} \approx 2.5 \times 10^{14}~{\rm neutrinos/m}^2 \, .
\end{equation}
That is, every square meter on the Earth's surface was peppered with
250 trillion neutrinos from the supernova. The detector has 2.14~kton
of water ($N_{\rm target} \sim 1.28
\times 10^{33}$ free target nucleons) and so using the average cross section
for weak interactions we have
\begin{equation}
\frac{1}{3} \ \frac{\rm Number}{\rm Surface \ Area} \ \sigma_{\rm weak} \ N_{\rm target}
= \frac{1}{3} \cdot 2.5 \times 10^{14} \cdot 5 \times 10^{-48}  \left(\frac{E_\nu}{{\rm
      MeV}} \right)^2 1.28 \times 10^{33} \sim 46~{\rm electron \
  neutrinos} \, .
\end{equation}
When the discovery of the supernova was first announced, Bahcall, Dar,
and Piran (DBP) immediately realized the possibility that Kamiokande
could have detected the neutrinos from it. They locked themselves in
their office, took the phone off the hook, did essentially the
calculation that you have just done, and sent a paper off tol Nature,
all within 24 hours. They wanted to make a prediction about the
neutrinos, untainted by any news that the neutrinos actually were
found. Indeed, a few days later, the news of IMB~\cite{Bionta:1987qt}
and Kamiokande~\cite{Hirata:1987hu} detection came out. The two
detectors in deep mines recorded a total of 19 neutrino interactions
over a span of 13 seconds. The BDP paper was published on 1987 March
12 (the supernova itself went off on February 23), and has the
following understated but triumphal final sentence: ``Note added in
proof: Since this paper was received on 2 March, the neutrino burst
was found by the Kamiokande experimental group, with properties
generally consistent with the calculated
expectations''~\cite{Bahcall:1987fz}. Making a rough correction for the 60\%
efficiency reduces the expected number of events to 28, within about a factor of
2 of the actual detection.\\

9.4~Substituting $M_\star$ and $M_{\rm BH}$ in (\ref{PGWs}) we obtain $P = 2.4 \times 10^{47}$~W.
At this emission rate the neutron star will fall into the black hole in 
$t \approx 2.9$~s.

\newpage

\twocolumngrid

\appendix

\section{Properties of the ellipse}

An ellipse is the set of all points for which the sum of the distances from two fixed points (foci) is constant, see Fig.~\ref{fig:ellipse}. Additionally to its major axis
$a$, an ellipse is characterized either by its minor axis $b$ or its
eccentricity $e$. The latter two quantities can be connected by
considering the two points at the end of the minor axis $b$, for which
$r=r' =a$ and $r^2 =b^2+(ae)^2$ or
\begin{equation}
b^2 = a^2 (1-e^2) \, .
\label{elipse-rel}
\end{equation}
Any point of an ellipse can be specified by the distance $r$ to one of
its focal points and an angle $\vartheta$ that is measured
counter-clockwise beginning from the major axis. From 
Fig.~\ref{fig:ellipse}, using $\cos (\pi - \vartheta) = - \cos \vartheta$,
we have
\begin{equation}
{r'}^{\,2} = r^2 + (ae + a)^2 + 2 r (2ae+a) \cos \vartheta \, .
\end{equation}
Eliminating $r'$ with the help of $r + r' = 2a$ and solving for $r$ we
obtain
\begin{equation}
r = \frac{a(1-e^2)}{1 + e \cos \vartheta} \, .
\label{elipse-eq}
\end{equation}

\begin{figure}
 \postscript{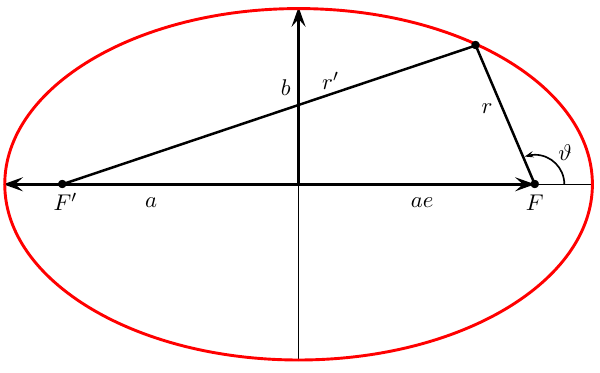}{0.95}
 \caption{An ellipse is defined by the condition $r + r' = 2a$, which describes the
set of points with a constant sum $2a$ of the distances $r$ and $r'$
to the two focal points $F$ and $F'$~\cite{Kachelriess}.}
\label{fig:ellipse}
\end{figure}

\section{Geometry of radiation}
\label{appA}

The solid angle $\Omega$ is the two dimensional analog of the
conventional one dimensional angle $\vartheta$. Just as the angle
$\vartheta$ is defined as the distance along a circle divided by the
radius of that circle, so the solid angle $\Omega$ is analogously
defined as the area on the surface of a sphere divided by the radius
squared of that sphere. The units for $\vartheta$􏰂 and $\Omega$ are
radians (r) and steradians (sr), respectively; although it should be noted
that both of these measures of angle have no actual dimensions. Since
the total surface area of a sphere of radius $R$ is $4\pi R^2$, the
total solid angle in one sphere is $4 \pi~{\rm sr}$.

Consider  a differential area $dA$ on the surface of a sphere, in the
form of a thin ring centered about the symmetry axis. This ring can
be thought of as the intersection of the spherical surface with two
cones, one of half-angle $\vartheta$, and other of half-angle $\vartheta + d
\vartheta$. The width
of this ring is $R d\vartheta$, and the radius of the ring is $R \sin \vartheta$. The
differential solid angle is then
\begin{equation}
d\Omega = \frac{dA}{R^2} = \frac{(2 \pi R \sin \vartheta) (R d \vartheta)}{R^2} = 2 \pi
\sin \vartheta d \vartheta.
\end{equation}
The solid angle inside a cone of half-angle $\vartheta_c$ can be
determined by integrating 
\begin{eqnarray}
\Omega & = & \int d \Omega = \left. \int_0^\pi 2 \pi \sin \vartheta \, d
\vartheta = - 2 \pi \cos \vartheta \right|^{\vartheta_c}_0 \nonumber \\ & = & 2 \pi (1 - \cos
\vartheta_c) \, .
\end{eqnarray}
It is often of interest to consider the small-angle approximation,
where $\vartheta_c \ll 1$. In this limit, $\cos \vartheta_c \approx 1 -
\vartheta_c^2/2$. Therefore, the solid angle of a cone with small half-angle
$\vartheta_c$ is $\Omega \approx \pi \theta_c^2$.

\section{Conservation of mass and  momentum}
\label{appB}

Consider a fluid with local density $\rho(t, x, y, x)$ and local
velocity $\vec u(t, x, y, z)$. Consider a control volume $V$ (not necessarily
small, not necessarily rectangular) which has boundary $S$. The total
mass in this volume is
\begin{equation}
M = \int \rho dV \, .
\label{eflow1}
\end{equation}
The rate-of-change of this mass is just
\begin{equation}
\frac{\partial M}{\partial t} = \int \frac{\partial \rho}{\partial t}
\, dV \, .
\label{eflow2}
\end{equation}
The only way such change can occur is by stuff flowing across the
boundary, so
\begin{equation}
\frac{\partial M}{\partial t} = \int \rho \vec u \cdot d \vec S \, .
\label{eflow3}
\end{equation}
We can change the surface integral into a volume integral using Green's theorem, to obtain
\begin{equation}
\frac{\partial M}{\partial t} = - \int \vec \nabla \cdot (\rho \vec u)
dV \, . 
\label{eflow4}
\end{equation}
Note that (\ref{eflow2}) and (\ref{eflow4}) must be equal no matter what volume $V$ we choose, so the
integrals must be pointwise equal. This gives us an expression for the
local conservation of mass
\begin{equation}
\frac{\partial \rho}{\partial t} + \vec \nabla \cdot ( \rho \vec u) =
0 \, ,
\label{eflow5}
\end{equation} 
which is sometimes called {\it continuity equation}.

We can go through the same process for momentum instead of mass. We
use $\Pi$ to represent momentum, to avoid conflict with $P$ which represents
pressure. The total momentum in the control volume is:
\begin{equation}
\Pi_i = \int \rho  \ u_i \ dV \,,
\label{eflow6}
\end{equation}
where the index $i$ runs over the three components of the momentum. The rate-of-change
thereof is just
\begin{equation}
\frac{\partial \Pi_i}{\partial t}  = \int \frac{\partial (\rho
  u_i)}{\partial t} \, dV \, .
\label{eflow7}
\end{equation}
We (temporarily) assume that there are no applied forces (i.e. no gravity)
and no pressure (e.g. a fluid of non-interacting dust particles). We
also assume viscous forces are negligible. Then, the only way a
momentum-change can occur is by momentum flowing across the boundary,
\begin{equation}
\frac{\partial \Pi_i}{\partial t} = \int (\rho u_i) \vec u \cdot dS =
\int (\rho u_i u^j d_j S) \, .
\label{eflow8}
\end{equation}
We are expressing dot products using the Einstein summation
convention, i.e. implied summation over repeated dummy indices, such
as $j$ in the previous expression. We can change the surface integral
into a volume integral using Green's theorem, to obtain
\begin{equation}
\frac{\partial \Pi_i}{\partial t} = - \int \nabla_j (\rho u_i u^j) dV
\, .
\label{eflow9}
\end{equation}
Note that (\ref{eflow7}) and (\ref{eflow9}) must be equal no matter what volume $V$ we choose, so
the integrands must be pointwise equal. This gives us an expression
for the local conservation of momentum,
\begin{equation}
\frac{\partial \Pi_i}{\partial t} = \frac{\partial (\rho u_i)}{\partial
  t} = - \nabla_j (\rho u_i u^j) \, .
\label{eflow10}
\end{equation}
We can understand this equation as follows: each component of the
momentum-density $\rho u_i$ (for each $i$ separately) obeys a local
conservation law. There are strong parallels between (\ref{eflow5}) and (\ref{eflow10}).  Note
that the $\nabla_j$ operator on the right-hand-side is differentiating
two velocities ($u_i$ and $u_j$) only one of which undergoes dot-product
summation (namely summation over $j$). Using vector component notation
(such as $\nabla_j u^j$) is a bit less elegant than using pure vector notation
(such as $\vec \nabla \cdot \vec u$) but in this case it makes things clearer.

We now consider the effect of pressure. It contributes a force on the
particles in the control volume, namely
\begin{equation}
F_i = \int P d_i S = - \int \nabla_i P dV \, .
\label{eflow11}
\end{equation}
A uniform gravitational field contributes another force, namely
\begin{equation}
F_i = \int \rho g_i dV \, .
\label{eflow12}
\end{equation}
These forces contribute to changing the momentum, by the second law of
motion:
\begin{equation}
\frac{\Pi'_i}{dt} = F_i \, .
\label{eflow13}
\end{equation}
Note the tricky notation: we write $d/dt$ rather than
$\partial/\partial t$, and $\Pi'$ rather than $\Pi$, to remind
ourselves that the three laws of motion apply to particles, not to the
control volume itself. The rate-of-change of $\Pi$, the momentum in the
control volume, contains the Newtonian contributions, (\ref{eflow11}) and
(\ref{eflow12}) via (\ref{eflow13}), plus the flow contributions (\ref{eflow10}).

Combining all the contributions we obtain the main result, Euler's
equation of motion:
\begin{equation}
\frac{\partial (\rho u_i)}{\partial t} + \nabla_j (\rho u_iu^j) = -
\nabla_i P + \rho g_i \, .
\label{eflow14}
\end{equation}

One sometimes encounters other ways of expressing the same equation of
motion. Rather than emphasizing the momentum, we might want to
emphasize the velocity. This is not a conserved quantity, but
sometimes it is easier to visualize and/or easier to measure. If we
expand the left-hand-side we have
\begin{equation}
\rho \frac{\partial u_i}{\partial t} + u_i \frac{\partial
  \rho}{\partial t} + u_i \nabla_j (\rho u^j) + \rho u_j \nabla_j u_i
= \rho g_i - \nabla_i P,
\label{eflow15}
\end{equation}
where the second and third terms cancel because of conservation of
mass (\ref{eflow5}), leaving us with
\begin{equation}
\rho \frac{\partial u_i}{\partial t}  + \rho u_j \nabla_j u_i
= - \nabla_i P + \rho g_i \, .
\label{eflow16}
\end{equation}
Converting from component notation to vector notation, we obtain
\begin{equation}
\rho \frac{\partial \vec u}{\partial t} + \rho \left( \vec u \cdot \vec \nabla \right)
  \vec u = -  \vec \nabla P + \rho \vec g \, .
\label{eflow17}
\end{equation}
If we now consider a plane-parallel ($\partial/\partial y = 0$,
$\partial/\partial z =0$, $\partial/\partial x = d/dx$)
steady-state
($\partial /\partial t =0$) flow and we ignore gravity, (\ref{eflow5}) and (\ref{eflow17}) become
\begin{equation}
\frac{d}{dx} (\rho u ) = 0 \,,
\label{eu3}
\end{equation}
and 
\begin{equation}
u \frac{du}{dx} = - \frac{1}{\rho} \frac{dP}{dx} \, ;
\label{eu4}
\end{equation}
respectively. (\ref{eu3}) immediately gives
\begin{equation}
\rho u ={\rm constant} \rightarrow \rho_1 u_1 = \rho_2 u_2 \, .
\end{equation}
Using
\begin{eqnarray}
\frac{d}{dx} (\rho u^2) & = & 2 \rho u \frac{du}{dx} + u^2
\frac{d\rho}{dx} \nonumber \\
& = & \rho u \frac{du}{dx} + u \left(\rho \frac{du}{dx} + u
\frac{d\rho}{dx} \right) \nonumber \\
 & = & \rho u \frac{du}{dx} + u \frac{d}{dx} (\rho u) \nonumber \\
 & = & \rho u \frac{du}{dx}
 \end{eqnarray}
(\ref{eu4}) can be rewritten as
\begin{equation}
\rho u \frac{du}{dx}  + \frac{dP}{dx} = \frac{d}{dx} (\rho u^2 + P) =
0 \, .
\end{equation}
This leads to
\begin{equation}
\rho u^2 + P = {\rm constant} \rightarrow \rho_1 u_1^2 + P_1 = \rho_2
u_2^2 + P_2 \,.
\end{equation}

\section{Kruskal coordinates} 
\label{appC}

One elegant coordinate substitution is the replacement of $r$ and $t$
by the Kruskal coordinates $x$ and $y$, which are defined by
the following two equations~\cite{Kruskal:1959vx}
\begin{equation}
xy = \left(\frac{r}{2M} - 1\right) e^{r/(2M)} 
\label{kruskal1}
\end{equation}
and 
\begin{equation}
x/y = e^{t/(2M)} \, .
\label{kruskal2}
\end{equation}
The angular coordinates $\theta$ and $\phi$ are kept the same.
Hereafter, we adopt geometrodynamic units $G = c =1$.
By taking the $\ln$ of (\ref{kruskal1}) and
(\ref{kruskal2}), and partially differentiating with respect to $x$ and $y$, we read off
\begin{equation}
\frac{dx}{x} + \frac{dy}{y} = \frac{dr}{r - 2M} + \frac{dr}{2M} =
\frac{dr}{2M (1 - 2M/r)} 
\end{equation}
and
\begin{equation}
\frac{dx}{x} - \frac{dy}{y} = \frac{dt}{2M} \, .
\end{equation}
The Schwarzschild metric (\ref{Sch-metric}) is now given by
\begin{eqnarray}
ds^2 & = & - 16 M^2 \left( 1 - \frac{2M}{r} \right) \frac{dx}{x} \frac{dy}{y}  -
  r^2 d \Omega^2 \nonumber \\
 & = & - \frac{32 M^3}{r} e^{-r/(2M)} \, dx \, dy - r^2 d\Omega^2 \, .
\end{eqnarray}
Note that, in the last expression, the zero and the pole at $r = 2M$
have cancelled out. The function $r(x, y)$ can be obtained by
inverting the algebraic expression (\ref{kruskal1}) and is regular in the entire
region $x y > -1$. In particular, nothing special seems to happen on
the two lines $x = 0$ and $y = 0$. Apparently, there is no physical
singularity or curvature singularity at $r \to 2M$. We do notice that the
line $x = 0$, $\theta$ and $\phi$ both constant, is lightlike, since two neighboring
points on that line obey $dx = d\theta = d\phi = 0$, and this implies that $ds^2 =
0$, regardless the value of $dy$. Likewise, the line $y = 0$ is
lightlike. Indeed, we can also read off from the original expression
(\ref{Sch-metric}) that if $r = 2M$, the lines with constant $\theta$ and $\phi$ are lightlike,
as $ds^2 = 0$ regardless the value of $dt$. The line $y = 0$ is called the
future horizon and the line $x = 0$ is the past horizon.

\begin{figure}
 \postscript{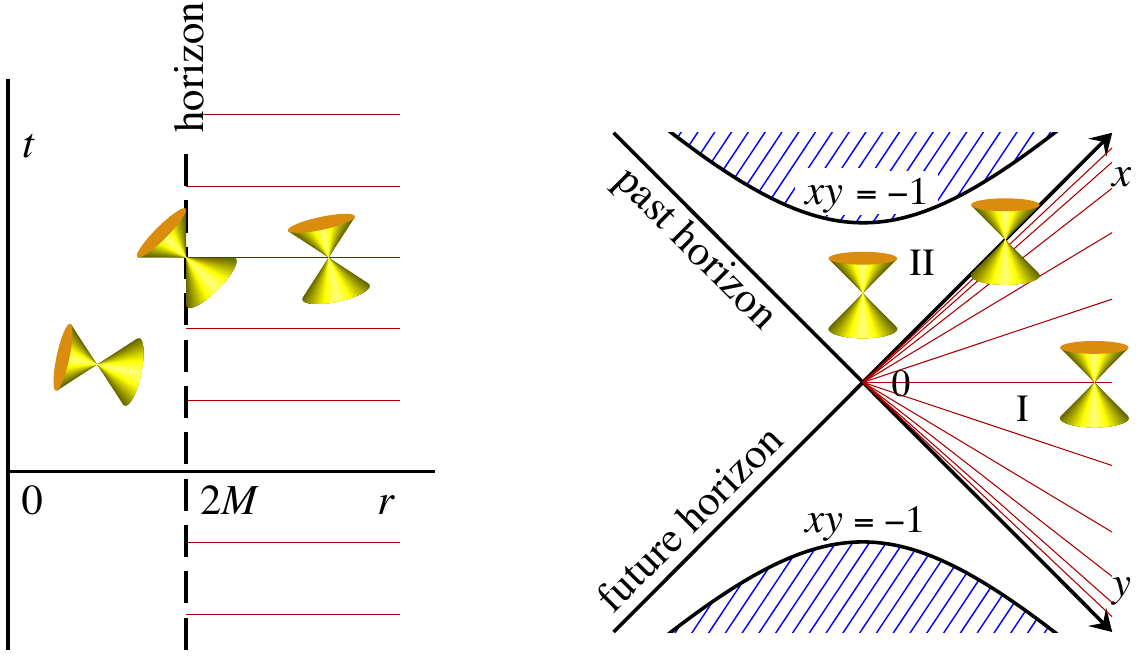}{0.95}
 \caption{{\bf Left.}The black hole in the Schwarzschild coordinates
   $(t,r)$. The event horizon is at $r = 2M$. {\bf Right.} Kruskal
   coordinates. Here, the coordinates of the horizon are at $x = 0$
   and at $y = 0$. The orientation of the local lightcones is
   indicated. Thin red lines are the time = constant lines in the
   physical part of spacetime~\cite{tHooft}.}
\label{fig:kruskal}
\end{figure}

An other important point to highlight is that (\ref{kruskal2})
attaches a real value for the time $t$ when $x$ and $y$ both have the
same sign, such as is the case in the region marked I in Fig.~\ref{fig:kruskal}, but
if $xy < 0$, as in region II,  the coordinate $t$ gets an imaginary
part. This means that region II is not part of our universe. Actually,
$t$ does not serve as a time coordinate there, but as a space
coordinate, since there, $dt^2$ enters with a negative sign in the metric
(\ref{Sch-metric}). $r$ is then the time coordinate.  

Even if we restrict ourselves to the regions where $t$ is real, we
find that, in general, every point $(r, t)$ in the physical region of
spacetime is mapped onto two points in the $(x, y)$ plane: the points
$(x, y)$ and $(-x, -y)$ are mapped onto the same point $(r, r)$. This
leads to the picture of a black hole being a ``wormhole'' connecting our
universe to another universe, or perhaps another region of the
spacetime of our universe~\cite{Einstein:1935tc}. However, there are no timelike or light
like paths connecting these two universes~\cite{Fuller:1962zza}. If this is a wormhole at
all, it is a purely spacelike one.

\section{Geometry of $\bm{S^3}$ and $\bm{H^3}$}
\label{appD}

Herein we provide a geometric interpretation of the hyper-sphere $S^3$
and the hyperbolic hyper-plane $H^3$. The explanation given herein
will build upon the content of the exquisite book by Kolb and Turner~\cite{Kolb:1990vq}.

We begin by studing the familiar two dimensional surfaces.
To visualize the two sphere it is convenient to introduce an extra
fictitious spatial dimension and to embed this two-dimensional curve
space in a three-dimensional Euclidean space with cartesian
coordinates $x_1, x_2, x_3$. The equation of the two sphere $S^2$ of radius
$R$ is
\begin{equation}
x_1^2 + x_2^2 + x_3 ^2 = R^2 \, .
\label{KT1}
\end{equation}
The line element in the three-dimensional Euclidean space is
\begin{equation}
ds^2 = dx_1^2 + dx_2^2 + dx_3^2 \, .
\label{Euclidean3}
\end{equation}
If $x_3$ is taken as the fictitious third spatial coordinate, it can
be eliminated from $ds^2$ by the use of (\ref{KT1})
\begin{equation}
ds^2 = dx_1^2 + dx_2^2 + \frac{(x_1 dx_1 + x_2 dx_2)^2}{R^2 - x_1^2 -
  x_2^2} \, .
\label{KT3}
\end{equation}
Now, introduce the coordinates $\varrho$ and $\theta$ defined in terms of $x_1$ and
$x_2$ by 
\begin{equation}
x_1 = \varrho \cos \theta \quad {\rm and} \quad x_2 = \varrho \sin
\theta \, .
\end{equation}
Physically, $\varrho$ and $\theta$ correspond to polar coordinates
in the $x_3$-plane; $x_3^2 = R^2 - \varrho^2$. In terms of the new
coordinates, (\ref{KT3}) becomes
\begin{equation}
ds^2 = \frac{R^2 d\varrho^2}{R^2 - \varrho^2} + \varrho^2 d \theta^2
\, .
\label{KT12}
\end{equation}
Note the similarity between this metric and the spatial hypersurface
with $k=1$ in (\ref{FRWmetric}).

Another convenient coordinate system for the two sphere is that
specified by the usual polar and azimuthal angles ($\theta, \phi$) of
spherical coordinates, related to $x_i$ by 
\begin{equation}
x_1 = R \sin \theta \cos \phi, \quad x_2 = R \sin \theta \sin \phi,
\quad x_3 = R \cos \theta \, . 
\end{equation}
In terms of these coordinates, (\ref{Euclidean3}) becomes
\begin{equation}
ds^2 = R^2 \left[ d\theta^2 + \sin^2 \theta d\phi \right] \, .
\label{KT14}
\end{equation}
 This form makes manifest the fact that the space is the two sphere of
 radius $R$.

The equivalent formulas for a space of constant negative curvature can
be obtained with the replacement $R \to iR$ in (\ref{KT1}). The metric
corresponding to the form of (\ref{KT12}) for the negative curvature case is
\begin{equation}
ds^2 = \frac{R^2 d\varrho^2}{R^2 + \varrho^2} + \varrho^2 d \theta^2
\, .
\end{equation}
and the metric in the form corresponding to (\ref{KT14}) is
\begin{equation}
ds^2 = R^2 \left[ d \theta + \sinh^2 \theta d \phi^2 \right] \, .
\end{equation}
The embedding of the hyperbolic plane $H^2$ in an Euclidean space
requires three fictitious extra dimensions, and such an embedding is
of little use in visualizing the geometry. While $H^2$
cannot be globally embedded in $\mathbb{R}^3$, it can be partailly
represented by the pseudosphere (\ref{pseudosphere}).

The generalization of the two-dimensional models discussed above to
three spatial dimensions is trivial. For the three sphere $S^3$ a fictitious
fourth spatial dimension is introduced and in cartesian coordinates
the three sphere is defined by $R^2 = x_1^2 + x_2^2 + x_3^2 +
x_4^2$. The spatial metric of a four-dimeniosnal Euclidean space is $ds^2
= dx_1^2 + dx_2^2 + dx_3^2 + dx_4^2$. The fictitious coordinate can be
removed to give
\begin{eqnarray}
ds^2 &= & dx_1^2 + dx_2^2 + dx_3^2 \nonumber \\
&+ & \frac{(x_1 dx_1 + x_2 dx_2 + x_3
  dx_3)^2}{R^2 - x_1^2 - x_2^2 - x_3^2 } \, .
\end{eqnarray}
In terms of coordinates $x_1 = \varrho \sin \theta \cos \phi$, $x_2 =
\varrho \sin \theta \sin \phi$, $x_3 = \varrho \cos \theta$, the
metric is given by the spatial part of (\ref{FRWmetric}) with $k=1$.  In terms of a
coordinate system that employs the 3 angular coordinates $(\chi,
\theta, \phi)$ of a four-dimensional spherical coordinate system, $x_1
= R \sin \chi \sin \theta \cos \phi$, $x_2 = R \sin \chi \sin \theta
\sin \phi$, $x_3 = R \sin \chi \cos \theta$, $x_4 = R \cos \chi$, the
metric is given by
\begin{equation}
ds^2 = R^2 \left[ d\chi^2 + \sin^2 \chi (d \theta^2 + \sin^2 \theta
  d\phi^2) \right] \, .
\end{equation}
The substitution $\chi = r/R$ leads to the spatial part of (\ref{FRWsmetric})
with $k=1$.

As in the two-dimensional example, the three-dimensional open model is
obtained by the replacement $R \to iR$, which gives the metric in the
form (\ref{FRWmetric}) with $k=-1$, or in the form (\ref{FRWsmetric}) with $\sin \chi \to \sinh
\chi$. Again the space is unbounded and $R$ sets the curvature
scale. Embedding $H^3$ in an Euclidean space requires four fictitious
extra dimensions.

\section{Dirac Delta Function}
\label{appE}

Dirac's delta function is defined by the following property
\begin{equation}
\delta (t) = \left\{ \begin{array}{cc} 0 & t\neq 0 \\
\infty & t = 0 \end{array} \right. \,,
\end{equation}
with
\begin{equation}
\int_{t_1}^{t_2} dt \, \delta (t) = 1 
\end{equation}
if $0 \in [t_1,t_2]$ (and zero otherwise). It is ``infinitely peaked'' at $t = 0$, with the
total area of unity. You can view this function as a limit of Gaussian
\begin{equation}
\delta (t) = \lim_{\sigma \to 0} \frac{1}{\sqrt{2 \pi} \ \sigma}
e^{-t^2/(2 \sigma^2)} \,,
\end{equation}
or a Lorentzian
\begin{equation}
\delta (t) = \lim_{\epsilon \to 0} \frac{1}{\pi} \frac{\epsilon}{t^2 +
  \epsilon^2} \, .
\end{equation}

The important property of the delta function is the following relation
\begin{equation}
\int dt \ f(t) \ \delta(t) = f(0) \,,
\end{equation}
which is valid for any {\it test function} $f(t)$ that is bounded and
differentiable to any order, and which vanishes outside a finite
range. This is easy to see. First of all, $\delta (t)$ vanishes
everywhere except $t = 0$. Therefore, it does not matter what values
the function $f(t)$ takes except at $t = 0$. You can then say $f (t)
\, \delta (t) = f (0) \, \delta (t)$. Then $f (0)$ can be pulled
outside the integral because it does not depend on $t$, and you obtain
the right-hand-side. This equation can easily be generalized to
\begin{equation}
\int dt \, f(t) \, \delta (t - t_0) = f(t_0) \, .
\end{equation}

Mathematically, the delta function is not a function, because it is
``too singular.'' Instead, it is said to be a ``distribution.'' It is a
generalized idea of functions, but can be used only inside
integrals. In fact, 􏰯 $\int dt \, \delta (t)$ can be regarded as an ``operator'' which
pulls the value of a {\it test function} at zero. Put it this way, it sounds
perfectly legitimate and well-defined. But as long as it is understood
that the delta function is eventually integrated, we can use it as if
it is a function.

The step (Heaviside) function,
\begin{equation}
\Theta (x) = \left\{ \begin{array}{cc} 1 & x \geq 0 \\
0 & x<0 \end{array} \right. \,,
\end{equation}
is the ``primitive'' (at least in symbolyc form) of the $\delta
(x)$. Equivalently, $\Theta'(x)$, has the symbolic limit $\delta
(x)$, as we show next. For any given {\it test function} $f(x)$, integration
by parts leads to
\begin{eqnarray}
\int_{-\infty}^{+\infty} \Theta'(x) f (x) dx & = & -
\int_{-\infty}^{+\infty} \Theta (x) f'(x) dx \nonumber \\
& = & - \int_0^\infty f'(x)
dx = f(0) \ ; 
  \end{eqnarray}
therefore $\Theta'(x) = \delta (x)$.


\begin{thebibliography}{99}

\bibitem{Galileo:1610} G. Galilei, {\it Sidereus Nuncius}, (T. Baglioni,
  Republic of Venice, 1610).


\bibitem{Galileo} G. Galilei, {\it Dialogues concerning two sciences}, (1638). 
 Reprinted on {\it On the Shoulders of Giants: The Great Works of Physics and Astronomy}, (Ed. S. Hawking, Running Press, Philadelphia, 2002) ISBN 0-7624-1348-4; p.399.

\bibitem{Copernicus} N. Copernicus, {\em De revolutionibus orbium coelestium}, (1543). Reprinted on {\it On the Shoulders of Giants: The Great Works of Physics and Astronomy}, (Ed. S. Hawking, Running Press, Philadelphia, 2002) ISBN 0-7624-1348-4; p.7.


\bibitem{Kepler:1619} J. Kepler, {\it Harmonices Mundi}  (Johann
  Planck, Linz, Austria, 1619). Reprinted on {\it On the Shoulders of Giants: The Great Works of Physics and Astronomy}, (Ed. S. Hawking, Running Press, Philadelphia, 2002) ISBN 0-7624-1348-4; p.635.

\bibitem{Newton:1687} I. Newton, Philosophi\ae Naturalis Principia Mathematica, (1687). Reprinted on On the Shoulders of Giants: The Great Works of Physics and Astronomy, (Ed. S. Hawk- ing, Running Press, Philadelphia, 2002) ISBN 0-7624- 1348-4; p.733.


\bibitem{Beringer:1900zz} 
  J.~Beringer {\it et al.} [Particle Data Group Collaboration],
  Phys.\ Rev.\ D {\bf 86}, 010001 (2012).
  doi:10.1103/PhysRevD.86.010001




\bibitem{Wright:1750} T. Wright, {\it An Original Theory of New Hypothesis of the Universe}, (H. Chapelle, London, 1750).


\bibitem{Messier:1781} C. Messier, {\it  Catalogue des N\'ebuleuses \&
    des amas d'\'Etoiles}, 1781. K. G.  Jones, {\it Messier's nebulae
    and star clusters},  (Cambridge University Press, 1991) ISBN 0-521-37079-5.

\bibitem{Kant:1755} I. Kant, {\it Allgemeine Naturgeschichte und Theorie
    des Himmels}, (Germany, 1755).


\bibitem{Hubble:b} E. Hubble, {\em The Realm of Nebulae}, (Yale
  University Press, New Haven, 1936; reprinted by Dover Publications,
  Inc., New York, 1958).





  
\bibitem{Planck:1901tja}   M. Planck,
Verh. d.  deutsch. phys. Ges. {\bf 2}, 202 (1900);
Verh. d. deutsch. phys. Ges. {\bf 2}, 237 (1900);
  Annalen Phys.\  {\bf 4}, 553 (1901).


\bibitem{Rybicki:1979} 
G. B. Rybicki and A. P. Lightman,
{\it Radiative Processes in Astrophysics},
(John Wiley \& Sons, Massachusetts, 1979) ISBN 978-0-471-82759-7.





\bibitem{Anchordoqui:2015uww} 
  L.~A.~Anchordoqui,
  arXiv:1512.04361 [physics.pop-ph].


\bibitem{Stefan:1879} 
J. Stefan, 
Wiener Ber.  {\bf 79}, 391 (1879).

\bibitem{Boltzmann:1884} 
L. Boltzmann, 
Annalen Phys. {\bf 22}, 291 (1884).

\bibitem{Kachelriess} M. Kachelriess, {\it A concise introduction to
    astrophysics}, lectures given at the Institutt for fysikk
NTNU, 2011.



\bibitem{Wien:1894} 
W. Wien, 
Annalen  Phys. {\bf 52}, 132 (1894).



\bibitem{FA} For further details see e.g., H. Karttunen, P. Kr\"oger,
  H. Oja, M. Poutanen, K. J. Donner, {\em Fundamental Astronomy}, (4th
  Edition, Springer-Verlag Berlin Heidelberg New York, 2003).




\bibitem{HR} E. Hertzsprung, Astron. Nachr. {\bf 196}, 201 (1913); 
H. N. Russell, Science {\bf 37}, 651 (1913). 


\bibitem{Shirley} Y. L. Shirley, {\it Fundamentals of Astronomy}
  lectures given at the University of Arizona, 2010.



\bibitem{Doppler} C. Doppler, 
Abh. K\"onigl. B\"ohm. Ges. Wiss. {\bf 2}, 465 (1843).



\bibitem{Weinberg:1977ji} 
  S.~Weinberg,
  {\it The First Three Minutes: A Modern View of the Origin of the Universe},
  (BasicBooks, New York, 1993) ISBN 0-465-02437-8.


\bibitem{Fraunhofer} J. Fraunhofer,
{\it Determination of the refractive and color-dispersing power of different types of glass, in relation to the improvement of achromatic telescopes}, Memoirs of the Royal Academy of Sciences in Munich {\bf 5}, 193 (1814-1815); see especially pages 202-205 and the plate following page 226.

\bibitem{Huggins} W. Huggins, 
Philos. Trans. Roy. Soc. London {\bf 158},  529 (1968)  doi:10.1098/rstl.1868.0022




\bibitem{Sarkar} S. Sarkar, {\it Lecture Notes on Special Relativity},
lectures given at  Oxford University, 2002.


\bibitem{Lorentz:1904} H. A. Lorentz,
Proc. R. Neth. Acad. Arts Sci. {\bf 6}, 809 (1904).


\bibitem{Anchordoqui:2015xca} 
  L.~A.~Anchordoqui,
  arXiv:1509.08868 [physics.pop-ph].





\bibitem{Mazeh:2000pa} 
  T.~Mazeh {\it et al.},
  Astrophys.\ J.\  {\bf 532}, L55 (2000)
  doi:10.1086/312558
  [astro-ph/0001284].

\bibitem{Queloz:2000xp} 
  D.~Queloz, A.~Eggenberger, M.~Mayor, C.~Perrier, J.~L.~Beuzit, D.~Naef, J.~P.~Sivan and S.~Udry,
  Astron.\ Astrophys.\  {\bf 359}, L13 (2000)
  [astro-ph/0006213].

\bibitem{Wittenmyer:2005nz} 
  R.~A.~Wittenmyer {\it et al.},
  Astrophys.\ J.\  {\bf 632}, 1157 (2005)
  doi:10.1086/433176
  [astro-ph/0504579].





\bibitem{Henry:2000gr} 
  G.~W.~Henry, G.~W.~Marcy, R.~P.~Butler and S.~S.~Vogt,
  Astrophys.\ J.\  {\bf 529}, L41 (2000).
  doi:10.1086/312458



\bibitem{Charbonneau:1999nk} 
  D.~Charbonneau, T.~M.~Brown, D.~W.~Latham and M.~Mayor,
  Astrophys.\ J.\  {\bf 529}, L45 (2000)
  doi:10.1086/312457
  [astro-ph/9911436].



\bibitem{Bethe:1939bt}
  H.~A.~Bethe,
  Phys.\ Rev.\  {\bf 55}, 434 (1939).



\bibitem{Chandra} S. Chandrasekhar, Mon. Not. Roy. Astron. Soc. {\bf
    95}, 207 (1935).


\bibitem{Longair:1994wu} 
  M.~S.~Longair,
  (Cambridge University Press, UK, 2011)  ISBN 978-0-521-75618-1.




\bibitem{Oppenheimer:1939ne}
  J.~R.~Oppenheimer and G.~M.~Volkoff,
  Phys.\ Rev.\  {\bf 55}, 374 (1939).


\bibitem{Sedov} L. I. Sedov,
J. App. Math. Mech. {\bf 10}, 241 (1946).

\bibitem{Taylor:1950a}
G. I. Taylor
Proc. Roy. Soc. {\bf 201}, 159  (1950)
doi:10.1098/rspa.1950.0049.

\bibitem{Taylor:1950b}  G. I. Taylor,
Proc. Roy. Soc. {\bf 201}, 175  (1950) doi:10.1098/rspa.1950.0050.

\bibitem{Hewish} A. Hewish, S. J. Bell, J. D. H. Pilkington,
P. F. Scott, and R. A. Collins
Nature {\bf 217}, 709 (1968) doi:10.1038/217709a0.



\bibitem{Gold} T. Gold, 
Nature {\bf 218}, 731 (1968) doi:10.1038/218731a0.

\bibitem{Boynton} P. E. Boynton, E. J. Groth III, R. B. Partridge, and
  D. T. Wilkinson, 
Astrophys. J. {\bf 157} L 197 (1969).

\bibitem{Oppenheimer:1939ue}
  J.~R.~Oppenheimer and H.~Snyder,
  Phys.\ Rev.\  {\bf 56}, 455 (1939).


\bibitem{Penrose:1964wq}
  R.~Penrose,
  Phys.\ Rev.\ Lett.\  {\bf 14}, 57 (1965).

\bibitem{Hawking:1965mf}
  S.~Hawking,
  Phys.\ Rev.\ Lett.\  {\bf 15}, 689 (1965).

\bibitem{Hawking:1966sx}
  S.~Hawking,
  Proc.\ Roy.\ Soc.\ Lond.\  A {\bf 294}, 511 (1966).

\bibitem{Hawking:1966jv}
  S.~Hawking,
  Proc.\ Roy.\ Soc.\ Lond.\  A {\bf 295}, 490 (1966);

\bibitem{Hawking:1967ju}
  S.~Hawking,
  Proc.\ Roy.\ Soc.\ Lond.\  A {\bf 300}, 187 (1967).

\bibitem{Hawking:1969sw}
  S.~W.~Hawking and R.~Penrose,
  Proc.\ Roy.\ Soc.\ Lond.\  A {\bf 314}, 529 (1970).




\bibitem{Gauss} K. F. Gauss, {\it General investigations of curved
    surfaces of 1827 and 1825}, (C. S. Robinson \& Co., University
  Press Princeton, N. J., 1902).

\bibitem{Bolyai} J. B\'olyai, {\it Appendix: Explaining the absolute true of
    space}, published as an appendix to the essay by his father
  F. B\'olyai {\it An attempt to introduce  youth to the
  fundamentals of pure science, elementary and advanced, by a clear and
  proper method} (Maros V\'as\'arhely, Transilvania, 1832).

\bibitem{Lobachevsky} N. I Lobachevsky, 
Kasanski Vestnik (Kazan Messenger),
  Feb-Mar, 178 (1829); April, 228 (1829);  Nov-Dec, 227 (1829);
  Mar-Apr, 251 (1830; Jul-Aug 571 (1830).


\bibitem{Bessel} F. W. Bessel; communicated by J. F. W. Herschel
Mon. Not. Roy. Astron. Soc. {\bf 6}, 136 (1844). 



\bibitem{Clark:1863} A. Clark, communicated by T. H. Safford, {\it The observed motions of the  companion of Sirius} (Cambridge: Welch, Bigelow, and Company, MA, 1863). 

\bibitem{Adams:1915} W. S. Adams,
Publications of the Astronomical Society of the Pacific {\bf 27}, 236
(1915)  doi:10.1086/122440.



\bibitem{Pauli} 
W. Pauli, 
Z. Phys. 31:765 (1925). 

\bibitem{Heisenberg:1927zz} 
  W.~Heisenberg,
  Z.\ Phys.\  {\bf 43}, 172 (1927).

\bibitem{Einstein:1905ve} 
  A.~Einstein,
  Annalen Phys.\  {\bf 17}, 891 (1905)
  [Annalen Phys.\  {\bf 14}, 194 (2005)].




\bibitem{Minkowski} H. Minkowski, 
Physikalische Zeitschrift {\bf 10}, 104  (1909).



\bibitem{Schwarzschild:1916uq} 
  K.~Schwarzschild,
  Sitzungsber.\ Preuss.\ Akad.\ Wiss.\ Berlin (Math.\ Phys.\ ) {\bf 1916}, 189 (1916)
  [physics/9905030].


\bibitem{Einstein:1916vd} 
  A.~Einstein,
  Annalen Phys.\  {\bf 49}, 769 (1916)
  [Annalen Phys.\  {\bf 14}, 517 (2005)].
  doi:10.1002/andp.200590044

\bibitem{Kretschmann} E. Kretschmann, Annalen Phys. {\bf 53}, 575  (1917).

\bibitem{Weinberg:1972kfs} 
S. Weinberg, {\em  Gravtitation and
    Cosmology} (John Wiley \& Sons, New York, 1972) ISBN 0-471-92567-5


\bibitem{Misner:1974qy} 
  C.~W.~Misner, K.~S.~Thorne and J.~A.~Wheeler,
  {\it Gravitation,}
  (W. H. Freeman, San Francisco, 1973) ISBN 978-0-7167-0344-0.


\bibitem{Dyson} F. W. Dyson, A. S. Eddington, and C. Davidson,
Phil. Trans.  Roy. Soc. {\bf 220A}, 291 (1920). 


\bibitem{Adams:1982hh} D. Adams, {\it Hitchhiker's Guide to the
    Galaxy: Life, the Universe and Everything}, (Harmony Books, NY,
  1982) ISBN 0-345-39182-9; see chapter 17.

\bibitem{Cameron} A. G. W. Cameron, Nature {\bf 229}, 178 (1971); 

\bibitem{Wilson} R. E. Wilson,
Astrophys. J. {\bf 170}, 529 (1971).

\bibitem{Giacconi} R. Giacconi, P. Gorenstein, H. Gursky, J. R Waters, Astrophys. J. {\bf 148}, L119 (1967)

\bibitem{Oda} M. Oda, P. Gorenstein, H.  Gursky, E. Kellogg, E. Schreier,
H. Tananbaum, R. Giacconi,  Astrophys. J. {\bf 166}, L1 (1971).


\bibitem{Webster} B. L. Webster and P. Murdin, Nature, {\bf 235}, 37 (1972).

\bibitem{Bolton} C. T. Bolton, Nature {\bf 235}, 271 (1972).

\bibitem{Petterson} J. A. Petterson, Astrophys. J. {\bf 224}, 625 (1978).


\bibitem{Stirling:2001xb} 
  A.~M.~Stirling, R.~E.~Spencer, C.~de la Force, M.~A.~Garrett, R.~P.~Fender and R.~N.~Ogley,
  Mon.\ Not.\ Roy.\ Astron.\ Soc.\  {\bf 327}, 1273 (2001)
  doi:10.1046/j.1365-8711.2001.04821.x
  [astro-ph/0107192].




\bibitem{Thomson} J. J. Thomson, {\it Conduction of electricity
    through gases} (Cambridge University Press, Cambridge, 1906).


\bibitem{Eddington} A. S. Eddington, {\it The internal constitution of the stars} (Cambridge University Press, Cambridge, 1926).


\bibitem{Biteau:2013nua} 
  J.~Biteau,
PhD thesis, 2013, pastel-00822242.  


\bibitem{Urry:1995mg} 
  C.~M.~Urry and P.~Padovani,
  Publ.\ Astron.\ Soc.\ Pac.\  {\bf 107}, 803 (1995)
  doi:10.1086/133630
  [astro-ph/9506063].

\bibitem{Dermer:2016jmw} 
  C.~D.~Dermer and B.~Giebels,
  arXiv:1602.06592 [astro-ph.HE].


\bibitem{Piner:2005px} 
  B.~G.~Piner, D.~Bhattarai, P.~G.~Edwards and D.~L.~Jones,
  Astrophys.\ J.\  {\bf 640}, 196 (2006)
  doi:10.1086/500006
  [astro-ph/0511664].



\bibitem{Cheseaux} J. P. L. de Cheseaux, {\em Trait\'e de la Com\`ete}
  (Lausanne, 1774), pp. 223 ff; reprinted in {\em The Bowl of Night},
  by F. P. Dickson (MIT Press, Cambridge, 1968) Appendix II.

\bibitem{Olbers} H. W. M. Olbers, {\em Bode's Jahrbuch}, 111 (1826); reprinted by Dickson, {\em op. cit.}, Appendix I.



\bibitem{Hubble} E. Hubble, Proc. Nat. Acad. Sci. {\bf 15}, 168 (1929).


\bibitem{Freedman:2000cf} 
  W.~L.~Freedman {\it et al.} [HST Collaboration],
  Astrophys.\ J.\  {\bf 553}, 47 (2001)
  doi:10.1086/320638
  [astro-ph/0012376].

\bibitem{SDSS} {\tt
  http://www.sdss.org/iotw/archive.html}


\bibitem{Chyba} C. F. Chyba, J. R. Gott, and A. Spitkovsky, {\it The
    Universe}, lectures given at Princeton University, 2009 - 2010.


\bibitem{Milne} E. A. Milne, 
 Z. Astrophysik {\bf 6}, 1 (1933).


\bibitem{Ryden:2003yy} 
  B.~Ryden,
  {\it Introduction to cosmology,}
(Addison-Wesley,  San Francisco, USA,  2003) ISBN 978-0805389128

\bibitem{Einstein:1917ce}
  A.~Einstein,
  Sitzungsber.\ Preuss.\ Akad.\ Wiss.\ Berlin (Math.\ Phys.\ ) 
{\bf 1917}, 142 (1917).


\bibitem{Friedmann:1} A. Friedmann, Z. Phys. {\bf 10}, 377 (1922).

\bibitem{Friedmann:2}  A. Friedmann, Z. Phys. {\bf 21}, 326 (1924).



\bibitem{Robertson:1} H. P. Robertson, Astrophys. J. {\bf 82}, 284
  (1935).

\bibitem{Robertson:2} H. P. Robertson, Astrophys. J. {\bf 83}, 187, 257 (1936).

\bibitem{Walker} A. G. Walker, Proc. Lond. Math. Soc. (2), {\bf 42} 90 (1936).


\bibitem{Mukhanov:2005sc} 
  V.~Mukhanov,
  {\it Physical Foundations of Cosmology,}
(Cambridge University Press,  UK, 2005) ISBN: 978-0-521-56398-7




\bibitem{Pogson:1856wh} 
  N.~Pogson,
  Mon.\ Not.\ Roy.\ Astron.\ Soc.\  {\bf 17}, 12 (1856).


\bibitem{Riess:1998cb} 
  A.~G.~Riess {\it et al.} [Supernova Search Team Collaboration],
  Astron.\ J.\  {\bf 116}, 1009 (1998)
  doi:10.1086/300499
  [astro-ph/9805201].


\bibitem{Perlmutter:1998np} 
  S.~Perlmutter {\it et al.} [Supernova Cosmology Project Collaboration],
  Astrophys.\ J.\  {\bf 517}, 565 (1999)
  doi:10.1086/307221
  [astro-ph/9812133].

\bibitem{Hamuy:1993}
M. Hamuy {\it et al}, 
Astron. J. {\bf 106}, 2392 (1993). 


\bibitem{Hamuy:1995in} 
  M.~Hamuy, M.~M.~Phillips, J.~Maza, N.~B.~Suntzeff, R.~A.~Schommer and R.~Aviles,
  Astron.\ J.\  {\bf 109}, 1 (1995).
  doi:10.1086/117251

\bibitem{Perlmutter:2003} 
  S.~Perlmutter,
Phys. Today, April 2003.




\bibitem{Bahcall:1999xn} 
  N.~A.~Bahcall, J.~P.~Ostriker, S.~Perlmutter and P.~J.~Steinhardt,
  Science {\bf 284}, 1481 (1999)
  doi:10.1126/science.284.5419.1481
  [astro-ph/9906463].




\bibitem{Penzias:1965wn} 
  A.~A.~Penzias and R.~W.~Wilson,
  Astrophys.\ J.\  {\bf 142}, 419 (1965).


\bibitem{Dicke:1965zz} 
  R.~H.~Dicke, P.~J.~E.~Peebles, P.~G.~Roll and D.~T.~Wilkinson,
  Astrophys.\ J.\  {\bf 142}, 414 (1965).
  doi:10.1086/148306


\bibitem{Smoot:1997xt} 
  G.~F.~Smoot,
  astro-ph/9705101.


\bibitem{Mather:1993ij} 
  J.~C.~Mather {\it et al.},
  Astrophys.\ J.\  {\bf 420}, 439 (1994).



\bibitem{Bennett:2012zja} 
  C.~L.~Bennett {\it et al.} [WMAP Collaboration],
  Astrophys.\ J.\ Suppl.\  {\bf 208}, 20 (2013)
  [arXiv:1212.5225 [astro-ph.CO]].



\bibitem{Adam:2015rua} 
  R.~Adam {\it et al.} [Planck Collaboration],
  arXiv:1502.01582 [astro-ph.CO].

\bibitem{Weinberg:2008zzc} 
  S.~Weinberg,
  {\it Cosmology,}
  (Oxford University  Press, UK, 2008) 
ISBN 978-0-19-852682-7.


\bibitem{Dodelson:2003ft} 
  S.~Dodelson,
  {\it Modern Cosmology,}
  (Academic Press, Elsevier, Amsterdam, 2003) ISBN 978-0-12-219141-1.



\bibitem{Anchordoqui:2013} 
L. A. Anchordoqui and T. C. Paul,
{\it Mathematical models of physics problems}
(Nova, New York, 2013) ISBN  978-1-62618-600-2.

\bibitem{Denton:2014nga} 
  P.~B.~Denton, L. A. Anchordoqui, A. A. Berlind, M. Richardson, and T. J.
  Weiler (for the JEM-EUSO Collaboration),
  J.\ Phys.\ Conf.\ Ser.\  {\bf 531}, 012004 (2014)
  doi:10.1088/1742-6596/531/1/012004
  [arXiv:1401.5757 [astro-ph.IM]].

\bibitem{Hinshaw:2008kr} 
  G.~Hinshaw {\it et al.} [WMAP Collaboration],
  Astrophys.\ J.\ Suppl.\  {\bf 180}, 225 (2009)
  doi:10.1088/0067-0049/180/2/225
  [arXiv:0803.0732 [astro-ph]].




\bibitem{Lineweaver:1996tk} 
  C.~H.~Lineweaver,
  ASP Conf.\ Ser.\  {\bf 126}, 185 (1997)
  [astro-ph/9702042].


\bibitem{Rubin:1970zza}
  V.~C.~Rubin and W.~K.~Ford, Jr.,
  Astrophys.\ J.\  {\bf 159}, 379 (1970).
  doi:10.1086/150317


\bibitem{Rubin:1980zd}
  V.~C.~Rubin, N.~Thonnard and W.~K.~Ford, Jr.,
  Astrophys.\ J.\  {\bf 238}, 471 (1980).
  doi:10.1086/158003


\bibitem{Rubin:1985ze} 
  V.~C.~Rubin, D.~Burstein, W.~K.~Ford, Jr. and N.~Thonnard,
  Astrophys.\ J.\  {\bf 289}, 81 (1985).
  doi:10.1086/162866



\bibitem{Zwicky:1933gu} 
  F.~Zwicky,
  Helv.\ Phys.\ Acta {\bf 6}, 110 (1933).




\bibitem{Clowe:2006eq} 
  D.~Clowe, M.~Bradac, A.~H.~Gonzalez, M.~Markevitch, S.~W.~Randall, C.~Jones and D.~Zaritsky,
  Astrophys.\ J.\  {\bf 648}, L109 (2006)
  doi:10.1086/508162
  [astro-ph/0608407].

\bibitem{Feng:2010gw} 
  J.~L.~Feng,
  Ann.\ Rev.\ Astron.\ Astrophys.\  {\bf 48}, 495 (2010)
  doi:10.1146/annurev-astro-082708-101659
  [arXiv:1003.0904 [astro-ph.CO]].




\bibitem{Ade:2015xua} 
  P.~A.~R.~Ade {\it et al.} [Planck Collaboration],
  arXiv:1502.01589 [astro-ph.CO].


\bibitem{Riess:2011yx} 
  A.~G.~Riess {\it et al.},
  Astrophys.\ J.\  {\bf 730}, 119 (2011)
  Erratum: [Astrophys.\ J.\  {\bf 732}, 129 (2011)]
  doi:10.1088/0004-637X/732/2/129, 10.1088/0004-637X/730/2/119
  [arXiv:1103.2976 [astro-ph.CO]].


\bibitem{Knop:2003iy}
  R.~A.~Knop {\it et al.}  [Supernova Cosmology Project Collaboration],
  Astrophys.\ J.\  {\bf 598}, 102 (2003)
  [arXiv:astro-ph/0309368].

\bibitem{Allen:2002sr}
  S.~W.~Allen, R.~W.~Schmidt and A.~C.~Fabian,
  Mon.\ Not.\ Roy.\ Astron.\ Soc.\  {\bf 334}, L11 (2002)
  [arXiv:astro-ph/0205007].



\bibitem{Lange:2000iq}
  A.~E.~Lange {\it et al.}  [Boomerang Collaboration],
  Phys.\ Rev.\  D {\bf 63}, 042001 (2001)
  [arXiv:astro-ph/0005004].

\bibitem{Balbi:2000tg}
  A.~Balbi {\it et al.},
  Astrophys.\ J.\  {\bf 545}, L1 (2000)
  [Erratum-ibid.\  {\bf 558}, L145 (2001)]
  [arXiv:astro-ph/0005124].





\bibitem{Carroll:1991mt}
  S.~M.~Carroll, W.~H.~Press and E.~L.~Turner,
  Ann.\ Rev.\ Astron.\ Astrophys.\  {\bf 30}, 499 (1992).


\bibitem{Meyer:1986pw}
  B.~S.~Meyer and D.~N.~Schramm,
  Astrophys.\ J.\  {\bf 311}, 406 (1986).

\bibitem{Aldering:2002dp}
  G.~Aldering {\it et al.}  [SNAP Collaboration],
  arXiv:astro-ph/0209550.




\bibitem{Halzen:1984mc} 
  F.~Halzen and A.~D.~Martin,
  {\it Quarks and leptons: An introductory course In modern particle physics,}
(John Wiley \& Sons, New York, 1984) ISBN 0-471-88741-2 


\bibitem{Barger:1987nn} 
  V.~D.~Barger and R.~J.~N.~Phillips,
  {\it Collider physics},
Front. Phys. {\bf 71}, 1 (1991) ISBN 0-201-14945-1



\bibitem{Quigg:2013ufa}
  C.~Quigg,
  {\it Gauge Theories of the strong, weak, and electromagnetic
    interactions},
Front. Phys. {\bf 56}, 1 (1983) ISBN 978-0805360202



\bibitem{Anchordoqui:2009eg} 
  L.~Anchordoqui and F.~Halzen,
  arXiv:0906.1271 [physics.ed-ph].



\bibitem{Agashe:2014kda} 
  K.~A.~Olive {\it et al.} [Particle Data Group Collaboration],
  Chin.\ Phys.\ C {\bf 38}, 090001 (2014).
  doi:10.1088/1674-1137/38/9/090001



\bibitem{Fritzsch:1973pi} 
  H.~Fritzsch, M.~Gell-Mann and H.~Leutwyler,
  Phys.\ Lett.\ B {\bf 47}, 365 (1973).
  


\bibitem{GellMann:1961ky} 
  M.~Gell-Mann,
  CTSL-20, TID-12608.


\bibitem{Ne'eman:1961cd} 
  Y.~Ne'eman,
  Nucl.\ Phys.\  {\bf 26}, 222 (1961).


\bibitem{GellMann:1964nj} 
  M.~Gell-Mann,
  Phys.\ Lett.\  {\bf 8}, 214 (1964).


\bibitem{Bose:1924s} S. N. Bose, 
Z. Phys. 26, 178 (1924).

\bibitem{Einstein:1925s} A. Einstein, 
 [Sitzungsber.\ Preuss.\ Akad.\ Wiss.\ Berlin (Math.\ Phys.\ ) {\bf 22},
261 (1924); {\bf 1}, 3 (1925);  {\bf 3}, 18 (1925).



\bibitem{Fermi:1926s} E. Fermi, 
 Rend. Lincei {\bf 3}, 145 (1926); Z. Phys. {\bf 36}, 902 (1926).

\bibitem{Dirac:1926s} P. A. M. Dirac, 
Proc. R. Soc. Lond. Ser. A {\bf 112}, 661 (1926).

\bibitem{Gross:1973id} 
  D.~J.~Gross and F.~Wilczek,
  Phys.\ Rev.\ Lett.\  {\bf 30}, 1343 (1973).


\bibitem{Politzer:1973fx} 
  H.~D.~Politzer,
  Phys.\ Rev.\ Lett.\  {\bf 30}, 1346 (1973).




\bibitem{Schwinger:1948yk} 
  J.~S.~Schwinger,
  Phys.\ Rev.\  {\bf 74}, 1439 (1948).

\bibitem{Schwinger:1948yj} 
  J.~S.~Schwinger,
  Phys.\ Rev.\  {\bf 75}, 651 (1948).
  doi:10.1103/PhysRev.75.651






\bibitem{Tomonaga:1946zz} 
  S.~Tomonaga,
  Prog.\ Theor.\ Phys.\  {\bf 1}, 27 (1946).


\bibitem{Feynman:1948ur} 
  R.~P.~Feynman,
  Rev.\ Mod.\ Phys.\  {\bf 20}, 367 (1948).


\bibitem{Dyson:1949bp} 
  F.~J.~Dyson,
  Phys.\ Rev.\  {\bf 75}, 486 (1949).



\bibitem{Dyson:1949ha} 
  F.~J.~Dyson,
  Phys.\ Rev.\  {\bf 75}, 1736 (1949).



\bibitem{Feynman:1950ir} 
  R.~P.~Feynman,
  Phys.\ Rev.\  {\bf 80}, 440 (1950).
  doi:10.1103/PhysRev.80.440




\bibitem{Schwinger:1948iu} 
  J.~S.~Schwinger,
  Phys.\ Rev.\  {\bf 73}, 416 (1948).







\bibitem{Glashow:1961tr} 
  S.~L.~Glashow,
  Nucl.\ Phys.\  {\bf 22}, 579 (1961).


\bibitem{Weinberg:1967tq} 
  S.~Weinberg,
  Phys.\ Rev.\ Lett.\  {\bf 19}, 1264 (1967).


\bibitem{Salam:1968rm} 
  A.~Salam,
  Conf.\ Proc.\ C {\bf 680519}, 367 (1968).


\bibitem{Higgs:1964pj} 
  P.~W.~Higgs,
  Phys.\ Rev.\ Lett.\  {\bf 13}, 508 (1964).



\bibitem{Englert:1964et} 
  F.~Englert and R.~Brout,
  Phys.\ Rev.\ Lett.\  {\bf 13}, 321 (1964).



\bibitem{ATLAS:2012ae} 
  G.~Aad {\it et al.} [ATLAS Collaboration],
  Phys.\ Lett.\ B {\bf 710}, 49 (2012)
  [arXiv:1202.1408 [hep-ex]].


\bibitem{Chatrchyan:2012tx} 
  S.~Chatrchyan {\it et al.} [CMS Collaboration],
 Phys.\ Lett.\ B {\bf 710}, 26 (2012)
  [arXiv:1202.1488 [hep-ex]].


\bibitem{ATLAS} 
  ATLAS Collaboration,
  {\it Search for resonances decaying to photon pairs in 3.2 fb$^{-1}$ of $pp$ collisions at $\sqrt{s}$ = 13 TeV with the ATLAS detector},
  ATLAS-CONF-2015-081.


\bibitem{CMS:2015dxe} 
  CMS Collaboration,
  {\it Search for new physics in high mass diphoton events in proton-proton
  collisions at 13~TeV},
  CMS-PAS-EXO-15-004.



\bibitem{Strumia:2016wys} 
  A.~Strumia,
  arXiv:1605.09401 [hep-ph]; and references therein.




\bibitem{Delmastro} M. Delmastro [on behalf of the ATLAS Collaboration],
  {\it Diphoton searches in ATLAS},
51st Rencontres de Moriond (Electroweak session) 17 May 2016, La Thuile (Italy).

\bibitem{Musella} P. Musella [on behalf of the CMS Collaboration],
  {\it Search for high mass diphoton resonances at CMS},
51st Rencontres de Moriond (Electroweak session) 17 May 2016, La Thuile (Italy).

\bibitem{CMS:2016owr} 
  CMS Collaboration,
   {\it  Search for new physics in high mass diphoton events in $3.3~\mathrm{fb}^{-1}$ of proton-proton collisions at $\sqrt{s}=13~\mathrm{TeV}$ and combined interpretation of searches at $8~\mathrm{TeV}$ and $13~\mathrm{TeV}$},
  CMS-PAS-EXO-16-018.



\bibitem{Kolb:1990vq} 
  E.~W.~Kolb and M.~S.~Turner,
  {\it The Early Universe,}
  Front.\ Phys.\  {\bf 69}, 1 (1990).
ISBN 0-201-11603-0  


\bibitem{Olive:2010mh} 
  K.~A.~Olive,
  CERN Yellow Report CERN-2010-002, 149-196
  [arXiv:1005.3955 [hep-ph]].


\bibitem{Guth:1980zm} 
  A.~H.~Guth,
  Phys.\ Rev.\ D {\bf 23}, 347 (1981).
  doi:10.1103/PhysRevD.23.347


\bibitem{Baumann:2009ds} 
  D.~Baumann,
  doi:10.1142/9789814327183$_-$0010
  arXiv:0907.5424 [hep-th].


\bibitem{Riotto:2002yw} 
  A.~Riotto,
  hep-ph/0210162.


\bibitem{Lineweaver:2003ie} 
  C.~H.~Lineweaver,
  astro-ph/0305179.




\bibitem{Sakharov:1967dj} 
  A.~D.~Sakharov,
  Pisma Zh.\ Eksp.\ Teor.\ Fiz.\  {\bf 5}, 32 (1967)
  [JETP Lett.\  {\bf 5}, 24 (1967)]
  [Sov.\ Phys.\ Usp.\  {\bf 34}, 392 (1991)]
  [Usp.\ Fiz.\ Nauk {\bf 161}, 61 (1991)].
  doi:10.1070/PU1991v034n05ABEH002497


\bibitem{Brust:2013xpv} 
  C.~Brust, D.~E.~Kaplan and M.~T.~Walters,
  JHEP {\bf 1312}, 058 (2013)
  doi:10.1007/JHEP12(2013)058
  [arXiv:1303.5379 [hep-ph]].

\bibitem{Bazavov:2009zn} 
  A.~Bazavov {\it et al.},
  Phys.\ Rev.\ D {\bf 80}, 014504 (2009)
  doi:10.1103/PhysRevD.80.014504
  [arXiv:0903.4379 [hep-lat]].

\bibitem{Anchordoqui:2011nh} 
  L.~A.~Anchordoqui and H.~Goldberg,
  Phys.\ Rev.\ Lett.\  {\bf 108}, 081805 (2012)
  doi:10.1103/PhysRevLett.108.081805
  [arXiv:1111.7264 [hep-ph]].




\bibitem{Laine:2006cp} 
  M.~Laine and Y.~Schroder,
  Phys.\ Rev.\ D {\bf 73}, 085009 (2006)
  doi:10.1103/PhysRevD.73.085009
  [hep-ph/0603048].



\bibitem{Steigman:2012nb} 
  G.~Steigman, B.~Dasgupta and J.~F.~Beacom,
  Phys.\ Rev.\ D {\bf 86}, 023506 (2012)
  doi:10.1103/PhysRevD.86.023506
  [arXiv:1204.3622 [hep-ph]].



\bibitem{Anchordoqui:2013wwa} 
  L.~A.~Anchordoqui, H.~Goldberg and B.~Vlcek,
  arXiv:1305.0146 [astro-ph.CO].






\bibitem{Alpher:1953zz} 
  R.~A.~Alpher, J.~W.~Follin and R.~C.~Herman,
  Phys.\ Rev.\  {\bf 92}, 1347 (1953).
  doi:10.1103/PhysRev.92.1347

\bibitem{Zeldovich:65} Ya.~B.~Zel'dovich, Adv. Astron. Astrophys. {\bf
    3}, 241 (1965).

\bibitem{Zeldovich:67} Ya.~B.~Zel'dovich,  Sov. Phys. Usp. {\bf 9}, 602 (1967).

\bibitem{Lee:1977ua} 
  B.~W.~Lee and S.~Weinberg,
  Phys.\ Rev.\ Lett.\  {\bf 39}, 165 (1977).
  doi:10.1103/PhysRevLett.39.165




\bibitem{Steigman:1977kc} 
  G.~Steigman, D.~N.~Schramm and J.~E.~Gunn,
  Phys.\ Lett.\ B {\bf 66}, 202 (1977).
  doi:10.1016/0370-2693(77)90176-9

\bibitem{Steigman:1986nh} 
  G.~Steigman, K.~A.~Olive, D.~N.~Schramm and M.~S.~Turner,
  Phys.\ Lett.\ B {\bf 176}, 33 (1986).
  doi:10.1016/0370-2693(86)90920-2


\bibitem{Dicus:1982bz} 
  D.~A.~Dicus, E.~W.~Kolb, A.~M.~Gleeson, E.~C.~G.~Sudarshan, V.~L.~Teplitz and M.~S.~Turner,
  Phys.\ Rev.\ D {\bf 26}, 2694 (1982).
  doi:10.1103/PhysRevD.26.2694


\bibitem{Dodelson:1992km} 
  S.~Dodelson and M.~S.~Turner,
  Phys.\ Rev.\ D {\bf 46}, 3372 (1992).
  doi:10.1103/PhysRevD.46.3372



\bibitem{Mangano:2001iu} 
  G.~Mangano, G.~Miele, S.~Pastor and M.~Peloso,
  Phys.\ Lett.\ B {\bf 534}, 8 (2002)
  doi:10.1016/S0370-2693(02)01622-2
  [astro-ph/0111408].


\bibitem{Mangano:2005cc} 
  G.~Mangano, G.~Miele, S.~Pastor, T.~Pinto, O.~Pisanti and P.~D.~Serpico,
  Nucl.\ Phys.\ B {\bf 729}, 221 (2005)
  doi:10.1016/j.nuclphysb.2005.09.041
  [hep-ph/0506164].




\bibitem{Sarkar:1995dd} 
  S.~Sarkar,
  Rept.\ Prog.\ Phys.\  {\bf 59}, 1493 (1996)
  doi:10.1088/0034-4885/59/12/001
  [hep-ph/9602260].


\bibitem{Olive:1999ij} 
  K.~A.~Olive, G.~Steigman and T.~P.~Walker,
  Phys.\ Rept.\  {\bf 333}, 389 (2000)
  doi:10.1016/S0370-1573(00)00031-4
  [astro-ph/9905320].

\bibitem{Gamow:1946eb} 
  G.~Gamow,
  Phys.\ Rev.\  {\bf 70}, 572 (1946).
  doi:10.1103/PhysRev7.0.572


\bibitem{Alpher:1948ve} 
  R.~A.~Alpher, H.~Bethe and G.~Gamow,
  Phys.\ Rev.\  {\bf 73}, 803 (1948).
  doi:10.1103/PhysRev.73.803

\bibitem{Gamow:1949zz} 
  G.~Gamow,
  Rev.\ Mod.\ Phys.\  {\bf 21}, 367 (1949).
  doi:10.1103/RevModPhys.21.367






\bibitem{Izotov:2007ed} 
  Y.~I.~Izotov, T.~X.~Thuan and G.~Stasinska,
  Astrophys.\ J.\  {\bf 662}, 15 (2007)
  doi:10.1086/513601
  [astro-ph/0702072 [ASTRO-PH]].


\bibitem{Peimbert:2007vm} 
  M.~Peimbert, V.~Luridiana and A.~Peimbert,
  Astrophys.\ J.\  {\bf 666}, 636 (2007)
  doi:10.1086/520571
  [astro-ph/0701580].


\bibitem{Steigman:2007xt} 
  G.~Steigman,
  Ann.\ Rev.\ Nucl.\ Part.\ Sci.\  {\bf 57}, 463 (2007)
  doi:10.1146/annurev.nucl.56.080805.140437
  [arXiv:0712.1100 [astro-ph]].


\bibitem{Simha:2008zj} 
  V.~Simha and G.~Steigman,
  JCAP {\bf 0806}, 016 (2008)
  doi:10.1088/1475-7516/2008/06/016
  [arXiv:0803.3465 [astro-ph]].

\bibitem{Izotov:2010ca} 
  Y.~I.~Izotov and T.~X.~Thuan,
  Astrophys.\ J.\  {\bf 710}, L67 (2010)
  doi:10.1088/2041-8205/710/1/L67
  [arXiv:1001.4440 [astro-ph.CO]].



\bibitem{Aver:2010wd} 
  E.~Aver, K.~A.~Olive and E.~D.~Skillman,
  JCAP {\bf 1103}, 043 (2011)
  doi:10.1088/1475-7516/2011/03/043
  [arXiv:1012.2385 [astro-ph.CO]].


\bibitem{Aver:2010wq} 
  E.~Aver, K.~A.~Olive and E.~D.~Skillman,
  JCAP {\bf 1005}, 003 (2010)
  doi:10.1088/1475-7516/2010/05/003
  [arXiv:1001.5218 [astro-ph.CO]].



\bibitem{Izotov:2014fga} 
  Y.~I.~Izotov, T.~X.~Thuan and N.~G.~Guseva,
  Mon.\ Not.\ Roy.\ Astron.\ Soc.\  {\bf 445}, no. 1, 778 (2014)
  doi:10.1093/mnras/stu1771
  [arXiv:1408.6953 [astro-ph.CO]].


\bibitem{Ade:2013zuv} 
  P.~A.~R.~Ade {\it et al.} [Planck Collaboration],
  Astron.\ Astrophys.\  {\bf 571}, A16 (2014)
  doi:10.1051/0004-6361/201321591
  [arXiv:1303.5076 [astro-ph.CO]].


\bibitem{Anchordoqui:2012qu} 
  L.~A.~Anchordoqui, H.~Goldberg and G.~Steigman,
  Phys.\ Lett.\ B {\bf 718}, 1162 (2013)
  doi:10.1016/j.physletb.2012.12.019
  [arXiv:1211.0186 [hep-ph]].



\bibitem{Hawking:rv}
S.~W.~Hawking,
Nature {\bf 248}, 30 (1974).

\bibitem{Hawking:sw}
S.~W.~Hawking,
Commun.\ Math.\ Phys.\  {\bf 43} (1975) 199.


\bibitem{Hartle:tp} 
J.~B.~Hartle and S.~W.~Hawking,
Phys.\ Rev.\ D {\bf 13} (1976) 2188.



\bibitem{Parker:1975jm}
  L.~Parker,
  Phys.\ Rev.\ D {\bf 12}, 1519 (1975).

\bibitem{Wald:1975kc}
  R.~M.~Wald,
  Commun.\ Math.\ Phys.\  {\bf 45}, 9 (1975).

\bibitem{Hawking:1976ra}
  S.~W.~Hawking,
  Phys.\ Rev.\ D {\bf 14}, 2460 (1976).





\bibitem{Page:df}
D.~N.~Page,
Phys.\ Rev.\ D {\bf 13} (1976) 198.


\bibitem{Page:1976wx}
  D.~N.~Page and S.~W.~Hawking,
  Astrophys.\ J.\  {\bf 206}, 1 (1976).

\bibitem{Hawking:1976de}
  S.~W.~Hawking,
  Phys.\ Rev.\  D {\bf 13}, 191 (1976).





 


\bibitem{ArkaniHamed:1998rs} 
  N.~Arkani-Hamed, S.~Dimopoulos and G.~R.~Dvali,
  Phys.\ Lett.\ B {\bf 429}, 263 (1998)
  doi:10.1016/S0370-2693(98)00466-3
  [hep-ph/9803315].




\bibitem{Banks:1999gd} 
  T.~Banks and W.~Fischler,
  hep-th/9906038.


\bibitem{Dimopoulos:2001hw} 
  S.~Dimopoulos and G.~L.~Landsberg,
  Phys.\ Rev.\ Lett.\  {\bf 87}, 161602 (2001)
  doi:10.1103/PhysRevLett.87.161602
  [hep-ph/0106295].


\bibitem{Giddings:2001bu} 
  S.~B.~Giddings and S.~D.~Thomas,
  Phys.\ Rev.\ D {\bf 65}, 056010 (2002)
  doi:10.1103/PhysRevD.65.056010
  [hep-ph/0106219].



\bibitem{Feng:2001ib} 
  J.~L.~Feng and A.~D.~Shapere,
  Phys.\ Rev.\ Lett.\  {\bf 88}, 021303 (2002)
  doi:10.1103/PhysRevLett.88.021303
  [hep-ph/0109106].


\bibitem{Anchordoqui:2001cg} 
  L.~A.~Anchordoqui, J.~L.~Feng, H.~Goldberg and A.~D.~Shapere,
  Phys.\ Rev.\ D {\bf 65}, 124027 (2002)
  doi:10.1103/PhysRevD.65.124027
  [hep-ph/0112247].


\bibitem{Emparan:2000rs} 
  R.~Emparan, G.~T.~Horowitz and R.~C.~Myers,
  Phys.\ Rev.\ Lett.\  {\bf 85}, 499 (2000)
  doi:10.1103/PhysRevLett.85.499
  [hep-th/0003118].


\bibitem{Anchordoqui:2003ug} 
  L.~A.~Anchordoqui, J.~L.~Feng, H.~Goldberg and A.~D.~Shapere,
  Phys.\ Lett.\ B {\bf 594}, 363 (2004)
  doi:10.1016/j.physletb.2004.05.051
  [hep-ph/0311365].


\bibitem{Anchordoqui:2002cp} 
  L.~Anchordoqui and H.~Goldberg,
  Phys.\ Rev.\ D {\bf 67}, 064010 (2003)
  doi:10.1103/PhysRevD.67.064010
  [hep-ph/0209337].



\bibitem{Myers:1986un} 
  R.~C.~Myers and M.~J.~Perry,
  Annals Phys.\  {\bf 172}, 304 (1986).
  doi:10.1016/0003-4916(86)90186-7

\bibitem{Chamblin:2003wg} 
  A.~Chamblin, F.~Cooper and G.~C.~Nayak,
  Phys.\ Rev.\ D {\bf 69}, 065010 (2004)
  doi:10.1103/PhysRevD.69.065010
  [hep-ph/0301239].


\bibitem{Antoniadis:1998ig} 
  I.~Antoniadis, N.~Arkani-Hamed, S.~Dimopoulos and G.~R.~Dvali,
  Phys.\ Lett.\ B {\bf 436}, 257 (1998)
  doi:10.1016/S0370-2693(98)00860-0
  [hep-ph/9804398].


\bibitem{Harris:2003eg} 
  C.~M.~Harris and P.~Kanti,
  JHEP {\bf 0310}, 014 (2003)
  doi:10.1088/1126-6708/2003/10/014
  [hep-ph/0309054].












 \bibitem{Kraushaar} W. L. Kraushaar, G. W. Clark, G. P. Garmire,
  R. Borken, P. Higbie, and C. Leong, and T. Thorsos,
Astrophys. J. {\bf 177}, 341 (1972). 

\bibitem{Fichtel} C. E. Fichtel, R. C. Hartman, D. A. Kniffen,
  D. J. Thomson, H. Ogelman, M. E. Ozel, T. turner, and G. F. Bignami
Astrophys. J. {\bf 198}, 163 (1975).



\bibitem{Sreekumar:1997un} 
  P.~Sreekumar {\it et al.} [EGRET Collaboration],
  Astrophys.\ J.\  {\bf 494}, 523 (1998)
  doi:10.1086/305222
  [astro-ph/9709257].


\bibitem{Hartman:1999fc}
  R.~C.~Hartman {\it et al.}  [EGRET Collaboration],
  Astrophys.\ J.\ Suppl.\  {\bf 123}, 79 (1999).

\bibitem{Ackermann:2014usa} 
  M.~Ackermann {\it et al.} [Fermi-LAT Collaboration],
  Astrophys.\ J.\  {\bf 799}, 86 (2015)
  doi:10.1088/0004-637X/799/1/86
  [arXiv:1410.3696 [astro-ph.HE]].



\bibitem{Aartsen:2014muf} 
  M.~G.~Aartsen {\it et al.} [IceCube Collaboration],
  Phys.\ Rev.\ D {\bf 91}, no. 2, 022001 (2015)
  doi:10.1103/PhysRevD.91.022001
  [arXiv:1410.1749 [astro-ph.HE]].



\bibitem{Anchordoqui:2014rca} 
  L.~A.~Anchordoqui, H.~Goldberg, T.~C.~Paul, L.~H.~M.~da Silva and B.~J.~Vlcek,
  Phys.\ Rev.\ D {\bf 90}, no. 12, 123010 (2014)
  doi:10.1103/PhysRevD.90.123010
  [arXiv:1410.0348 [astro-ph.HE]].




\bibitem{Felix} For further details see e.g., F. A. Aharonian, {\em
    Very high energy cosmic gamma radiation: A critical window on the
    extreme universe,} (Singapore: World Scientific Publishing, 2004)
ISBN 981-02-4573-4.

\bibitem{Learned:2000sw}
J.~G.~Learned and K.~Mannheim,
Ann.\ Rev.\ Nucl.\ Part.\ Sci.\  {\bf 50}, 679 (2000).



\bibitem{Hess} V. F. Hess, Phys. Z. {\bf 13}, 1804 (1912).

\bibitem{Auger:1938} P. Auger, R. Maze, T. Grivet-Meyer, 
Comptes Rendus {\bf 206}, 1721 (1938).

\bibitem{Auger:1939} P. Auger, P. Ehrenfest, R. Maze, J. Daudin,
  Robley, and A. Fr\'eon,
Rev. Mod. Phys. {\bf 11}, 288 (1939).



\bibitem{Bird:1994uy}
D.~J.~Bird {\it et al.},
Astrophys.\ J.\  {\bf 441}, 144 (1995).

\bibitem{ThePierreAuger:2015rma} 
  A.~Aab {\it et al.} [Pierre Auger Collaboration],
  Nucl.\ Instrum.\ Meth.\ A {\bf 798}, 172 (2015)
  doi:10.1016/j.nima.2015.06.058
  [arXiv:1502.01323 [astro-ph.IM]].




\bibitem{Abraham:2010zz} 
  J.~Abraham {\it et al.} [Pierre Auger Collaboration],
  Nucl.\ Instrum.\ Meth.\ A {\bf 613}, 29 (2010)
  doi:10.1016/j.nima.2009.11.018
  [arXiv:1111.6764 [astro-ph.IM]].



\bibitem{Abraham:2009pm} 
  J.~Abraham {\it et al.} [Pierre Auger Collaboration],
  Nucl.\ Instrum.\ Meth.\ A {\bf 620}, 227 (2010)
  doi:10.1016/j.nima.2010.04.023
  [arXiv:0907.4282 [astro-ph.IM]].



\bibitem{Anchordoqui:2004xb} 
  L.~Anchordoqui, M.~T.~Dova, A.~G.~Mariazzi, T.~McCauley, T.~C.~Paul, S.~Reucroft and J.~Swain,
  Annals Phys.\  {\bf 314}, 145 (2004)
  doi:10.1016/j.aop.2004.07.003
  [hep-ph/0407020].



\bibitem{Torres:2004hk} 
  D.~F.~Torres and L.~A.~Anchordoqui,
  Rept.\ Prog.\ Phys.\  {\bf 67}, 1663 (2004)
  doi:10.1088/0034-4885/67/9/R03
  [astro-ph/0402371].

\bibitem{Antoni:2005wq} 
  T.~Antoni {\it et al.} [KASCADE Collaboration],
  Astropart.\ Phys.\  {\bf 24}, 1 (2005)
  doi:10.1016/j.astropartphys.2005.04.001
  [astro-ph/0505413].

\bibitem{Bird:1993yi} 
  D.~J.~Bird {\it et al.} [HiRes Collaboration],
  Phys.\ Rev.\ Lett.\  {\bf 71}, 3401 (1993).
  doi:10.1103/PhysRevLett.71.3401

\bibitem{Abbasi:2007sv} 
  R.~U.~Abbasi {\it et al.} [HiRes Collaboration],
  Phys.\ Rev.\ Lett.\  {\bf 100}, 101101 (2008)
  doi:10.1103/PhysRevLett.100.101101
  [astro-ph/0703099].

\bibitem{Abraham:2008ru} 
  J.~Abraham {\it et al.} [Pierre Auger Collaboration],
  Phys.\ Rev.\ Lett.\  {\bf 101}, 061101 (2008)
  doi:10.1103/PhysRevLett.101.061101
  [arXiv:0806.4302 [astro-ph]].



\bibitem{Apel:2012tda} 
  W.~D.~Apel {\it et al.},
  Astropart.\ Phys.\  {\bf 36}, 183 (2012).
  doi:10.1016/j.astropartphys.2012.05.023

\bibitem{Aartsen:2013wda} 
  M.~G.~Aartsen {\it et al.} [IceCube Collaboration],
  Phys.\ Rev.\ D {\bf 88}, no. 4, 042004 (2013)
  doi:10.1103/PhysRevD.88.042004
  [arXiv:1307.3795 [astro-ph.HE]].


\bibitem{AbuZayyad:2000ay} 
  T.~Abu-Zayyad {\it et al.} [HiRes-MIA Collaboration],
  Astrophys.\ J.\  {\bf 557}, 686 (2001)
  doi:10.1086/322240
  [astro-ph/0010652].

\bibitem{Bergman:2007kn} 
  D.~R.~Bergman and J.~W.~Belz,
  J.\ Phys.\ G {\bf 34}, R359 (2007)
  doi:10.1088/0954-3899/34/10/R01
  [arXiv:0704.3721 [astro-ph]].

\bibitem{Kampert:2012mx} 
  K.~H.~Kampert and M.~Unger,
  Astropart.\ Phys.\  {\bf 35}, 660 (2012)
  doi:10.1016/j.astropartphys.2012.02.004
  [arXiv:1201.0018 [astro-ph.HE]].

\bibitem{Abreu:2011ve} 
  P.~Abreu {\it et al.} [Pierre Auger Collaboration],
  Astropart.\ Phys.\  {\bf 34}, 627 (2011)
  doi:10.1016/j.astropartphys.2010.12.007
  [arXiv:1103.2721 [astro-ph.HE]].


\bibitem{Auger:2012an}
  P.~Abreu {\it et al.} [Pierre Auger Collaboration],
  Astrophys.\ J.\ Suppl.\  {\bf 203} (2012) 34
  doi:10.1088/0067-0049/203/2/34
  [arXiv:1210.3736 [astro-ph.HE]].

\bibitem{ThePierreAuger:2014nja} 
  A.~Aab {\it et al.} [Pierre Auger Collaboration],
  Astrophys.\ J.\  {\bf 802}, no. 2, 111 (2015)
  doi:10.1088/0004-637X/802/2/111
  [arXiv:1411.6953 [astro-ph.HE]].

\bibitem{Aab:2014kda} 
  A.~Aab {\it et al.} [Pierre Auger Collaboration],
  Phys.\ Rev.\ D {\bf 90}, no. 12, 122005 (2014)
  doi:10.1103/PhysRevD.90.122005
  [arXiv:1409.4809 [astro-ph.HE]].

\bibitem{Aab:2014aea} 
  A.~Aab {\it et al.} [Pierre Auger Collaboration],
  Phys.\ Rev.\ D {\bf 90}, no. 12, 122006 (2014)
  doi:10.1103/PhysRevD.90.122006
  [arXiv:1409.5083 [astro-ph.HE]].

\bibitem{Berezinsky:2002nc} 
  V.~Berezinsky, A.~Z.~Gazizov and S.~I.~Grigorieva,
  Phys.\ Rev.\ D {\bf 74}, 043005 (2006)
  doi:10.1103/PhysRevD.74.043005
  [hep-ph/0204357].

\bibitem{Ahlers:2012az} 
  M.~Ahlers, L.~A.~Anchordoqui and A.~M.~Taylor,
  Phys.\ Rev.\ D {\bf 87}, no. 2, 023004 (2013)
  doi:10.1103/PhysRevD.87.023004
  [arXiv:1209.5427 [astro-ph.HE]].




\bibitem{Aloisio:2013hya} 
  R.~Aloisio, V.~Berezinsky and P.~Blasi,
  JCAP {\bf 1410}, no. 10, 020 (2014)
  doi:10.1088/1475-7516/2014/10/020
  [arXiv:1312.7459 [astro-ph.HE]].

\bibitem{Giacinti:2015hva} 
  G.~Giacinti, M.~Kachelrieß and D.~V.~Semikoz,
  Phys.\ Rev.\ D {\bf 91}, no. 8, 083009 (2015)
  doi:10.1103/PhysRevD.91.083009
  [arXiv:1502.01608 [astro-ph.HE]].



\bibitem{Unger:2015laa} 
  M.~Unger, G.~R.~Farrar and L.~A.~Anchordoqui,
  Phys.\ Rev.\ D {\bf 92}, no. 12, 123001 (2015)
  doi:10.1103/PhysRevD.92.123001
  [arXiv:1505.02153 [astro-ph.HE]].

\bibitem{AbuZayyad:2012ru} 
  T.~Abu-Zayyad {\it et al.} [Telescope Array Collaboration],
  Astrophys.\ J.\  {\bf 768}, L1 (2013)
  doi:10.1088/2041-8205/768/1/L1
  [arXiv:1205.5067 [astro-ph.HE]].

\bibitem{Abraham:2010mj} 
  J.~Abraham {\it et al.} [Pierre Auger Collaboration],
  Phys.\ Lett.\ B {\bf 685}, 239 (2010)
  doi:10.1016/j.physletb.2010.02.013
  [arXiv:1002.1975 [astro-ph.HE]].

\bibitem{Greisen:1966jv} 
  K.~Greisen,
  Phys.\ Rev.\ Lett.\  {\bf 16}, 748 (1966).
  doi:10.1103/PhysRevLett.16.748


\bibitem{Zatsepin:1966jv} 
  G.~T.~Zatsepin and V.~A.~Kuzmin,
  JETP Lett.\  {\bf 4}, 78 (1966)
  [Pisma Zh.\ Eksp.\ Teor.\ Fiz.\  {\bf 4}, 114 (1966)].



\bibitem{Aloisio:2009sj} 
  R.~Aloisio, V.~Berezinsky and A.~Gazizov,
  Astropart.\ Phys.\  {\bf 34}, 620 (2011)
  doi:10.1016/j.astropartphys.2010.12.008
  [arXiv:0907.5194 [astro-ph.HE]].


\bibitem{Anchordoqui:2003vc} 
  L.~A.~Anchordoqui, H.~Goldberg, F.~Halzen and T.~J.~Weiler,
  Phys.\ Lett.\ B {\bf 593}, 42 (2004)
  doi:10.1016/j.physletb.2004.04.054
  [astro-ph/0311002].


\bibitem{Anchordoqui:2001nt} 
  L.~A.~Anchordoqui, H.~Goldberg and T.~J.~Weiler,
  Phys.\ Rev.\ Lett.\  {\bf 87}, 081101 (2001)
  doi:10.1103/PhysRevLett.87.081101
  [astro-ph/0103043].

\bibitem{Anchordoqui:2011gy} 
  L.~A.~Anchordoqui,
  doi:10.5170/CERN-2013-003.303
  arXiv:1104.0509 [hep-ph].







\bibitem{Schramm:1987ra} 
  D.~N.~Schramm,
  Comments Nucl.\ Part.\ Phys.\  {\bf 17}, no. 5, 239 (1987).


\bibitem{Anchordoqui:2009nf} 
  L.~A.~Anchordoqui and T.~Montaruli,
  Ann.\ Rev.\ Nucl.\ Part.\ Sci.\  {\bf 60}, 129 (2010)
  doi:10.1146/annurev.nucl.012809.104551
  [arXiv:0912.1035 [astro-ph.HE]].

\bibitem{Achterberg:2006md} 
  A.~Achterberg {\it et al.} [IceCube Collaboration],
  Astropart.\ Phys.\  {\bf 26}, 155 (2006)
  doi:10.1016/j.astropartphys.2006.06.007
  [astro-ph/0604450].

\bibitem{Abbasi:2008aa} 
  R.~Abbasi {\it et al.} [IceCube Collaboration],
  Nucl.\ Instrum.\ Meth.\ A {\bf 601}, 294 (2009)
  doi:10.1016/j.nima.2009.01.001
  [arXiv:0810.4930 [physics.ins-det]].


\bibitem{IceCube:2012nn} 
  R.~Abbasi {\it et al.} [IceCube Collaboration],
  Nucl.\ Instrum.\ Meth.\ A {\bf 700}, 188 (2013)
  doi:10.1016/j.nima.2012.10.067
  [arXiv:1207.6326 [astro-ph.IM]].

\bibitem{Beresinsky:1969qj} 
  V.~S.~Berezinsky and G.~T.~Zatsepin,
  Phys.\ Lett.\ B {\bf 28}, 423 (1969).
  doi:10.1016/0370-2693(69)90341-4


\bibitem{Aartsen:2013bka} 
  M.~G.~Aartsen {\it et al.} [IceCube Collaboration],
  Phys.\ Rev.\ Lett.\  {\bf 111}, 021103 (2013)
  doi:10.1103/PhysRevLett.111.021103
  [arXiv:1304.5356 [astro-ph.HE]].


\bibitem{Schonert:2008is} 
  S.~Schonert, T.~K.~Gaisser, E.~Resconi and O.~Schulz,
  Phys.\ Rev.\ D {\bf 79}, 043009 (2009)
  doi:10.1103/PhysRevD.79.043009
  [arXiv:0812.4308 [astro-ph]].


\bibitem{Gaisser:2014bja} 
  T.~K.~Gaisser, K.~Jero, A.~Karle and J.~van Santen,
  Phys.\ Rev.\ D {\bf 90}, no. 2, 023009 (2014)
  doi:10.1103/PhysRevD.90.023009
  [arXiv:1405.0525 [astro-ph.HE]].


\bibitem{Aartsen:2013jdh} 
  M.~G.~Aartsen {\it et al.} [IceCube Collaboration],
  Science {\bf 342}, 1242856 (2013)
  doi:10.1126/science.1242856
  [arXiv:1311.5238 [astro-ph.HE]].





\bibitem{Anchordoqui:2013qsi} 
  L.~A.~Anchordoqui, H.~Goldberg, M.~H.~Lynch, A.~V.~Olinto, T.~C.~Paul and T.~J.~Weiler,
  Phys.\ Rev.\ D {\bf 89}, no. 8, 083003 (2014)
  doi:10.1103/PhysRevD.89.083003
  [arXiv:1306.5021 [astro-ph.HE]].



\bibitem{Aartsen:2014gkd} 
  M.~G.~Aartsen {\it et al.} [IceCube Collaboration],
  Phys.\ Rev.\ Lett.\  {\bf 113}, 101101 (2014)
  doi:10.1103/PhysRevLett.113.101101
  [arXiv:1405.5303 [astro-ph.HE]].



\bibitem{Aartsen:2015knd} 
  M.~G.~Aartsen {\it et al.} [IceCube Collaboration],
  Astrophys.\ J.\  {\bf 809}, no. 1, 98 (2015)
  doi:10.1088/0004-637X/809/1/98
  [arXiv:1507.03991 [astro-ph.HE]].



\bibitem{Aartsen:2015zva} 
  M.~G.~Aartsen {\it et al.} [IceCube Collaboration],
  arXiv:1510.05223 [astro-ph.HE].


\bibitem{Aartsen:2015ivb} 
  M.~G.~Aartsen {\it et al.} [IceCube Collaboration],
  Phys.\ Rev.\ Lett.\  {\bf 114}, no. 17, 171102 (2015)
  doi:10.1103/PhysRevLett.114.171102
  [arXiv:1502.03376 [astro-ph.HE]].





\bibitem{Neronov:2015osa} 
  A.~Neronov and D.~V.~Semikoz,
  Astropart.\ Phys.\  {\bf 75}, 60 (2016)
  doi:10.1016/j.astropartphys.2015.11.002
  [arXiv:1509.03522 [astro-ph.HE]].

\bibitem{Neronov:2016bnp} 
  A.~Neronov and D.~Semikoz,
  Phys.\ Rev.\ D {\bf 93}, no. 12, 123002 (2016)
  doi:10.1103/PhysRevD.93.123002
  [arXiv:1603.06733 [astro-ph.HE]].


\bibitem{Razzaque:2013uoa} 
  S.~Razzaque,
  Phys.\ Rev.\ D {\bf 88}, 081302 (2013)
  doi:10.1103/PhysRevD.88.081302
  [arXiv:1309.2756 [astro-ph.HE]].


\bibitem{Bai:2014kba} 
  Y.~Bai, A.~J.~Barger, V.~Barger, R.~Lu, A.~D.~Peterson and J.~Salvado,
  Phys.\ Rev.\ D {\bf 90}, no. 6, 063012 (2014)
  doi:10.1103/PhysRevD.90.063012
  [arXiv:1407.2243 [astro-ph.HE]].


\bibitem{Anchordoqui:2016dcp} 
  L.~A.~Anchordoqui,
  arXiv:1606.01816 [astro-ph.HE].



\bibitem{Anchordoqui:2013dnh} 
  L.~A.~Anchordoqui {\it et al.},
  JHEAp {\bf 1-2}, 1 (2014)
  doi:10.1016/j.jheap.2014.01.001
  [arXiv:1312.6587 [astro-ph.HE]].






\bibitem{Thorne:1980rt} 
  K.~S.~Thorne,
  Rev.\ Mod.\ Phys.\  {\bf 52}, 285 (1980).
  doi:10.1103/RevModPhys.52.285



\bibitem{Thorne:1980ru} 
  K.~S.~Thorne,
  Rev.\ Mod.\ Phys.\  {\bf 52}, 299 (1980).
  doi:10.1103/RevModPhys.52.299


\bibitem{Peters:1963ux} 
  P.~C.~Peters and J.~Mathews,
  Phys.\ Rev.\  {\bf 131}, 435 (1963).
  doi:10.1103/PhysRev.131.435

\bibitem{Hulse:1974eb}
  R.~A.~Hulse and J.~H.~Taylor,
  Astrophys.\ J.\  {\bf 195}, L51 (1975).

\bibitem{Abramovici:1992ah}
  A.~Abramovici {\it et al.},
  Science {\bf 256}, 325 (1992).




\bibitem{Abbott:2016blz} 
  B.~P.~Abbott {\it et al.} [LIGO Scientific and Virgo Collaborations],
  Phys.\ Rev.\ Lett.\  {\bf 116}, no. 6, 061102 (2016)
  doi:10.1103/PhysRevLett.116.061102
  [arXiv:1602.03837 [gr-qc]].


\bibitem{Aartsen:2014mfp} 
  M.~G.~Aartsen {\it et al.} [IceCube and LIGO Scientific and VIRGO Collaborations],
  Phys.\ Rev.\ D {\bf 90}, no. 10, 102002 (2014)
  doi:10.1103/PhysRevD.90.102002
  [arXiv:1407.1042 [astro-ph.HE]].




\bibitem{Aartsen:2015dml} 
  M.~G.~Aartsen {\it et al.} [IceCube and Pierre Auger and Telescope Array Collaborations],
  JCAP {\bf 1601}, no. 01, 037 (2016)
  doi:10.1088/1475-7516/2016/01/037
  [arXiv:1511.09408 [astro-ph.HE]].


\bibitem{Fermi-LAT:2016qqr} 
  M.~Ackermann {\it et al.} [Fermi-LAT Collaboration],
  arXiv:1602.04488 [astro-ph.HE].




\bibitem{Adrian-Martinez:2016xgn} 
  S.~Adrian-Martinez {\it et al.} [ANTARES and IceCube and LIGO Scientific and Virgo Collaborations],
  arXiv:1602.05411 [astro-ph.HE].


\bibitem{Aartsen:2014njl} 
  M.~G.~Aartsen {\it et al.} [IceCube Collaboration],
  arXiv:1412.5106 [astro-ph.HE].

\bibitem{Neronov:2016zou} 
  A.~Neronov, D.~V.~Semikoz, L.~A.~Anchordoqui, J.~Adams and A.~V.~Olinto,
  arXiv:1606.03629 [astro-ph.IM].


\bibitem{Aab:2016vlz} 
  A.~Aab {\it et al.} [Pierre Auger Collaboration],
  arXiv:1604.03637 [astro-ph.IM].








\bibitem{Bradley:1727} J. Bradley
Phil. Trans. {\bf 35}, 637, (1727 - 1728)  doi:10.1098/rstl.1727.0064.



\bibitem{michael} M. Unger, {\it High energy
    astrophysics}, lectures given at New York University,  2015.


\bibitem{Ma} C.-P. Ma, {\it Astro 161}, lectures given at the University
  of California, Berkeley.

\bibitem{Peacock} J. A. Peacock, {\it Cosmological Physics}, (Cambridge
University Press, U.K., 1999) ISBN: 0-521-42270-1.


\bibitem{Denton:2014hfa} 
  P.~B.~Denton and T.~J.~Weiler,
  Astrophys.\ J.\  {\bf 802}, no. 1, 25 (2015)
  doi:10.1088/0004-637X/802/1/25
  [arXiv:1409.0883 [astro-ph.HE]].



\bibitem{Giddings:2008gr} 
  S.~B.~Giddings and M.~L.~Mangano,
  Phys.\ Rev.\ D {\bf 78}, 035009 (2008)
  doi:10.1103/PhysRevD.78.035009
  [arXiv:0806.3381 [hep-ph]].



\bibitem{Giddings:2008pi} 
  S.~B.~Giddings and M.~L.~Mangano,
  arXiv:0808.4087 [hep-ph].

\bibitem{Hillas:1985is} 
  A.~M.~Hillas,
  Ann.\ Rev.\ Astron.\ Astrophys.\  {\bf 22}, 425 (1984).
  doi:10.1146/annurev.aa.22.090184.002233



\bibitem{Bionta:1987qt}
  R.~M.~Bionta {\it et al.},
  Phys.\ Rev.\ Lett.\  {\bf 58}, 1494 (1987).


\bibitem{Hirata:1987hu}
  K.~Hirata {\it et al.}  [KAMIOKANDE-II Collaboration],
  Phys.\ Rev.\ Lett.\  {\bf 58}, 1490 (1987);

\bibitem{Bahcall:1987fz} 
  J.~N.~Bahcall, A.~Dar and T.~Piran,
  Nature {\bf 326}, 135 (1987).
  doi:10.1038/326135a0



\bibitem{Kruskal:1959vx} 
  M.~D.~Kruskal,
  Phys.\ Rev.\  {\bf 119}, 1743 (1960).
  doi:10.1103/PhysRev.119.1743


\bibitem{tHooft}  G.~'t Hooft, {\it Introduction to the theory of black holes}, lectures given at Utrecht University, 2009.


\bibitem{Einstein:1935tc} 
  A.~Einstein and N.~Rosen,
  Phys.\ Rev.\  {\bf 48}, 73 (1935).
  doi:10.1103/PhysRev.48.73


\bibitem{Fuller:1962zza} 
  R.~W.~Fuller and J.~A.~Wheeler,
  Phys.\ Rev.\  {\bf 128}, 919 (1962).
  doi:10.1103/PhysRev.128.919






\end{thebibliography}
\end{document}